\documentclass[twocolumn,tighten]{aastex631}
\usepackage[]{txfonts}
\usepackage{natbib, twoopt}
\usepackage{float}
\usepackage{graphicx}
\usepackage{epstopdf}
\usepackage{enumitem}
\usepackage[caption=false]{subfig}
\usepackage{color}\definecolor{AliceBlue}{rgb}{0.94,0.97,1.00}
\definecolor{AntiqueWhite1}{rgb}{1.00,0.94,0.86}
\definecolor{AntiqueWhite2}{rgb}{0.93,0.87,0.80}
\definecolor{AntiqueWhite3}{rgb}{0.80,0.75,0.69}
\definecolor{AntiqueWhite4}{rgb}{0.55,0.51,0.47}
\definecolor{AntiqueWhite}{rgb}{0.98,0.92,0.84}
\definecolor{BlanchedAlmond}{rgb}{1.00,0.92,0.80}
\definecolor{BlueViolet}{rgb}{0.54,0.17,0.89}
\definecolor{CadetBlue1}{rgb}{0.60,0.96,1.00}
\definecolor{CadetBlue2}{rgb}{0.56,0.90,0.93}
\definecolor{CadetBlue3}{rgb}{0.48,0.77,0.80}
\definecolor{CadetBlue4}{rgb}{0.33,0.53,0.55}
\definecolor{CadetBlue}{rgb}{0.37,0.62,0.63}
\definecolor{CornflowerBlue}{rgb}{0.39,0.58,0.93}
\definecolor{DarkBlue}{rgb}{0.00,0.00,0.55}
\definecolor{DarkCyan}{rgb}{0.00,0.55,0.55}
\definecolor{DarkGoldenrod1}{rgb}{1.00,0.73,0.06}
\definecolor{DarkGoldenrod2}{rgb}{0.93,0.68,0.05}
\definecolor{DarkGoldenrod3}{rgb}{0.80,0.58,0.05}
\definecolor{DarkGoldenrod4}{rgb}{0.55,0.40,0.03}
\definecolor{DarkGoldenrod}{rgb}{0.72,0.53,0.04}
\definecolor{DarkGray}{rgb}{0.66,0.66,0.66}
\definecolor{DarkGreen}{rgb}{0.00,0.39,0.00}
\definecolor{DarkGrey}{rgb}{0.66,0.66,0.66}
\definecolor{DarkKhaki}{rgb}{0.74,0.72,0.42}
\definecolor{DarkMagenta}{rgb}{0.55,0.00,0.55}
\definecolor{DarkOliveGreen1}{rgb}{0.79,1.00,0.44}
\definecolor{DarkOliveGreen2}{rgb}{0.74,0.93,0.41}
\definecolor{DarkOliveGreen3}{rgb}{0.64,0.80,0.35}
\definecolor{DarkOliveGreen4}{rgb}{0.43,0.55,0.24}
\definecolor{DarkOliveGreen}{rgb}{0.33,0.42,0.18}
\definecolor{DarkOrange1}{rgb}{1.00,0.50,0.00}
\definecolor{DarkOrange2}{rgb}{0.93,0.46,0.00}
\definecolor{DarkOrange3}{rgb}{0.80,0.40,0.00}
\definecolor{DarkOrange4}{rgb}{0.55,0.27,0.00}
\definecolor{DarkOrange}{rgb}{1.00,0.55,0.00}
\definecolor{DarkOrchid1}{rgb}{0.75,0.24,1.00}
\definecolor{DarkOrchid2}{rgb}{0.70,0.23,0.93}
\definecolor{DarkOrchid3}{rgb}{0.60,0.20,0.80}
\definecolor{DarkOrchid4}{rgb}{0.41,0.13,0.55}
\definecolor{DarkOrchid}{rgb}{0.60,0.20,0.80}
\definecolor{DarkRed}{rgb}{0.55,0.00,0.00}
\definecolor{DarkSalmon}{rgb}{0.91,0.59,0.48}
\definecolor{DarkSeaGreen1}{rgb}{0.76,1.00,0.76}
\definecolor{DarkSeaGreen2}{rgb}{0.71,0.93,0.71}
\definecolor{DarkSeaGreen3}{rgb}{0.61,0.80,0.61}
\definecolor{DarkSeaGreen4}{rgb}{0.41,0.55,0.41}
\definecolor{DarkSeaGreen}{rgb}{0.56,0.74,0.56}
\definecolor{DarkSlateBlue}{rgb}{0.28,0.24,0.55}
\definecolor{DarkSlateGray1}{rgb}{0.59,1.00,1.00}
\definecolor{DarkSlateGray2}{rgb}{0.55,0.93,0.93}
\definecolor{DarkSlateGray3}{rgb}{0.47,0.80,0.80}
\definecolor{DarkSlateGray4}{rgb}{0.32,0.55,0.55}
\definecolor{DarkSlateGray}{rgb}{0.18,0.31,0.31}
\definecolor{DarkSlateGrey}{rgb}{0.18,0.31,0.31}
\definecolor{DarkTurquoise}{rgb}{0.00,0.81,0.82}
\definecolor{DarkViolet}{rgb}{0.58,0.00,0.83}
\definecolor{DeepPink1}{rgb}{1.00,0.08,0.58}
\definecolor{DeepPink2}{rgb}{0.93,0.07,0.54}
\definecolor{DeepPink3}{rgb}{0.80,0.06,0.46}
\definecolor{DeepPink4}{rgb}{0.55,0.04,0.31}
\definecolor{DeepPink}{rgb}{1.00,0.08,0.58}
\definecolor{DeepSkyBlue1}{rgb}{0.00,0.75,1.00}
\definecolor{DeepSkyBlue2}{rgb}{0.00,0.70,0.93}
\definecolor{DeepSkyBlue3}{rgb}{0.00,0.60,0.80}
\definecolor{DeepSkyBlue4}{rgb}{0.00,0.41,0.55}
\definecolor{DeepSkyBlue}{rgb}{0.00,0.75,1.00}
\definecolor{DimGray}{rgb}{0.41,0.41,0.41}
\definecolor{DimGrey}{rgb}{0.41,0.41,0.41}
\definecolor{DodgerBlue1}{rgb}{0.12,0.56,1.00}
\definecolor{DodgerBlue2}{rgb}{0.11,0.53,0.93}
\definecolor{DodgerBlue3}{rgb}{0.09,0.45,0.80}
\definecolor{DodgerBlue4}{rgb}{0.06,0.31,0.55}
\definecolor{DodgerBlue}{rgb}{0.12,0.56,1.00}
\definecolor{FloralWhite}{rgb}{1.00,0.98,0.94}
\definecolor{ForestGreen}{rgb}{0.13,0.55,0.13}
\definecolor{GhostWhite}{rgb}{0.97,0.97,1.00}
\definecolor{GreenYellow}{rgb}{0.68,1.00,0.18}
\definecolor{HotPink1}{rgb}{1.00,0.43,0.71}
\definecolor{HotPink2}{rgb}{0.93,0.42,0.65}
\definecolor{HotPink3}{rgb}{0.80,0.38,0.56}
\definecolor{HotPink4}{rgb}{0.55,0.23,0.38}
\definecolor{HotPink}{rgb}{1.00,0.41,0.71}
\definecolor{IndianRed1}{rgb}{1.00,0.42,0.42}
\definecolor{IndianRed2}{rgb}{0.93,0.39,0.39}
\definecolor{IndianRed3}{rgb}{0.80,0.33,0.33}
\definecolor{IndianRed4}{rgb}{0.55,0.23,0.23}
\definecolor{IndianRed}{rgb}{0.80,0.36,0.36}
\definecolor{LavenderBlush1}{rgb}{1.00,0.94,0.96}
\definecolor{LavenderBlush2}{rgb}{0.93,0.88,0.90}
\definecolor{LavenderBlush3}{rgb}{0.80,0.76,0.77}
\definecolor{LavenderBlush4}{rgb}{0.55,0.51,0.53}
\definecolor{LavenderBlush}{rgb}{1.00,0.94,0.96}
\definecolor{LawnGreen}{rgb}{0.49,0.99,0.00}
\definecolor{LemonChiffon1}{rgb}{1.00,0.98,0.80}
\definecolor{LemonChiffon2}{rgb}{0.93,0.91,0.75}
\definecolor{LemonChiffon3}{rgb}{0.80,0.79,0.65}
\definecolor{LemonChiffon4}{rgb}{0.55,0.54,0.44}
\definecolor{LemonChiffon}{rgb}{1.00,0.98,0.80}
\definecolor{LightBlue1}{rgb}{0.75,0.94,1.00}
\definecolor{LightBlue2}{rgb}{0.70,0.87,0.93}
\definecolor{LightBlue3}{rgb}{0.60,0.75,0.80}
\definecolor{LightBlue4}{rgb}{0.41,0.51,0.55}
\definecolor{LightBlue}{rgb}{0.68,0.85,0.90}
\definecolor{LightCoral}{rgb}{0.94,0.50,0.50}
\definecolor{LightCyan1}{rgb}{0.88,1.00,1.00}
\definecolor{LightCyan2}{rgb}{0.82,0.93,0.93}
\definecolor{LightCyan3}{rgb}{0.71,0.80,0.80}
\definecolor{LightCyan4}{rgb}{0.48,0.55,0.55}
\definecolor{LightCyan}{rgb}{0.88,1.00,1.00}
\definecolor{LightGoldenrod1}{rgb}{1.00,0.93,0.55}
\definecolor{LightGoldenrod2}{rgb}{0.93,0.86,0.51}
\definecolor{LightGoldenrod3}{rgb}{0.80,0.75,0.44}
\definecolor{LightGoldenrod4}{rgb}{0.55,0.51,0.30}
\definecolor{LightGoldenrodYellow}{rgb}{0.98,0.98,0.82}
\definecolor{LightGoldenrod}{rgb}{0.93,0.87,0.51}
\definecolor{LightGray}{rgb}{0.83,0.83,0.83}
\definecolor{LightGreen}{rgb}{0.56,0.93,0.56}
\definecolor{LightGrey}{rgb}{0.83,0.83,0.83}
\definecolor{LightPink1}{rgb}{1.00,0.68,0.73}
\definecolor{LightPink2}{rgb}{0.93,0.64,0.68}
\definecolor{LightPink3}{rgb}{0.80,0.55,0.58}
\definecolor{LightPink4}{rgb}{0.55,0.37,0.40}
\definecolor{LightPink}{rgb}{1.00,0.71,0.76}
\definecolor{LightSalmon1}{rgb}{1.00,0.63,0.48}
\definecolor{LightSalmon2}{rgb}{0.93,0.58,0.45}
\definecolor{LightSalmon3}{rgb}{0.80,0.51,0.38}
\definecolor{LightSalmon4}{rgb}{0.55,0.34,0.26}
\definecolor{LightSalmon}{rgb}{1.00,0.63,0.48}
\definecolor{LightSeaGreen}{rgb}{0.13,0.70,0.67}
\definecolor{LightSkyBlue1}{rgb}{0.69,0.89,1.00}
\definecolor{LightSkyBlue2}{rgb}{0.64,0.83,0.93}
\definecolor{LightSkyBlue3}{rgb}{0.55,0.71,0.80}
\definecolor{LightSkyBlue4}{rgb}{0.38,0.48,0.55}
\definecolor{LightSkyBlue}{rgb}{0.53,0.81,0.98}
\definecolor{LightSlateBlue}{rgb}{0.52,0.44,1.00}
\definecolor{LightSlateGray}{rgb}{0.47,0.53,0.60}
\definecolor{LightSlateGrey}{rgb}{0.47,0.53,0.60}
\definecolor{LightSteelBlue1}{rgb}{0.79,0.88,1.00}
\definecolor{LightSteelBlue2}{rgb}{0.74,0.82,0.93}
\definecolor{LightSteelBlue3}{rgb}{0.64,0.71,0.80}
\definecolor{LightSteelBlue4}{rgb}{0.43,0.48,0.55}
\definecolor{LightSteelBlue}{rgb}{0.69,0.77,0.87}
\definecolor{LightYellow1}{rgb}{1.00,1.00,0.88}
\definecolor{LightYellow2}{rgb}{0.93,0.93,0.82}
\definecolor{LightYellow3}{rgb}{0.80,0.80,0.71}
\definecolor{LightYellow4}{rgb}{0.55,0.55,0.48}
\definecolor{LightYellow}{rgb}{1.00,1.00,0.88}
\definecolor{LimeGreen}{rgb}{0.20,0.80,0.20}
\definecolor{MediumAquamarine}{rgb}{0.40,0.80,0.67}
\definecolor{MediumBlue}{rgb}{0.00,0.00,0.80}
\definecolor{MediumOrchid1}{rgb}{0.88,0.40,1.00}
\definecolor{MediumOrchid2}{rgb}{0.82,0.37,0.93}
\definecolor{MediumOrchid3}{rgb}{0.71,0.32,0.80}
\definecolor{MediumOrchid4}{rgb}{0.48,0.22,0.55}
\definecolor{MediumOrchid}{rgb}{0.73,0.33,0.83}
\definecolor{MediumPurple1}{rgb}{0.67,0.51,1.00}
\definecolor{MediumPurple2}{rgb}{0.62,0.47,0.93}
\definecolor{MediumPurple3}{rgb}{0.54,0.41,0.80}
\definecolor{MediumPurple4}{rgb}{0.36,0.28,0.55}
\definecolor{MediumPurple}{rgb}{0.58,0.44,0.86}
\definecolor{MediumSeaGreen}{rgb}{0.24,0.70,0.44}
\definecolor{MediumSlateBlue}{rgb}{0.48,0.41,0.93}
\definecolor{MediumSpringGreen}{rgb}{0.00,0.98,0.60}
\definecolor{MediumTurquoise}{rgb}{0.28,0.82,0.80}
\definecolor{MediumVioletRed}{rgb}{0.78,0.08,0.52}
\definecolor{MidnightBlue}{rgb}{0.10,0.10,0.44}
\definecolor{MintCream}{rgb}{0.96,1.00,0.98}
\definecolor{MistyRose1}{rgb}{1.00,0.89,0.88}
\definecolor{MistyRose2}{rgb}{0.93,0.84,0.82}
\definecolor{MistyRose3}{rgb}{0.80,0.72,0.71}
\definecolor{MistyRose4}{rgb}{0.55,0.49,0.48}
\definecolor{MistyRose}{rgb}{1.00,0.89,0.88}
\definecolor{NavajoWhite1}{rgb}{1.00,0.87,0.68}
\definecolor{NavajoWhite2}{rgb}{0.93,0.81,0.63}
\definecolor{NavajoWhite3}{rgb}{0.80,0.70,0.55}
\definecolor{NavajoWhite4}{rgb}{0.55,0.47,0.37}
\definecolor{NavajoWhite}{rgb}{1.00,0.87,0.68}
\definecolor{NavyBlue}{rgb}{0.00,0.00,0.50}
\definecolor{OldLace}{rgb}{0.99,0.96,0.90}
\definecolor{OliveDrab1}{rgb}{0.75,1.00,0.24}
\definecolor{OliveDrab2}{rgb}{0.70,0.93,0.23}
\definecolor{OliveDrab3}{rgb}{0.60,0.80,0.20}
\definecolor{OliveDrab4}{rgb}{0.41,0.55,0.13}
\definecolor{OliveDrab}{rgb}{0.42,0.56,0.14}
\definecolor{OrangeRed1}{rgb}{1.00,0.27,0.00}
\definecolor{OrangeRed2}{rgb}{0.93,0.25,0.00}
\definecolor{OrangeRed3}{rgb}{0.80,0.22,0.00}
\definecolor{OrangeRed4}{rgb}{0.55,0.15,0.00}
\definecolor{OrangeRed}{rgb}{1.00,0.27,0.00}
\definecolor{PaleGoldenrod}{rgb}{0.93,0.91,0.67}
\definecolor{PaleGreen1}{rgb}{0.60,1.00,0.60}
\definecolor{PaleGreen2}{rgb}{0.56,0.93,0.56}
\definecolor{PaleGreen3}{rgb}{0.49,0.80,0.49}
\definecolor{PaleGreen4}{rgb}{0.33,0.55,0.33}
\definecolor{PaleGreen}{rgb}{0.60,0.98,0.60}
\definecolor{PaleTurquoise1}{rgb}{0.73,1.00,1.00}
\definecolor{PaleTurquoise2}{rgb}{0.68,0.93,0.93}
\definecolor{PaleTurquoise3}{rgb}{0.59,0.80,0.80}
\definecolor{PaleTurquoise4}{rgb}{0.40,0.55,0.55}
\definecolor{PaleTurquoise}{rgb}{0.69,0.93,0.93}
\definecolor{PaleVioletRed1}{rgb}{1.00,0.51,0.67}
\definecolor{PaleVioletRed2}{rgb}{0.93,0.47,0.62}
\definecolor{PaleVioletRed3}{rgb}{0.80,0.41,0.54}
\definecolor{PaleVioletRed4}{rgb}{0.55,0.28,0.36}
\definecolor{PaleVioletRed}{rgb}{0.86,0.44,0.58}
\definecolor{PapayaWhip}{rgb}{1.00,0.94,0.84}
\definecolor{PeachPuff1}{rgb}{1.00,0.85,0.73}
\definecolor{PeachPuff2}{rgb}{0.93,0.80,0.68}
\definecolor{PeachPuff3}{rgb}{0.80,0.69,0.58}
\definecolor{PeachPuff4}{rgb}{0.55,0.47,0.40}
\definecolor{PeachPuff}{rgb}{1.00,0.85,0.73}
\definecolor{PowderBlue}{rgb}{0.69,0.88,0.90}
\definecolor{RosyBrown1}{rgb}{1.00,0.76,0.76}
\definecolor{RosyBrown2}{rgb}{0.93,0.71,0.71}
\definecolor{RosyBrown3}{rgb}{0.80,0.61,0.61}
\definecolor{RosyBrown4}{rgb}{0.55,0.41,0.41}
\definecolor{RosyBrown}{rgb}{0.74,0.56,0.56}
\definecolor{RoyalBlue1}{rgb}{0.28,0.46,1.00}
\definecolor{RoyalBlue2}{rgb}{0.26,0.43,0.93}
\definecolor{RoyalBlue3}{rgb}{0.23,0.37,0.80}
\definecolor{RoyalBlue4}{rgb}{0.15,0.25,0.55}
\definecolor{RoyalBlue}{rgb}{0.25,0.41,0.88}
\definecolor{SaddleBrown}{rgb}{0.55,0.27,0.07}
\definecolor{SandyBrown}{rgb}{0.96,0.64,0.38}
\definecolor{SeaGreen1}{rgb}{0.33,1.00,0.62}
\definecolor{SeaGreen2}{rgb}{0.31,0.93,0.58}
\definecolor{SeaGreen3}{rgb}{0.26,0.80,0.50}
\definecolor{SeaGreen4}{rgb}{0.18,0.55,0.34}
\definecolor{SeaGreen}{rgb}{0.18,0.55,0.34}
\definecolor{SkyBlue1}{rgb}{0.53,0.81,1.00}
\definecolor{SkyBlue2}{rgb}{0.49,0.75,0.93}
\definecolor{SkyBlue3}{rgb}{0.42,0.65,0.80}
\definecolor{SkyBlue4}{rgb}{0.29,0.44,0.55}
\definecolor{SkyBlue}{rgb}{0.53,0.81,0.92}
\definecolor{SlateBlue1}{rgb}{0.51,0.44,1.00}
\definecolor{SlateBlue2}{rgb}{0.48,0.40,0.93}
\definecolor{SlateBlue3}{rgb}{0.41,0.35,0.80}
\definecolor{SlateBlue4}{rgb}{0.28,0.24,0.55}
\definecolor{SlateBlue}{rgb}{0.42,0.35,0.80}
\definecolor{SlateGray1}{rgb}{0.78,0.89,1.00}
\definecolor{SlateGray2}{rgb}{0.73,0.83,0.93}
\definecolor{SlateGray3}{rgb}{0.62,0.71,0.80}
\definecolor{SlateGray4}{rgb}{0.42,0.48,0.55}
\definecolor{SlateGray}{rgb}{0.44,0.50,0.56}
\definecolor{SlateGrey}{rgb}{0.44,0.50,0.56}
\definecolor{SpringGreen1}{rgb}{0.00,1.00,0.50}
\definecolor{SpringGreen2}{rgb}{0.00,0.93,0.46}
\definecolor{SpringGreen3}{rgb}{0.00,0.80,0.40}
\definecolor{SpringGreen4}{rgb}{0.00,0.55,0.27}
\definecolor{SpringGreen}{rgb}{0.00,1.00,0.50}
\definecolor{SteelBlue1}{rgb}{0.39,0.72,1.00}
\definecolor{SteelBlue2}{rgb}{0.36,0.67,0.93}
\definecolor{SteelBlue3}{rgb}{0.31,0.58,0.80}
\definecolor{SteelBlue4}{rgb}{0.21,0.39,0.55}
\definecolor{SteelBlue}{rgb}{0.27,0.51,0.71}
\definecolor{VioletRed1}{rgb}{1.00,0.24,0.59}
\definecolor{VioletRed2}{rgb}{0.93,0.23,0.55}
\definecolor{VioletRed3}{rgb}{0.80,0.20,0.47}
\definecolor{VioletRed4}{rgb}{0.55,0.13,0.32}
\definecolor{VioletRed}{rgb}{0.82,0.13,0.56}
\definecolor{WhiteSmoke}{rgb}{0.96,0.96,0.96}
\definecolor{YellowGreen}{rgb}{0.60,0.80,0.20}
\definecolor{aliceblue}{rgb}{0.94,0.97,1.00}
\definecolor{antiquewhite}{rgb}{0.98,0.92,0.84}
\definecolor{aquamarine1}{rgb}{0.50,1.00,0.83}
\definecolor{aquamarine2}{rgb}{0.46,0.93,0.78}
\definecolor{aquamarine3}{rgb}{0.40,0.80,0.67}
\definecolor{aquamarine4}{rgb}{0.27,0.55,0.45}
\definecolor{aquamarine}{rgb}{0.50,1.00,0.83}
\definecolor{azure1}{rgb}{0.94,1.00,1.00}
\definecolor{azure2}{rgb}{0.88,0.93,0.93}
\definecolor{azure3}{rgb}{0.76,0.80,0.80}
\definecolor{azure4}{rgb}{0.51,0.55,0.55}
\definecolor{azure}{rgb}{0.94,1.00,1.00}
\definecolor{beige}{rgb}{0.96,0.96,0.86}
\definecolor{bisque1}{rgb}{1.00,0.89,0.77}
\definecolor{bisque2}{rgb}{0.93,0.84,0.72}
\definecolor{bisque3}{rgb}{0.80,0.72,0.62}
\definecolor{bisque4}{rgb}{0.55,0.49,0.42}
\definecolor{bisque}{rgb}{1.00,0.89,0.77}
\definecolor{black}{rgb}{0.00,0.00,0.00}
\definecolor{blanchedalmond}{rgb}{1.00,0.92,0.80}
\definecolor{blue1}{rgb}{0.00,0.00,1.00}
\definecolor{blue2}{rgb}{0.00,0.00,0.93}
\definecolor{blue3}{rgb}{0.00,0.00,0.80}
\definecolor{blue4}{rgb}{0.00,0.00,0.55}
\definecolor{blueviolet}{rgb}{0.54,0.17,0.89}
\definecolor{blue}{rgb}{0.00,0.00,1.00}
\definecolor{brown1}{rgb}{1.00,0.25,0.25}
\definecolor{brown2}{rgb}{0.93,0.23,0.23}
\definecolor{brown3}{rgb}{0.80,0.20,0.20}
\definecolor{brown4}{rgb}{0.55,0.14,0.14}
\definecolor{brown}{rgb}{0.65,0.16,0.16}
\definecolor{burlywood1}{rgb}{1.00,0.83,0.61}
\definecolor{burlywood2}{rgb}{0.93,0.77,0.57}
\definecolor{burlywood3}{rgb}{0.80,0.67,0.49}
\definecolor{burlywood4}{rgb}{0.55,0.45,0.33}
\definecolor{burlywood}{rgb}{0.87,0.72,0.53}
\definecolor{cadetblue}{rgb}{0.37,0.62,0.63}
\definecolor{chartreuse1}{rgb}{0.50,1.00,0.00}
\definecolor{chartreuse2}{rgb}{0.46,0.93,0.00}
\definecolor{chartreuse3}{rgb}{0.40,0.80,0.00}
\definecolor{chartreuse4}{rgb}{0.27,0.55,0.00}
\definecolor{chartreuse}{rgb}{0.50,1.00,0.00}
\definecolor{chocolate1}{rgb}{1.00,0.50,0.14}
\definecolor{chocolate2}{rgb}{0.93,0.46,0.13}
\definecolor{chocolate3}{rgb}{0.80,0.40,0.11}
\definecolor{chocolate4}{rgb}{0.55,0.27,0.07}
\definecolor{chocolate}{rgb}{0.82,0.41,0.12}
\definecolor{coral1}{rgb}{1.00,0.45,0.34}
\definecolor{coral2}{rgb}{0.93,0.42,0.31}
\definecolor{coral3}{rgb}{0.80,0.36,0.27}
\definecolor{coral4}{rgb}{0.55,0.24,0.18}
\definecolor{coral}{rgb}{1.00,0.50,0.31}
\definecolor{cornflowerblue}{rgb}{0.39,0.58,0.93}
\definecolor{cornsilk1}{rgb}{1.00,0.97,0.86}
\definecolor{cornsilk2}{rgb}{0.93,0.91,0.80}
\definecolor{cornsilk3}{rgb}{0.80,0.78,0.69}
\definecolor{cornsilk4}{rgb}{0.55,0.53,0.47}
\definecolor{cornsilk}{rgb}{1.00,0.97,0.86}
\definecolor{cyan1}{rgb}{0.00,1.00,1.00}
\definecolor{cyan2}{rgb}{0.00,0.93,0.93}
\definecolor{cyan3}{rgb}{0.00,0.80,0.80}
\definecolor{cyan4}{rgb}{0.00,0.55,0.55}
\definecolor{cyan}{rgb}{0.00,1.00,1.00}
\definecolor{darkblue}{rgb}{0.00,0.00,0.55}
\definecolor{darkcyan}{rgb}{0.00,0.55,0.55}
\definecolor{darkgoldenrod}{rgb}{0.72,0.53,0.04}
\definecolor{darkgray}{rgb}{0.66,0.66,0.66}
\definecolor{darkgreen}{rgb}{0.00,0.39,0.00}
\definecolor{darkgrey}{rgb}{0.66,0.66,0.66}
\definecolor{darkkhaki}{rgb}{0.74,0.72,0.42}
\definecolor{darkmagenta}{rgb}{0.55,0.00,0.55}
\definecolor{darkolive}{rgb}{0.33,0.42,0.18}
\definecolor{darkorange}{rgb}{1.00,0.55,0.00}
\definecolor{darkorchid}{rgb}{0.60,0.20,0.80}
\definecolor{darkred}{rgb}{0.55,0.00,0.00}
\definecolor{darksalmon}{rgb}{0.91,0.59,0.48}
\definecolor{darksea}{rgb}{0.56,0.74,0.56}
\definecolor{darkslate}{rgb}{0.18,0.31,0.31}
\definecolor{darkslate}{rgb}{0.18,0.31,0.31}
\definecolor{darkslate}{rgb}{0.28,0.24,0.55}
\definecolor{darkturquoise}{rgb}{0.00,0.81,0.82}
\definecolor{darkviolet}{rgb}{0.58,0.00,0.83}
\definecolor{deeppink}{rgb}{1.00,0.08,0.58}
\definecolor{deepsky}{rgb}{0.00,0.75,1.00}
\definecolor{dimgray}{rgb}{0.41,0.41,0.41}
\definecolor{dimgrey}{rgb}{0.41,0.41,0.41}
\definecolor{dodgerblue}{rgb}{0.12,0.56,1.00}
\definecolor{firebrick1}{rgb}{1.00,0.19,0.19}
\definecolor{firebrick2}{rgb}{0.93,0.17,0.17}
\definecolor{firebrick3}{rgb}{0.80,0.15,0.15}
\definecolor{firebrick4}{rgb}{0.55,0.10,0.10}
\definecolor{firebrick}{rgb}{0.70,0.13,0.13}
\definecolor{floralwhite}{rgb}{1.00,0.98,0.94}
\definecolor{forestgreen}{rgb}{0.13,0.55,0.13}
\definecolor{gainsboro}{rgb}{0.86,0.86,0.86}
\definecolor{ghostwhite}{rgb}{0.97,0.97,1.00}
\definecolor{gold1}{rgb}{1.00,0.84,0.00}
\definecolor{gold2}{rgb}{0.93,0.79,0.00}
\definecolor{gold3}{rgb}{0.80,0.68,0.00}
\definecolor{gold4}{rgb}{0.55,0.46,0.00}
\definecolor{goldenrod1}{rgb}{1.00,0.76,0.15}
\definecolor{goldenrod2}{rgb}{0.93,0.71,0.13}
\definecolor{goldenrod3}{rgb}{0.80,0.61,0.11}
\definecolor{goldenrod4}{rgb}{0.55,0.41,0.08}
\definecolor{goldenrod}{rgb}{0.85,0.65,0.13}
\definecolor{gold}{rgb}{1.00,0.84,0.00}
\definecolor{gray0}{rgb}{0.00,0.00,0.00}
\definecolor{gray100}{rgb}{1.00,1.00,1.00}
\definecolor{gray10}{rgb}{0.10,0.10,0.10}
\definecolor{gray11}{rgb}{0.11,0.11,0.11}
\definecolor{gray12}{rgb}{0.12,0.12,0.12}
\definecolor{gray13}{rgb}{0.13,0.13,0.13}
\definecolor{gray14}{rgb}{0.14,0.14,0.14}
\definecolor{gray15}{rgb}{0.15,0.15,0.15}
\definecolor{gray16}{rgb}{0.16,0.16,0.16}
\definecolor{gray17}{rgb}{0.17,0.17,0.17}
\definecolor{gray18}{rgb}{0.18,0.18,0.18}
\definecolor{gray19}{rgb}{0.19,0.19,0.19}
\definecolor{gray1}{rgb}{0.01,0.01,0.01}
\definecolor{gray20}{rgb}{0.20,0.20,0.20}
\definecolor{gray21}{rgb}{0.21,0.21,0.21}
\definecolor{gray22}{rgb}{0.22,0.22,0.22}
\definecolor{gray23}{rgb}{0.23,0.23,0.23}
\definecolor{gray24}{rgb}{0.24,0.24,0.24}
\definecolor{gray25}{rgb}{0.25,0.25,0.25}
\definecolor{gray26}{rgb}{0.26,0.26,0.26}
\definecolor{gray27}{rgb}{0.27,0.27,0.27}
\definecolor{gray28}{rgb}{0.28,0.28,0.28}
\definecolor{gray29}{rgb}{0.29,0.29,0.29}
\definecolor{gray2}{rgb}{0.02,0.02,0.02}
\definecolor{gray30}{rgb}{0.30,0.30,0.30}
\definecolor{gray31}{rgb}{0.31,0.31,0.31}
\definecolor{gray32}{rgb}{0.32,0.32,0.32}
\definecolor{gray33}{rgb}{0.33,0.33,0.33}
\definecolor{gray34}{rgb}{0.34,0.34,0.34}
\definecolor{gray35}{rgb}{0.35,0.35,0.35}
\definecolor{gray36}{rgb}{0.36,0.36,0.36}
\definecolor{gray37}{rgb}{0.37,0.37,0.37}
\definecolor{gray38}{rgb}{0.38,0.38,0.38}
\definecolor{gray39}{rgb}{0.39,0.39,0.39}
\definecolor{gray3}{rgb}{0.03,0.03,0.03}
\definecolor{gray40}{rgb}{0.40,0.40,0.40}
\definecolor{gray41}{rgb}{0.41,0.41,0.41}
\definecolor{gray42}{rgb}{0.42,0.42,0.42}
\definecolor{gray43}{rgb}{0.43,0.43,0.43}
\definecolor{gray44}{rgb}{0.44,0.44,0.44}
\definecolor{gray45}{rgb}{0.45,0.45,0.45}
\definecolor{gray46}{rgb}{0.46,0.46,0.46}
\definecolor{gray47}{rgb}{0.47,0.47,0.47}
\definecolor{gray48}{rgb}{0.48,0.48,0.48}
\definecolor{gray49}{rgb}{0.49,0.49,0.49}
\definecolor{gray4}{rgb}{0.04,0.04,0.04}
\definecolor{gray50}{rgb}{0.50,0.50,0.50}
\definecolor{gray51}{rgb}{0.51,0.51,0.51}
\definecolor{gray52}{rgb}{0.52,0.52,0.52}
\definecolor{gray53}{rgb}{0.53,0.53,0.53}
\definecolor{gray54}{rgb}{0.54,0.54,0.54}
\definecolor{gray55}{rgb}{0.55,0.55,0.55}
\definecolor{gray56}{rgb}{0.56,0.56,0.56}
\definecolor{gray57}{rgb}{0.57,0.57,0.57}
\definecolor{gray58}{rgb}{0.58,0.58,0.58}
\definecolor{gray59}{rgb}{0.59,0.59,0.59}
\definecolor{gray5}{rgb}{0.05,0.05,0.05}
\definecolor{gray60}{rgb}{0.60,0.60,0.60}
\definecolor{gray61}{rgb}{0.61,0.61,0.61}
\definecolor{gray62}{rgb}{0.62,0.62,0.62}
\definecolor{gray63}{rgb}{0.63,0.63,0.63}
\definecolor{gray64}{rgb}{0.64,0.64,0.64}
\definecolor{gray65}{rgb}{0.65,0.65,0.65}
\definecolor{gray66}{rgb}{0.66,0.66,0.66}
\definecolor{gray67}{rgb}{0.67,0.67,0.67}
\definecolor{gray68}{rgb}{0.68,0.68,0.68}
\definecolor{gray69}{rgb}{0.69,0.69,0.69}
\definecolor{gray6}{rgb}{0.06,0.06,0.06}
\definecolor{gray70}{rgb}{0.70,0.70,0.70}
\definecolor{gray71}{rgb}{0.71,0.71,0.71}
\definecolor{gray72}{rgb}{0.72,0.72,0.72}
\definecolor{gray73}{rgb}{0.73,0.73,0.73}
\definecolor{gray74}{rgb}{0.74,0.74,0.74}
\definecolor{gray75}{rgb}{0.75,0.75,0.75}
\definecolor{gray76}{rgb}{0.76,0.76,0.76}
\definecolor{gray77}{rgb}{0.77,0.77,0.77}
\definecolor{gray78}{rgb}{0.78,0.78,0.78}
\definecolor{gray79}{rgb}{0.79,0.79,0.79}
\definecolor{gray7}{rgb}{0.07,0.07,0.07}
\definecolor{gray80}{rgb}{0.80,0.80,0.80}
\definecolor{gray81}{rgb}{0.81,0.81,0.81}
\definecolor{gray82}{rgb}{0.82,0.82,0.82}
\definecolor{gray83}{rgb}{0.83,0.83,0.83}
\definecolor{gray84}{rgb}{0.84,0.84,0.84}
\definecolor{gray85}{rgb}{0.85,0.85,0.85}
\definecolor{gray86}{rgb}{0.86,0.86,0.86}
\definecolor{gray87}{rgb}{0.87,0.87,0.87}
\definecolor{gray88}{rgb}{0.88,0.88,0.88}
\definecolor{gray89}{rgb}{0.89,0.89,0.89}
\definecolor{gray8}{rgb}{0.08,0.08,0.08}
\definecolor{gray90}{rgb}{0.90,0.90,0.90}
\definecolor{gray91}{rgb}{0.91,0.91,0.91}
\definecolor{gray92}{rgb}{0.92,0.92,0.92}
\definecolor{gray93}{rgb}{0.93,0.93,0.93}
\definecolor{gray94}{rgb}{0.94,0.94,0.94}
\definecolor{gray95}{rgb}{0.95,0.95,0.95}
\definecolor{gray96}{rgb}{0.96,0.96,0.96}
\definecolor{gray97}{rgb}{0.97,0.97,0.97}
\definecolor{gray98}{rgb}{0.98,0.98,0.98}
\definecolor{gray99}{rgb}{0.99,0.99,0.99}
\definecolor{gray9}{rgb}{0.09,0.09,0.09}
\definecolor{gray}{rgb}{0.75,0.75,0.75}
\definecolor{green1}{rgb}{0.00,1.00,0.00}
\definecolor{green2}{rgb}{0.00,0.93,0.00}
\definecolor{green3}{rgb}{0.00,0.80,0.00}
\definecolor{green4}{rgb}{0.00,0.55,0.00}
\definecolor{greenyellow}{rgb}{0.68,1.00,0.18}
\definecolor{green}{rgb}{0.00,1.00,0.00}
\definecolor{grey0}{rgb}{0.00,0.00,0.00}
\definecolor{grey100}{rgb}{1.00,1.00,1.00}
\definecolor{grey10}{rgb}{0.10,0.10,0.10}
\definecolor{grey11}{rgb}{0.11,0.11,0.11}
\definecolor{grey12}{rgb}{0.12,0.12,0.12}
\definecolor{grey13}{rgb}{0.13,0.13,0.13}
\definecolor{grey14}{rgb}{0.14,0.14,0.14}
\definecolor{grey15}{rgb}{0.15,0.15,0.15}
\definecolor{grey16}{rgb}{0.16,0.16,0.16}
\definecolor{grey17}{rgb}{0.17,0.17,0.17}
\definecolor{grey18}{rgb}{0.18,0.18,0.18}
\definecolor{grey19}{rgb}{0.19,0.19,0.19}
\definecolor{grey1}{rgb}{0.01,0.01,0.01}
\definecolor{grey20}{rgb}{0.20,0.20,0.20}
\definecolor{grey21}{rgb}{0.21,0.21,0.21}
\definecolor{grey22}{rgb}{0.22,0.22,0.22}
\definecolor{grey23}{rgb}{0.23,0.23,0.23}
\definecolor{grey24}{rgb}{0.24,0.24,0.24}
\definecolor{grey25}{rgb}{0.25,0.25,0.25}
\definecolor{grey26}{rgb}{0.26,0.26,0.26}
\definecolor{grey27}{rgb}{0.27,0.27,0.27}
\definecolor{grey28}{rgb}{0.28,0.28,0.28}
\definecolor{grey29}{rgb}{0.29,0.29,0.29}
\definecolor{grey2}{rgb}{0.02,0.02,0.02}
\definecolor{grey30}{rgb}{0.30,0.30,0.30}
\definecolor{grey31}{rgb}{0.31,0.31,0.31}
\definecolor{grey32}{rgb}{0.32,0.32,0.32}
\definecolor{grey33}{rgb}{0.33,0.33,0.33}
\definecolor{grey34}{rgb}{0.34,0.34,0.34}
\definecolor{grey35}{rgb}{0.35,0.35,0.35}
\definecolor{grey36}{rgb}{0.36,0.36,0.36}
\definecolor{grey37}{rgb}{0.37,0.37,0.37}
\definecolor{grey38}{rgb}{0.38,0.38,0.38}
\definecolor{grey39}{rgb}{0.39,0.39,0.39}
\definecolor{grey3}{rgb}{0.03,0.03,0.03}
\definecolor{grey40}{rgb}{0.40,0.40,0.40}
\definecolor{grey41}{rgb}{0.41,0.41,0.41}
\definecolor{grey42}{rgb}{0.42,0.42,0.42}
\definecolor{grey43}{rgb}{0.43,0.43,0.43}
\definecolor{grey44}{rgb}{0.44,0.44,0.44}
\definecolor{grey45}{rgb}{0.45,0.45,0.45}
\definecolor{grey46}{rgb}{0.46,0.46,0.46}
\definecolor{grey47}{rgb}{0.47,0.47,0.47}
\definecolor{grey48}{rgb}{0.48,0.48,0.48}
\definecolor{grey49}{rgb}{0.49,0.49,0.49}
\definecolor{grey4}{rgb}{0.04,0.04,0.04}
\definecolor{grey50}{rgb}{0.50,0.50,0.50}
\definecolor{grey51}{rgb}{0.51,0.51,0.51}
\definecolor{grey52}{rgb}{0.52,0.52,0.52}
\definecolor{grey53}{rgb}{0.53,0.53,0.53}
\definecolor{grey54}{rgb}{0.54,0.54,0.54}
\definecolor{grey55}{rgb}{0.55,0.55,0.55}
\definecolor{grey56}{rgb}{0.56,0.56,0.56}
\definecolor{grey57}{rgb}{0.57,0.57,0.57}
\definecolor{grey58}{rgb}{0.58,0.58,0.58}
\definecolor{grey59}{rgb}{0.59,0.59,0.59}
\definecolor{grey5}{rgb}{0.05,0.05,0.05}
\definecolor{grey60}{rgb}{0.60,0.60,0.60}
\definecolor{grey61}{rgb}{0.61,0.61,0.61}
\definecolor{grey62}{rgb}{0.62,0.62,0.62}
\definecolor{grey63}{rgb}{0.63,0.63,0.63}
\definecolor{grey64}{rgb}{0.64,0.64,0.64}
\definecolor{grey65}{rgb}{0.65,0.65,0.65}
\definecolor{grey66}{rgb}{0.66,0.66,0.66}
\definecolor{grey67}{rgb}{0.67,0.67,0.67}
\definecolor{grey68}{rgb}{0.68,0.68,0.68}
\definecolor{grey69}{rgb}{0.69,0.69,0.69}
\definecolor{grey6}{rgb}{0.06,0.06,0.06}
\definecolor{grey70}{rgb}{0.70,0.70,0.70}
\definecolor{grey71}{rgb}{0.71,0.71,0.71}
\definecolor{grey72}{rgb}{0.72,0.72,0.72}
\definecolor{grey73}{rgb}{0.73,0.73,0.73}
\definecolor{grey74}{rgb}{0.74,0.74,0.74}
\definecolor{grey75}{rgb}{0.75,0.75,0.75}
\definecolor{grey76}{rgb}{0.76,0.76,0.76}
\definecolor{grey77}{rgb}{0.77,0.77,0.77}
\definecolor{grey78}{rgb}{0.78,0.78,0.78}
\definecolor{grey79}{rgb}{0.79,0.79,0.79}
\definecolor{grey7}{rgb}{0.07,0.07,0.07}
\definecolor{grey80}{rgb}{0.80,0.80,0.80}
\definecolor{grey81}{rgb}{0.81,0.81,0.81}
\definecolor{grey82}{rgb}{0.82,0.82,0.82}
\definecolor{grey83}{rgb}{0.83,0.83,0.83}
\definecolor{grey84}{rgb}{0.84,0.84,0.84}
\definecolor{grey85}{rgb}{0.85,0.85,0.85}
\definecolor{grey86}{rgb}{0.86,0.86,0.86}
\definecolor{grey87}{rgb}{0.87,0.87,0.87}
\definecolor{grey88}{rgb}{0.88,0.88,0.88}
\definecolor{grey89}{rgb}{0.89,0.89,0.89}
\definecolor{grey8}{rgb}{0.08,0.08,0.08}
\definecolor{grey90}{rgb}{0.90,0.90,0.90}
\definecolor{grey91}{rgb}{0.91,0.91,0.91}
\definecolor{grey92}{rgb}{0.92,0.92,0.92}
\definecolor{grey93}{rgb}{0.93,0.93,0.93}
\definecolor{grey94}{rgb}{0.94,0.94,0.94}
\definecolor{grey95}{rgb}{0.95,0.95,0.95}
\definecolor{grey96}{rgb}{0.96,0.96,0.96}
\definecolor{grey97}{rgb}{0.97,0.97,0.97}
\definecolor{grey98}{rgb}{0.98,0.98,0.98}
\definecolor{grey99}{rgb}{0.99,0.99,0.99}
\definecolor{grey9}{rgb}{0.09,0.09,0.09}
\definecolor{grey}{rgb}{0.75,0.75,0.75}
\definecolor{honeydew1}{rgb}{0.94,1.00,0.94}
\definecolor{honeydew2}{rgb}{0.88,0.93,0.88}
\definecolor{honeydew3}{rgb}{0.76,0.80,0.76}
\definecolor{honeydew4}{rgb}{0.51,0.55,0.51}
\definecolor{honeydew}{rgb}{0.94,1.00,0.94}
\definecolor{hotpink}{rgb}{1.00,0.41,0.71}
\definecolor{indianred}{rgb}{0.80,0.36,0.36}
\definecolor{ivory1}{rgb}{1.00,1.00,0.94}
\definecolor{ivory2}{rgb}{0.93,0.93,0.88}
\definecolor{ivory3}{rgb}{0.80,0.80,0.76}
\definecolor{ivory4}{rgb}{0.55,0.55,0.51}
\definecolor{ivory}{rgb}{1.00,1.00,0.94}
\definecolor{khaki1}{rgb}{1.00,0.96,0.56}
\definecolor{khaki2}{rgb}{0.93,0.90,0.52}
\definecolor{khaki3}{rgb}{0.80,0.78,0.45}
\definecolor{khaki4}{rgb}{0.55,0.53,0.31}
\definecolor{khaki}{rgb}{0.94,0.90,0.55}
\definecolor{lavenderblush}{rgb}{1.00,0.94,0.96}
\definecolor{lavender}{rgb}{0.90,0.90,0.98}
\definecolor{lawngreen}{rgb}{0.49,0.99,0.00}
\definecolor{lemonchiffon}{rgb}{1.00,0.98,0.80}
\definecolor{lightblue}{rgb}{0.68,0.85,0.90}
\definecolor{lightcoral}{rgb}{0.94,0.50,0.50}
\definecolor{lightcyan}{rgb}{0.88,1.00,1.00}
\definecolor{lightgoldenrod}{rgb}{0.93,0.87,0.51}
\definecolor{lightgoldenrod}{rgb}{0.98,0.98,0.82}
\definecolor{lightgray}{rgb}{0.83,0.83,0.83}
\definecolor{lightgreen}{rgb}{0.56,0.93,0.56}
\definecolor{lightgrey}{rgb}{0.83,0.83,0.83}
\definecolor{lightpink}{rgb}{1.00,0.71,0.76}
\definecolor{lightsalmon}{rgb}{1.00,0.63,0.48}
\definecolor{lightsea}{rgb}{0.13,0.70,0.67}
\definecolor{lightsky}{rgb}{0.53,0.81,0.98}
\definecolor{lightslate}{rgb}{0.47,0.53,0.60}
\definecolor{lightslate}{rgb}{0.47,0.53,0.60}
\definecolor{lightslate}{rgb}{0.52,0.44,1.00}
\definecolor{lightsteel}{rgb}{0.69,0.77,0.87}
\definecolor{lightyellow}{rgb}{1.00,1.00,0.88}
\definecolor{limegreen}{rgb}{0.20,0.80,0.20}
\definecolor{linen}{rgb}{0.98,0.94,0.90}
\definecolor{magenta1}{rgb}{1.00,0.00,1.00}
\definecolor{magenta2}{rgb}{0.93,0.00,0.93}
\definecolor{magenta3}{rgb}{0.80,0.00,0.80}
\definecolor{magenta4}{rgb}{0.55,0.00,0.55}
\definecolor{magenta}{rgb}{1.00,0.00,1.00}
\definecolor{maroon1}{rgb}{1.00,0.20,0.70}
\definecolor{maroon2}{rgb}{0.93,0.19,0.65}
\definecolor{maroon3}{rgb}{0.80,0.16,0.56}
\definecolor{maroon4}{rgb}{0.55,0.11,0.38}
\definecolor{maroon}{rgb}{0.69,0.19,0.38}
\definecolor{mediumaquamarine}{rgb}{0.40,0.80,0.67}
\definecolor{mediumblue}{rgb}{0.00,0.00,0.80}
\definecolor{mediumorchid}{rgb}{0.73,0.33,0.83}
\definecolor{mediumpurple}{rgb}{0.58,0.44,0.86}
\definecolor{mediumsea}{rgb}{0.24,0.70,0.44}
\definecolor{mediumslate}{rgb}{0.48,0.41,0.93}
\definecolor{mediumspring}{rgb}{0.00,0.98,0.60}
\definecolor{mediumturquoise}{rgb}{0.28,0.82,0.80}
\definecolor{mediumviolet}{rgb}{0.78,0.08,0.52}
\definecolor{midnightblue}{rgb}{0.10,0.10,0.44}
\definecolor{mintcream}{rgb}{0.96,1.00,0.98}
\definecolor{mistyrose}{rgb}{1.00,0.89,0.88}
\definecolor{moccasin}{rgb}{1.00,0.89,0.71}
\definecolor{navajowhite}{rgb}{1.00,0.87,0.68}
\definecolor{navyblue}{rgb}{0.00,0.00,0.50}
\definecolor{navy}{rgb}{0.00,0.00,0.50}
\definecolor{oldlace}{rgb}{0.99,0.96,0.90}
\definecolor{olivedrab}{rgb}{0.42,0.56,0.14}
\definecolor{orange1}{rgb}{1.00,0.65,0.00}
\definecolor{orange2}{rgb}{0.93,0.60,0.00}
\definecolor{orange3}{rgb}{0.80,0.52,0.00}
\definecolor{orange4}{rgb}{0.55,0.35,0.00}
\definecolor{orangered}{rgb}{1.00,0.27,0.00}
\definecolor{orange}{rgb}{1.00,0.65,0.00}
\definecolor{orchid1}{rgb}{1.00,0.51,0.98}
\definecolor{orchid2}{rgb}{0.93,0.48,0.91}
\definecolor{orchid3}{rgb}{0.80,0.41,0.79}
\definecolor{orchid4}{rgb}{0.55,0.28,0.54}
\definecolor{orchid}{rgb}{0.85,0.44,0.84}
\definecolor{palegoldenrod}{rgb}{0.93,0.91,0.67}
\definecolor{palegreen}{rgb}{0.60,0.98,0.60}
\definecolor{paleturquoise}{rgb}{0.69,0.93,0.93}
\definecolor{paleviolet}{rgb}{0.86,0.44,0.58}
\definecolor{papayawhip}{rgb}{1.00,0.94,0.84}
\definecolor{peachpuff}{rgb}{1.00,0.85,0.73}
\definecolor{peru}{rgb}{0.80,0.52,0.25}
\definecolor{pink1}{rgb}{1.00,0.71,0.77}
\definecolor{pink2}{rgb}{0.93,0.66,0.72}
\definecolor{pink3}{rgb}{0.80,0.57,0.62}
\definecolor{pink4}{rgb}{0.55,0.39,0.42}
\definecolor{pink}{rgb}{1.00,0.75,0.80}
\definecolor{plum1}{rgb}{1.00,0.73,1.00}
\definecolor{plum2}{rgb}{0.93,0.68,0.93}
\definecolor{plum3}{rgb}{0.80,0.59,0.80}
\definecolor{plum4}{rgb}{0.55,0.40,0.55}
\definecolor{plum}{rgb}{0.87,0.63,0.87}
\definecolor{powderblue}{rgb}{0.69,0.88,0.90}
\definecolor{purple1}{rgb}{0.61,0.19,1.00}
\definecolor{purple2}{rgb}{0.57,0.17,0.93}
\definecolor{purple3}{rgb}{0.49,0.15,0.80}
\definecolor{purple4}{rgb}{0.33,0.10,0.55}
\definecolor{purple}{rgb}{0.63,0.13,0.94}
\definecolor{red1}{rgb}{1.00,0.00,0.00}
\definecolor{red2}{rgb}{0.93,0.00,0.00}
\definecolor{red3}{rgb}{0.80,0.00,0.00}
\definecolor{red4}{rgb}{0.55,0.00,0.00}
\definecolor{red}{rgb}{1.00,0.00,0.00}
\definecolor{rosybrown}{rgb}{0.74,0.56,0.56}
\definecolor{royalblue}{rgb}{0.25,0.41,0.88}
\definecolor{saddlebrown}{rgb}{0.55,0.27,0.07}
\definecolor{salmon1}{rgb}{1.00,0.55,0.41}
\definecolor{salmon2}{rgb}{0.93,0.51,0.38}
\definecolor{salmon3}{rgb}{0.80,0.44,0.33}
\definecolor{salmon4}{rgb}{0.55,0.30,0.22}
\definecolor{salmon}{rgb}{0.98,0.50,0.45}
\definecolor{sandybrown}{rgb}{0.96,0.64,0.38}
\definecolor{seagreen}{rgb}{0.18,0.55,0.34}
\definecolor{seashell1}{rgb}{1.00,0.96,0.93}
\definecolor{seashell2}{rgb}{0.93,0.90,0.87}
\definecolor{seashell3}{rgb}{0.80,0.77,0.75}
\definecolor{seashell4}{rgb}{0.55,0.53,0.51}
\definecolor{seashell}{rgb}{1.00,0.96,0.93}
\definecolor{sienna1}{rgb}{1.00,0.51,0.28}
\definecolor{sienna2}{rgb}{0.93,0.47,0.26}
\definecolor{sienna3}{rgb}{0.80,0.41,0.22}
\definecolor{sienna4}{rgb}{0.55,0.28,0.15}
\definecolor{sienna}{rgb}{0.63,0.32,0.18}
\definecolor{skyblue}{rgb}{0.53,0.81,0.92}
\definecolor{slateblue}{rgb}{0.42,0.35,0.80}
\definecolor{slategray}{rgb}{0.44,0.50,0.56}
\definecolor{slategrey}{rgb}{0.44,0.50,0.56}
\definecolor{snow1}{rgb}{1.00,0.98,0.98}
\definecolor{snow2}{rgb}{0.93,0.91,0.91}
\definecolor{snow3}{rgb}{0.80,0.79,0.79}
\definecolor{snow4}{rgb}{0.55,0.54,0.54}
\definecolor{snow}{rgb}{1.00,0.98,0.98}
\definecolor{springgreen}{rgb}{0.00,1.00,0.50}
\definecolor{steelblue}{rgb}{0.27,0.51,0.71}
\definecolor{tan1}{rgb}{1.00,0.65,0.31}
\definecolor{tan2}{rgb}{0.93,0.60,0.29}
\definecolor{tan3}{rgb}{0.80,0.52,0.25}
\definecolor{tan4}{rgb}{0.55,0.35,0.17}
\definecolor{tan}{rgb}{0.82,0.71,0.55}
\definecolor{thistle1}{rgb}{1.00,0.88,1.00}
\definecolor{thistle2}{rgb}{0.93,0.82,0.93}
\definecolor{thistle3}{rgb}{0.80,0.71,0.80}
\definecolor{thistle4}{rgb}{0.55,0.48,0.55}
\definecolor{thistle}{rgb}{0.85,0.75,0.85}
\definecolor{tomato1}{rgb}{1.00,0.39,0.28}
\definecolor{tomato2}{rgb}{0.93,0.36,0.26}
\definecolor{tomato3}{rgb}{0.80,0.31,0.22}
\definecolor{tomato4}{rgb}{0.55,0.21,0.15}
\definecolor{tomato}{rgb}{1.00,0.39,0.28}
\definecolor{turquoise1}{rgb}{0.00,0.96,1.00}
\definecolor{turquoise2}{rgb}{0.00,0.90,0.93}
\definecolor{turquoise3}{rgb}{0.00,0.77,0.80}
\definecolor{turquoise4}{rgb}{0.00,0.53,0.55}
\definecolor{turquoise}{rgb}{0.25,0.88,0.82}
\definecolor{violetred}{rgb}{0.82,0.13,0.56}
\definecolor{violet}{rgb}{0.93,0.51,0.93}
\definecolor{wheat1}{rgb}{1.00,0.91,0.73}
\definecolor{wheat2}{rgb}{0.93,0.85,0.68}
\definecolor{wheat3}{rgb}{0.80,0.73,0.59}
\definecolor{wheat4}{rgb}{0.55,0.49,0.40}
\definecolor{wheat}{rgb}{0.96,0.87,0.70}
\definecolor{whitesmoke}{rgb}{0.96,0.96,0.96}
\definecolor{white}{rgb}{1.00,1.00,1.00}
\definecolor{yellow1}{rgb}{1.00,1.00,0.00}
\definecolor{yellow2}{rgb}{0.93,0.93,0.00}
\definecolor{yellow3}{rgb}{0.80,0.80,0.00}
\definecolor{yellow4}{rgb}{0.55,0.55,0.00}
\definecolor{yellowgreen}{rgb}{0.60,0.80,0.20}
\definecolor{yellow}{rgb}{1.00,1.00,0.00}
\usepackage{amsfonts}
\usepackage{bm}
\usepackage{longtable}
\usepackage{multirow}
\usepackage{pifont}
\newcommand{\cmark}{\ding{51}}%
\usepackage{gensymb}

\usepackage{booktabs} 
 
\newcommand{\gskfont}{
}
\newcommand{\jcafont}{
  \bfseries 
  \color{green}
}
\newcommand{\vpfont}{
  \bfseries 
  \color{magenta}
}
\newcommand{\yxfont}{
  \bfseries 
  \color{blue}
}

\DeclareTextFontCommand{\gsk}{\gskfont}
\DeclareTextFontCommand{\vp}{\vpfont}
\DeclareTextFontCommand{\yx}{\yxfont}
\DeclareTextFontCommand{\jca}{\jcafont}

\newcommand{\radyn}{\texttt{RADYN}}
\newcommand{\radynfp}{\texttt{RADYN+FP}}
\newcommand{\fpc}{\texttt{FP}}
\newcommand{\rhpar}{\texttt{RH15D}}

\shorttitle{Evolution of Heating Rates at Flare Ribbon Fronts}
\shortauthors{Kerr, Polito, Xu, Allred}

\begin{document}


	\title{Solar Flare Ribbon Fronts II: Evolution of heating rates in individual flare footpoints}
	\author[0000-0001-5316-914X]{Graham~S. Kerr}
	\email{graham.s.kerr@nasa.gov}
	\email{kerrg@cua.edu}
	\affil{NASA Goddard Space Flight Center, Heliophysics Science Division, Code 671, 8800 Greenbelt Rd., Greenbelt, MD 20771, USA}
 	\affil{Department of Physics, Catholic University of America, 620 Michigan Avenue, Northeast, Washington, DC 20064, USA}
	
        \author[0000-0002-4980-7126]{Vanessa Polito}
        \affiliation{Lockheed Martin Solar and Astrophysics Laboratory, Building 203, 3251 Hanover Street, Palo Alto, CA 94304, USA}
        \affiliation{{Department of Physics Oregon State University, 301 Weniger Hall, Corvallis, 97331, OR, USA}}
        
        \author{Yan Xu}
        \affil{Institute for Space Weather Sciences, New Jersey Institute of Technology, 323 Martin Luther King Boulevard, Newark, NJ 07102-1982}
         \affil{Big Bear Solar Observatory, New Jersey Institute of Technology, 40386 North Shore Lane, Big Bear City, CA 92314-9672, USA}
	
	\author[0000-0003-4227-6809]{Joel~C. Allred}
	\affil{NASA Goddard Space Flight Center, Heliophysics Science Division, Code 671, 8800 Greenbelt Rd., Greenbelt, MD 20771, USA}

	\date{Received / Accepted}
	
	
	\begin{abstract}	
Solar flare ribbon fronts appear ahead of the bright structures that normally characterise solar flares, and can persist for an extended period of time in spatially localised patches before transitioning to `regular' bright ribbons. They likely represent the initial onset of flare energy deposition into the chromosphere. Chromospheric spectra (e.g. \ion{He}{1} 10830~\AA\ and the \ion{Mg}{2} near-UV lines) from ribbon fronts exhibit properties rather different to typical flare behaviour. In prior numerical modelling efforts we were unable to reproduce the long lifetime of ribbon fronts. Here we present a series of numerical experiments that are rather simple but which have important implications. We inject a very low flux of nonthermal electrons ($F = 5\times10^{8}$~erg~s$^{-1}$~cm$^{-2}$) into the chromosphere for 100~s before ramping up to standard flare energy fluxes $(F = 10^{10-11}$~erg~s$^{-1}$~cm$^{-2}$). Synthetic spectra not only sustained their ribbon front-like properties for significantly longer, in the case of harder nonthermal electron spectra the ribbon front behaviour persisted for the entirety of this weak-heating phase. Lengthening or shortening the duration of the weak-heating phase commensurately lengthened or shortened the ribbon front lifetimes. Ribbon fronts transitioned to regular bright ribbons when the upper chromosphere became sufficiently hot and dense, which happened faster for softer nonthermal electron spectra. Thus, the lifetime of flare ribbon fronts are a direct measure of the duration over which a relatively low flux of high energy electrons precipitates to the chromosphere prior to the bombardment of a much larger energy flux.
	 	\end{abstract}

\section{Introduction}\label{sec:intro}

The liberation of magnetic energy following magnetic reconnection in solar and stellar atmospheres drives eruptive events (flares, coronal mass ejections, jets etc.,) on a range of scales.  Oftentimes particles are accelerated during these dynamic events, with mounting evidence that nonthermal particles are present even in small events such as microflares and nanoflares \citep[e.g.][]{2020ApJ...891L..34G,2021MNRAS.507.3936C,2018ApJ...856..178P,2023FrASS..1014901P}. Diagnosing the various processes at play during solar (stellar) eruptive events (SEEs) are active areas of inquiry, and high resolution observations of solar flares are key to understanding the physics of magnetic energy release, energy transport, and energy dissipation. Dissipation of this energy ultimately leads to the enhancement of the radiative output across various parts of the electromagnetic spectrum, providing windows throughout the flaring atmospheres. 

A dramatic manifestation of flare energy release is the extended flare ribbon structures that appear in optical, infrared, and UV radiation radiation \citep[][]{2011SSRv..159...19F}. These ribbons form at the base of flare loops in the chromosphere and transition region (TR), and as such are also referred to as flare footpoints. Hard X-ray and radio observations suggest populations of nonthermal electrons in compact footpoints co-temporal and co-spatial with flare ribbons, and the properties of those HXRs have been used to infer the energy distribution of those energetic electrons \citep[see reviews by][]{2011SSRv..159..107H,2011SSRv..159..301K}. These HXR footpoints are pointed to as evidence for the standard flare model, in which electrons accelerated in the corona following magnetic reconnection bombard the chromosphere and transition region where they lose their energy due to Coulomb collisions. Plasma heating, ionisation and mass flows result. Of course, it is likely that other energy transport mechanisms are present also, including accelerated protons and heavy ions, thermal conduction following direct \textsl{in-situ} heating, and magnetohydrodynamic waves \citep[see discussions in Section 3 of][]{2023FrASS...960862K}. 

High spatial resolution (sub-arcsecond) observations afforded by modern ground-based and space-based observatories have revealed that not only do flare ribbons contain sub-structure, they can be accompanied by a narrow leading edge, or `ribbon front' which exhibits properties that differ from the main bright ribbon and trailing edge. Flare ribbons undergo an apparent propagation over the field of view caused by sequential reconnection of different loops. The brightest parts of the ribbon are thought to be sites into which large energy fluxes from those newly reconnected loops are deposited. Those locations cool but remain brighter than the pre-flare for sometime creating a hazy region behind the propagating bright region that has been referred to as the trailing ribbon. Observations from the Goode Solar Telescope at Big Bear Solar Observatory \citep[GST/BBSO;][]{2012ASPC..463..357G} and the Interface Region Imaging Spectrograph \citep[IRIS;][]{2014SoPh..289.2733D} now suggest that in some sections of flare ribbons there is a very narrow leading edge ahead of the bright ribbon \citep[e.g.][]{2016ApJ...819...89X,2022ApJ...924L..18X,2018ApJ...861...62P,2021ApJ...915...77P,2023ApJ...944..104P,2023ApJS..268...62W}. Those ribbon fronts represent the initial deposition of energy into the chromosphere. For a cartoon of these regions see Figure 10 of \cite{2023ApJ...944..104P}.

\cite{2016ApJ...819...89X} identified narrow, 350-500~km wide, structures along ahead of propagating ribbons in two flares. \ion{He}{1} 10830~\AA\ filtergrams of this ribbon front showed a \textsl{negative} contrast, which then transitioned to the expected positive contrast in the bright ribbon phase of each source's lifetime. The ribbon front phase lasted several dozen seconds. Of note is that only one ribbon in each flare exhibited ribbon fronts. Subsequent observations \citep{2022ApJ...924L..18X} with coarse spectral coverage (but unfortunately lacking temporal information) confirmed that dimming occurs over the full line. 

At the same locations where \ion{He}{1} 10830~\AA\ enhanced absorption occurred \cite{2016ApJ...819...89X} found that the \ion{Mg}{2} k line, observed by IRIS, also exhibited characteristics that differed from the typical bright ribbon spectra. The optically thick \ion{Mg}{2} k line has a central reversal such that its core (referred to as the k3 component) is flanked by two emission peaks (referred to as the k2r and k2v components for the red and blue peak, or together just k2). Along the ribbon fronts, the \ion{Mg}{2} central reversals deepened, their k3 cores were blueshifted, their emission peaks were asymmetric (k2r was stronger than k2v), the k2 peak separation increased, and the lines became extremely broad. In contrast, the more typical flare behaviour is that the lines become almost single peaked, or have only a shallow reversal, and are redshifted or have red-wing asymmetries \citep[e.g.][]{2015A&A...582A..50K,2018ApJ...861...62P}. Using machine learning techniques, \cite{2018ApJ...861...62P}, \cite{2021ApJ...912..121P} and \cite{2021ApJ...915...77P} analysed 33 M and X class flares, discovering that those ribbon front-like behaviours occurred in most if not all flares in their sample. Another property of ribbon fronts was that the \ion{Mg}{2} subordinate triplet (forming alongside the h \& k resonance lines) were in emission rather than absorption. The approximately 1-3 minute time period for the profiles with deeper reversals to become single peaked  was comparable the lifetimes of the \ion{He}{1} 10830~\AA\ negative ribbons.

These \ion{Mg}{2} near-UV (NUV) spectral characteristics were placed on a more quantitative footing by \cite{2023ApJ...944..104P}, who analysed four flares in detail (their metrics are defined in Section~\ref{sec:linemetrics}). They also found that ribbon front locations did not contain explosive chromospheric evaporation or strong TR spectral line intensities. These ribbon front spectra typically evolved into bright ribbon spectra. From one of the events with higher cadence data it was noted that from the onset of the ribbon front it took approximately $45$~s for explosive evaporation to begin. In other locations of that same flare, some ribbon fronts failed to evolve into the explosive evaporation regime, or to exhibit intense transition region emission. Maps of ribbon front locations indicated that they do not appear uniformly along the flare ribbons. Rather, they can be spatially localised and spotty, consistent with the fact that the \ion{He}{1} 10830~\AA\ ribbon front dimming was confined to one ribbon of the two-ribbon flares. \cite{2023ApJS..268...62W} analysed three additional flares in detail, including noting that ribbon fronts were observed in \ion{Mg}{2} with deep reversals and small blueshifts. Notably, they report that some of the ribbon front characteristics, the deep central reversals, were not solely confined to the ribbon fronts and could sporadically appear in the bright ribbon segments also. 

Numerical modelling of flare ribbon fronts has revealed that radiation hydrodynamic (RHD) loop simulations of electron-beam driven flares\footnote{Flare energy was delivered for either $t=10$~s at a constant rate, or was delivered more gradually over $t=20$~s in a triangular profile so that the total energy was fixed between both scenarios.} can successfully reproduce ribbon front characteristics of \ion{He}{1} 10830~\AA\ \citep[][]{2020ApJ...897L...6H,2021ApJ...912..153K}, and \ion{Mg}{2} NUV \citep[][]{2023ApJ...944..104P}. \cite{2020ApJ...897L...6H} demonstrated that the observed pattern of \ion{He}{1} 10830~\AA\ negative contrast followed by positive contrast was caused by an overall dimming of the line before it went into emission once the chromosphere became hot. In order to produce this period of enhanced absorption a population of nonthermal particles must be present \citep[][]{2021ApJ...912..153K}, which cause nonthermal collisional ionisation of He. Recombinations can then overpopulate orthohelium (the multiplet state responsible for \ion{He}{1} 10830~\AA\ and \ion{He}{1} D3) which then absorbs more photospheric radiation, producing a deeper absorption line. Experiments using only thermal conduction from a flare-heated corona, or of electron beam driven flares that omitted nonthermal collisional He ionisation, were unable to produce the dimming at flare onset. Nonthermal electron distributions with a smaller energy flux, delivered more gradually, were able to produce slightly longer lived ribbon fronts. Those with a harder electron energy distribution (in this case, a larger value of low-energy cutoff) drove more pronounced dimmings. 

Using the simulations from \cite{2021ApJ...912..153K},  \cite{2023ApJ...944..104P} synthesised \ion{Mg}{2} NUV spectra, and obtained the same metrics as their observational analysis of IRIS flares. The synthetic spectra that bore characteristics consistent with ribbon front behaviours originated from simulations with smaller, more gently injected energy fluxes with larger low-energy cutoffs. That is, the same energy input parameters can produce \ion{He}{1} and \ion{Mg}{2} ribbon front behaviours, from which we inferred that two distinct heating regimes exist. Ribbon fronts were caused by injection of a weak flux of electrons, while bright ribbon regions are locations bombarded a much larger energy flux (which can then drive strong evaporative upflows). The `best-fit' experiments used triangular energy flux profiles, that peaked $>1\times10^{9}$~erg~s$^{-1}$~cm$^{-2}$. Additional experiments that modified the heating profile (extending the rise time to the peak value or extending the total duration of the flare) demonstrated that simply increasing the time over which a modest energy flux was deposited did not drive subsequent explosive evaporation. Likewise, the transition region (in this case \ion{O}{4} 1401~\AA\ was used as a proxy for the TR response) did not brighten in the latter stages of those simulations. So, the evolution from ribbon front to bright ribbon is not solely a matter of time, the energy flux itself likely has to increase. 

\begin{figure}
	\centering 
	\vbox{
	\hbox{
	\subfloat{\includegraphics[width = 0.5\textwidth, clip = true, trim = 0.cm 0.cm 0.cm 0.cm]{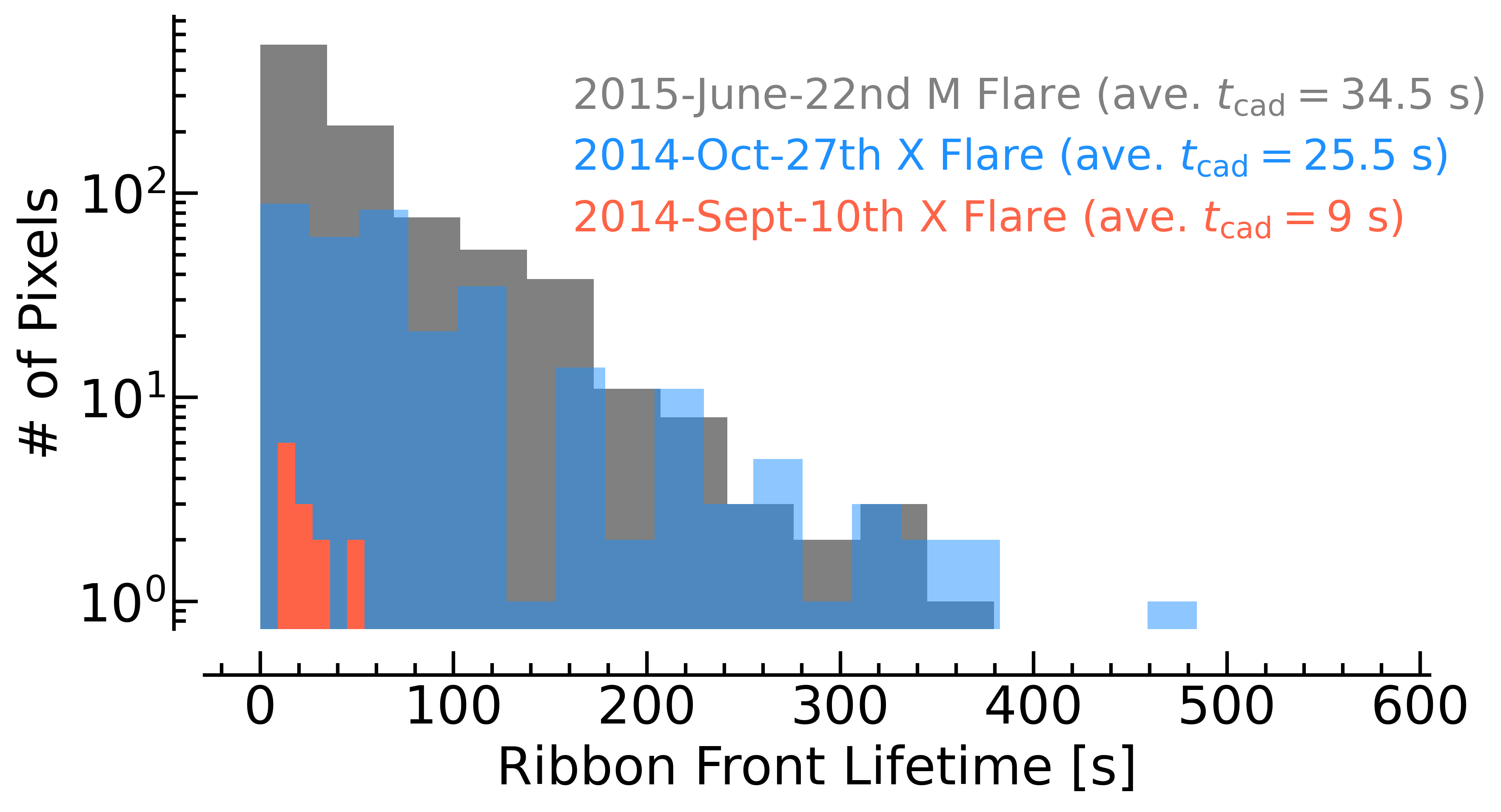}}	
	}
	}
	\caption{\textsl{Histograms of observed ribbon fronts lifetimes, obtained from the flares studied by \cite{2023ApJ...944..104P}. Since each flare had a different cadence, the bin size varies in each event. }}
	\label{fig:obsrflifetimes}
\end{figure}

Whilst chromospheric ribbon front behaviour seems well characterised, the Doppler shifts of the transition region seems more variable. Bright ribbon observations show that the TR generally exhibits redshifts \citep[e.g.][]{2015ApJ...811..139T,2016ApJ...816...89P}, with occasional observations of spatially localised blueshifts \citep[e.g.][]{Jeffreyeaav2794,2022ApJ...934...80L}. \cite{2023ApJ...944..104P} found that along ribbon fronts, both redshifts and blueshifts of \ion{O}{4} 1401~\AA\ could occur, though models only predicted blueshifted ribbon fronts. Adding an additional volumetric heating rate in the corona to mimic \textsl{in-situ} heating following reconnection was unable to produce redshifted \ion{O}{4} 1401~\AA\ alongside the \ion{Mg}{2} ribbon front behaviours. 

\cite{2023ApJ...944..104P} defined several metrics to define if a particular pixel was identified as a \ion{Mg}{2} ribbon front, including the depth of central reversal, separation of the \ion{Mg}{2} k2 emission peaks, the k2 peak asymmetry and the relative intensity of \ion{Mg}{2} triplet to \ion{Mg}{2} k line. Those pixels had variable lifetimes before transitioning to regular bright ribbon pixels. A histogram of those lifetimes is shown in Figure~\ref{fig:obsrflifetimes}, where we see a large range, from as short as a few seconds, up to many hundreds of seconds. Details of these flares can be found in \cite{2023ApJ...944..104P}, but we note here that the finite cadence of the observations means that there could be even shorter lifetimes. In that figure, since the cadence varies from flare-to-flare, the bin size on the x-axis differs. Another interesting thing to note, which can be gleaned from Figures 2-5 of \cite{2023ApJ...944..104P} is the varying number of sources along the ribbons that exhibit ribbon front behaviours within each flare. Of course, this is from three datasets only. An in-depth survey of IRIS flares may reveal a different distribution, but combined with the results of \cite{2018ApJ...861...62P}, \cite{2021ApJ...915...77P} \&  \cite{2021ApJ...912..121P} we can state with some confidence that ribbon front lifetimes can be exceed many dozens of seconds. 

Despite capturing the spectral properties of ribbon fronts, \cite{2021ApJ...912..153K} and \cite{2023ApJ...944..104P} were only able to maintain those characteristics for a few seconds, in stark contrast to the observed longer duration ribbon fronts. In this follow on study we aim to determine if we can extend the simulated ribbon front duration to be more consistent with the observations.


\section{Numerical Experiments}\label{sec:numex}
\gsk{\subsection{Motivating our long duration energy deposition experiments}\label{sec:motivation}}
\gsk{A major model-data discrepancy is that the observed ribbon front lifetimes can be up to one or two orders of magnitude longer than the synthetic ribbon front spectra which are only a few seconds. We found in \cite{2021ApJ...912..153K} that a simulation with a very weak injection of nonthermal electrons could result in \textsl{only} enhanced absorption of \ion{He}{1} 10830~\AA\ (i.e., without subsequent emission), and in \cite{2023ApJ...944..104P}  that two regimes of energy fluxes were required to explain ribbon fronts versus bright ribbons. Based on those points, here we perform a series of rather simple experiments aiming to achieve persistent ribbon front-like behaviour that subsequently transitions to bright ribbon fronts with explosive evaporation. Though seemingly simple, these experiments carry implications for energy release and particle acceleration processes in localised regions of the flare ribbons. They also present a somewhat contrarian view to the prevailing thought on energy injection timescales, that we discuss briefly here.}

\gsk{In our experiments we model flares in which we inject a weak flux of nonthermal electrons for $t_{\mathrm{dur}} = 100$~s, before raising this flux over a period of 100s to values required to produce the observed bright ribbon behaviour. We stress, though, that our primary focus was obtaining realistic durations of the ribbon front phase and so the temporal profile of the `main' heating phase was rather ad-hoc. This was designed to determine how quickly the flare footpoint would transition from ribbon front to bright ribbon, and ultimately produce chromospheric evaporation. Some further experiments were performed that varied the duration of the weak heating phase from $t_{\mathrm{dur}} = 25-150$~s.
}

\gsk{These timings (i.e. around 200s total heating) are rather long compared to what has become the general community's consensus, which more typically assumes bombardment by electron beams into individual locations on the order of a few seconds to tens of seconds \citep[see e.g. discussions in the review by][]{2022FrASS...960856K}. This is based on both observations of the chromospheric response (i.e. the rapid apparent motion of bright fare ribbons as new field lines reconnect) as well as hard X-ray footpoint motion, and the appearance of structure in hard X-ray lightcurves which can be interpreted as elementary bursts of energy injection \cite[see e.g. the recent example by][]{2024arXiv240210546C}.}

\gsk{Despite the body of evidence that suggests a relatively short duration of injection of energetic electrons into discrete areas, we are content that our experiments with long duration heating are appropriate to explore the properties of these ribbon front flare sources. This is especially true as prior experiments with short duration heating have been unable to reproduce ribbon front behaviours with the appropriate lifetimes. Our weak heating phase was observationally driven, based on lifetimes of ribbon front sources. The main heating phase, though arguably longer than the actual main heating phase in real flares, was intended to show that ribbon fronts transition to bright ribbons even if the increased energy flux is modest. Further, since \cite{2023ApJ...944..104P} found that not every ribbon front source ultimately produced chromospheric evaporation, the ramp-up in energy flux in our experiments will demonstrate that explosive evaporation occurs only when the energy flux becomes strong. As we will see in the remainder of this manuscript, our new experiments (1) successfully reproduce ribbon front lifetimes, (2) result in a transition to bright ribbon sources as soon as the energy flux is increased, (3) that when the energy flux in the main heating phase is large enough explosive evaporation is produced. }

\gsk{Current hard X-ray observations do not preclude the extended weak heating phase we propose. In the ribbon front phase of our model the nonthermal electron flux is orders of magnitude weaker than that of the main heating phase which produces the bright ribbons. Since the dynamic range of prior solar hard X-ray observatories has been small, the X-rays produced during ribbon front lifetimes would not be detectable in the presence of brighter X-ray sources elsewhere on the flare ribbons. Nor is the X-ray spatial resolution sufficient to isolate these small localised flare ribbon front sources from the rest of the bright ribbon. Finally, we note that although the general consensus is that the bright ribbon sources only have a few seconds to tens of seconds of nonthermal particle bombardment, there is healthy debate as to what keeps flare sources cooling slowly compared to the rapid cooling in flare loop models. There have been suggestions than some other form of energy is deposited for a longer period of time \citep[see e.g.][]{2016ApJ...820...14Q,2018ApJ...856...27Z,2022ApJ...926..164A}. Although our main heating phase is 100s of electron bombardment, this could in principle be a few seconds of nonthermal electrons, followed by the yet unknown source of energy flux. }

\subsection{Radiation hydrodynamic modelling of flares with \texttt{RADYN}}\label{sec:radyn}
\begin{figure}
	\centering 
	\vbox{
	\hbox{
	\subfloat{\includegraphics[width = 0.5\textwidth, clip = true, trim = 0.cm 0.cm 0.cm 0.cm]{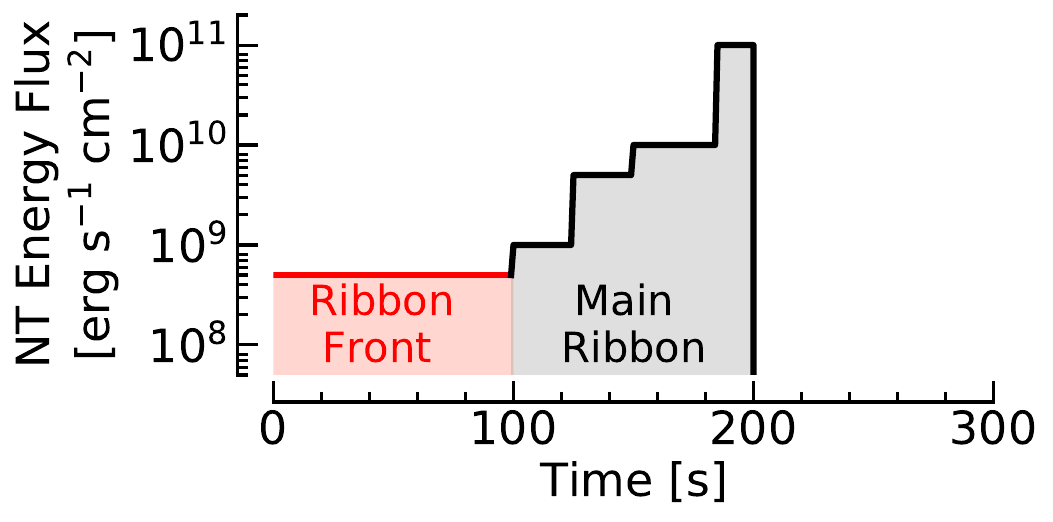}}	
	}
	}
	\caption{\textsl{The temporal evolution of the total energy flux of the injected nonthermal electron energy distribution in our simulation grid. The period designed to mimic the ribbon fronts are shown in red, and the period designed to mimic the more typical `main ribbon' times are shown in grey.}}
	\label{fig:fluxschematic}
\end{figure}

	To simulate the flaring atmosphere we employed the \radynfp\ numerical code, which is the combination of the field-aligned radiation hydrodynamic (RHD) code \radyn\ \citep[][]{1992ApJ...397L..59C,1995ApJ...440L..29C,1997ApJ...481..500C,1999ApJ...521..906A,2002ApJ...572..626C,2005ApJ...630..573A,2015ApJ...809..104A,2023A&A...673A.150C}, coupled with the nonthermal particle transport code \fpc\ \citep[][]{2020ApJ...902...16A}. \radyn\ solves the equations of hydrodynamics, plane-parallel non-local thermodynamic equilibrium (NLTE) radiation transfer, and non-equilibrium atomic level populations, including charge conservation. \fpc\ solves the propagation and thermalisation of a distribution of energetic particles through a specified atmosphere, which in our case is the evolving \radyn\ flares. \gsk{Feedback between each aspect of the problem is included self-consistently. Energy injected by the nonthermal particles affects the thermodynamic structure of the atmosphere, which varies the atomic level populations and emitted radiation. In turn this also affects the thermodynamic state through non-local radiative heating and cooling, which affects the energy deposition profile. The changing temperature of the atmosphere is included in the solution of the particle transport, which solves the full Coulomb collisional operator, making no cold- or warm-target assumption. Instead it is valid for both extremes and everything in between}. \fpc\ was merged with \radyn\ in 2021, replacing the prior imbedded Fokker-Planck based module from \cite{2015ApJ...809..104A}.

	\radyn\ has become a workhorse of the flare community given its ability to model the response of solar chromosphere and transition region at very high spatial resolution (sub-meter where necessary), and to included the important effects of NLTE radiation transport. We do not provide an exhaustive description of the code here, instead we direct readers to recent general descriptions of the code in \cite{2015ApJ...809..104A} and \cite{2023A&A...673A.150C}, and to recent detailed reviews of the ways in which \radyn\ has been used in tandem with observations from IRIS to explore flare physics \cite{2022FrASS...960856K,2023FrASS...960862K}. Some salient points regarding \radyn's capabilities and limitations when modelling He that are relevant to our study are noted below, the majority of which are similar to, and discussed in more detail in, \cite{2021ApJ...912..153K}.

	The model He atom employed in our experiments is a 9-level-with-continuum. \gsk{The upper and lower levels of \ion{He}{1} 10830~\AA\ included but without substructure of the $^{3}P$ state}.  Processes important for populating orthohelium are included. The photoionisation-recombination route is achieved by an irradiating EUV spectrum from the transition region and corona, which uses the local density and temperature of each grid cell when constructing an integrated downward-directed radiation field (note that irradiation by lines modelled in detail by \radyn, such as \ion{He}{2} 304~\AA\ are not included but we believe that this is only a minor effect). The nonthermal collisional ionisation-recombination route is modelled by adding to the rate equations nonthermal collisional ionisation of \ion{He}{1} and \ion{He}{2} \citep[following the recipes of][]{1985A&AS...60..425A}. The effects of the beam-neutralising return current are included when modelling the nonthermal particle distribution. Thermal collisional ionisation-recombination, direct collisional excitation routes, and  collisions between the upper and lower levels are included, which become more important when the flare chromosphere is sufficiently hot. 

\begin{figure*}[ht]
	\centering 
	\vbox{
 
	\hbox{
	\hspace{0.75in}
	\subfloat{\includegraphics[width = 0.75\textwidth, clip = true, trim = 0.cm 0.cm 0.cm 0.cm]{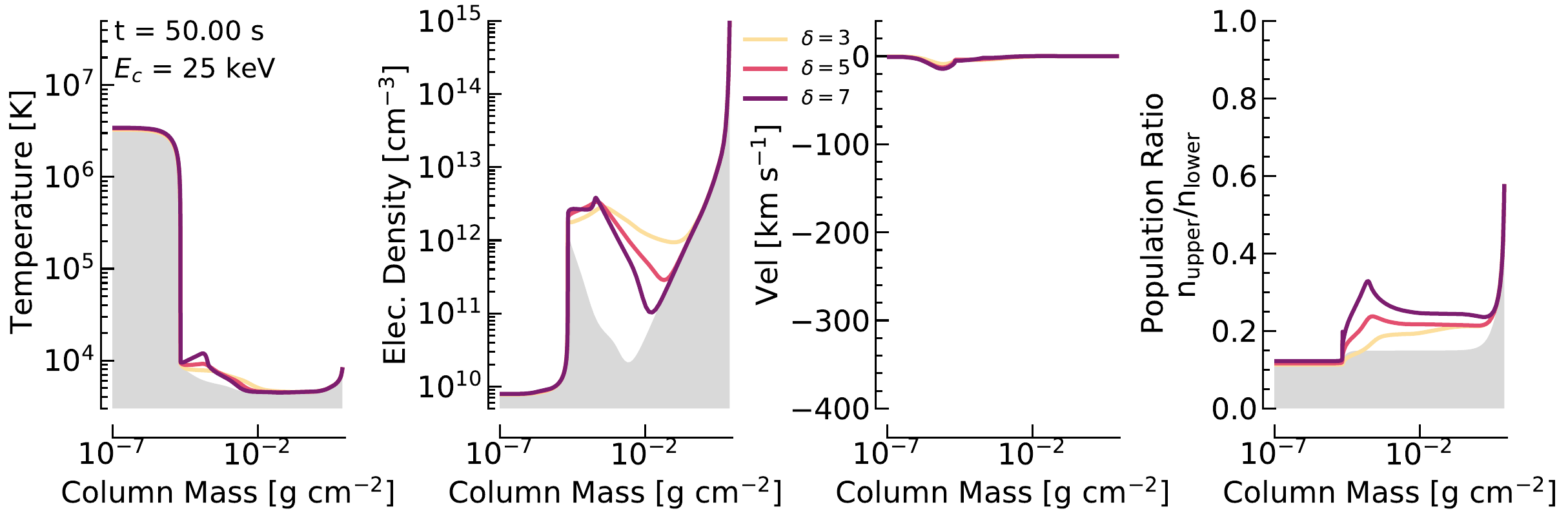}}
	}
	}
	\vbox{
	\hbox{
	\hspace{0.75in}
	\subfloat{\includegraphics[width = 0.75\textwidth, clip = true, trim = 0.cm 0.cm 0.cm 0.cm]{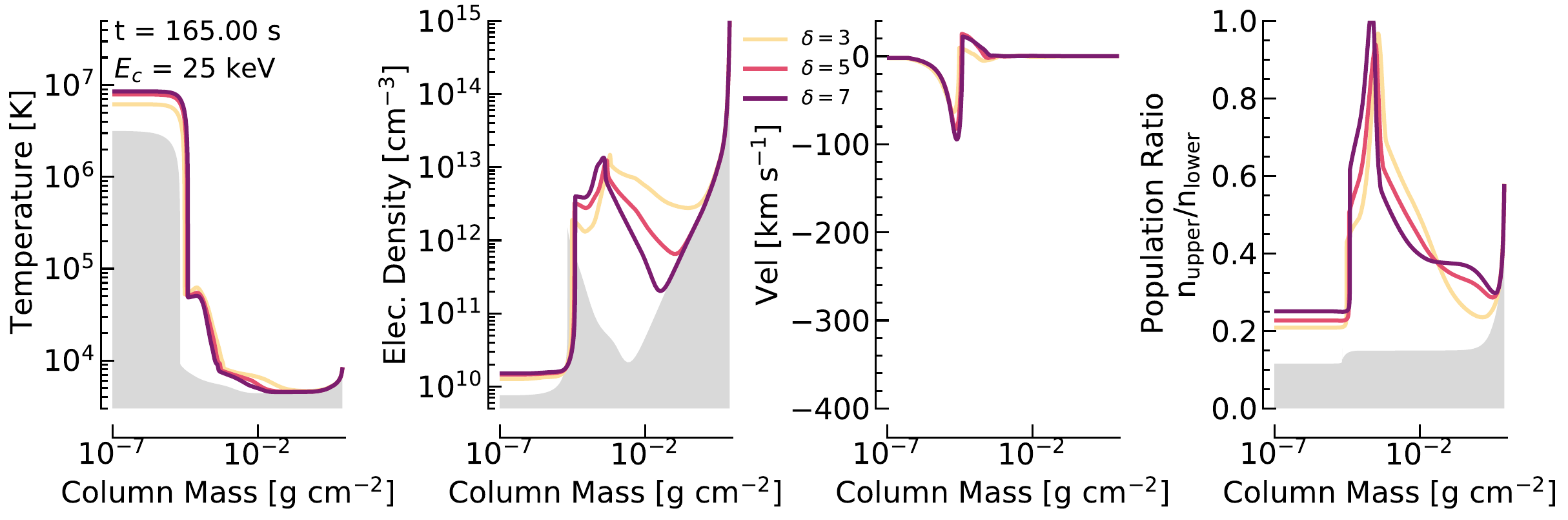}}
	}
	}
	\vbox{
	\hbox{
	\hspace{0.75in}
	\subfloat{\includegraphics[width = 0.75\textwidth, clip = true, trim = 0.cm 0.cm 0.cm 0.cm]{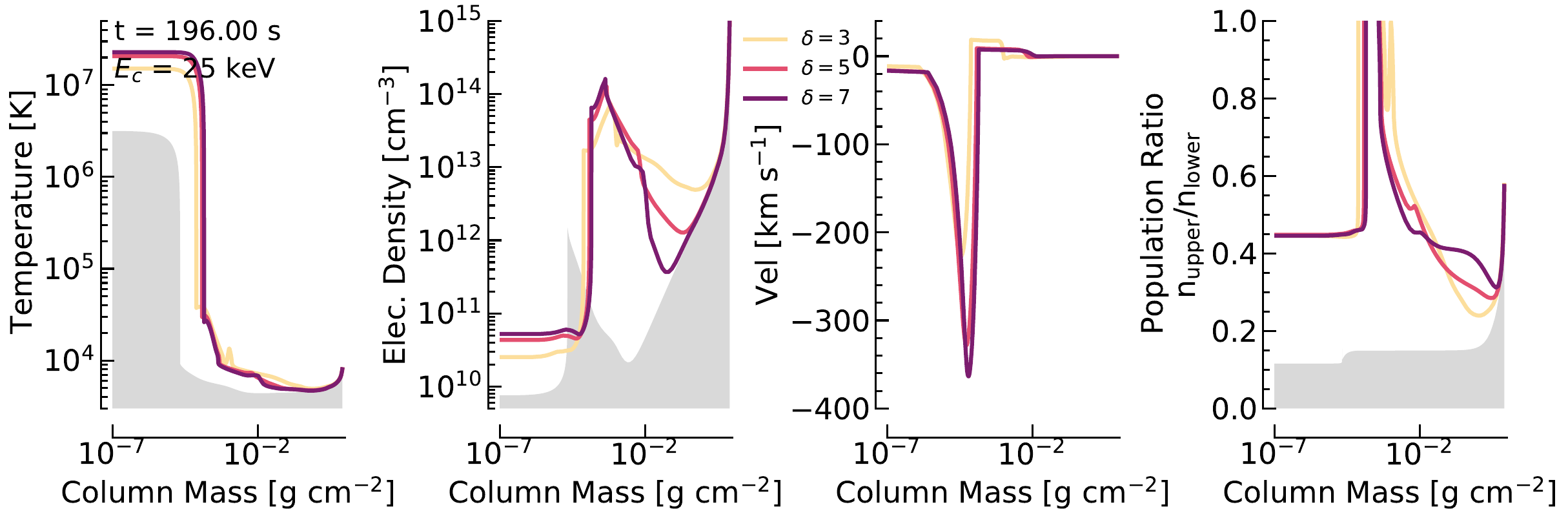}}
	}
	}
	\caption{\textsl{Illustrative examples of the flare atmospheric stratification at different phases of heating. The top row shows a snapshot during the weak-heating (ribbon front) phase, $t=50$~s. The middle row shows a time during the middle of the `main ribbon' phase, $t=165$~s. The bottom row shows a snapshot near the end of the strong-heating phase, $t=196$~s. The shaded area in each panel is the pre-flare atmospheric state, and each colour represents a different spectral index $\delta$, for a fixed $E_{c} = 25$~keV. From left-to-right we show the temperature, electron density, bulk velocity (upflows are negative) and the ratio of the populations of the upper-to-lower levels of \ion{He}{1} 10830~\AA. An animated version is available, starting at $t=0$~s and ending at $t=300$~s, with a real time duration of 60~s.}}
	\label{fig:atmos_examples}
\end{figure*}

	Our experiments used the same initial atmosphere as in \cite{2021ApJ...912..153K}, which is an 11Mm half-loop (with 191 grid points), with an apex temperature of $T=3.15$~MK and electron density $n_{e} = 7.6\times10^{9}$~cm$^{-3}$. Flare energy was injected in the form of a distribution of nonthermal electrons with spectral index $\delta = [3, 5, 7]$, and low-energy cutoff $E_{c} = [10, 15, 20, 25, 30]$~keV. The energy flux injected varied with time, but was the same for each experiment. Given the indications from \cite{2021ApJ...912..153K} that a weak energy flux may sustain a longer period of enhanced absorption, we set the duration of weak-heating to be $t_{\mathrm{dur}} = 100$~s \citep[similar to the observations of][]{2016ApJ...819...89X} with a constant energy flux of $F = 5\times10^{8}$~erg~s$^{-1}$~cm$^{-2}$, followed by an \textsl{ad-hoc} ramp up over the next $100$~s to more typical flare energy fluxes, reaching $F = 1\times10^{11}$~erg~s$^{-1}$~cm$^{-2}$ by $t=200$~s. A schematic of the evolution of the energy flux is shown in Figure~\ref{fig:fluxschematic}. Since we are primarily concerned with demonstrating the requirement of a weak-heating phase for some flare sources, and the difference between the synthetic spectra during their ribbon front phase and the main/bright ribbon phase, the particular details of strong-heating phase are not as important for our current experiments. We elected to slowly ramp up the energy injection in order to determine if there were any obvious signatures that appeared in the synthetic spectra as they transitioned from ribbon front-like to typical-flare spectra.
		
	In addition to this parameter study we modelled five flares with varying $t_{\mathrm{dur}} = [25, 50, 75, 125, 150]$~s, all with $\delta = 5$ and $E_{c} = 25$~keV. The subsequent heating of those additional flares then followed the gray shaded area of Figure~\ref{fig:fluxschematic}, just offset in time, such that the strong-heating phase injected the same total energy flux in each simulation. 

	Figure~\ref{fig:atmos_examples} shows three typical snapshots of the stratification of temperature, electron density, bulk velocity (upflows are negative), and the ratio of the populations of the upper-to-lower levels of \ion{He}{1} 10830~\AA\ (2p $^{3}$P versus 2s $^{3}$S), for $\delta = [3, 5, 7]$, $E_{c} = 25$~keV, and $t_{\mathrm{dur}} = 100$~s. One snapshot shows $t=50$~s, during our supposed weak-heating (ribbon front) phase, one snapshot shows $t=165$~s, midway through the ramp-up of the strong-heating phase, and the final snapshot shows $t=196$~s, when the injected energy flux was at its maximum (where we would expect bright ribbon behaviour). During the weak-heating phase gentle chromospheric evaporation is present, with velocities on the order $v\sim10-20$~km~s$^{-1}$. Modest temperature increases are present throughout the chromosphere, with softer nonthermal electron distributions (larger $\delta$; smaller $E_{c}$) producing somewhat warmer and denser upper chromospheres. Harder nonthermal electron distributions produce both temperature and electron density enhancements to greater depth. While there are enhanced populations of the \ion{He}{1} 2p $^{3}$P level relative to the \ion{He}{1} 2s $^{3}$S level in the upper chromosphere (where the line forms) for all flares, the ratio is notably larger for softer beams. During the strong-heating phase  the atmosphere evolves more dramatically, as one would expect, with orders of magnitude increases in electron density, explosive chromospheric evaporation and chromospheric condensations on the order a few 10s~km~s$^{-1}$.

\subsection{Post-processing flare atmospheres with \texttt{RH15D}}\label{sec:rh15d}
	To provide additional model validation, we complement the \ion{He}{1} 10830~\AA\ synthetic spectra from \radynfp\  with synthetic \ion{Mg}{2} NUV spectra from \rhpar\ \citep{2001ApJ...557..389U,2015A&A...574A...3P}. This code includes the effects of partial frequency redistribution (PRD), important for \ion{Mg}{2} even in the elevated density of the flare chromosphere \citep[][]{2013ApJ...772...89L,2019ApJ...883...57K}, and overlapping transitions. We used the hybrid PRD approximation of \cite{2012ApJ...749..136L}. \rhpar\ solves the multi-line, multi-species NLTE radiation transfer and atomic level population equations given an input atmosphere (in our case, the \radyn\ flare atmospheres processed at 1~s snapshots). The species modelled in NLTE are: H, \ion{C}{1}+\ion{C}{2}, \ion{O}{1}, \ion{Si}{1}+\ion{Si}{2} \& \ion{Mg}{2}, with several others included as LTE background opacity. Background opacity was also provided by modelling the many thousands of lines in the Kurucz linelists\footnote{\url{http://kurucz.harvard.edu/linelists.htm}}, between $\lambda = [20-8000]$~\AA. A constant value of microturbulence was added at each grid cell, with the value $v_{\mathrm{turb}} = 7$~km~s$^{-1}$ based on \cite{2015ApJ...809L..30C}, and the same as used in \citep{2023ApJ...944..104P}. This is somewhat smaller than the flare values observed by \cite{2024MNRAS.527.2523K}, but since those values were obtained at flare peak we did not know when the transition from background to flare values in the chromosphere should take place so we chose to keep this constant. While the \ion{Mg}{2} spectra will be more narrow than observed, this is a known problem \citep[e.g.][]{2024MNRAS.527.2523K} and does not affect our conclusions. 

	For \ion{Mg}{2} we used the 11-level-with-continuum atomic model of \cite{2013ApJ...772...89L}. Though \rhpar\ does not model non-equilibrium effects we have previously demonstrated that these likely have a minimal impact on the \ion{Mg}{2} solution \citep{2019ApJ...885..119K}\gsk{, aside from the initial switch on and cessation of flare energy injection. Relaxation timescales are $\tau_\mathrm{relax} < 0.1$~s during the bulk of the fare, rising to $\tau_\mathrm{relax} \sim 0.1-2$~s during the initial bombardment and first couple of seconds of the cooling phase. Omitting PRD effects for \ion{Mg}{2} were much more impactful.} Further, we keep the H populations and electron densities fixed, such that they include nonthermal and non-equilibrium effects, mitigating omitting these effects from other species. \cite{2019ApJ...879...19Z} improved the treatment of Stark damping for \ion{Mg}{2}, which we employ here. 


\section{Forward modelling flaring \ion{He}{1} 10830~\AA}\label{sec:heiflares}

\subsection{Sustained \ion{He}{1} 10830~\AA\ Dimming}

\begin{figure}
	\centering 
	\vbox{
	\hbox{
	\subfloat{\includegraphics[width = 0.5\textwidth, clip = true, trim = 0.cm 0.cm 0.cm 0.cm]{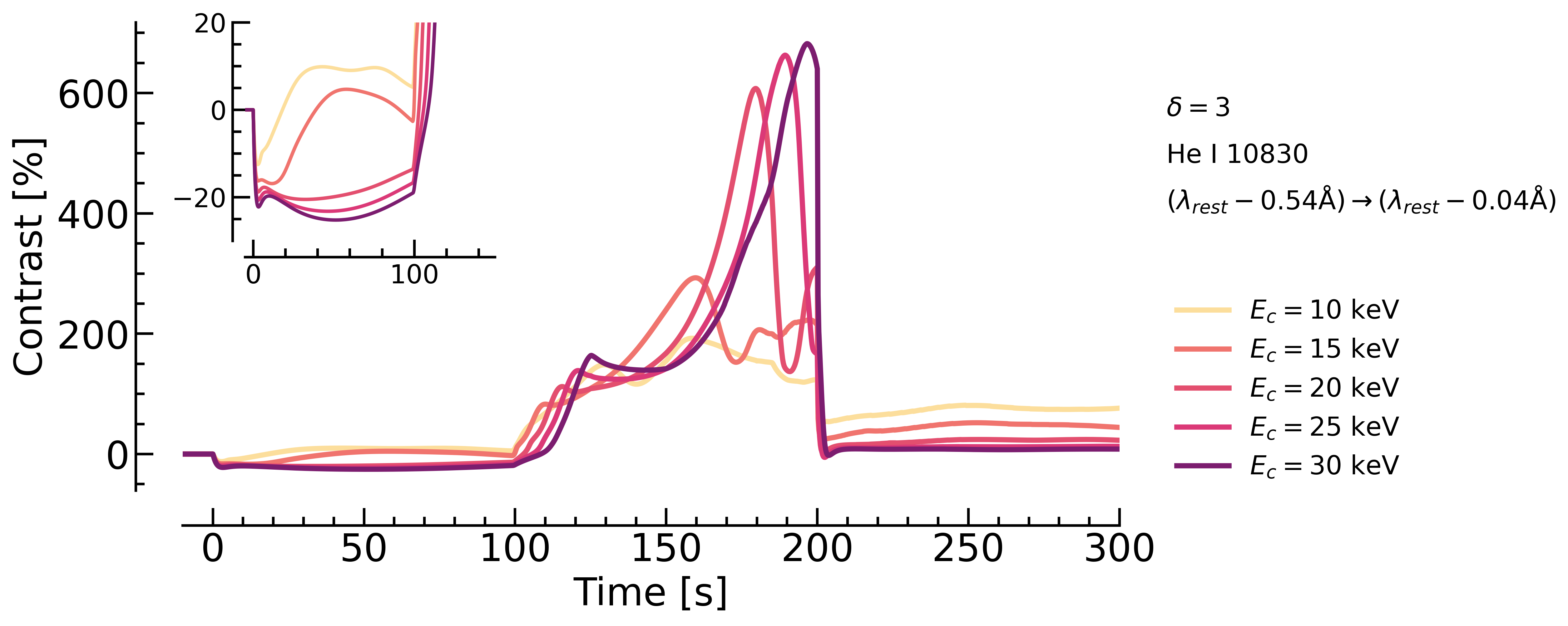}}
	}
	}
	\vbox{
	\hbox{
	\subfloat{\includegraphics[width = 0.5\textwidth, clip = true, trim = 0.cm 0.cm 0.cm 0.cm]{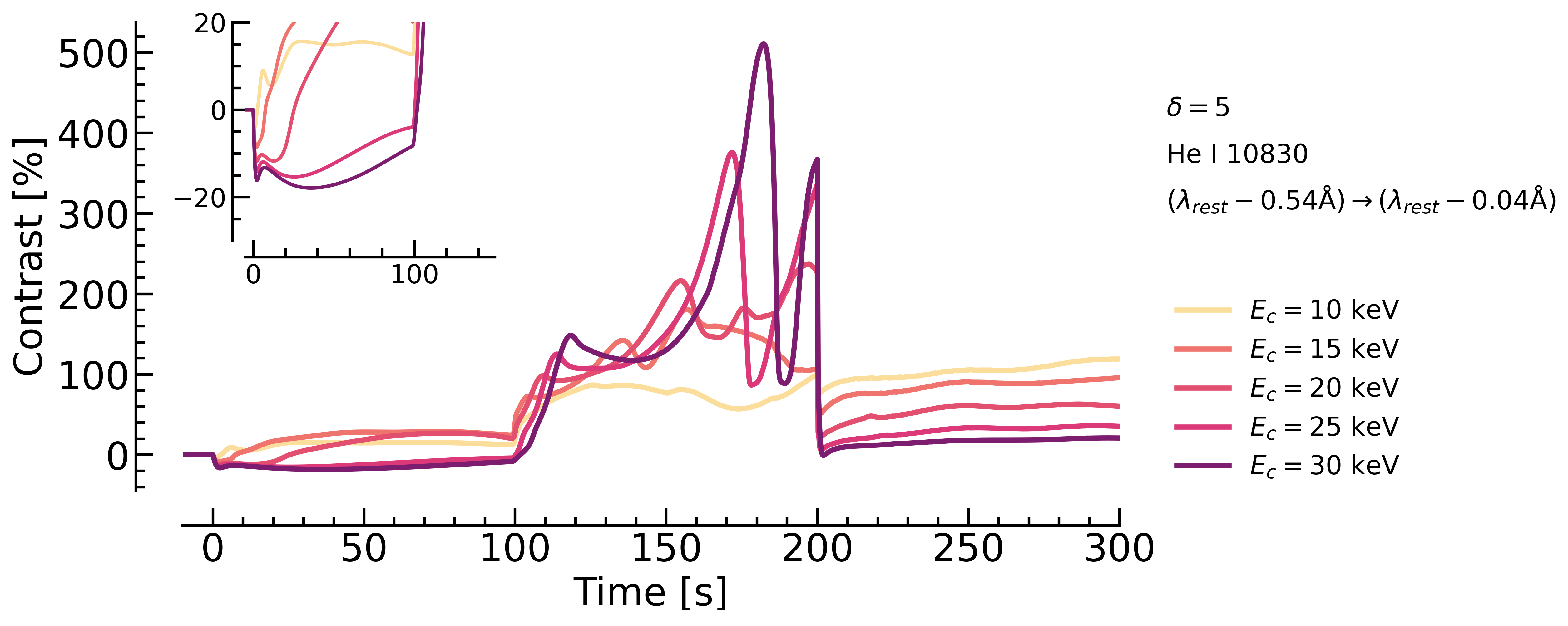}}
	}
	}
	\vbox{
	\hbox{
	\subfloat{\includegraphics[width = 0.5\textwidth, clip = true, trim = 0.cm 0.cm 0.cm 0.cm]{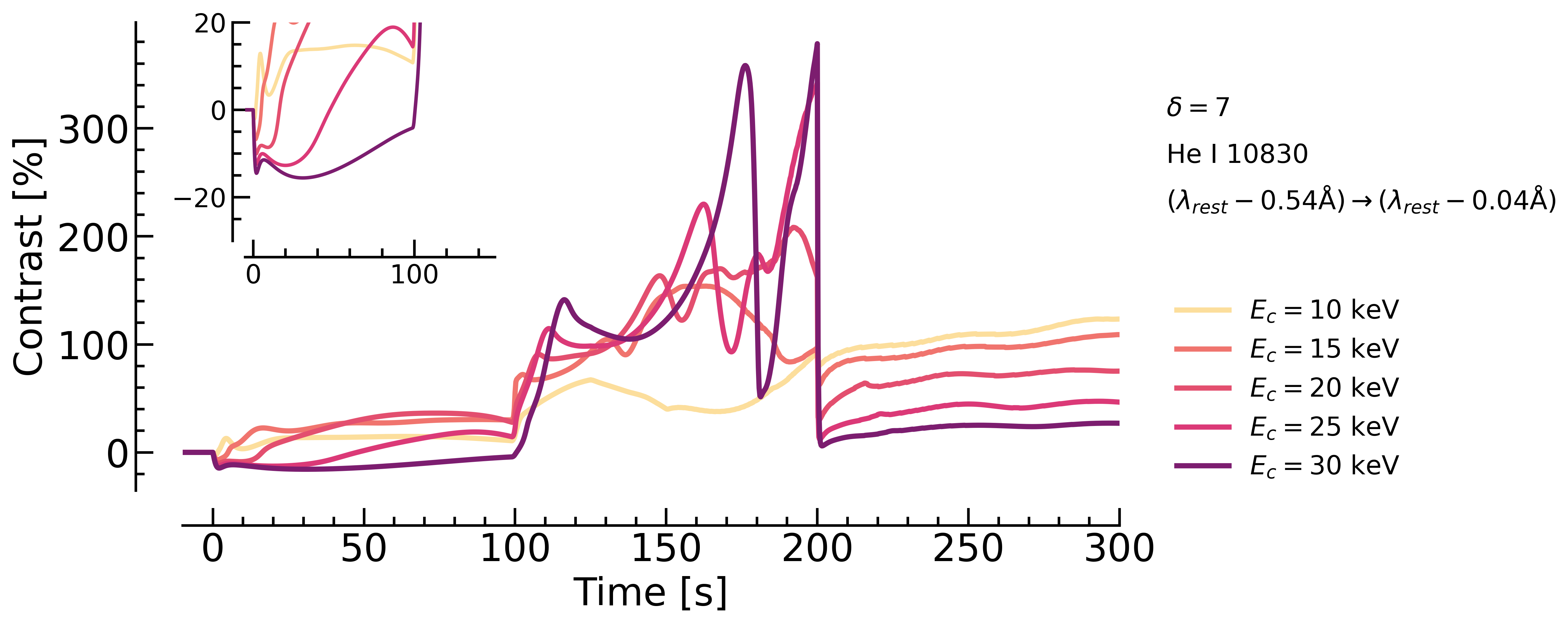}}
	}
	}
	\caption{\textsl{Lightcurves of \ion{He}{1} 10830~\AA, integrated between $[\lambda_{0}-0.04~\AA] \rightarrow [\lambda_{0}-0.54~\AA]$ from the simulations with $t_{\mathrm{dur}} = 100$~s. Each panel shows the effect of varying the low-energy cutoff $E_{c}$ (shown as coloured lines; darker is larger $E_{c}$) with $\delta = 3$ (top row), $\delta = 5$ (middle row), $\delta = 7$ (bottom row). Insets show zoomed in portions during the weak-heating phase. Harder nonthermal electron distributions maintain the lifetime of enhanced absorption throughout the $t_{\mathrm{dur}} = 100$~s weak-heating phase. Softer distributions results in a quicker transition to an emission profile.}}
	\label{fig:lcurves_bbso_grid}
\end{figure}

	All of our flare simulations exhibited the expected behaviour of enhanced \ion{He}{1} 10830~\AA\ absorption (dimming), followed by a subsequent brightening. However, unlike the experiments of \cite{2020ApJ...897L...6H} or \cite{2021ApJ...912..153K}, many of our flares exhibited \ion{He}{1} 10830~\AA\ dimming that persisted for dozens of seconds, up to $t\sim100$~s, with harder nonthermal electron distributions producing enhanced \ion{He}{1} 10830~\AA\ absorption for longer periods. That is, \textsl{the dimming persisted for the entirety of the weak-heating phase} in several of our flare simulations, and for durations similar to the observations of \cite{2016ApJ...819...89X}.
\begin{figure}[ht]
	\centering 
	\vbox{
	\hbox{
	\vspace{-.175in}
	\subfloat{\includegraphics[width = 0.3\textwidth, clip = true, trim = 0.cm 0.cm 0.cm 0.cm]{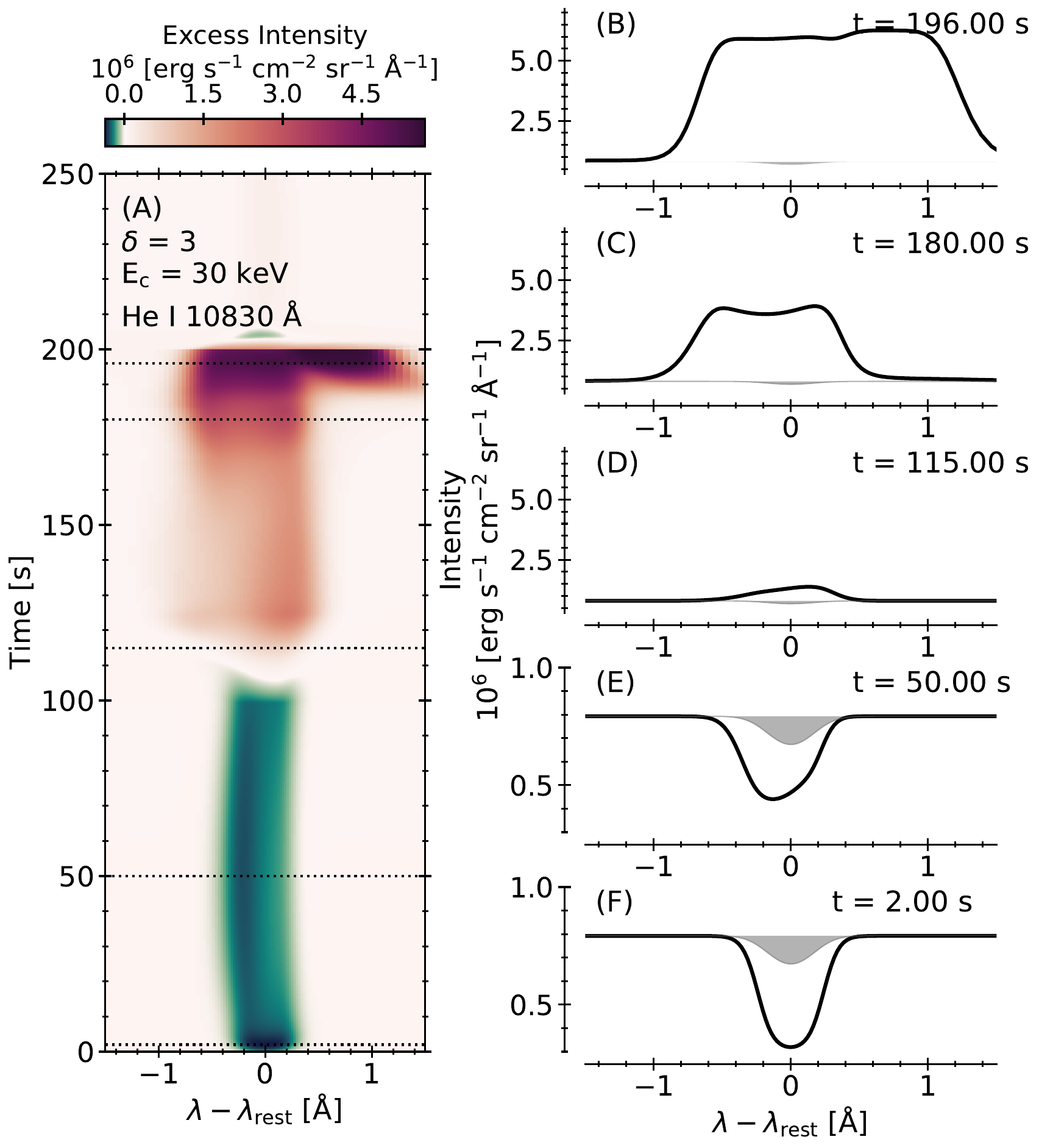}}
	}
	}
	\vbox{
	\hbox{
	\vspace{-.175in}
	\subfloat{\includegraphics[width = 0.3\textwidth, clip = true, trim = 0.cm 0.cm 0.cm 0.cm]{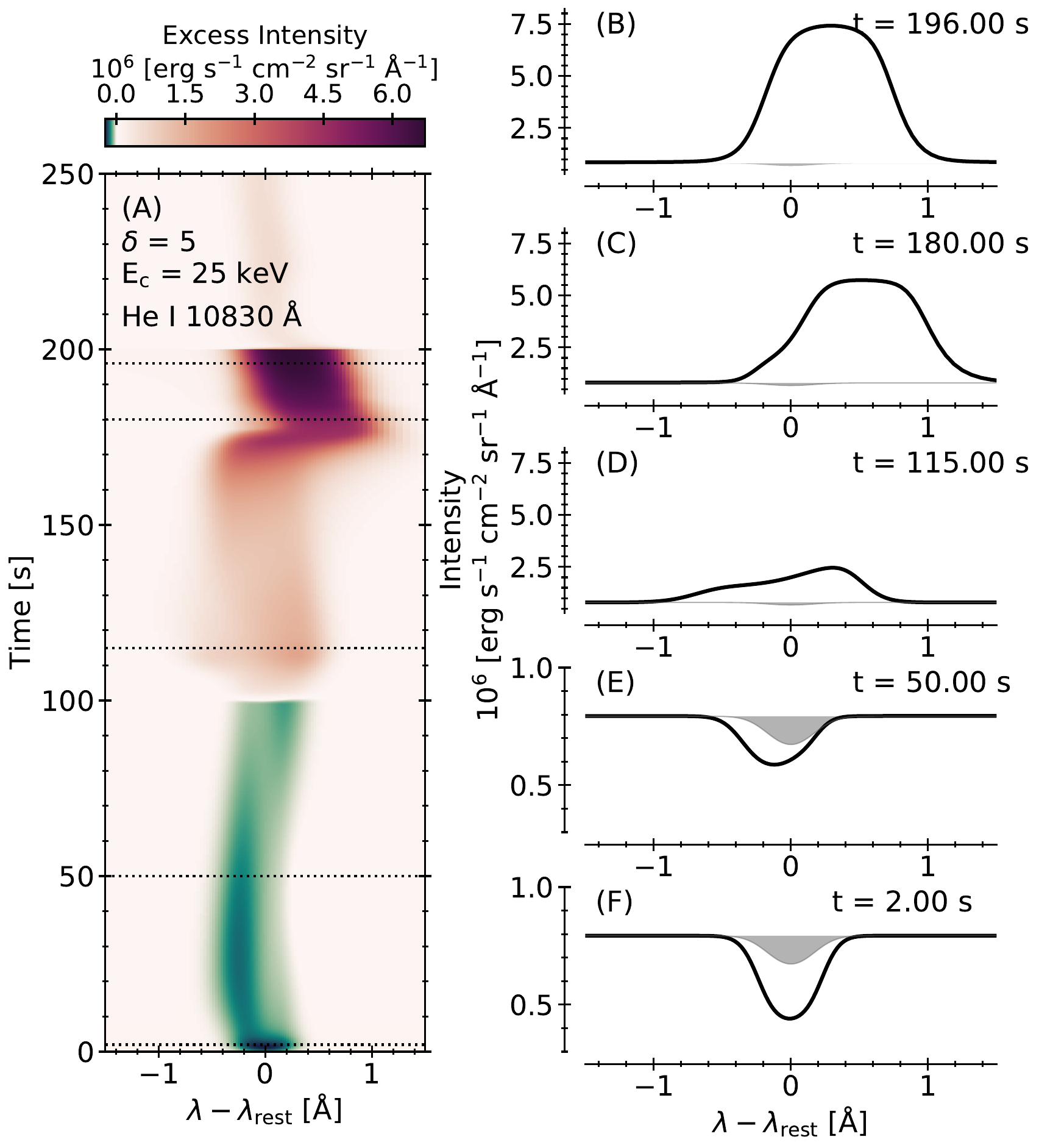}}
	}
	}
	\vbox{
	\hbox{
	\vspace{-.1in}
	\subfloat{\includegraphics[width = 0.3\textwidth, clip = true, trim = 0.cm 0.cm 0.cm 0.cm]{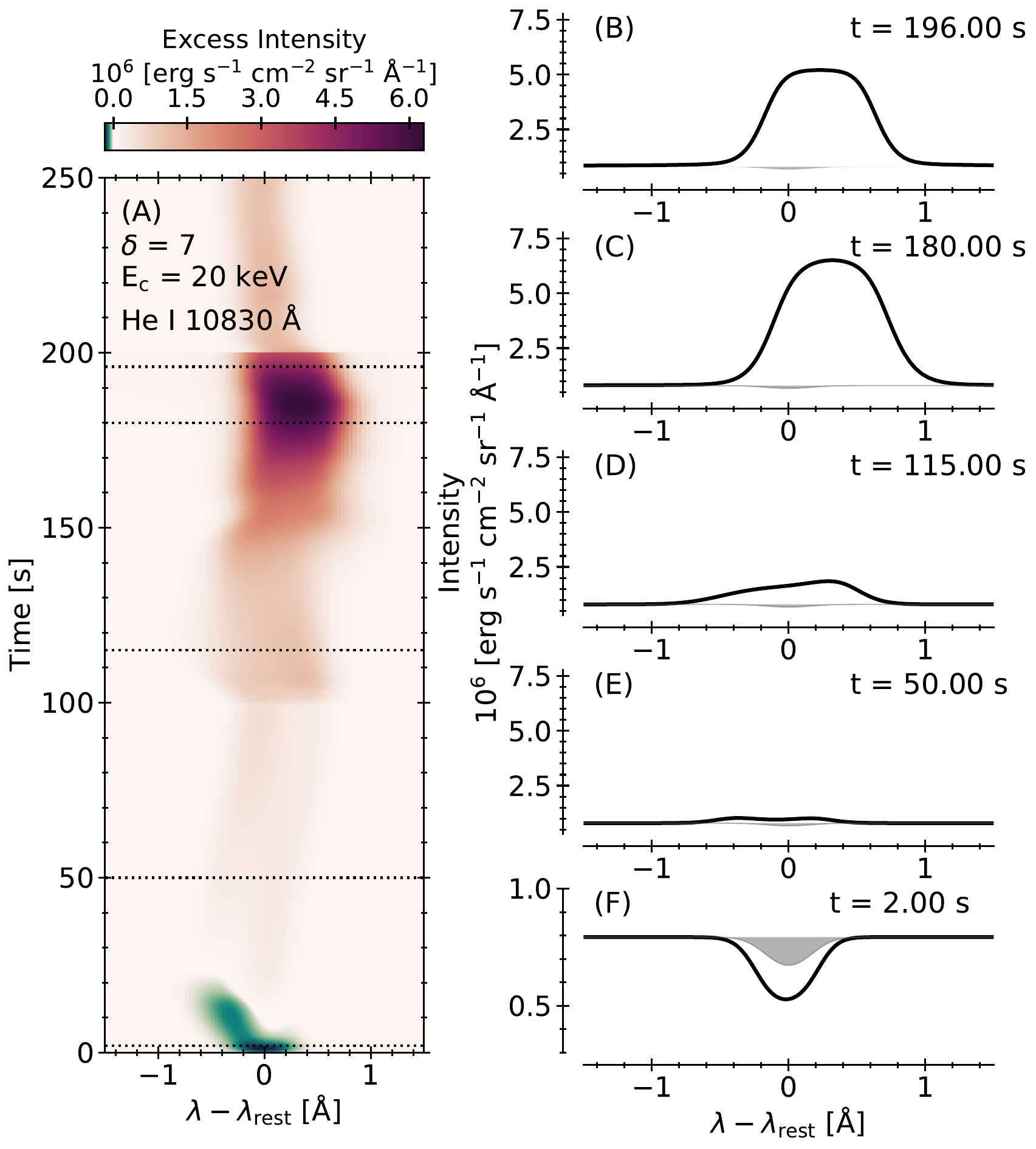}}
	}
	}
	\caption{\textsl{Examples of pre-flare subtracted \ion{He}{1} 10830~\AA\ line profiles as a function of time (images), with cutouts at particular times (which are not pre-flare subtracted), with $t=0$ profiles shaded grey. The top row is $\delta = 3$ $E_{c} = 30$~keV, the middle row is $\delta = 5$ $E_{c} = 25$~keV, and the bottom row is $\delta = 7$ $E_{c} = 20$~keV. Redder colours are positive (i.e. enhancements) and greener colours are are negative (i.e. dimmings), and note that the colorbar scales on each side of the zero line are not uniform.}}
	\label{fig:he10830_profiles_cutouts}
\end{figure}

	Lightcurves of the \ion{He}{1} 10830~\AA\ line, integrated in the range $[\lambda_{0}-0.04~\AA] \rightarrow [\lambda_{0}-0.54~\AA]$, mimicking the wavelength range of the passband in the \cite{2016ApJ...819...89X} BBSO observations, are shown in Figure~\ref{fig:lcurves_bbso_grid} for each of the 15 simulations with $t_{\mathrm{dur}} = 100$~s. The contrast of the wavelength-integrated flaring intensity ($I_{\mathrm{flare}}$) to the pre-flare intensity ($I_{{0}}$) was defined as: ($I_{\mathrm{flare}} - I_{{0}})/I_{{0}} \times 100$. It is immediately clear that the period of dimming is strongly dependent on the properties of the injected nonthermal electron beam, with harder nonthermal electron distributions resulting in longer duration dimming. Softer nonthermal distributions, though generally lasting longer than the few seconds of dimming in \cite{2020ApJ...897L...6H} and \cite{2021ApJ...912..153K}, transition to emission more rapidly. 

	The slow ramp-up to more typical flare energy fluxes results in more structure than the lightcurves shown in \cite{2021ApJ...912..153K}, some of which are reminiscent of the \cite{2016ApJ...819...89X} with a local minima (though the lines are still in emission) and multiple peaks over time. This seems to occur for flares with larger $E_{c}$ in our parameter grid, and is due to the presence of downward directed mass flows (chromospheric condensations) which Doppler shift the spectra outside of the selected window, towards redder wavelengths.

	Despite the fact that Doppler shifts result in structure in the lightcurves, the periods of enhanced absorption do occur over the full spectral line profiles, and do persist for the timescales indicated in Figure~\ref{fig:lcurves_bbso_grid}. Even though small Doppler motions are present during the dimming period, they are modest and insufficient to create an artificial dimming feature. This is illustrated in Figure~\ref{fig:he10830_profiles_cutouts}, which shows the \ion{He}{1} 10830~\AA\ profiles as a function of time from three simulations, with cutouts at a few snapshots. In that figure it is also clear that there can be a transitory time, where parts of the line profile are undergoing enhanced absorption, parts are similar to the pre-flare, and parts are enhanced relative to the pre-flare. Appendix~\ref{sec:helium_profiles_maingrid} shows the wavelength-time panels for each simulation with $t_{\mathrm{dur}} = 100$~s. 

\begin{figure}
	\centering 
	\vbox{
	\hbox{
	\subfloat{\includegraphics[width = 0.5\textwidth, clip = true, trim = 0.cm 0.cm 0.cm 0.cm]{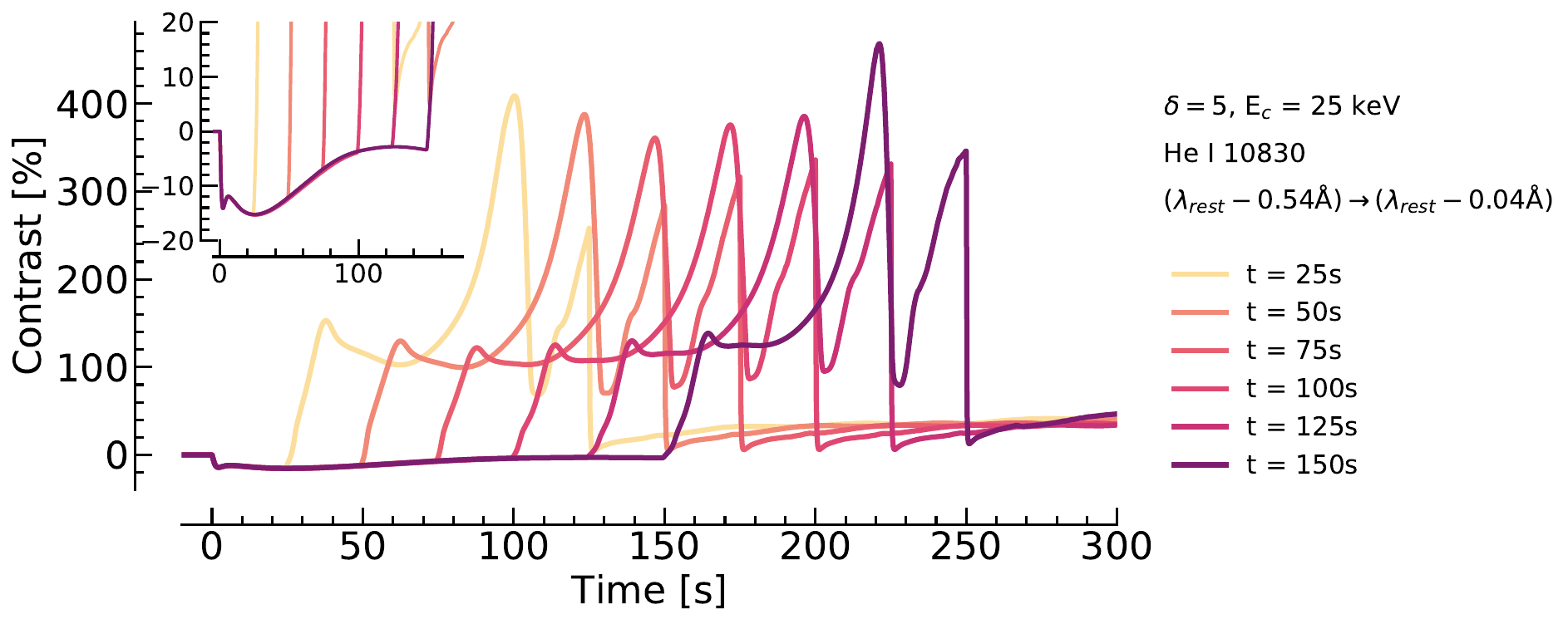}}	
	}
	}
	\caption{\textsl{Lightcurves of \ion{He}{1} 10830~\AA, integrated between $[\lambda_{0}-0.04~\AA] \rightarrow [\lambda_{0}-0.54~\AA]$ \citep[the wavelength range of the BBSO observations studied by][]{2016ApJ...819...89X}, from simulations with fixed $\delta = 5$ and $E_{c} = 25$~keV, but variable durations of weak-heating, $t_{\mathrm{dur}}$. The evolution of the injected energy flux follows the post-weak-heating phase (the grey shaded region of Figure~\ref{fig:fluxschematic}) such that each flare had the same 100s of stronger heating.  The inset shows a zoomed-in view of the weak-heating phase, where it is clear that shortening or lengthening the duration of the weak-heating phase commensurately shortens or lengthens the time taken for the line to go into emission.}}
	\label{fig:lcurves_bbso_varytdur}
\end{figure}

	The duration of the weak-heating phase has a direct impact on the lifetime of the \ion{He}{1} 10830~\AA\ ribbon front-like behaviour. Shortening $t_{\mathrm{dur}}$ reduces the time taken for the line to show a positive contrast, and lengthening $t_{\mathrm{dur}}$ increases the time taken for the line to transition from absorption to emission. In fact, for the harder nonthermal electron distributions, the lifetime of the dimming equals the duration of the weak-heating phase, as shown in the \ion{He}{1} 10830~\AA\ lightcurves in Figure~\ref{fig:lcurves_bbso_varytdur} where we vary $t_{\mathrm{dur}} = [25, 50, 75, 100, 125, 150]$~s. This implies that, in a rough sense, the length of the observed dimming period informs us about the duration of weak precipitation of nonthermal particles. There is of course some level of ambiguity as we saw for some intermediate cases (e.g. $\delta = 5$; $E_{c} = 20$~keV) that \ion{He}{1} 10830~\AA\ is driven into emission prior to the start of the strong-heating phase. 

\subsection{Flaring Plasma Conditions During Dimming Periods versus Emission Periods}\label{sec:he_plasmaprops}

\begin{figure}
	\centering 
	\vbox{
	\hbox{
	\subfloat{\includegraphics[width = 0.5\textwidth, clip = true, trim = 0.cm 0.cm 0.cm 0.cm]{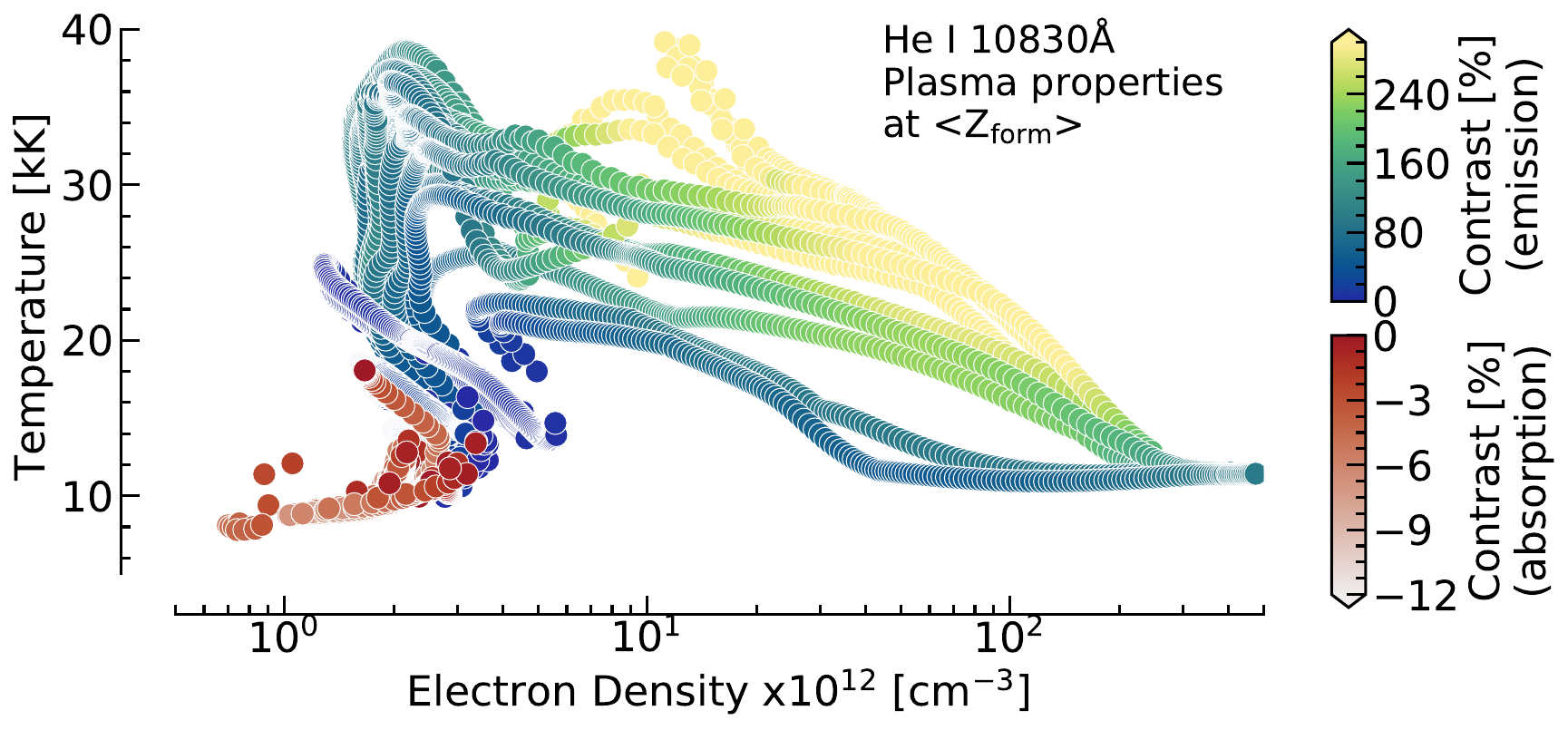}}	
	}
	}
	\vbox{
	\hbox{
	\subfloat{\includegraphics[width = 0.5\textwidth, clip = true, trim = 0.cm 0.cm 0.cm 0.cm]{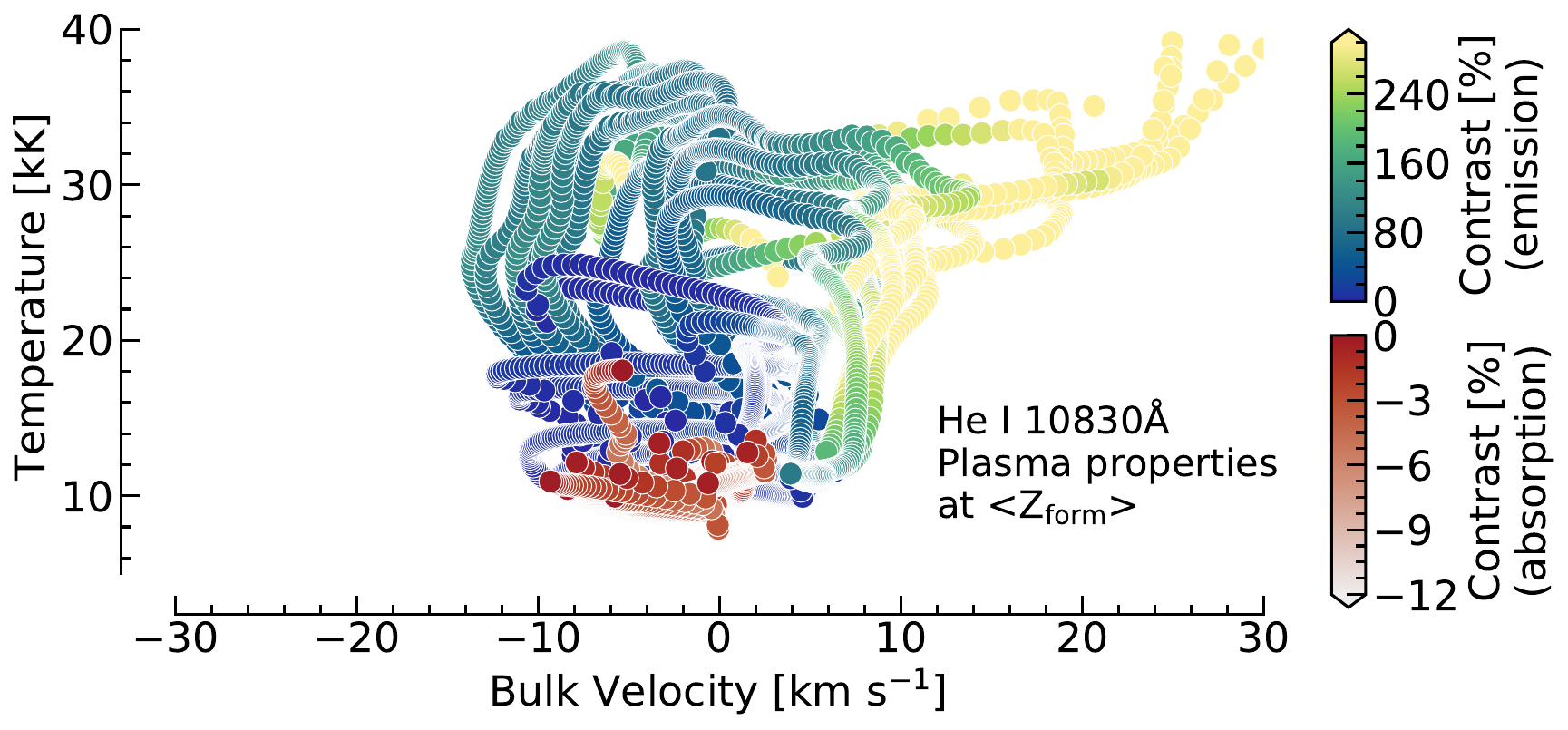}}	
	}
	}
	\vbox{
	\hbox{
	\subfloat{\includegraphics[width = 0.5\textwidth, clip = true, trim = 0.cm 0.cm 0.cm 0.cm]{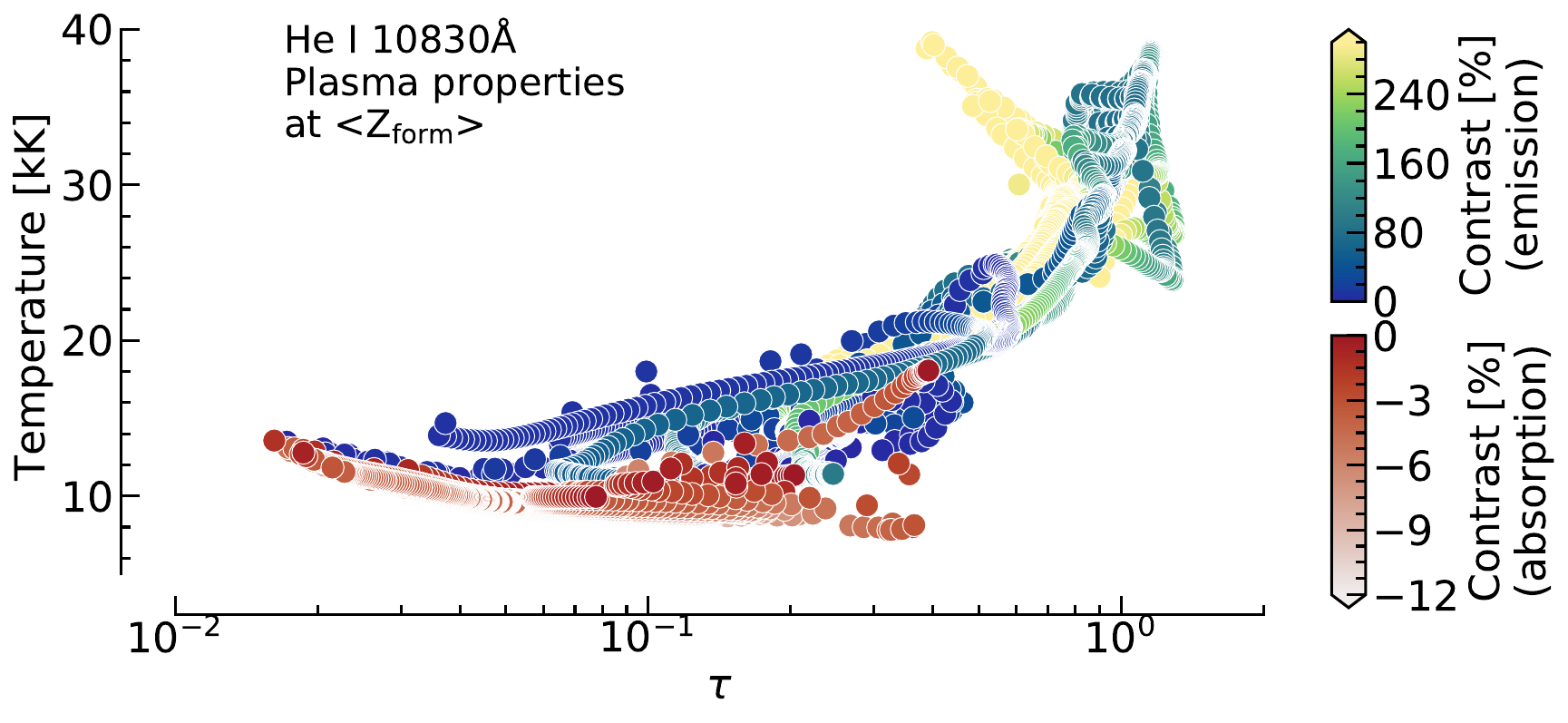}}	
	}
	}
	\caption{\textsl{The average plasma properties in the formation region of the \ion{He}{1} 10830~\AA\ line core, defined as the wavelength with largest optical depth. Colours represent the contrast of the line ($[\lambda_{0}-1.5~\AA] \rightarrow [\lambda_{0}+1.5~\AA]$), \gsk{at a particular instant}, where reds are enhanced absorption, and blues-yellows are emission. \gsk{Each circle represents an individual snapshot from a single simulation, and every simulation in our parameter survey is included.} The top panel shows temperature versus electron density. The middle panel shows temperature versus bulk velocity (upflows are negative). The bottom panel shows temperature versus optical depth $\tau$. Although there is some overlap in the low-positive contrast range, it is clear that a hot ($T>15-20$~kK) and/or dense ($n_{e} > 3.5\times10^{12}$~cm$^{-3}$) atmosphere drives the line into emission.}}
	\label{fig:plasmaprops}
\end{figure}

	Here we explore the formation properties of the line and the plasma conditions in the \ion{He}{1} 10830~\AA\ line forming region of the atmosphere, building upon some prior efforts in \cite{2020ApJ...897L...6H}, to understand when the line is driven into emission. That is, can we state confidently what the differences in plasma conditions are at the ribbon front versus the main bright ribbons? 
 
 	The contribution function to the emergent intensity, $C_{I}(\lambda,\mu,z)$, can be defined as: 
 
 \begin{equation}
 C_{I}(\lambda,\mu,z) = \frac{1}{\mu}~S_{\lambda}(z)~e^{\tau_{\lambda}(z)/\mu}~\chi_{\lambda}(z), 
 \end{equation}
 
 \noindent where $z$ is the height in the atmosphere, $\mu = \cos \theta$ ($\theta$ is the viewing angle between the line-of-sight and the normal), $S_{\lambda}$ is the source function, $\tau_{\lambda}$ is the optical depth, and $\chi_{\lambda}$ is the opacity. Integrating through some depth scale yields the emergent intensity. This allows us to understand, effectively, where in the atmosphere the intensity originates. 
 
	\cite{2020ApJ...897L...6H} averaged the plasma properties over a broad range of heights where $C_{I}$ was elevated. Here, we follow a similar, but more robust, approach where we carefully identify the range of heights that most significantly contribute to \ion{He}{1} 10830~\AA\ emission and perform a weighted average over that range \citep[following the approaches of][]{2017ApJ...836...12K,2019ApJ...879...19Z,2023ApJ...944..186M,2024MNRAS.527.2523K}. The normalised cumulative distribution function of $C_{I}$ was constructed, and the heights at which the bulk of the emission is formed was identified. For our purposes we defined this as being the heights bounding $10\% - 95\%$ of the line's contribution (similar results were obtained for $5\% - 95\%$, and $20\% - 95\%$). Then, the average plasma properties between those (time-varying) heights was measured, weighted by $C_{I}$, which can be sharply peaked. 

	Since $C_{I}$ is a function of wavelength we selected the line core as the reference point with which to asses plasma conditions, defined as the wavelength at which $z(\tau_{\lambda} = 1)$ was maximised. That is, the wavelength at which the line's optical depth was largest (this does not necessarily mean the line was optically thick throughout the flare). Doppler motions can shift the wavelength of the line core when they are co-spatial with regions of high orthohelium density. 

	Temperature, electron density, bulk velocity, and the optical depth in the line core emitting region were all measured as a function of time, and compared to the contrast of the wavelength-integrated intensity of the full line ($\lambda_{0} \pm 1.5$~\AA), shown in Figure~\ref{fig:plasmaprops}. Each panel shows how one of the properties varies with temperature, with colour representing the contrast of the line (the magnitude of the negative contrasts is generally smaller than when using the BBSO wavelength-integration range). \gsk{This figure is the aggregate of our whole simulation grid, with each symbol in effect representing a single snapshot from whichever simulation it happens to originate from. Here the intention is not to show a time evolution, which we glean from other figures and analysis (from which we also know that a negative contrast represents the notional ribbon front phase of each simulation). Rather, the intention is to understand what general atmospheric conditions lead to either a negative or positive contrast, by combining the full simulation dataset.  }

	During periods of enhanced absorption the atmosphere is generally cooler and less dense than periods of strong emission, with some overlap between smaller negative and smaller positive contrasts. The, rough, threshold between positive and negative \ion{He}{1} 10830~\AA\ contrast is $n_{e}\sim3.5\times10^{12}$~cm$^{-3}$ and $T\sim13.5$~kK \citep[which is essentially the same temperature as estimated by][but a larger electron density]{2020ApJ...897L...6H}. Of note is that we find that the \ion{He}{1} 10830~\AA\ line can be strongly in emission even if $T<13.5$~kK if the upper chromospheric electron density is sufficiently elevated, which in our simulations means $n_{e}>3\times10^{13}$~cm$^{-3}$. At $T>18$~kK we find only emission. 

	Opacity in the line core varies over time in the simulations, and over the particular evolution of each flare atmosphere, but generally it increased during the flare such that the optical depth approached or exceeded unity, meaning that optical depth effects play a role in line formation. This is true both during the periods of dimming or enhancement but $\tau > 0.5$ only when the line is strongly enhanced, which is intuitive given the larger densities when that is the case.

	Finally, we note that periods of dimming are generally associated with a gentle upflow of the line forming region, with $v_{z} \sim [-10, 0]$~km~s$^{-1}$. Downflows are present in some of the simulations, up to $v_{z} = 30$~km~s$^{-1}$.

	So, the ribbon front-like \ion{He}{1} 10830~\AA\ behaviour is present when the line forming region in the mid-upper chromosphere is still relatively cool, and is undergoing gentle evaporation, while the brighter flare ribbons are associated with a much warmer and denser upper chromosphere.

	\cite{2021ApJ...912..153K} noted the increased population of \ion{He}{1} 2s $^{3}$S through the chromosphere due to nonthermal collisional ionisation of He, that increases the opacity and produces the dimming. There is also an enhancement of \ion{He}{1} 2p $^{P}$ level, but during the dimming period the ratio of the upper-to-lower level in the upper chromosphere decreases since \ion{He}{1} 2s $^{3}$S is preferentially populated. At the time that the line switches to being enhanced this ratio reverses, and the line is driven into emission. This seemingly happens at $T>13.5$~kK and $n_{e}>3.5\times10^{12}$~cm$^{-3}$ (or a cooler plasma but with $n_{e}>> 3.5\times10^{12}$~cm$^{-3}$), and is at least partially due to increased collisional ionisation from the lower-to-upper level.


\section{Forward Modelling Flaring \ion{Mg}{2} NUV}\label{sec:mgiiflares}

\subsection{General \ion{Mg}{2} Behaviour}

\begin{figure}
	\centering 
	\vbox{
	\hbox{
	\subfloat{\includegraphics[width = 0.5\textwidth, clip = true, trim = 0.cm 0.cm 0.cm 0.cm]{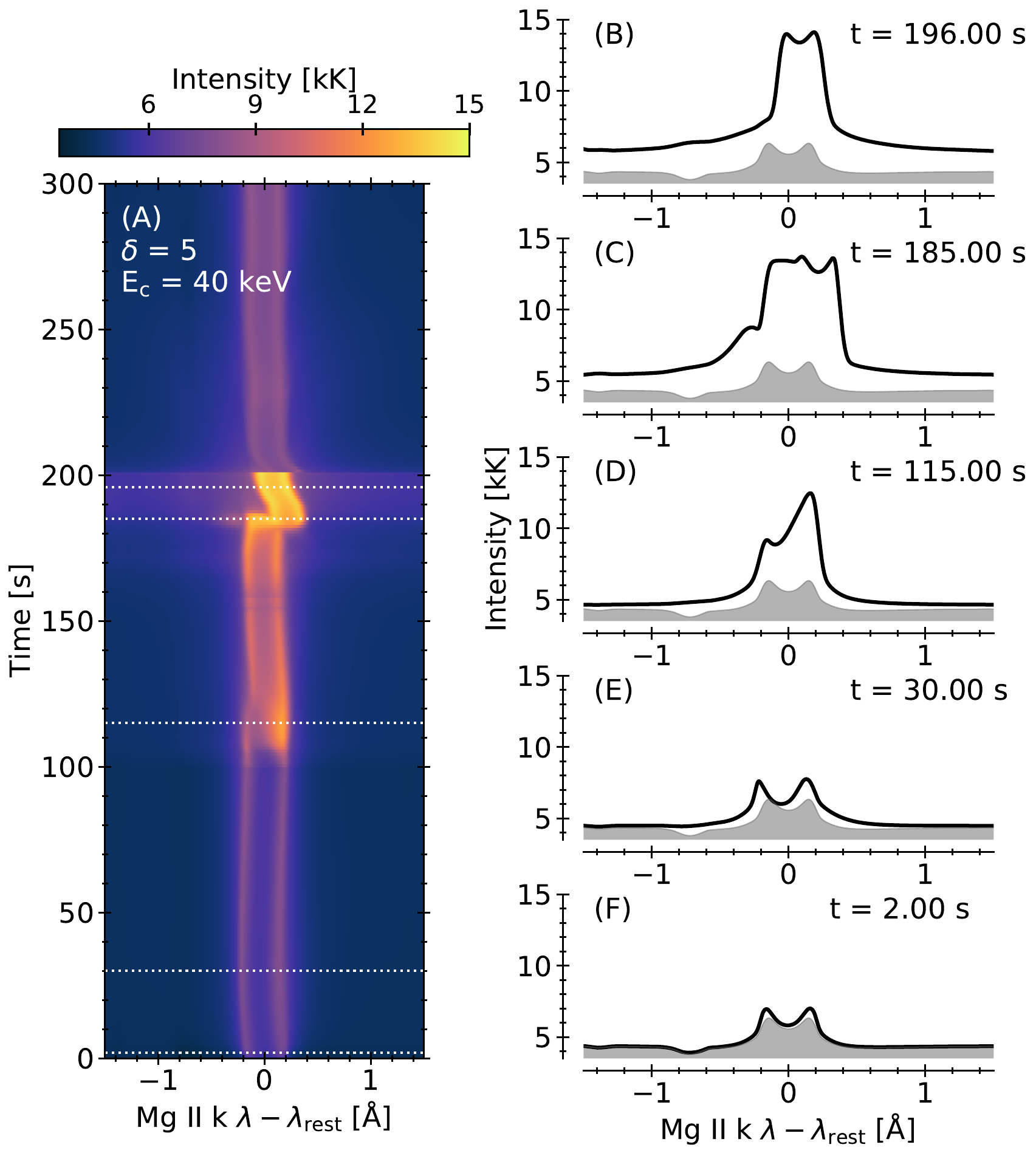}}
	}
	}	
	\caption{\textsl{Example \ion{Mg}{2} k line spectra from the $\delta = 5$, $E_{c} = 30$~keV simulation. The image shows the evolution of the spectra during the simulation, with cutouts at certain times, indicated by the horizontal white lines.}}
	\label{fig:mgii_spectra_ex}
\end{figure}

	In general the \ion{Mg}{2} spectra follow a common evolution through the weak-heating phase and the ramp-up to peak energy injection. Quantitative differences, and the rate of their evolution, in the various spectral line characteristics such as the intensity, depth of central reversals, and other properties (discussed in detail in Section~\ref{sec:linemetrics}) arise with varying the injected nonthermal electron beam parameters. Figure~\ref{fig:mgii_spectra_ex} shows the evolution of \ion{Mg}{2} line in the $\delta = 5$, $E_{c} = 30$~keV simulation. Appendix~\ref{sec:helium_profiles_maingrid} shows the temporal evolution for the other experiments, along with the \ion{Mg}{2} 2798~\AA\ subordinate blend

	Upon initial energy injection the \ion{Mg}{2} k line increases in intensity, and the depth of its central reversal increases.  The line core is slightly blueshifted, and the line becomes asymmetric with more intense red peak (k2r) relative to blue peak (k2v). As the energy flux injected increases the \ion{Mg}{2} k line intensity similarly increases, and the central reversal becomes shallower. At times the reversal almost fully fills in, particularly close to flare peak, and at other times it effectively merges with the k2v component such that the line appears with a broad blue shoulder. In response to chromospheric condensations the lines become redshifted or exhibit redshifted components alongside mostly stationary components. All the subordinate lines behave similarly to the \ion{Mg}{2} k line, though have a tendency to transition to being single-peaked once the weak-heating phase has ended. For some simulations (those with softer nonthermal electron distributions) the subordinate triplet appears single peaked even during the weak-heating phase. 

Our quantitative model-data comparison of \ion{Mg}{2} ribbon front behaviour follows the framework of \cite{2023ApJ...944..104P}, where we extract the wavelength positions of the k2r, k2v, and k3 components of the \ion{Mg}{2} k line, and use those to build quantitative metrics (described in more detail below), which can be compared to the results of the four observed solar flares  studied by \cite{2023ApJ...944..104P}. \gsk{The impacts of IRIS instrumental effects on the analysis that follows was investigated for a range of representative exposure times and cadences. While the main conclusions and trends did not change, the reduced spectral resolution and smearing in exposure time did result in some minor quantitative differences in comparison to the native \rhpar\ spectra. See Appendix~\ref{sec:irisquality} for an overview of the impact of IRIS instrumental effects on the synthetic spectra.} 
	
\subsection{\ion{Mg}{2} k Line Characteristics}\label{sec:linemetrics}
	
	For each \ion{Mg}{2} profile we extract several metrics that characterise their response to the flare during the various stages of flare energy injection. These metrics are the same as used by \cite{2023ApJ...944..104P}. They are the depth of the central reversal, $D_\mathrm{CR}$, the asymmetry of the k2 peaks, $A_\mathrm{k2}$ (where a positive value means more intense k2r versus k2v peak), the separation of the k2 peaks, $S_\mathrm{k2}$, and the Doppler shift of the k3 component (i.e. the Doppler shift of the line core), $v_{\mathrm{Dopp,k3}}$:

\begin{equation}\label{eq:dep}
D_\mathrm{CR} = - \frac{I_{\mathrm{k2v}} - I_{\mathrm{k3}}}{I_{\mathrm{k2v}} + I_{\mathrm{k3}}},
\end{equation}

\begin{equation}\label{eq:diff}
A_\mathrm{k2} = \frac{I_{\mathrm{k2r}} - I_{\mathrm{k2v}}}{I_{\mathrm{k2r}} + I_{\mathrm{k2v}}},
\end{equation}

\begin{equation}\label{eq:sep}
S_\mathrm{k2} = v_{\mathrm{k2r}} - v_{\mathrm{k2v}}.
\end{equation}

\noindent In those expressions $I$ variables are the intensities of the various components, and $v$ variables are the positions of the various component in velocity-space. 

\begin{figure*}
	\centering 
	\vbox{
	\hbox{
	\hspace{0.in}
	\subfloat{\includegraphics[width = 0.5\textwidth, clip = true, trim = 0.cm 0.cm 0.cm 0.cm]{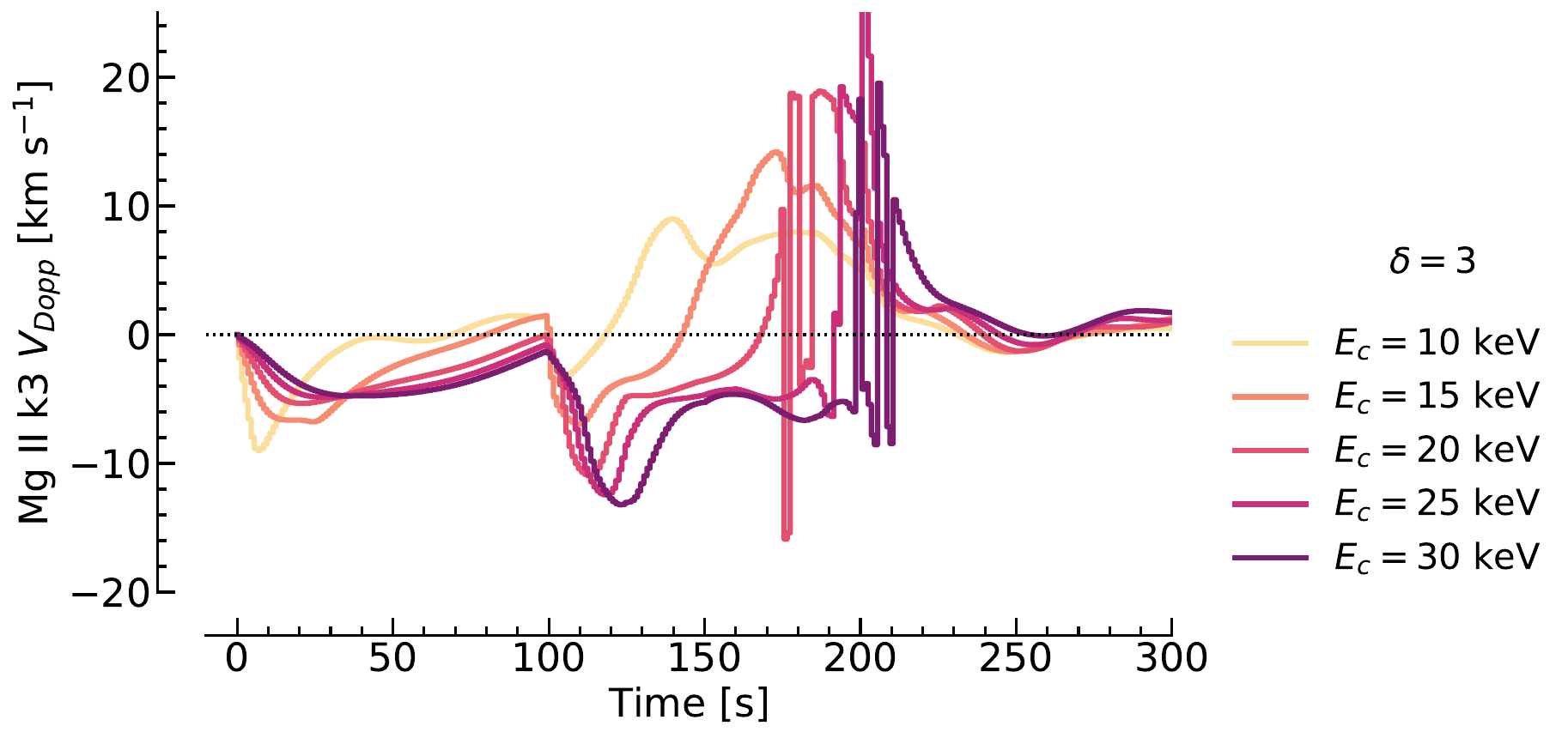}}
			\subfloat{\includegraphics[width = 0.5\textwidth, clip = true, trim = 0.cm 0.cm 0.cm 0.cm]{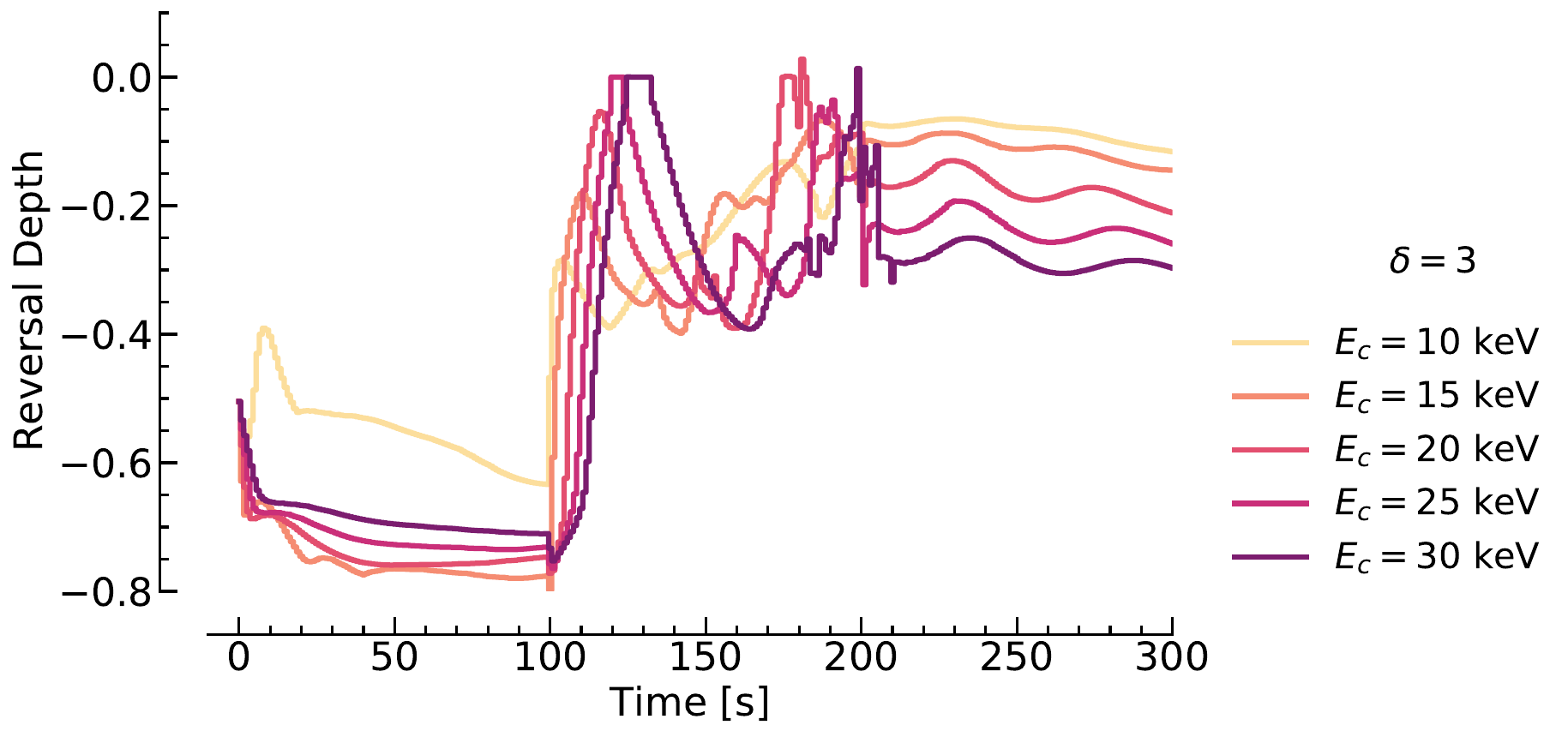}}
	}
	}
	\vbox{
	\hbox{
	\subfloat{\includegraphics[width = 0.5\textwidth, clip = true, trim = 0.cm 0.cm 0.cm 0.cm]{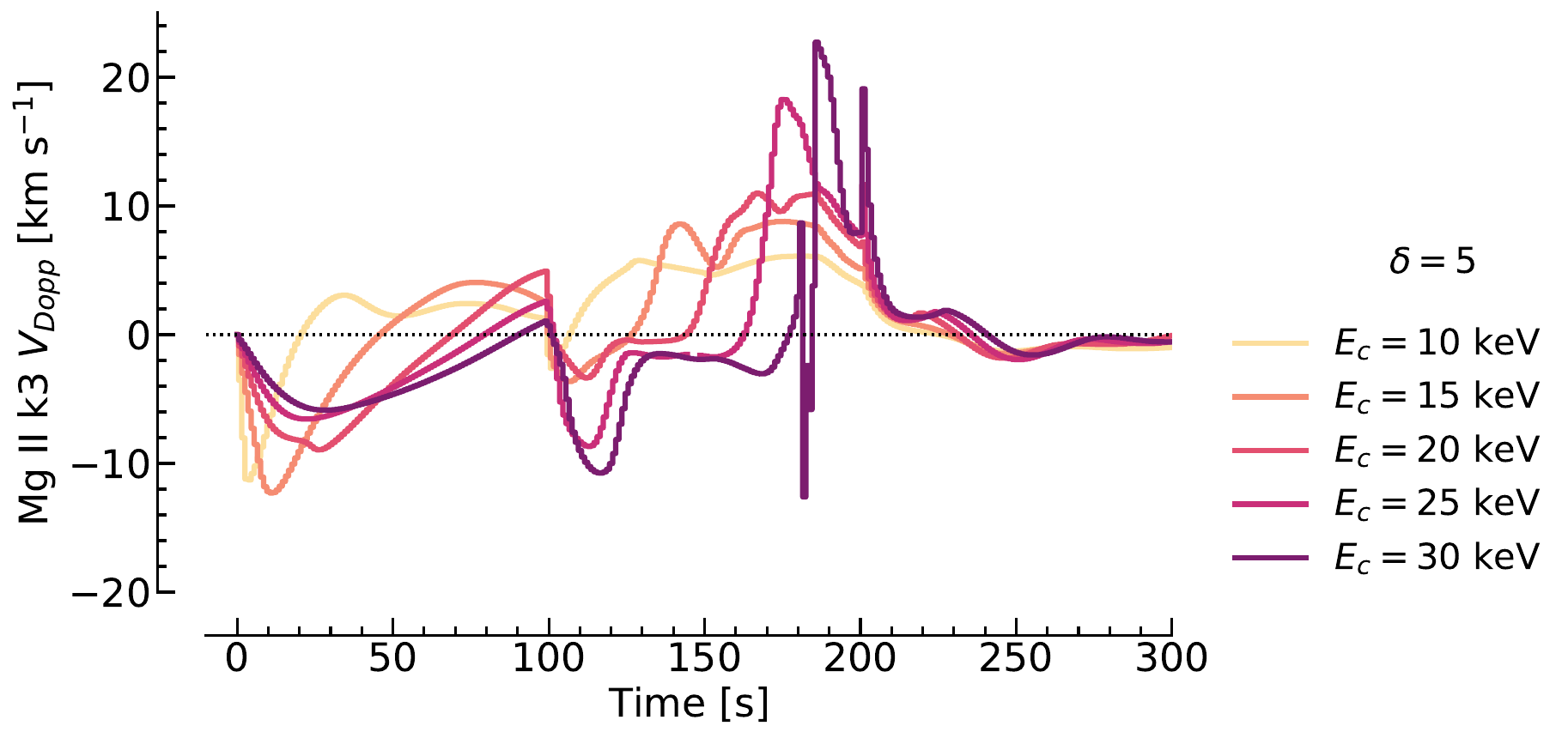}}	
		\subfloat{\includegraphics[width = 0.5\textwidth, clip = true, trim = 0.cm 0.cm 0.cm 0.cm]{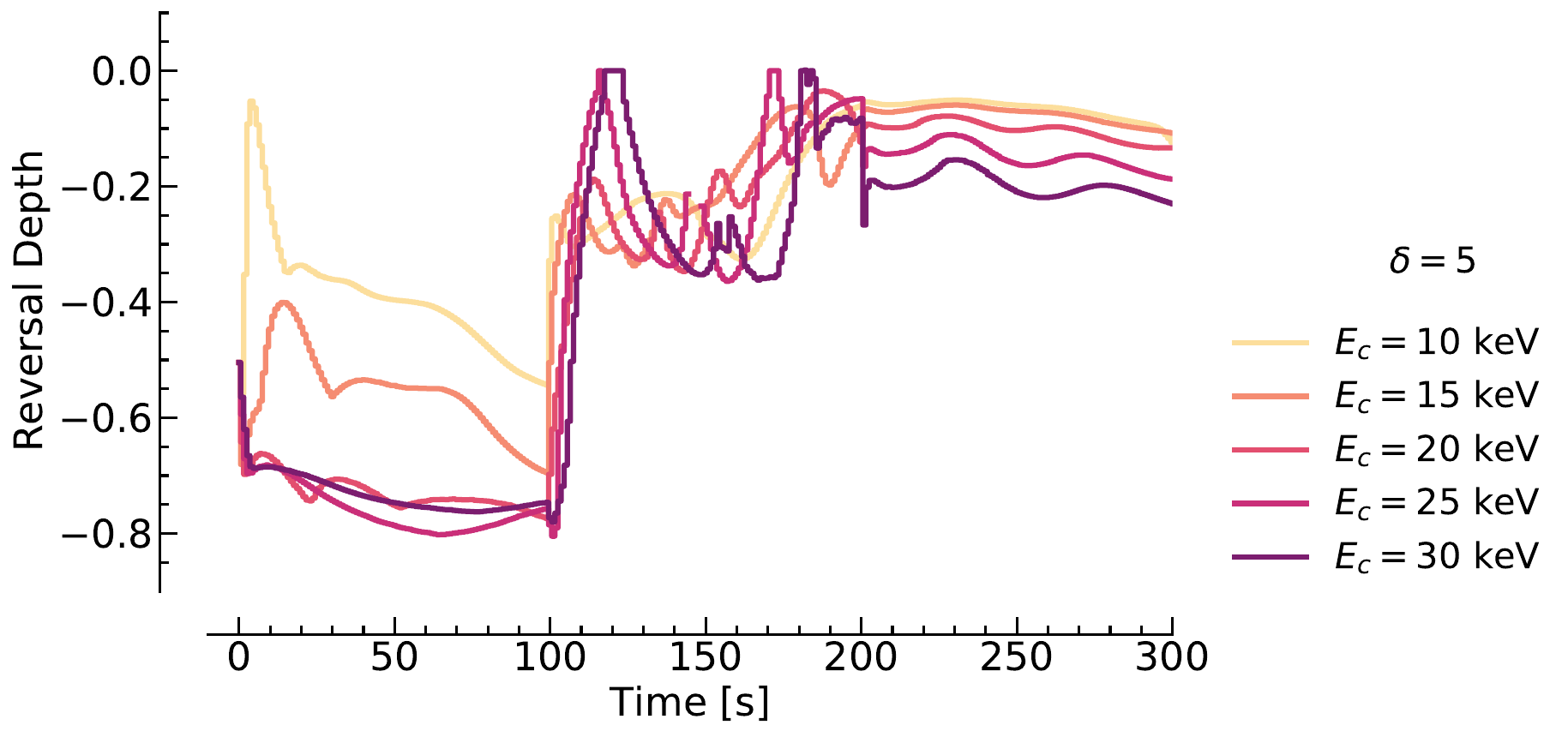}}
	}
	}
	\vbox{
	\hbox{
	\subfloat{\includegraphics[width = 0.5\textwidth, clip = true, trim = 0.cm 0.cm 0.cm 0.cm]{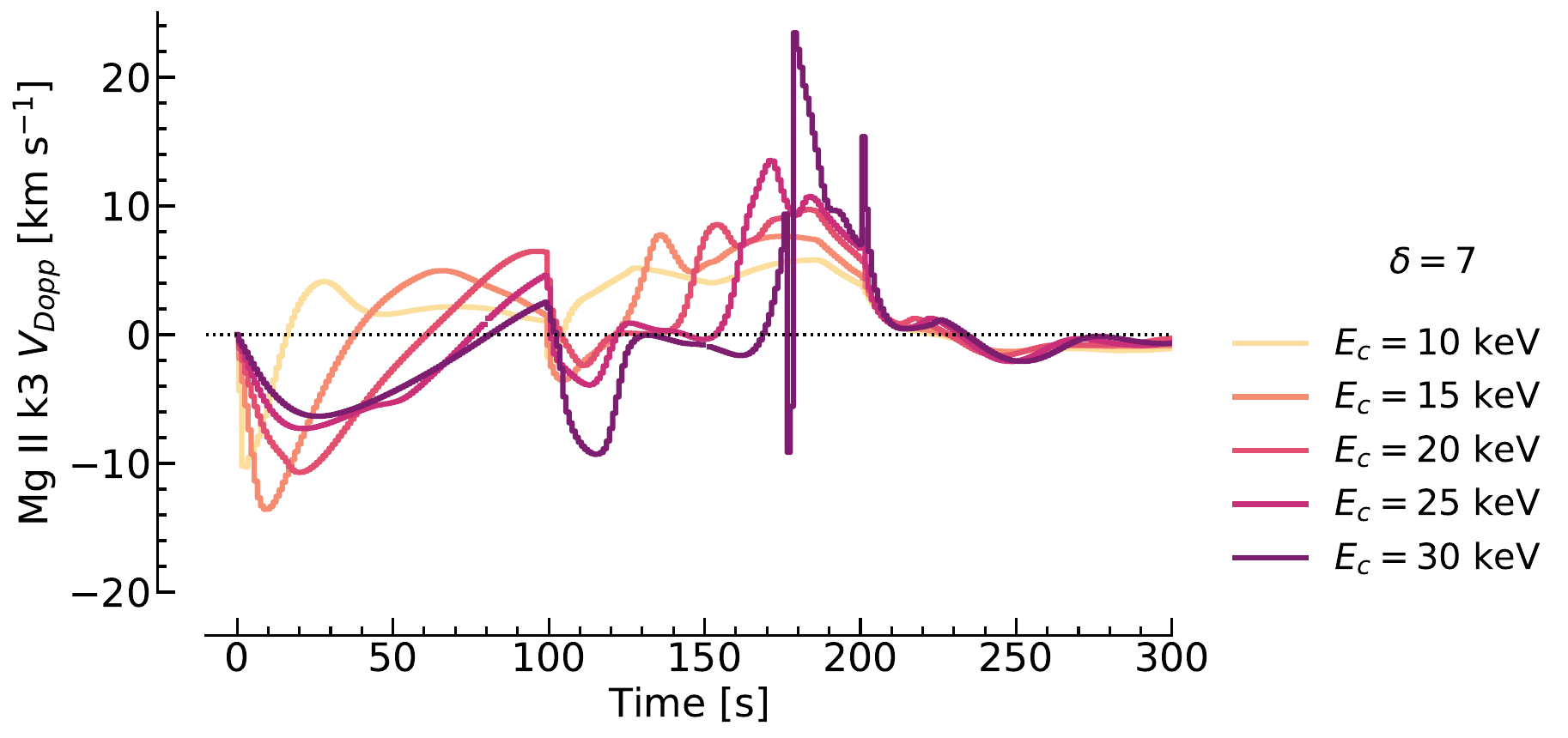}}
		\subfloat{\includegraphics[width = 0.5\textwidth, clip = true, trim = 0.cm 0.cm 0.cm 0.cm]{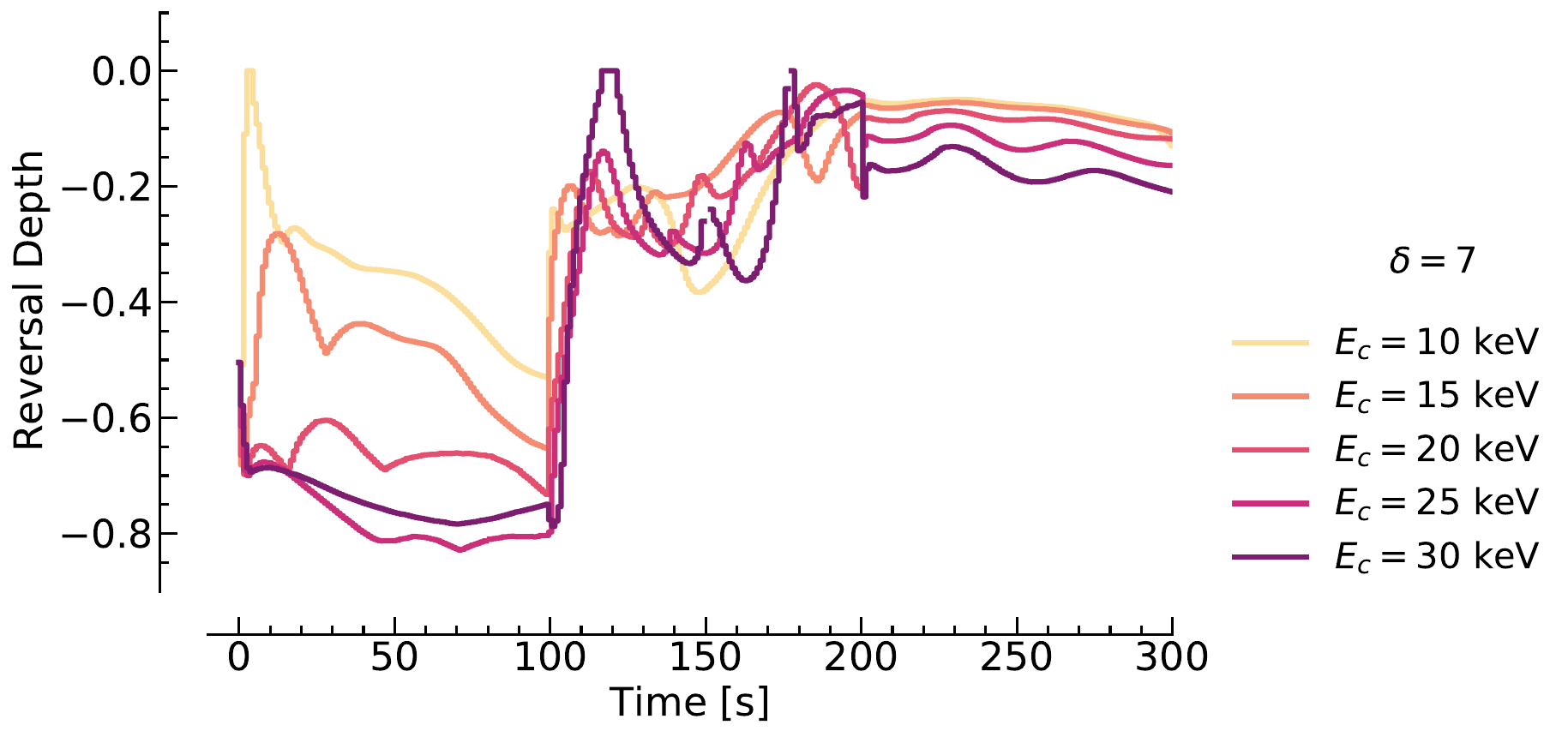}}
	}
	}
	\caption{\textsl{The evolution of Doppler shift of \ion{Mg}{2} k3 (first column), and the depth of the \ion{Mg}{2} k central reversal (second column), where colour indicates the low-energy cutoff. The top panel shows the the $\delta = 3$ simulations, the middle shows the $\delta = 5$ simulations and the bottom shows the $\delta = 7$ simulations.}}
	\label{fig:mgii_metrics}
\end{figure*}

\begin{figure*}
	\centering 
	\vbox{
	\hbox{
	\hspace{0.in}
	\subfloat{\includegraphics[width = 0.5\textwidth, clip = true, trim = 0.cm 0.cm 0.cm 0.cm]{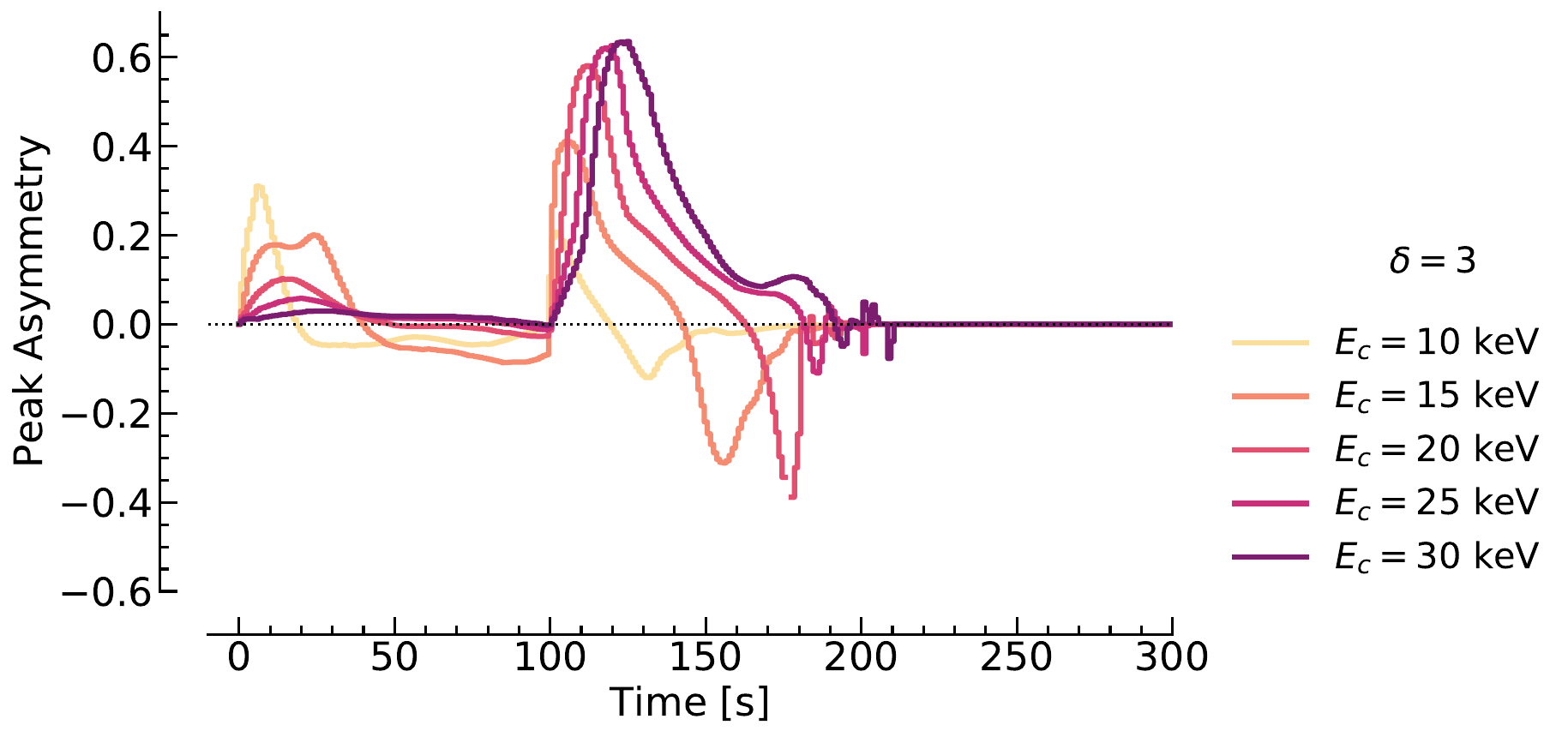}}
			\subfloat{\includegraphics[width = 0.5\textwidth, clip = true, trim = 0.cm 0.cm 0.cm 0.cm]{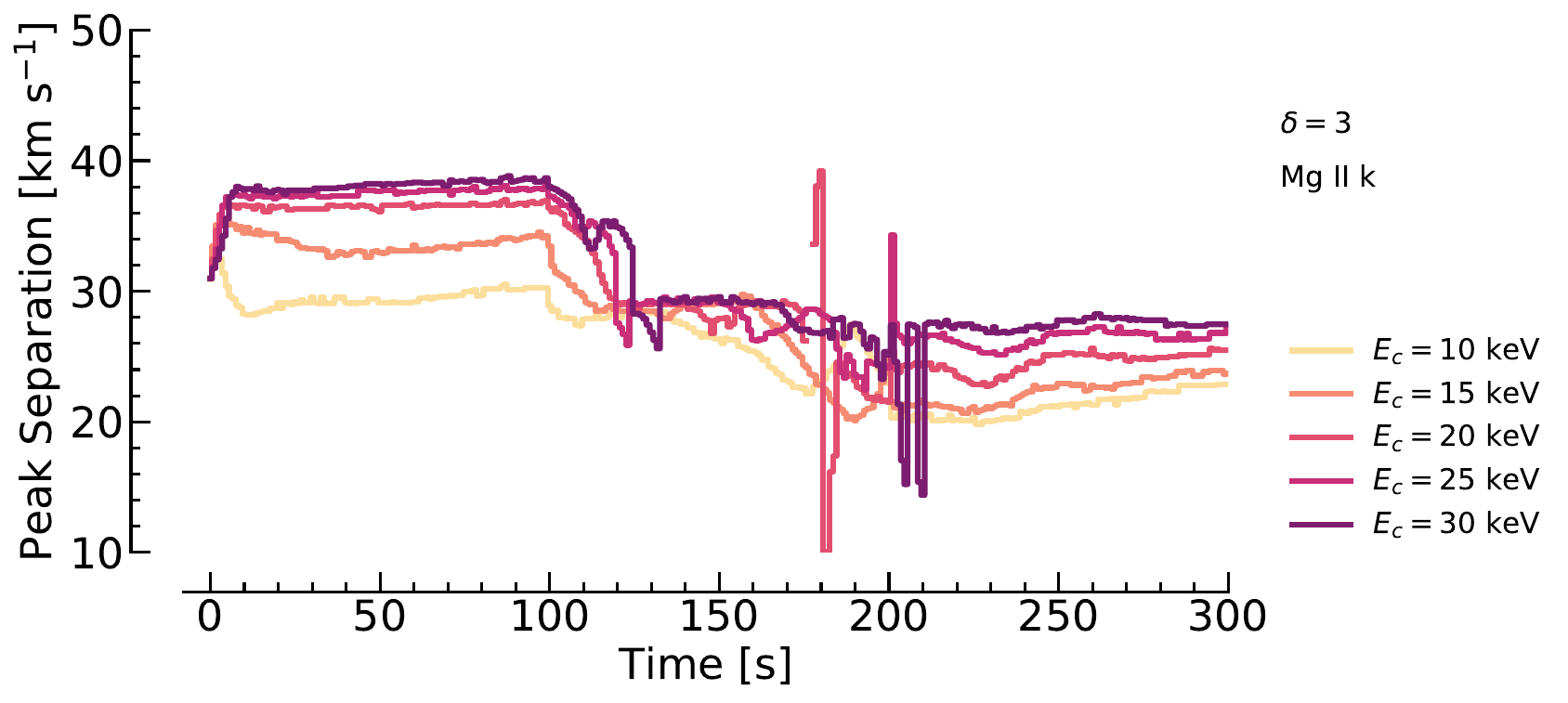}}
	}
	}
	\vbox{
	\hbox{
	\subfloat{\includegraphics[width = 0.5\textwidth, clip = true, trim = 0.cm 0.cm 0.cm 0.cm]{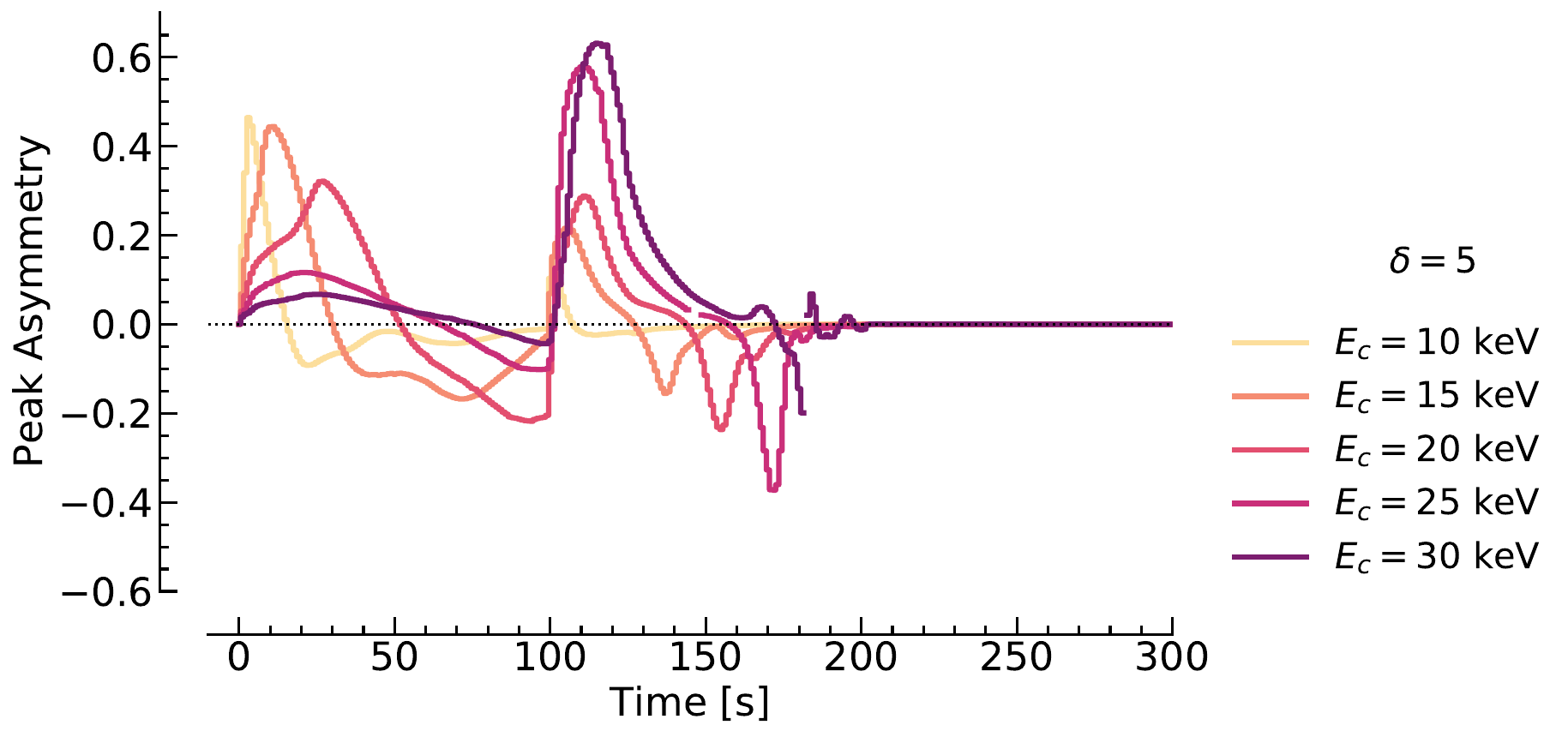}}	
		\subfloat{\includegraphics[width = 0.5\textwidth, clip = true, trim = 0.cm 0.cm 0.cm 0.cm]{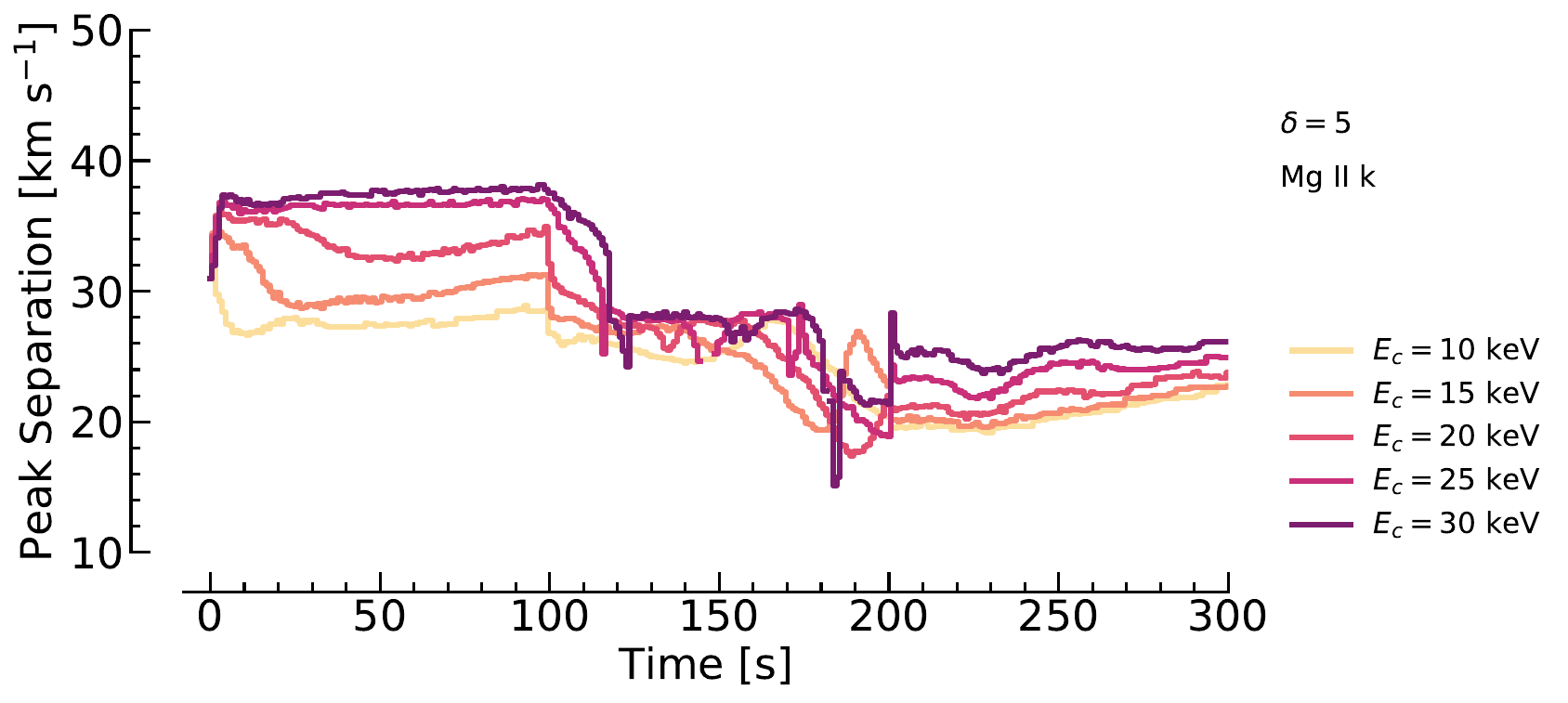}}
	}
	}
	\vbox{
	\hbox{
	\subfloat{\includegraphics[width = 0.5\textwidth, clip = true, trim = 0.cm 0.cm 0.cm 0.cm]{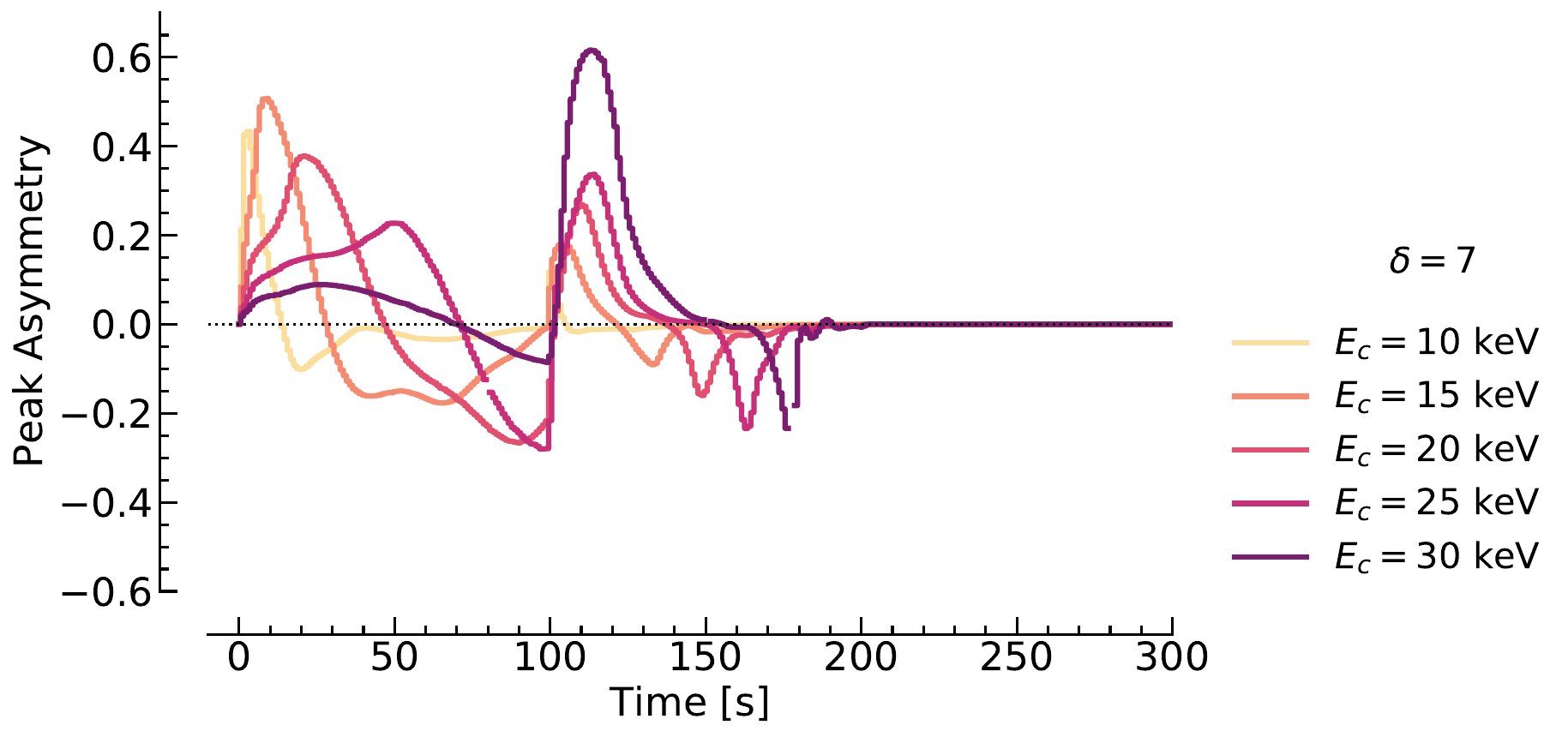}}
		\subfloat{\includegraphics[width = 0.5\textwidth, clip = true, trim = 0.cm 0.cm 0.cm 0.cm]{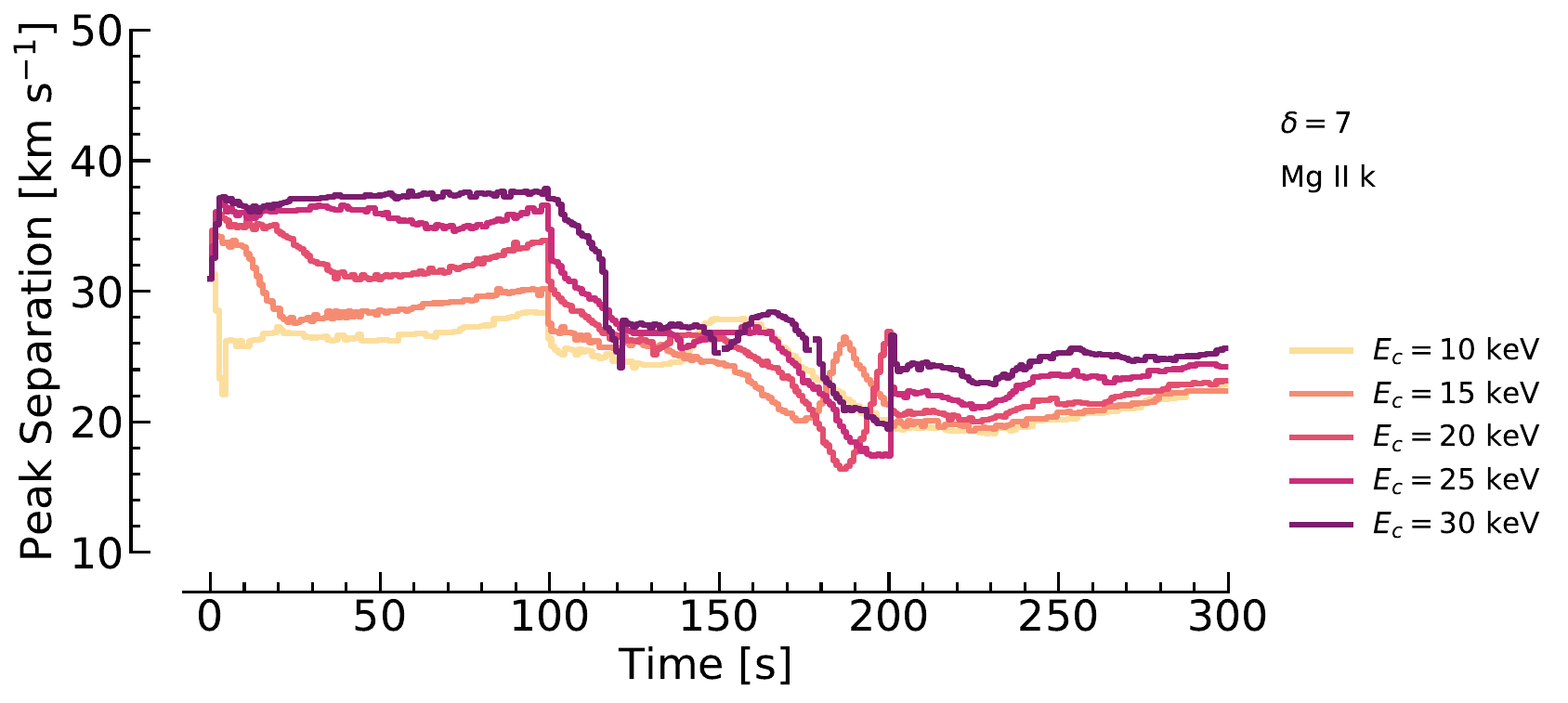}}
	}
	}
	\caption{\textsl{The evolution of asymmetry of the \ion{Mg}{2} k2 peaks (first column) and their separation (second column), where colour indicates the low-energy cutoff. The top panel shows the the $\delta = 3$ simulations, the middle shows the $\delta = 5$ simulations and the bottom shows the $\delta = 7$ simulations.}}
	\label{fig:mgii_metrics2}
\end{figure*}

	We find similar results as with \ion{He}{1} 10830~\AA, which is that the weak-heating phase of simulations with harder nonthermal electron distributions produce \ion{Mg}{2} spectra consistent with ribbon front observations. In fact, the same simulations that typify ribbon fronts in \ion{He}{1} 10830~\AA\ can be characterised as ribbon fronts in \ion{Mg}{2}. 

	Slight blueshifts of the line core (k3) were present, up to $\sim-(10-12)$~km~s$^{-1}$, which transition to redshifts by the peak of the strong-heating phase, albeit in a somewhat noisy manner in some simulations (Figure~\ref{fig:mgii_metrics}, left column). This is due to the presence of multiple components in the line, one slightly blueshifted or stationary and one redshifted. The algorithm to identify k3 can sometimes jump between these components. Observations generally are redshifted in the main ribbon, but this often manifests as an asymmetric red-wing not always as a shifted k3. These Doppler shifts are consistent with the observational analysis of ribbon front \cite{2023ApJ...944..104P} in which the distribution of k3 Doppler shifts peaked at $-10$~km~s$^{-1}$ (though some of those observed profiles could exhibit even larger blueshifts up to $-20$~km~s$^{-1}$). The magnitude of $V_{\mathrm{Dopp}}$ did not seem to depend strongly on $E_{c}$ or $\delta$, though a faster evolution back towards rest was associated with softer nonthermal distributions.

	Red peak (k2r) asymmetries were present during the weak-heating phase, though were only strong for a brief time (Figure~\ref{fig:mgii_metrics2}, left column). The harder nonthermal electron beams only produced mildly asymmetric profiles, with values consistent with the bulk of the observed distribution of $A_{\mathrm{k2}}\sim0-0.2$. The tail of the observed distribution of asymmetries with larger values of $A_\mathrm{k2}$ is seemingly suggestive of softer nonthermal electron beams. Most of the simulations actually began exhibiting blue peak asymmetries during the weak-heating phase. During the ramp-up to peak energy deposition $A_{\mathrm{k2}}$ could be either blue or red dominant.

	The k3 Doppler shift and the peak asymmetry are perhaps not as `clean' as \ion{He}{1} 10830~\AA, since the \ion{Mg}{2} spectra do not persist with ribbon front-like characteristics for the entirety of the weak-heating phase, beginning their transition to what might be considered more consistent with regular bright ribbon characteristics earlier. However, the k2 peak separation and the depth of the central reversal may offer a better diagnostic of the lifetime of any weak-heating phase.

	For the majority of the weak-heating phase the k2 peak separation increases from the pre-flare value, and remains fairly constant in the harder nonthermal electron beam simulations, with values $S_{\mathrm{k2}} \sim35-40$~km~s$^{-1}$ (Figure~\ref{fig:mgii_metrics2}, right column).  Their temporal evolution tracks the lifetime of the \ion{He}{1} 10830~\AA\ ribbon front-like behaviours pretty well, with the softer nonthermal electron beam simulations more rapidly transitioning to a smaller k2 peak separation. The same simulations that persist with enhanced absorption of \ion{He}{1} 10830~\AA\ until the start of the ramp-up of energy injection have a larger k2 peak separation throughout the weak-heating phase. The observations of \cite{2023ApJ...944..104P} suggest a broader range of values of $S_{\mathrm{k2}}$  than present in our simulations.

	The central reversal deepens in every simulation at the onset of energy deposition, but in the case of softer nonthermal electron distributions the reversal switches to being shallower than the pre-flare quite rapidly. A trend of deepening reversal depth is present in the harder nonthermal electron distributions, in which $D_\mathrm{CR}$ only reduces in magnitude once the energy flux injected begins to increase. During the latter half of the simulations the profiles  have very shallow central reversals and occasionally become single peaked. We might expect only single peaked profiles during the bright ribbon phase, but these have proven tricky to produce in RHD simulations, requiring large electron densities \citep[e.g.][]{2017ApJ...842...82R,2019ApJ...879...19Z}. It is possible that a more aggressive ramp-up time, and larger energy flux following the weak-heating phase could lead to those conditions but here we are primarily interested in the ribbon front period.

	Of the range of ribbon front-like behaviours, the increase of intensity accompanied by increased depth of the central reversal best typifies the duration of the weak-heating phase. 

	As noted in the previous section, the profiles as observed with IRIS could appear different due to instrumental effects. We repeated the analysis of \ion{Mg}{2} line characteristics for the profiles converted to IRIS quality. The trends identified do not change but certainly some of the finer scale temporal differences are not seen, and some of the noisier behaviour is reduced since multiple components (redshifted plus stationary) are not present as often. The magnitude of the Doppler shift of k3 is largely unaffected. There are some quantitative differences in the depth of the reversal since the profiles appear artificially shallower at IRIS resolution. Similarly the profiles may appear single peaked more often. The only notable difference in timing is that the $\delta = 5$, $E_{c} = 20$~keV simulation transitions from deeper-to-shallower central reversal sooner. Peak separation only differs by a few km~s$^{-1}$. Red peak asymmetries are accentuated somewhat but the temporal trend is unchanged. Summing over spectral pixels ($2\times2$ binning) does introduce short lived local maxima and minima. 

\subsection{Explaining the \ion{Mg}{2} Characteristics}
	\begin{figure*}
	\centering 
	\vbox{
	\hbox{
	\hspace{1in}
	\subfloat{\includegraphics[width = 0.35\textwidth, clip = true, trim = 0.cm 0.cm 0.cm 0.cm]{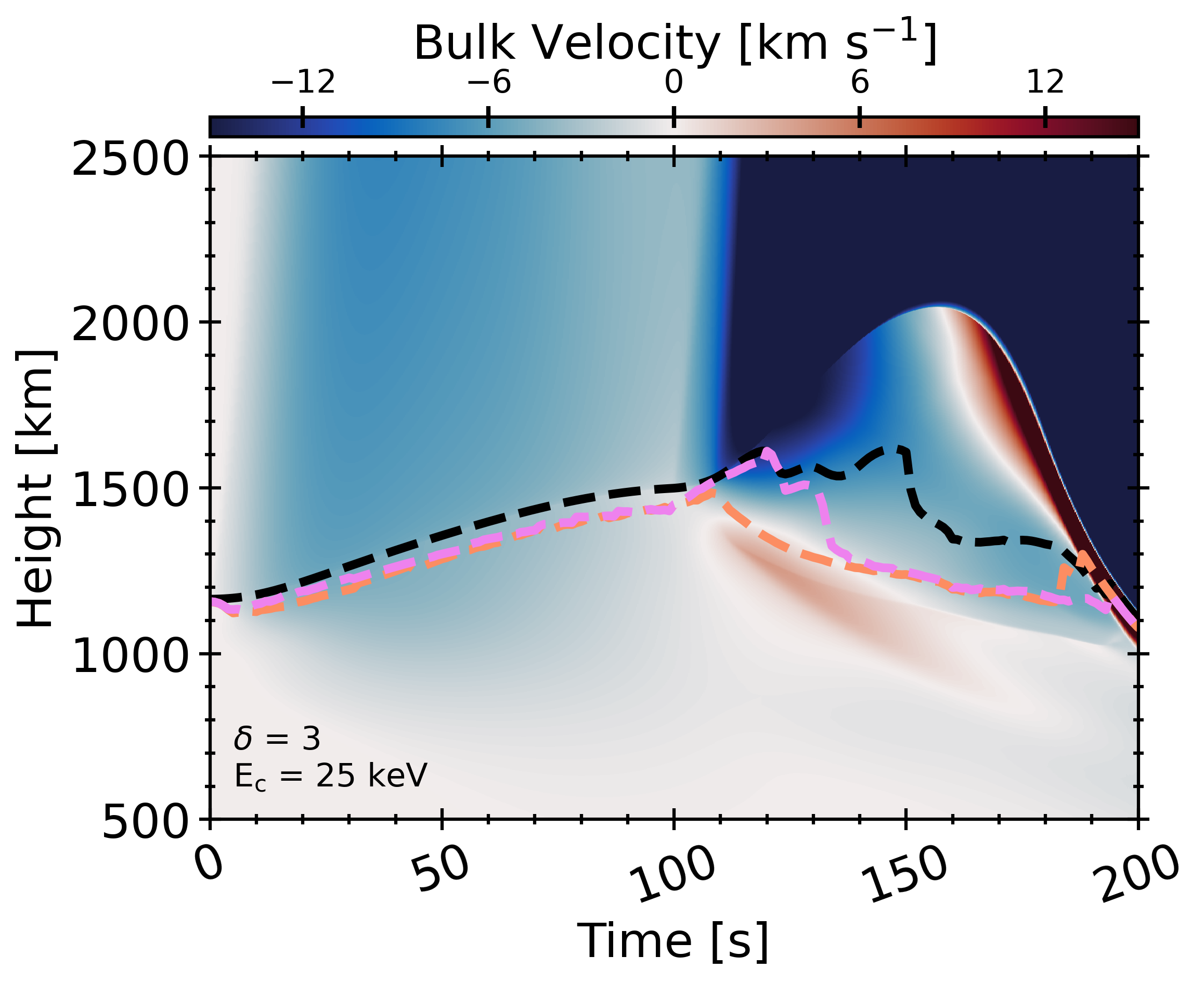}}
	\subfloat{\includegraphics[width = 0.35\textwidth, clip = true, trim = 0.cm 0.cm 0.cm 0.cm]{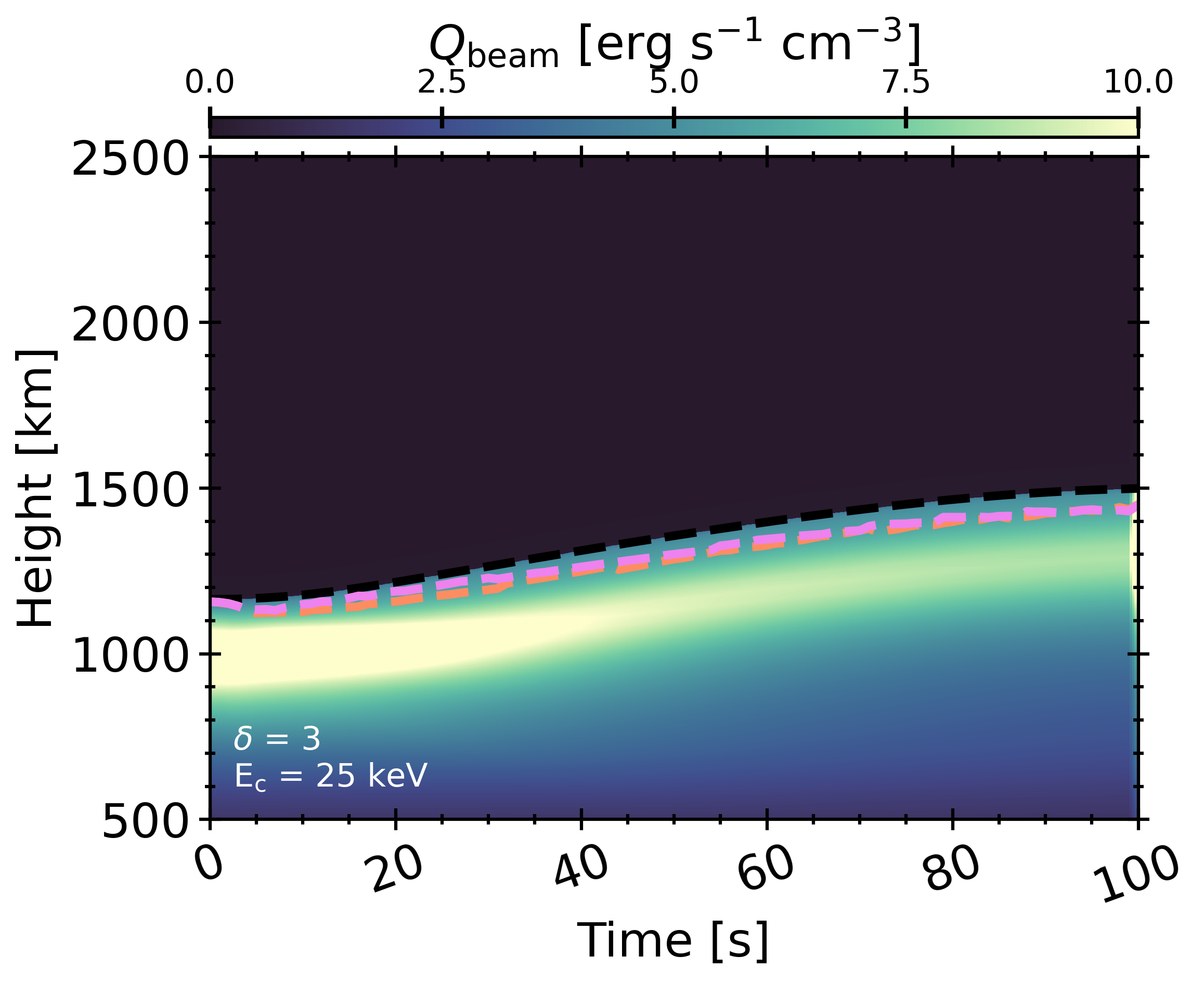}}
	}
	}
	\vbox{
	\hbox{
	\hspace{1in}
	\subfloat{\includegraphics[width = 0.35\textwidth, clip = true, trim = 0.cm 0.cm 0.cm 0.cm]{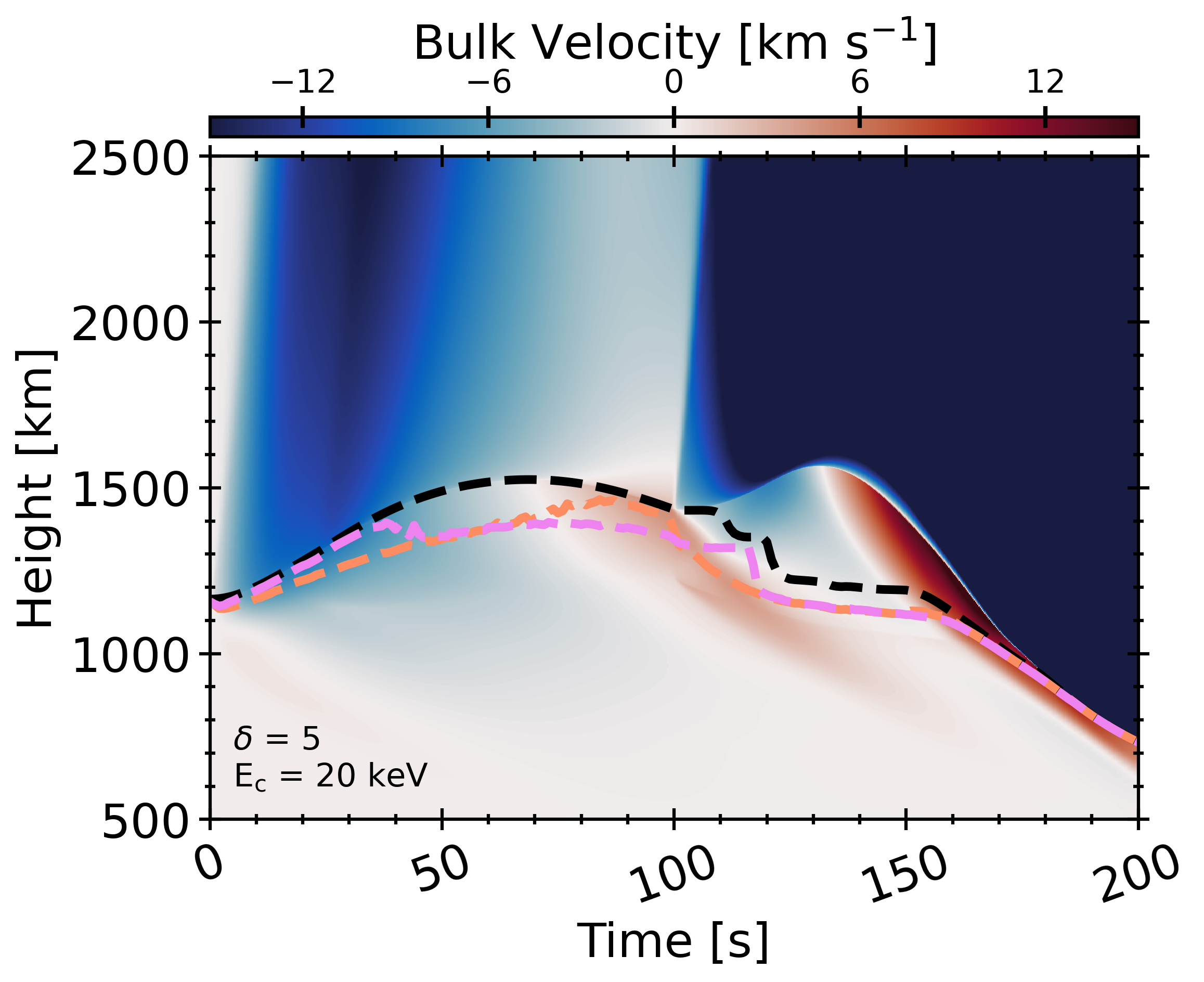}}	
	\subfloat{\includegraphics[width = 0.35\textwidth, clip = true, trim = 0.cm 0.cm 0.cm 0.cm]{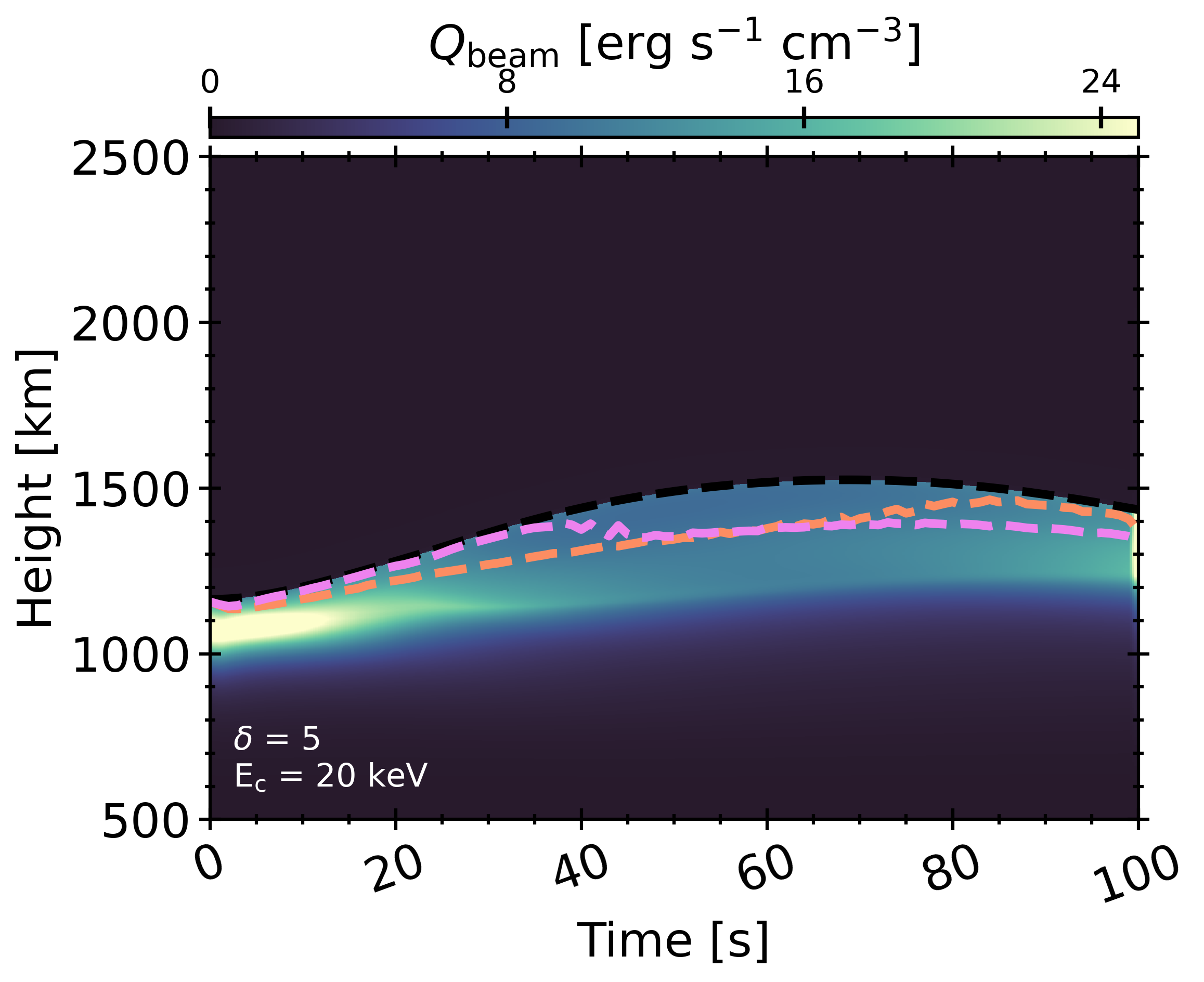}}
	}
	}
	\vbox{
	\hbox{
	\hspace{1in}
	\subfloat{\includegraphics[width = 0.35\textwidth, clip = true, trim = 0.cm 0.cm 0.cm 0.cm]{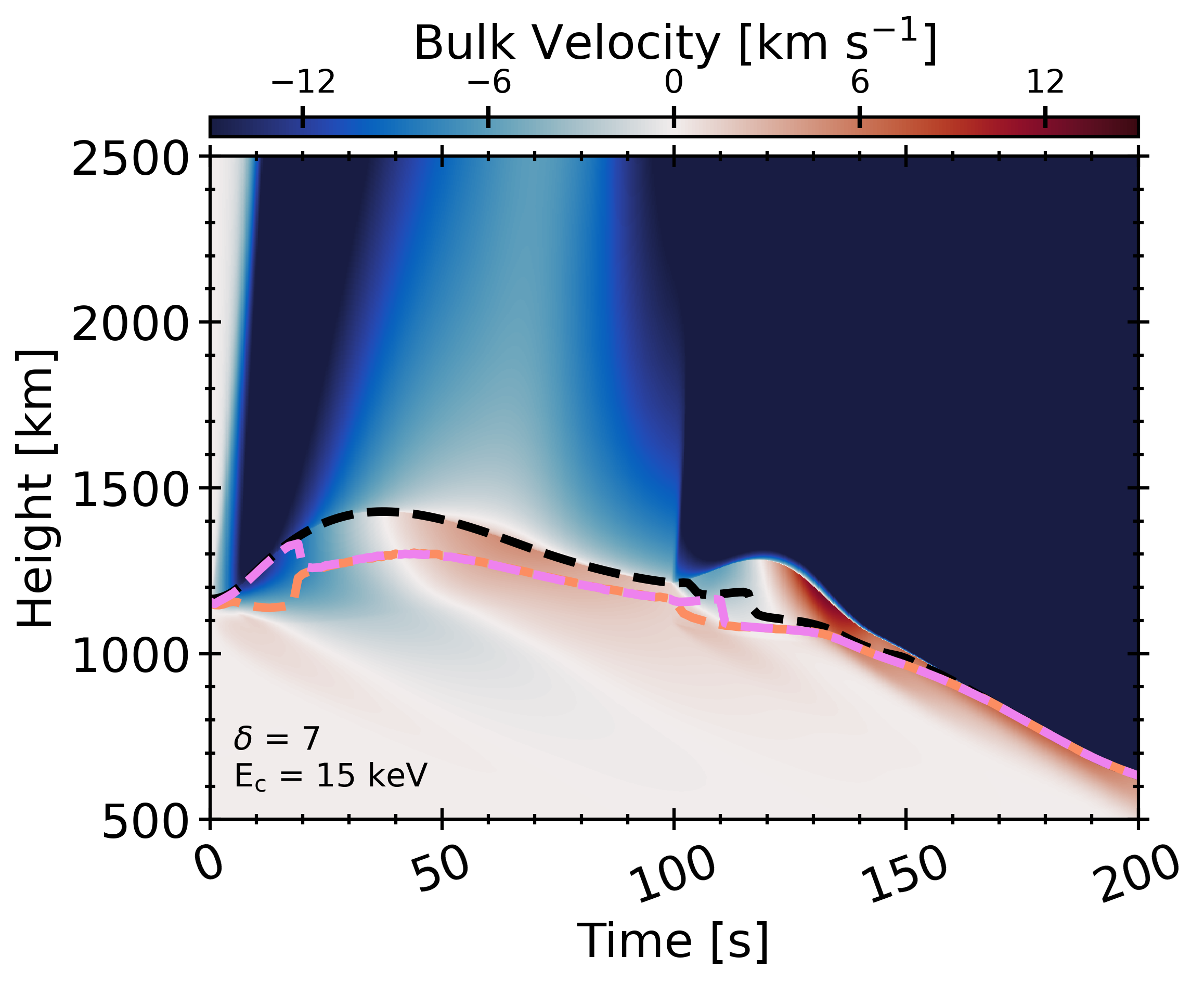}}
	\subfloat{\includegraphics[width = 0.35\textwidth, clip = true, trim = 0.cm 0.cm 0.cm 0.cm]{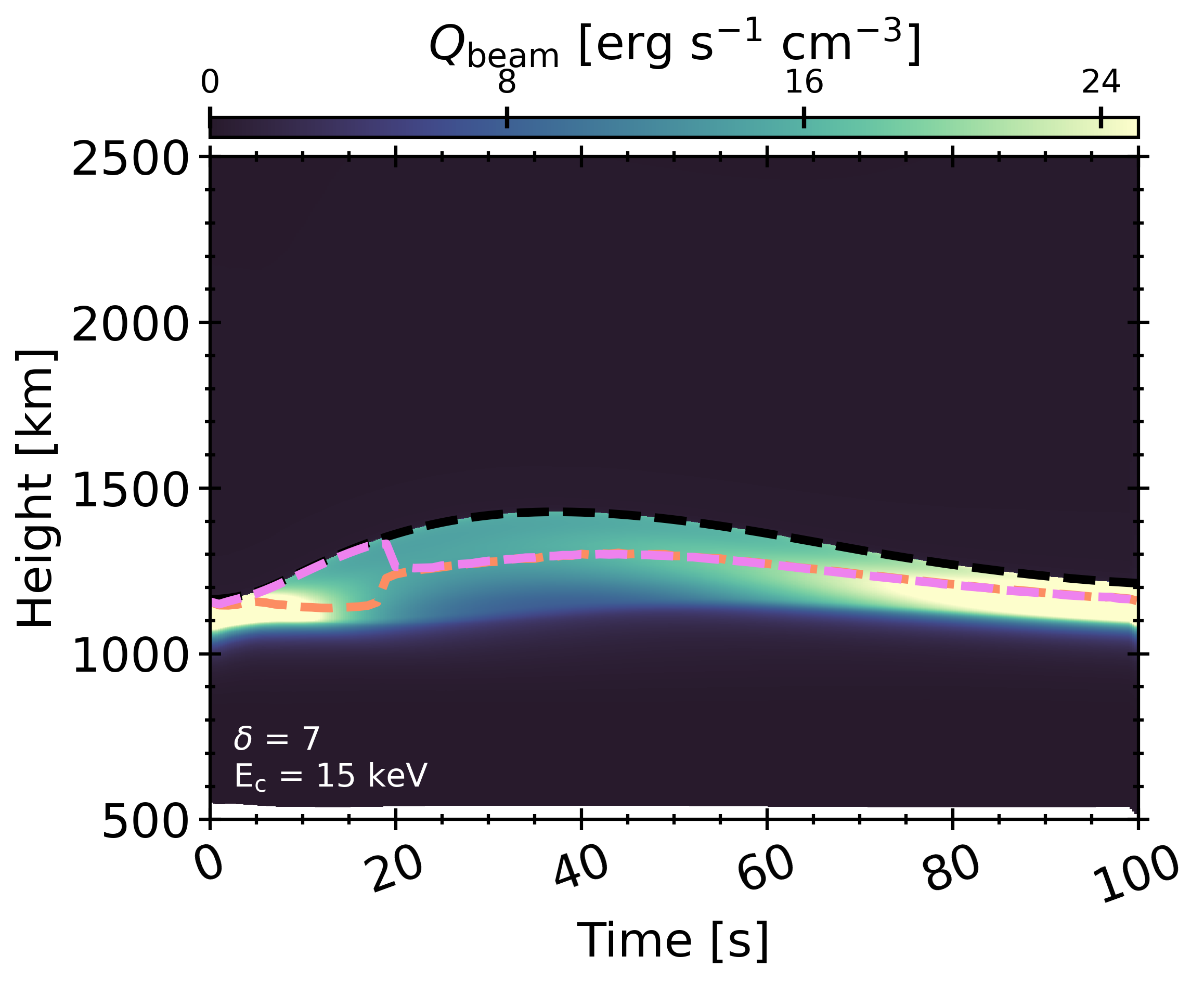}}
	}
	}
	\caption{\textsl{The atmospheric velocity (left column; upflows are negative), and nonthermal electron heating rate (right column) from three simulations, with the formation height of k3 (black dashed line), k2v (violet dashed line), and k2r (red dashed line) overlaid. The top row is a hard electron beam ($\delta = 3$, $E_{c} = 25$~keV), the middle row is an intermediate case ($\delta = 5$, $E_{c} = 20$~keV), and the bottom row is a soft electron beam ($\delta = 7$, $E_{c} = 15$~keV). Colourbars are saturated in order to more clearly show weaker values.}}
	\label{fig:k3shift_ex}
\end{figure*}
	The obvious question is, what do these ribbon front \ion{Mg}{2} profiles tell us about the evolution of the plasma, and why do we need a period of weak energy injection by hard nonthermal electron distributions to produce those conditions? Using a similar approach as in Section~\ref{sec:he_plasmaprops}, the plasma properties averaged over the formation height of k2v, k2r and k3 were obtained and contrasted during the different stages of energy deposition. 
	
	In this section we follow closely the foundational investigations of \ion{Mg}{2} profile characteristics by \cite{2013ApJ...772...89L}, \cite{2013ApJ...772...90L}, and \cite{2013ApJ...778..143P}. Those authors studied a quiet Sun 3D RMHD simulation, and cautioned that their results are indicative primarily of those conditions and that more active scenarios could differ significantly. For that reason, where appropriate, we confirm any inferences based on their conclusions by applying elements of their analysis to our flare simulations. An obvious caveat is that their conclusions were based on hundreds of thousands of spectra, whereas we have 4515 (15 1D RHD flare simulations each with 301 snapshots, representing 1s cadence). 
	
	Conspicuously absent from our exploration of the origin of the various ribbon front characteristics is the \ion{Mg}{2} line widths which can broaden to an even greater degree in ribbon fronts than in bright ribbon segments. This is because our flare models are still unable to reproduce the observed degree of broadening \citep[see e.g.][]{2024MNRAS.527.2523K}, indicating that we are missing some key physics. 

\subsubsection{What causes the \ion{Mg}{2} k3 Doppler shift?}
	Gentle (subsonic) chromospheric evaporation \citep{1985ApJ...289..414F,1985ApJ...289..425F,1985ApJ...289..434F,2015ApJ...808..177R} occurred during the weak-heating phase of the flares, which transitioned to explosive evaporation when the injected energy flux increased, illustrated by velocity maps in Figure~\ref{fig:k3shift_ex} (upflows are negative). This figure shows results from three simulations: one with a hard nonthermal electron distribution ($\delta = 3$, $E_{c} = 25$~keV), one with an `intermediate' nonthermal electron distribution ($\delta = 5$, $E_{c} = 20$~keV), and one with a soft nonthermal electron distribution ($\delta = 7$, $E_{c} = 15$~keV). Mean formation heights of the k2v, k2r, and k3 components overlaid.

\begin{figure}
	\centering 
	\vbox{
	\hbox{
	\hspace{0in}
	\subfloat{\includegraphics[width = 0.475\textwidth, clip = true, trim = 0.cm 0.cm 0.cm 0.cm]{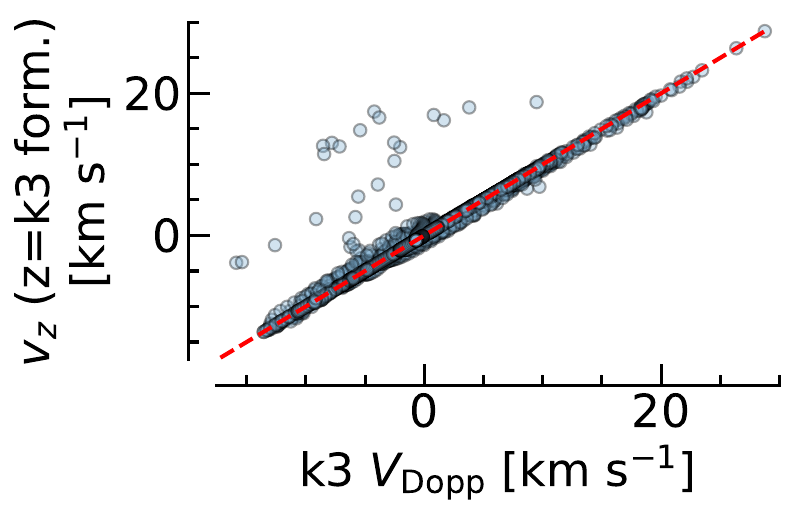}}
	}
	}
	\vbox{
	\hbox{
	\hspace{0in}
	\subfloat{\includegraphics[width = 0.475\textwidth, clip = true, trim = 0.cm 0.cm 0.cm 0.cm]{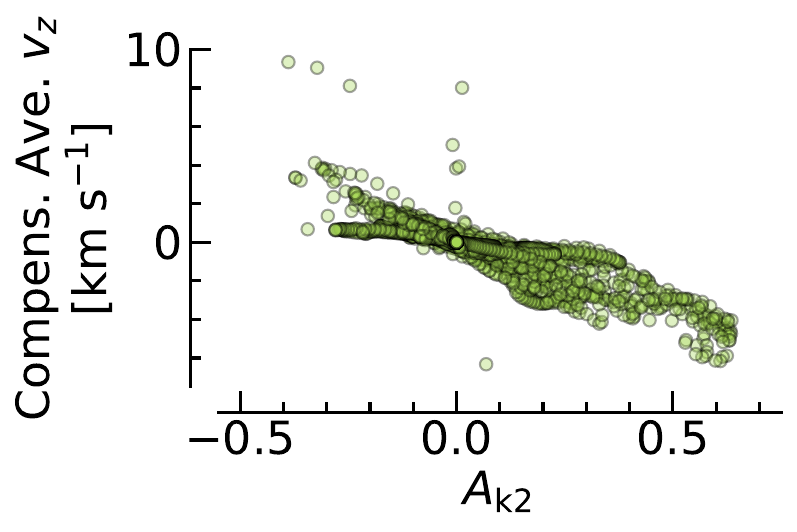}}	
	}
	}
	\vbox{
	\hbox{
	\hspace{0in}
	\subfloat{\includegraphics[width = 0.475\textwidth, clip = true, trim = 0.cm 0.cm 0.cm 0.cm]{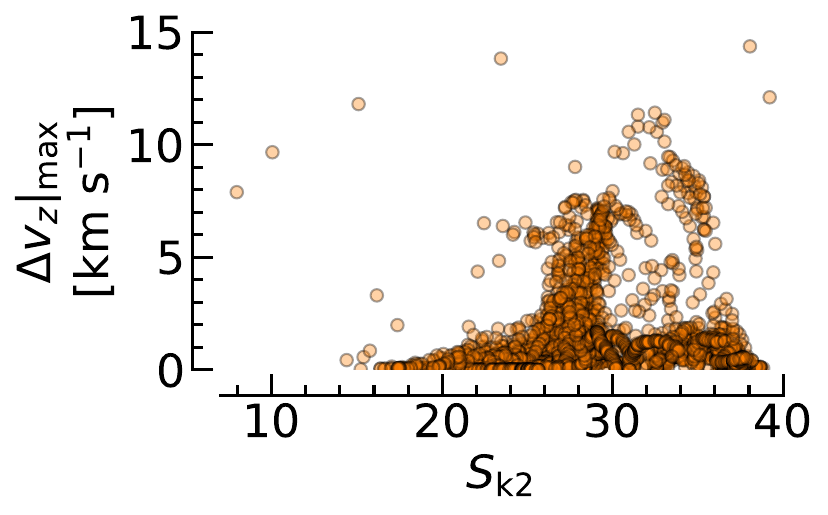}}
	}
	}
	\caption{\textsl{Relating \ion{Mg}{2} k line characteristics to atmospheric velocities. Every 1s snapshot from all 15 simulations is included, covering both ribbon front and stronger heating phases. The top panel shows a strong correlation between the Doppler shift of \ion{Mg}{2} k3 and the average plasma velocity within the k3 formation height. The red dashed line is $y=x$. The middle panel is shows how the k2 peak asymmetry varies with the `compensated average velocity' (average velocity in the inter-peak region less the plasma velocity at the k2 formation height).  The bottom panel shows how the k2 peak separation varies with the largest velocity differences present between in the inter-peak formation region. }}
	\label{fig:mgiicorrels}
\end{figure}

	When the nonthermal electron distribution is hard enough, during the weak-heating phase there are few low-energy electrons and energy deposition location peaks \textsl{below} the formation height of the \ion{Mg}{2} k line core, seen in the right hand column of Figure~\ref{fig:k3shift_ex}, which compares the mean formation heights to the nonthermal electron heating rate $Q_{\mathrm{beam}}$. The upflow originates fairly deep in the chromosphere, and so the k3 component therefore forms in that upflowing region and is blueshifted. Doppler shifts of k3 are very well correlated with the plasma velocity in the k3 formation region in the quiet Sun \citep{2013ApJ...772...90L,2013ApJ...778..143P}, and we show in Figure~\ref{fig:mgiicorrels} that this relationship holds true in flares. Almost every profile exhibits a k3 Doppler shift that is very close to the average plasma velocity at the k3 formation height (most points lie close to the $y=x$ line).

	Over time, as the density of the upper chromosphere increases due to evaporation, the spatial profile of energy deposition becomes more spread out such that the origin of the upflow drifts higher in altitude and the \ion{Mg}{2} blueshift subsides. This is why the Doppler shift of the line core does not show a behaviour that lasts the entirety of the weak-heating phase in those same simulations that persist with \ion{He}{1} 10830~\AA\ ribbon front characteristics. 

	When the nonthermal electron distribution is instead rather soft, there are sufficient numbers of low-energy electrons that the upper chromosphere/lower transition region is quickly perturbed, and in these cases the altitude of the origin of the upflow more rapidly drifts higher.

	Note that during the intermediate heating regime of some simulations the fluxes of lower-energy electrons are not strong enough to make the beams lose most of their energy in the upper chromosphere/lower transition region so that there can be a second period of blueshifted k3. At those times the whole chromosphere is, however, strongly heated so that the other ribbon front characteristics are not present. Once the energy fluxes approach the peak values we see the expected redshifts signifying downflowing plasma.

\subsubsection{What causes the \ion{Mg}{2} k2 Red Asymmetry?}
	
	\cite{2013ApJ...772...90L} demonstrated in their quiet Sun simulation that if there was an average large scale plasma motion in the inter-peak region\footnote{The region between the k2 and k3 formation heights, where the k2 formation height was defined as the mean of the k2v and k2r formation heights} then there would likely be a k2 peak asymmetry. They compared the asymmetry to the `compensated average velocity', which is the average plasma velocity in the inter-peak region minus the plasma velocity at the k2 formation height. This latter step is to have the values pass through the origin, since adding an offset to the whole atmosphere (a velocity at k2 formation height) just shifts the distribution without changing the relationship to the asymmetry. If the plasma was generally upflowing then the k2r peak would be stronger, and vice versa. For every snapshot in our 15 flare simulations we find the same relationship is present during the flares (middle panel of Figure~\ref{fig:mgiicorrels}). Note that we defined the formation heights as described earlier, but \cite{2013ApJ...772...90L} used the height at which optical depth unity was reached for each wavelength. Additionally, in our case downflows are positive and a positive $A_{\mathrm{k2}}$ means k2r is larger than k2v.   

	It may be somewhat counterintuitive, but when the plasma in the inter-peak region is upflowing the k2r peak becomes more intense compared to the k2v peak, and vice versa. This phenomenon has been discussed in detail in the context of \ion{Ca}{2} H2v bright grains \citep{1997ApJ...481..500C}, H$\alpha$ asymmetries in flares \citep{2015ApJ...813..125K}, and \ion{Mg}{2} k2 asymmetries in flares \citep{2016ApJ...827..101K}. In short, the absorption profile (also referred to as the extinction profile) of the line is shifted by plasma motions. If there is an average upflow in the inter-peak region then this absorption profile is shifted to the blue, and more blue-peak photons are absorbed compared to the red-peak which has a smaller opacity than the photons would experience in the stationary case. The situation is reversed in the case of an average downflow. 
	
	During the weak-heating phase the region between k2 and k3 is generally upflowing (Figure~\ref{fig:k3shift_ex}) and \ion{Mg}{2} profiles exhibit red peak asymmetries with stronger k2r peaks. For the same reason that the k3 blueshift does not persist for the entirety of the weak-heating phase, the spatially evolving mass motions in the mid-upper chromosphere impact the strength of $A_{\mathrm{k2}}$. Once the energy fluxes approach the peak values there are typically large average downflows present in the inter-peak formation region, driving the peaks to exhibit blue asymmetries.

\subsubsection{What Causes the Increased Peak Separation?}
	The separation of the k2 peaks is related to the velocity gradients in the mid-upper chromosphere since they lead to height-dependent wavelength shifts of the \ion{Mg}{2} k line's thermal absorption profile, which when integrated through height effectively widen's the overall absorption profile \citep{2013ApJ...772...90L}. A positive correlation between the maximum velocity difference in the inter-peak formation region ($\Delta v_{z}|_\mathrm{max}$) and the k2 peak separation was found by \cite{2013ApJ...772...90L}. They also identified a small population of \ion{Mg}{2} profiles which had large $S_{\mathrm{k2}}$ but small $\Delta v_{z}|_\mathrm{max}$, that arose due to a temperature rise in the lower chromosphere that naturally widened the absorption profile and lowered the k2 formation height. 
	
	In the bottom panel of Figure~\ref{fig:mgiicorrels} we show how $S_{\mathrm{k2}}$ varies with $\Delta v_{z}|_\mathrm{max}$, for every profile in our simulation grid. The scatter is reminiscent of the equivalent figures from the quiet Sun experiments of \cite{2013ApJ...772...90L} and \cite{2013ApJ...778..143P}, with a population of large $S_{\mathrm{k2}}$ at small $\Delta v_{z}|_\mathrm{max}$. This suggests to us that the source of increased $S_{\mathrm{k2}}$ in the weak-heating phase is due to mid-lower chromosphere temperature increases rather than velocity gradients in the mid-upper chromosphere. Consistent with this picture is the fact that the simulations that are more similar to flare ribbon fronts have longer periods in which the mid chromosphere is heated without very strong temperature increases in the upper chromosphere. That itself results from the harder, more deeply penetrating, nonthermal electron energy distributions. 
	
	When the chromosphere becomes strongly compressed during the peak of the strong-heating phase, and the upper chromospheric temperature is much more enhanced, $S_{\mathrm{k2}}$ tends to be smaller than the pre-flare (also true in observations), which is due to the fact that the k3 and k2 peaks form much closer in height and thus the velocity gradient in the inter-peak region is small. 
	
	Microturbulence would further broaden the absorption profile, which could be very variable in flares, and play a role in setting $S_{\mathrm{k2}}$. We keep microturbulence fixed throughout our simulations in the absence of solid constraints, but with values comparable to those found by \cite[][]{2024MNRAS.527.2523K} and \cite{2023FrASS..1033429S}. It is worth noting that \cite{Jeffreyeaav2794} observed an increase of \ion{Si}{4} resonance line nonthermal widths at the flare onset, prior to any strong intensity increases, so microturbulence could play some role in flare ribbon fronts. Study of the chromospheric \ion{O}{1} 1355.598~\AA, \ion{Fe}{2} 2814.445~\AA\ or \ion{Cl}{1} 1351.66~\AA\ ribbon front behaviour, in particular their nonthermal widths, would be very useful. 
	
\subsubsection{What Causes the Deeper \ion{Mg}{2} k Line Central Reversals?}
\begin{figure*}
	\centering 
	\vbox{
	\hbox{
	\hspace{0in}
	\subfloat{\includegraphics[width = 0.5\textwidth, clip = true, trim = 0.cm 0.cm 0.cm 0.cm]{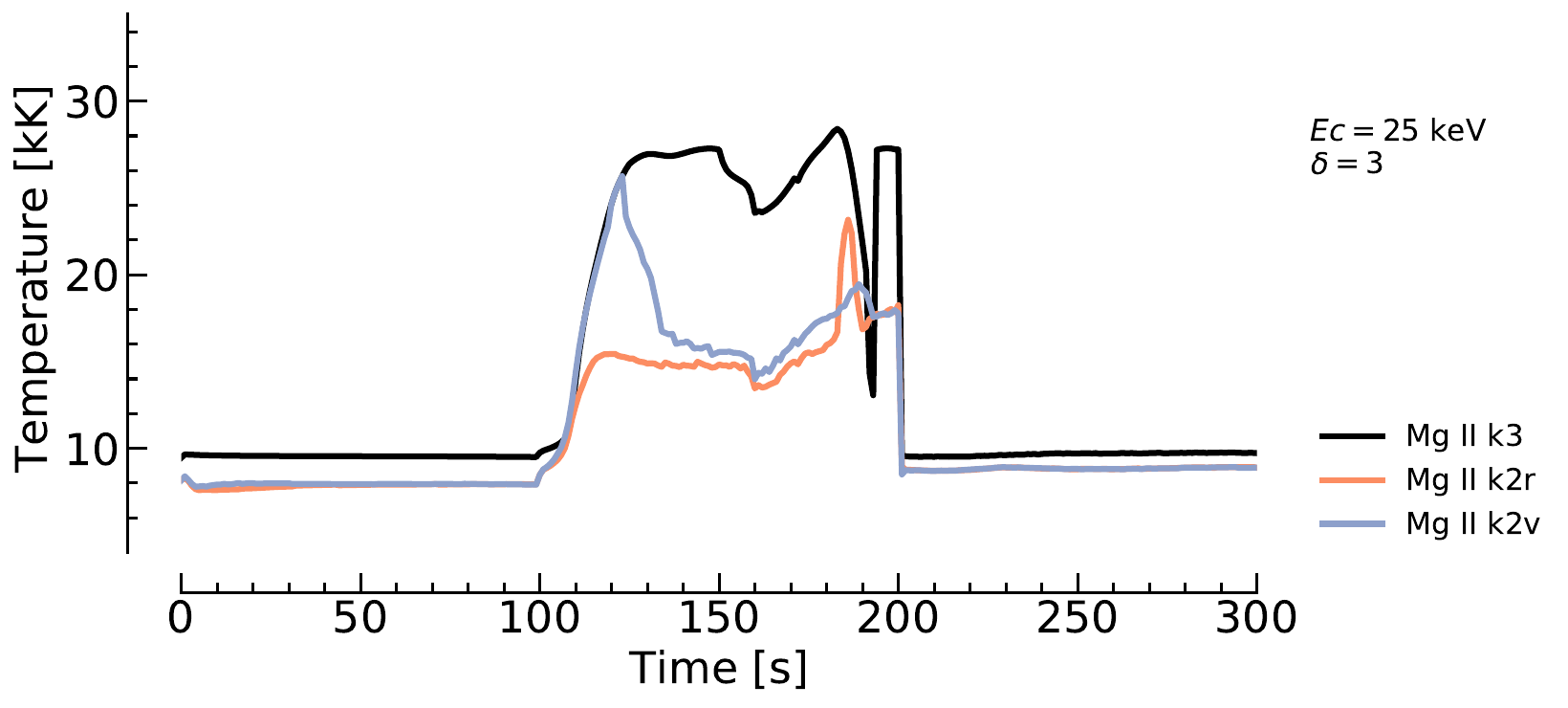}}
	\subfloat{\includegraphics[width = 0.5\textwidth, clip = true, trim = 0.cm 0.cm 0.cm 0.cm]{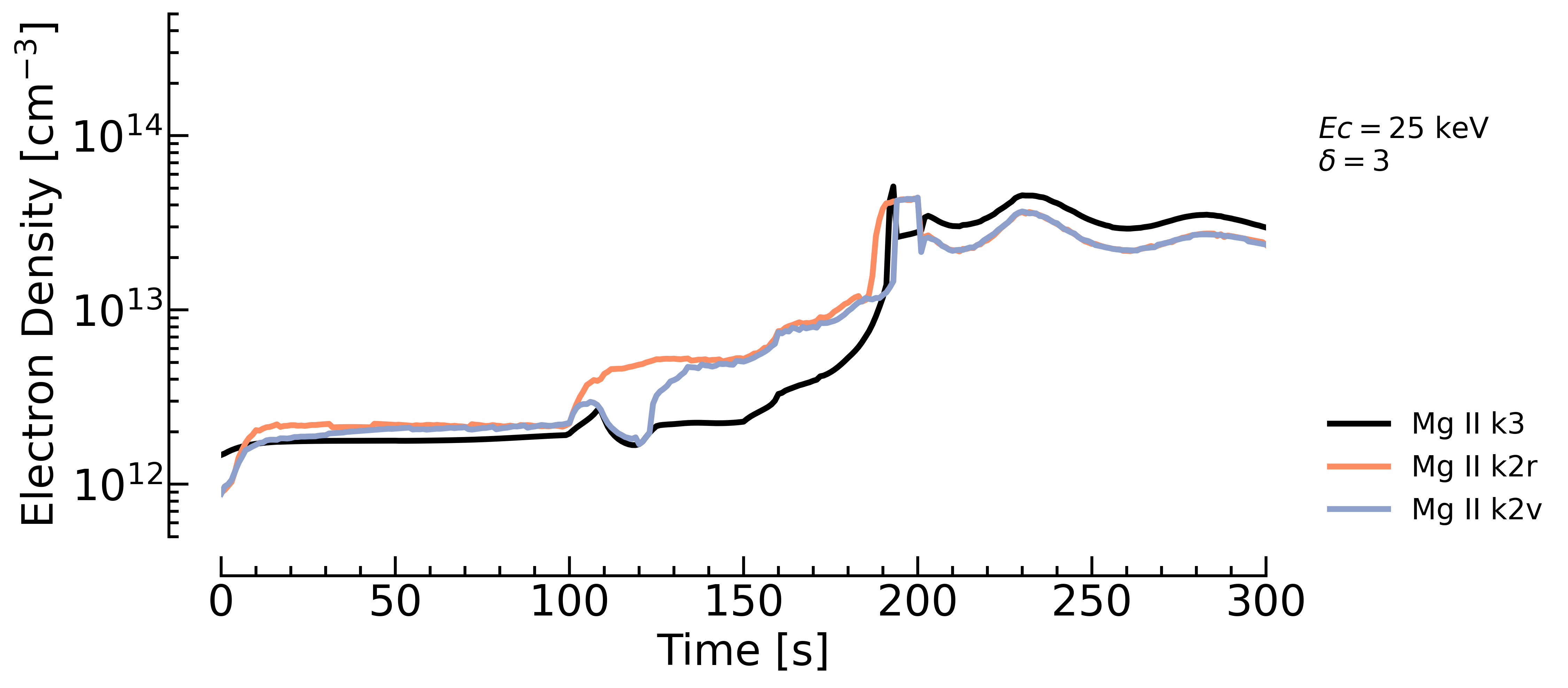}}
	}
	}
	\vbox{
	\hbox{
	\hspace{0in}
	\subfloat{\includegraphics[width = 0.5\textwidth, clip = true, trim = 0.cm 0.cm 0.cm 0.cm]{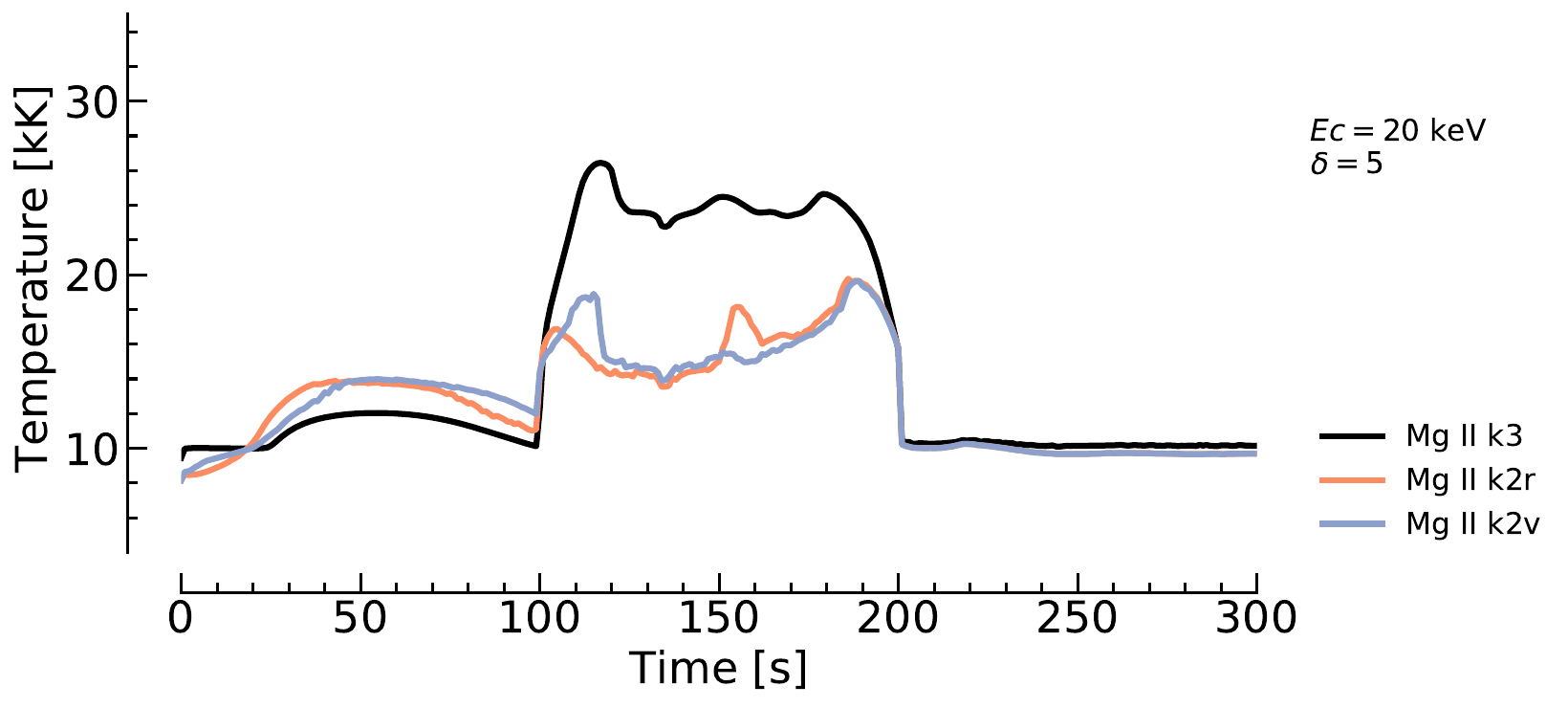}}	
	\subfloat{\includegraphics[width = 0.5\textwidth, clip = true, trim = 0.cm 0.cm 0.cm 0.cm]{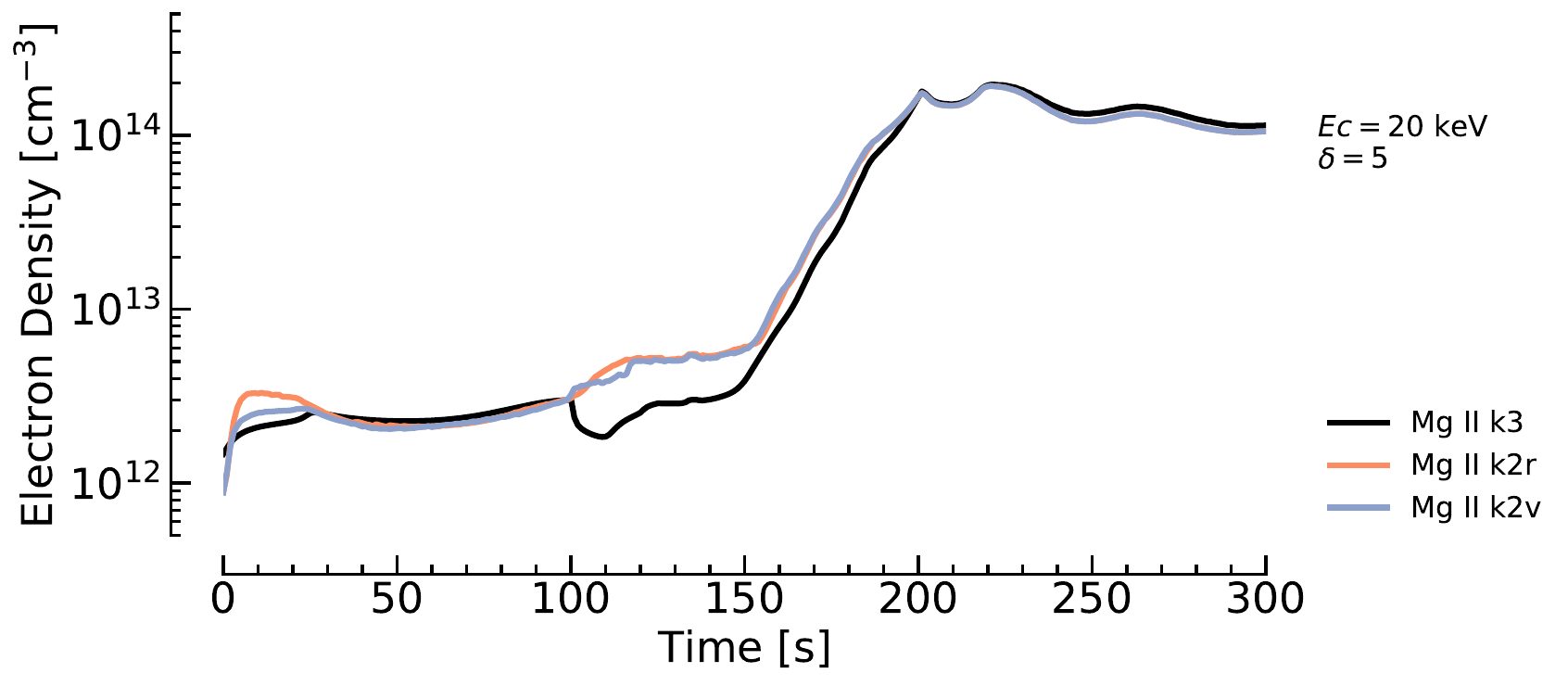}}
	}
	}
	\vbox{
	\hbox{
	\hspace{0in}
	\subfloat{\includegraphics[width = 0.5\textwidth, clip = true, trim = 0.cm 0.cm 0.cm 0.cm]{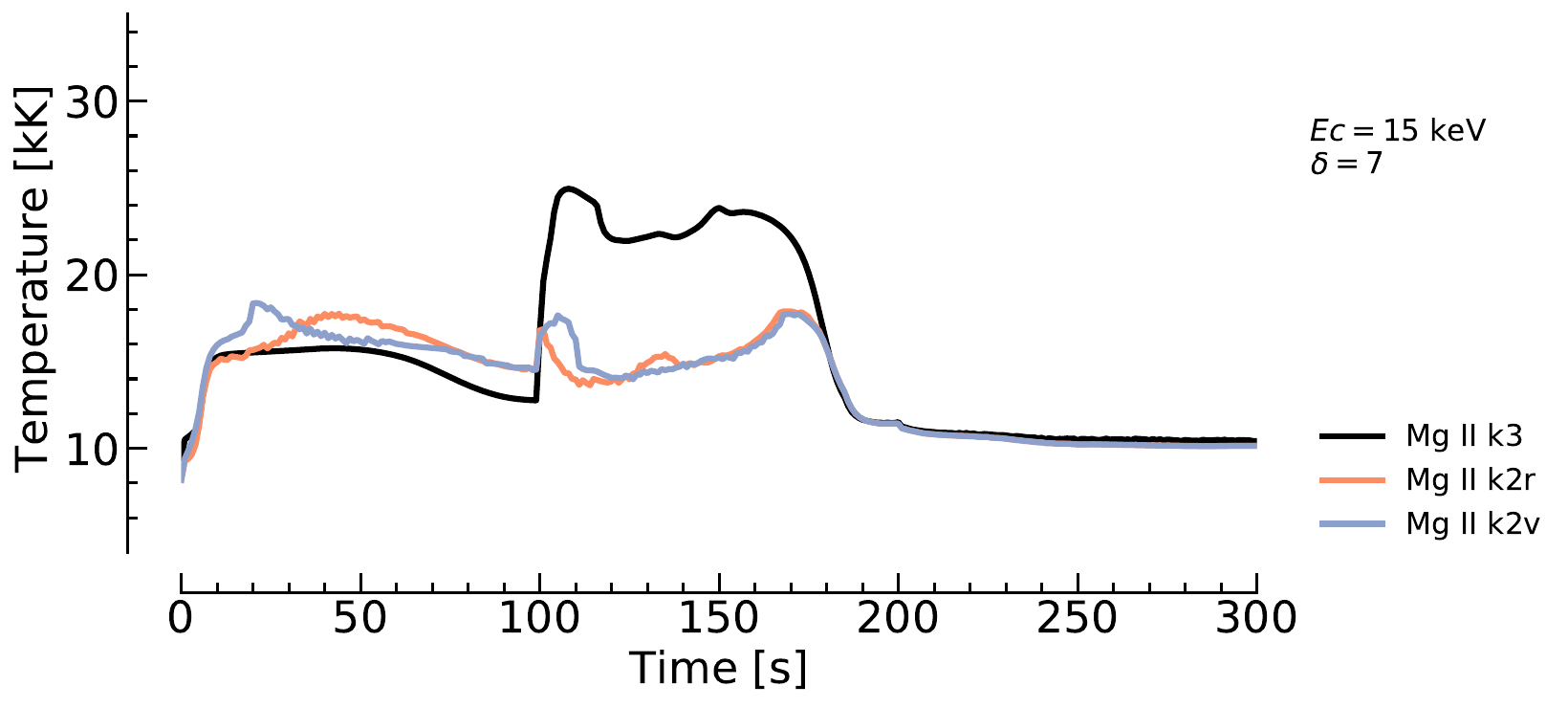}}
	\subfloat{\includegraphics[width = 0.5\textwidth, clip = true, trim = 0.cm 0.cm 0.cm 0.cm]{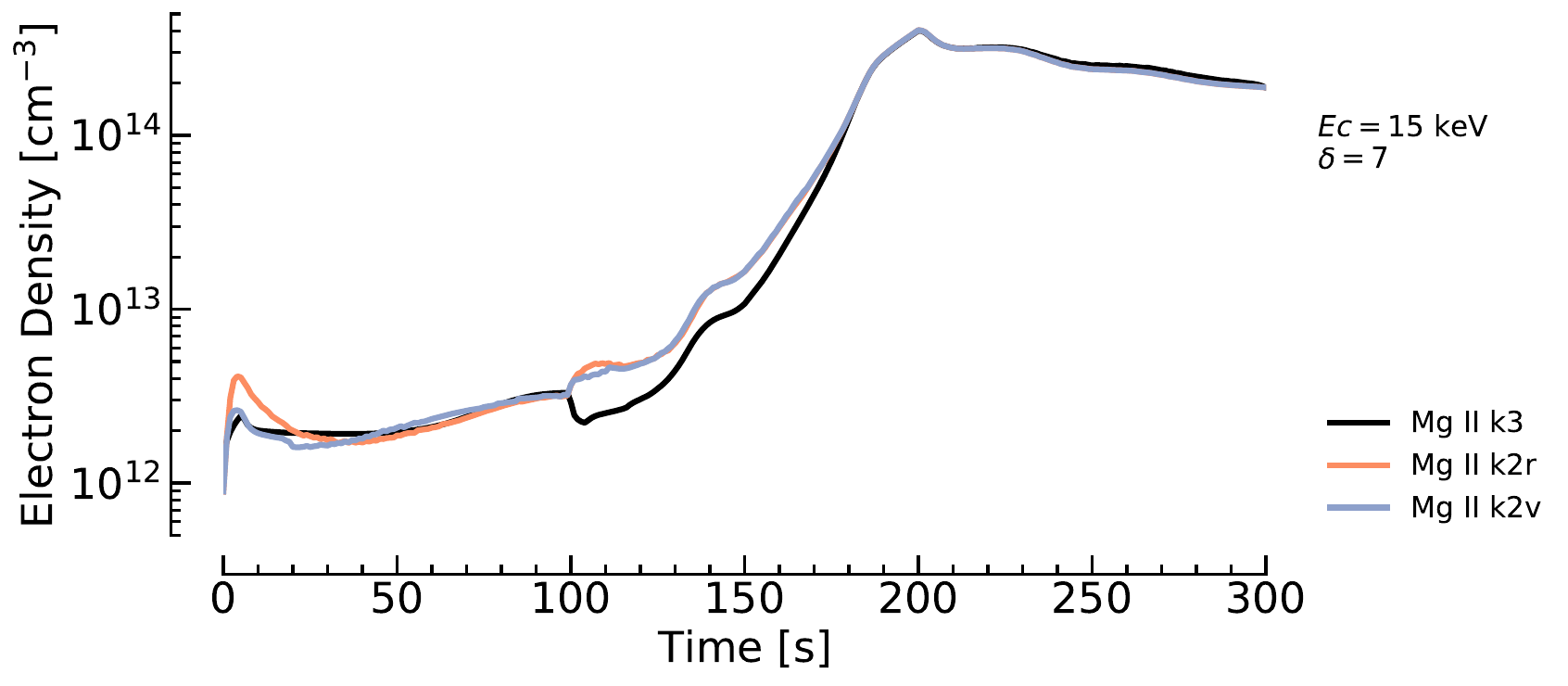}}
	}
	}
	\caption{\textsl{The plasma temperature (left column), and electron density (right column) averaged over the formation region of k3 (black), k2r (red), and k2v (blue) from three simulations. The top row is a hard electron beam ($\delta = 3$, $E_{c} = 25$~keV), the middle row is an intermediate case ($\delta = 5$, $E_{c} = 20$~keV), and the bottom row is a soft electron beam ($\delta = 7$, $E_{c} = 15$~keV).}}
	\label{fig:avtemps}
\end{figure*}
	The formation of the \ion{Mg}{2} lines is a complex interplay of modulation of the velocity, temperature, density and optical depth stratification. Figure~\ref{fig:k3shift_ex} illustrates that the mean formation height of the k2 and k3 line components vary in time, with the velocity structure playing a dominant role in the k3 formation height which increases in response to the gentle evaporation. In some simulations, those with larger upflows in the k3 formation region, the k2v formation height also increases, forming some distance from the k2r. \gsk{This is related to the fact that when the k3 component (the line core) is very Doppler shifted, in this case to bluer wavelengths, the absorption profile of the line is commensurately shifted. Significantly more k2v photons are absorbed as a result and the opacity shifts to higher in altitude. At the same time, the opacity is reduced in the red part of the line, such that the k2r formation height is deeper in the atmosphere.} More typically the k2 peaks form relatively close in altitude to each other.
	
	During the weak energy injection phase the mid-chromosphere is more perturbed than the upper chromosphere. As we alluded to in \cite{2023ApJ...944..104P}, but show explicitly here, this means that initially the k2 peaks experience much larger increases in the electron density than the k3 core.  This is particularly true of the simulations with the hardest nonthermal electron distributions. Temperature evolution is more complicated, as it can both increase or decrease (for example, the decrease in k2 formation height in the $\delta = 3$, $E_{c} = 25$~keV simulation results in a slight drop in formation temperature). Figure~\ref{fig:avtemps} shows the temporal evolution of the temperature and electron density in the formation region of k2r, k2v, and k3, for three exemplar simulations (the same as those shown in Figure~\ref{fig:k3shift_ex}). 
		
\begin{figure*}
	\centering 
	\vbox{
	\hbox{
	\hspace{0in}
	\subfloat{\includegraphics[width = 0.5\textwidth, clip = true, trim = 0.cm 0.cm 0.cm 0.cm]{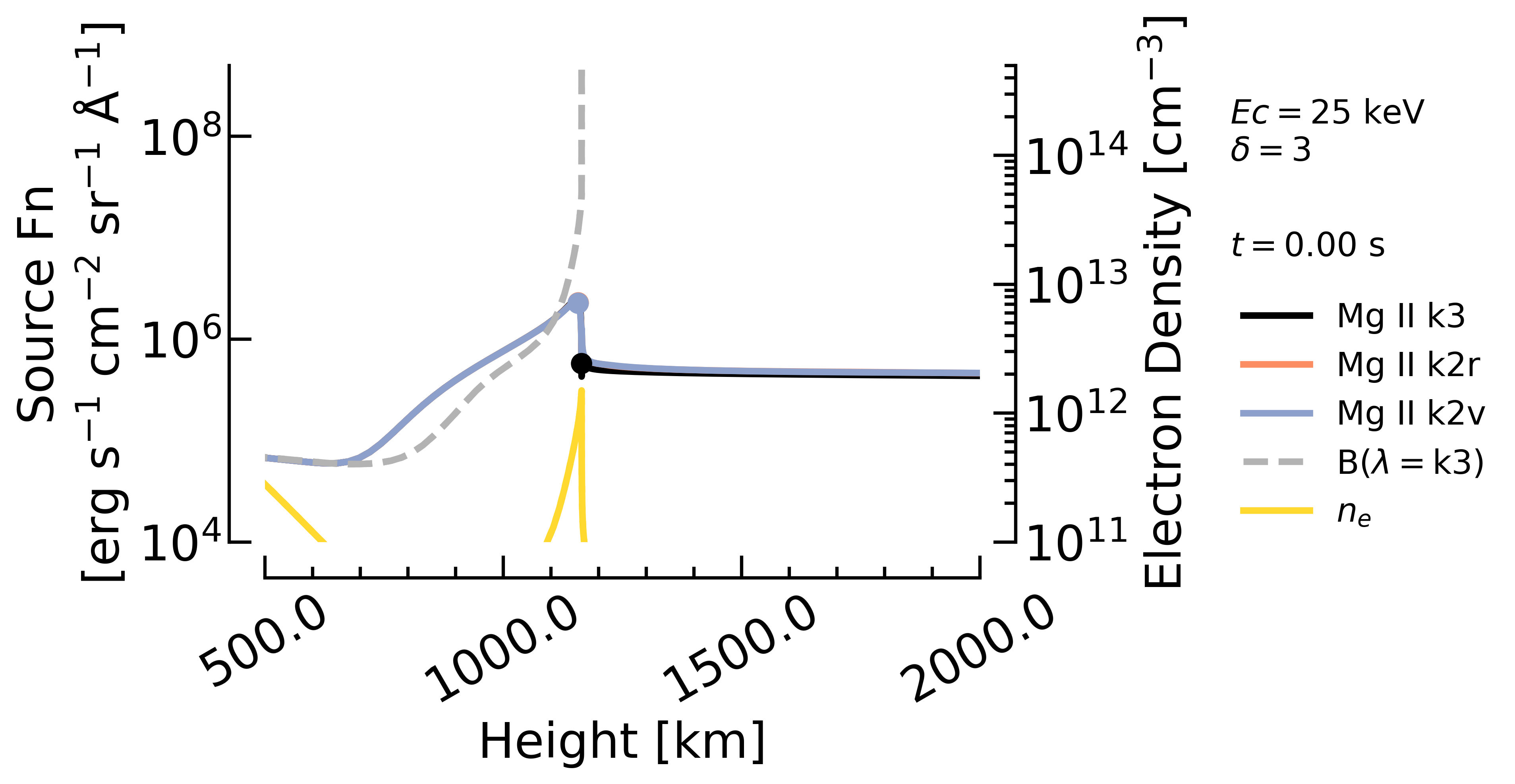}}
	\subfloat{\includegraphics[width = 0.5\textwidth, clip = true, trim = 0.cm 0.cm 0.cm 0.cm]{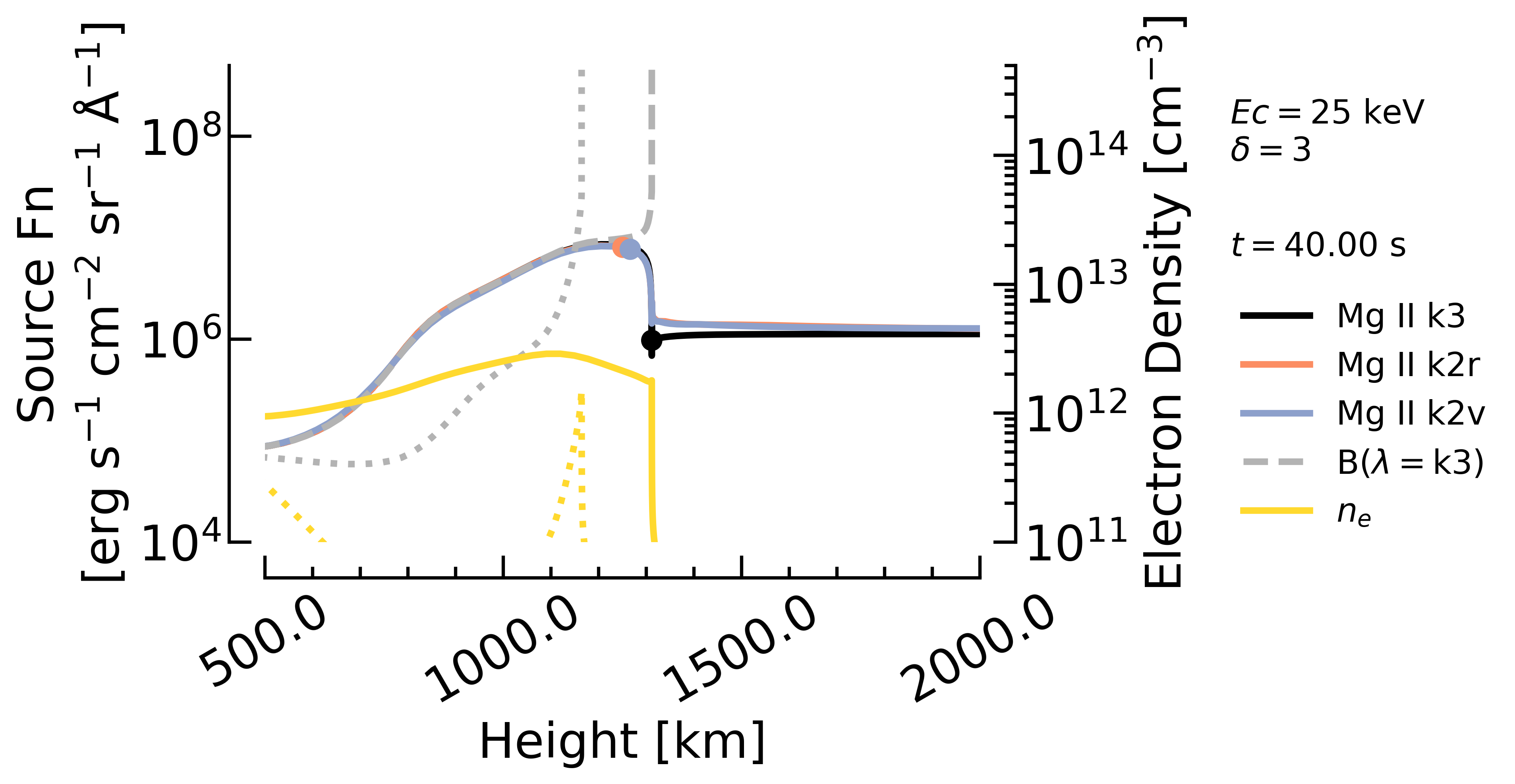}}
	}
	}
	\vbox{
	\hbox{
	\hspace{0in}
	\subfloat{\includegraphics[width = 0.5\textwidth, clip = true, trim = 0.cm 0.cm 0.cm 0.cm]{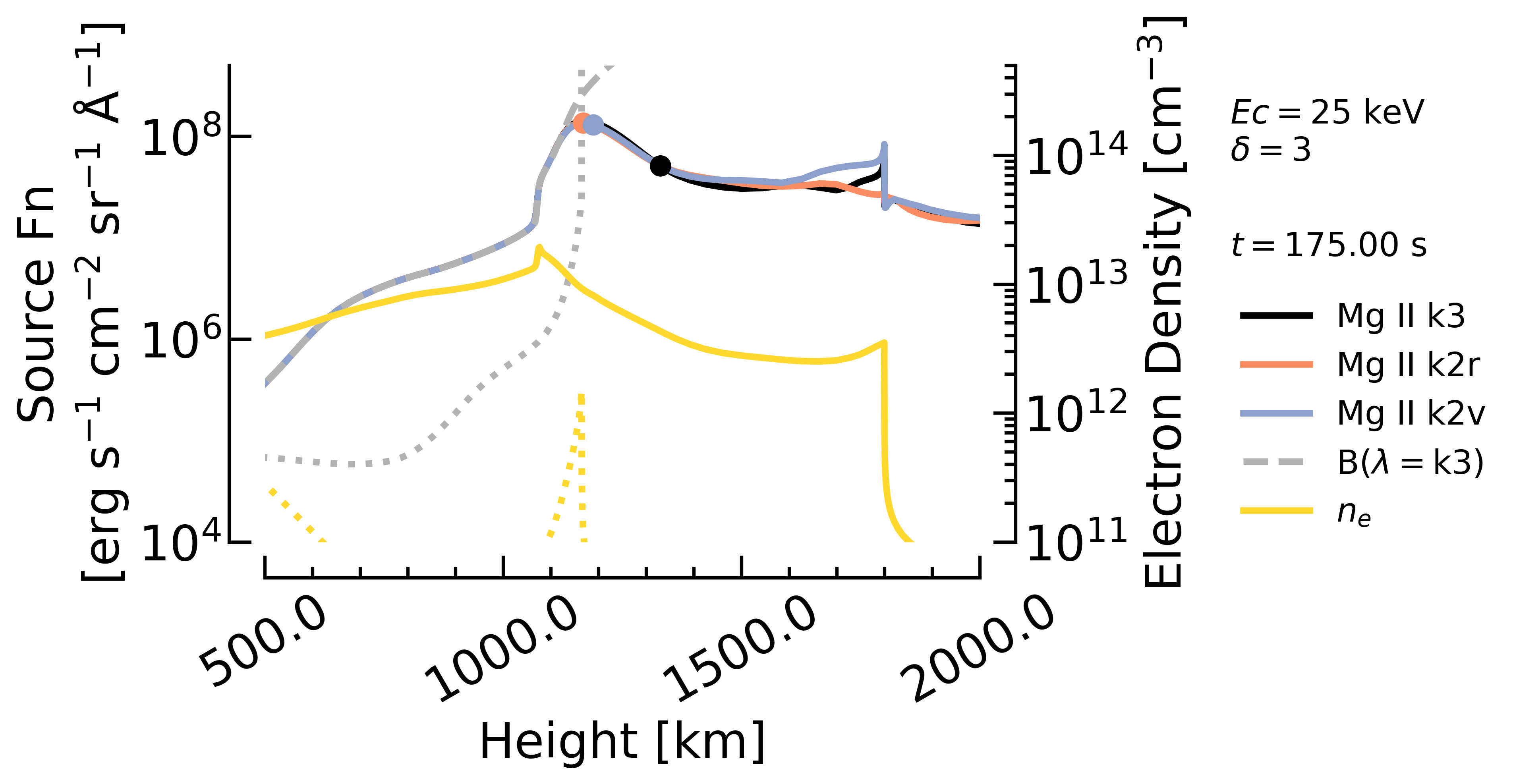}}	
	\subfloat{\includegraphics[width = 0.5\textwidth, clip = true, trim = 0.cm 0.cm 0.cm 0.cm]{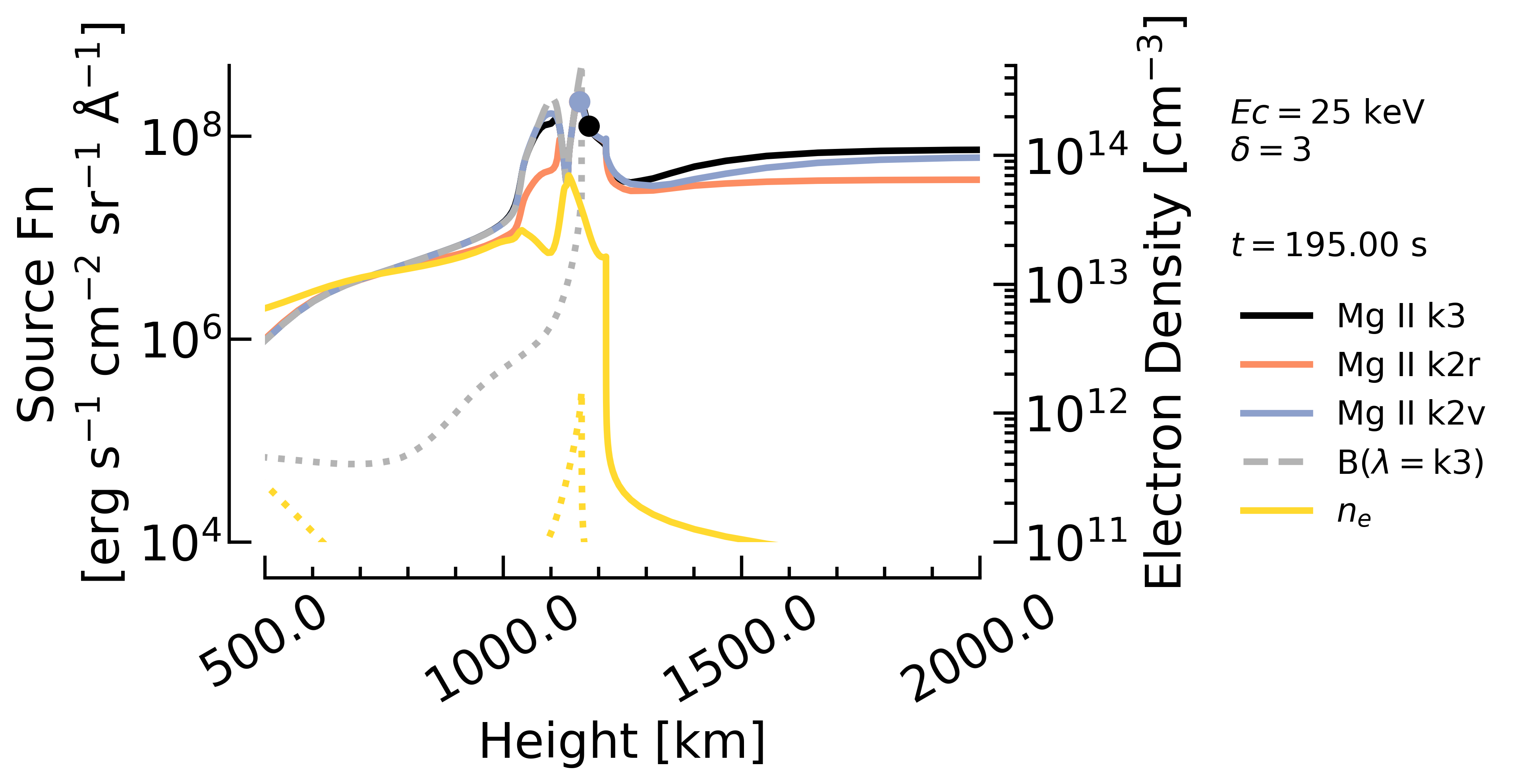}}
	}
	}
	\caption{\textsl{Source function stratification of k3 (black), k2v (blue) and k2r (red) in four snapshots of the $\delta = 3$, $E_{c} = 25$~keV simulation. The coloured symbols show the formation height of each component. Top left shows $t=0$~s, and top right shows $t=40$~s, during the ribbon front phase, where the difference in magnitude of the source functions has increased. Bottom panels show times during the bright ribbon phase. The dashed grey line in each panel is the Planck function at the wavelength of k3  (the dotted grey line is $t=0$~s). The yellow lines show the electron density stratification (the dotted yellow is $t=0$~s). }}
	\label{fig:sfnsnapshots}
\end{figure*}

	In response to the increased electron density and temperature the source function, $S_{\lambda}$, of each component is elevated. Larger electron densities also mean that the k2 $S_{\lambda}$ more closely tracks the background temperature rise of the chromosphere. Effectively, during the weak-heating phase the k2 peaks' $S_{\lambda}$ become larger, more so than the k3 $S_{\lambda}$ which does not show a commensurately large increase. The difference between the k2 and k3 $S_{\lambda}$ therefore rises, and since the emergent intensity is, roughly, related to $S_{\lambda}$ at the height of $\tau_{\lambda} = 1$ (the Eddington-Barbier relation), the intensity of k2 increases more than k3, resulting in the deepening of the \ion{Mg}{2} central reversal. Figure~\ref{fig:sfnsnapshots} shows $S_{\lambda}(z)$ in several snapshots from the $\delta = 3$, $E_{c} = 25$~keV simulation to illustrate these points. In that figure the heights of k2v, k2r and k3 formation are indicated, from which varying differences in $S_{\lambda}$ can be inferred. 
	
	Later in the flare once the injected energy flux has increased two effects happen that result in the central reversal becoming more shallow. The upper chromosphere conditions become sufficient to lessen the differences between k2 and k3 $S_{\lambda}$ since the k3 $S_{\lambda}$ also tracks the Planck function more closely. At the same time, the chromosphere becomes compressed and the formation height difference between k2 and k3 reduces substantially and with that the differences in k2 and k3 $S_{\lambda}$. 
	
\begin{figure}
	\centering 
	\vbox{
	\hbox{
	\hspace{0in}
	\subfloat{\includegraphics[width = 0.5\textwidth, clip = true, trim = 0.cm 0.cm 0.cm 0.cm]{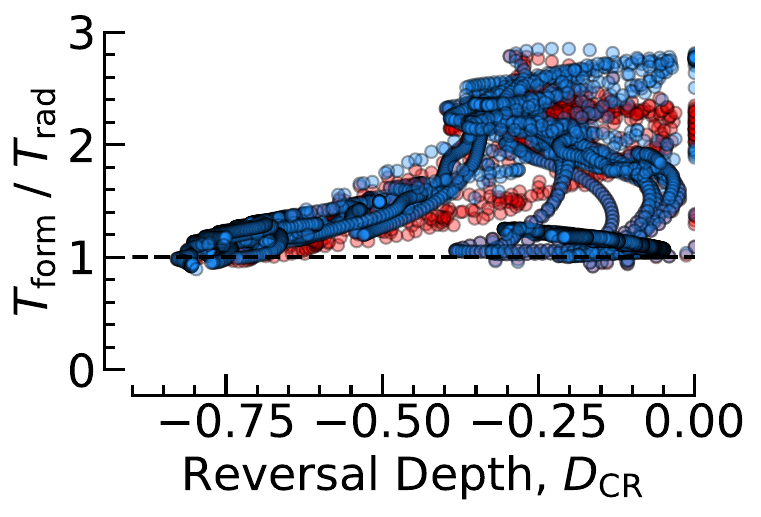}}
	}
	}
	\caption{\textsl{Ratio of the plasma temperature in the k2v (blue circles) or k2r (red circles) formation region to the k2v or k2r radiation temperature, as a function of the depth of the central reversal. The dashed line shows a ratio of unity, which indicates that the radiation temperature is reflective of the local plasma temperature.}}
	\label{fig:sfnratio}
\end{figure}

	\cite{2013ApJ...772...90L} found that in the quiet Sun scenario the intensity of the k2 peaks can constrain the plasma temperature in the mid chromosphere, since the temperature at their formation height is only around 500K or so larger than the intensity expressed as a radiation temperature. While this does not hold as strongly during flares, with the ratio of plasma temperature to radiation temperature being $T_{\mathrm{form}} / T_{\mathrm{rad}} \sim 1-3$, it is the case that when the central reversal has deepened, this ratio is closer to unity. So, during the ribbon front phase the radiation temperature of the k2 peaks can acts as a rough guide to the temperature at their formation height. Figure~\ref{fig:sfnratio} shows $T_{\mathrm{form}} / T_{\mathrm{rad}}$ for the k2v (blue circles) and k2r (red circles), as a function of $D_{\mathrm{CR}}$. This ratio lies between 1-1.5 for profiles with central reversals deeper than the pre-flare. 
	
\section{Discussion}
\begin{table}
\centering
\begin{tabular}{| c | c c c c c |} \toprule
    \textbf{Simulation} & \textbf{\ion{He}{1}} & $\boldsymbol{D_{\mathrm{CR}}}$ & $\boldsymbol{S_{\mathrm{k2}}}$ & $\boldsymbol{A_{\mathrm{k2}}}$ & $\boldsymbol{V_{\mathrm{k3,Dopp}}}$  \\ 
    
    \textbf{Harder to Softer} & \textbf{Dims} & \textbf{Incr.} &\textbf{Incr.}  & \textbf{Red} & \textbf{Blue}  \\

    \toprule
    $\boldsymbol{\delta = 3, E_{c} = 30}$~\textbf{keV} & \cmark               &  \cmark &  \cmark              &   \cmark              & \cmark               \\
    $\boldsymbol{\delta = 5, E_{c} = 30}$~\textbf{keV} & \cmark               &  \cmark &  \cmark              &  $\blacksquare$ &  \cmark              \\
    $\boldsymbol{\delta = 3, E_{c} = 25}$~\textbf{keV} & \cmark               &  \cmark &  \cmark              &  \cmark               & \cmark               \\
    $\boldsymbol{\delta = 3, E_{c} = 20}$~\textbf{keV} & \cmark               &  \cmark &  \cmark              &  $\blacksquare$ & \cmark               \\
    $\boldsymbol{\delta = 5, E_{c} = 25}$~\textbf{keV} & \cmark               &  \cmark &  \cmark              &  $\blacksquare$ & $\blacksquare$  \\
    $\boldsymbol{\delta = 7, E_{c} = 30}$~\textbf{keV} & \cmark               &  \cmark &  \cmark              &  $\blacksquare$ & $\blacksquare$  \\
    $\boldsymbol{\delta = 3, E_{c} = 15}$~\textbf{keV} & $\blacksquare$ &  \cmark &  \cmark              &  $\blacksquare$ & $\blacksquare$  \\
    $\boldsymbol{\delta = 3, E_{c} = 10}$~\textbf{keV} &                          &              &                           &                           &                            \\
    $\boldsymbol{\delta = 5, E_{c} = 20}$~\textbf{keV} & $\blacksquare$ &  \cmark &  \cmark              &  $\blacksquare$ & $\blacksquare$  \\
    $\boldsymbol{\delta = 7, E_{c} = 25}$~\textbf{keV} & $\blacksquare$ &  \cmark &  \cmark              &  $\blacksquare$ & $\blacksquare$  \\
    $\boldsymbol{\delta = 5, E_{c} = 15}$~\textbf{keV} &                          &              &                           &                           &                            \\
    $\boldsymbol{\delta = 7, E_{c} = 20}$~\textbf{keV} &                          &  \cmark &  $\blacksquare$ &  $\blacksquare$ & $\blacksquare$  \\
    $\boldsymbol{\delta = 5, E_{c} = 10}$~\textbf{keV} &                          &              &                           &                           &                            \\
    $\boldsymbol{\delta = 7, E_{c} = 15}$~\textbf{keV} &                          &              &                           &                           &                            \\
    $\boldsymbol{\delta = 7, E_{c} = 10}$~\textbf{keV} &                          &              &                           &                           &                           \\
   \bottomrule 
    \end{tabular}
    \caption{{{Persistence of ribbon front characteristics through the duration of the weak-heating phase. Simulations are ordered by the number of $E=50$~keV electrons in the distribution at $t=0$~s, from largest to smallest. A tick mark indicates that the characteristic is present for the duration of the weak-heating phase, and a black square ($\blacksquare$) indicates that it is present for the a notable ($> \sim 50\%$) portion of the weak-heating phase. An empty cell means that the characteristic is either not present or disappears fairly quickly.}}}
    \label{tab:rfsummary}
\end{table}

The combined results are summarised in Table~\ref{tab:rfsummary}, which notes which ribbon front characteristics persist for the whole duration of the weak-heating phase (tick marks), or for some significant portion of it ($>\sim50\%$; black squares): \ion{He}{1} 10830~\AA\ dimming, increasing reversal depth, increasing k2 separation, red k2 asymmetry, and blueshifted k3 line core. Simulations have been ordered by decreasing number of $50$~keV electrons in the distribution at $t=0$~s, a measure of hardest to softest electron beams. It is immediately obvious that the harder nonthermal electron distributions are more likely to have sustained ribbon fronts, and that three properties are better metrics of the lifetime of any weak-heating phase, which are \ion{He}{1} dimming, increasing $D_{\mathrm{CR}}$, and increasing $S_{\mathrm{k2}}$. 

Our prior study \citep{2023ApJ...944..104P} found that there was generally an observed evolution from ribbon front to bright ribbon characteristics, but that long duration weak-heating by itself could not produce this in simulations. Emission from the TR (\ion{O}{4} 1401~\AA) and hot flare plasma (\ion{Fe}{21} 1354.1~\AA) was synthesised (shown in Appendix~\ref{sec:tr_profiles_maingrid}), from which we found that most of the simulations that had persistent ribbon fronts in Table~\ref{tab:rfsummary} later produced strong TR emission and drive explosive chromospheric evaporation once the energy flux became large enough (i.e. after a certain time in the main heating phase). By increasing the energy flux after the weak-heating phase our new simulations could mimic the observed transition, such that the two heating regimes identified in \cite{2023ApJ...944..104P} are actually likely not independent. It was not necessary for a sustained ribbon front to exist prior to our simulations exhibiting explosive evaporation, with both blueshifted and strong \ion{Fe}{21} present in the softer simulations. Of note is that redshifted \ion{O}{4} emission was not produced in any of the ribbon front simulations, with redshifted TR only present later in the flare. Finally, if the main heating phase is modest (e.g. the first 50s or so of the main heating phase in our experiments) there is not explosive evaporation, which is consistent with the fact that not every ribbon front transitioned to a bright ribbon that also had explosive evaporation in \cite{2023ApJ...944..104P}. The post-ribbon front behaviour will vary depending on the characteristics of the post-weak heating phase. 

It is clear that the TR and hot flare plasma can be used as additional constraints of our model of an evolving energy flux, and we stress that our experiments here were overly simplified. The timing of the post-weak-heating phase was arbitrary, as was the magnitude of the energy flux. An allowance should also be made for $\delta$ and $E_{c}$ to vary over the lifetime of heating. Additional mechanisms of energy transport could also be present, for example heat flux from \textsl{in-situ} coronal heating, MHD waves, or the retracting loops \citep[e.g.][]{2023ApJ...944..147A}. Despite the simplifications made for our exploratory study, the general picture that has emerged is a compelling explanation for various observed ribbon behaviours. 

The lifetime of ribbon fronts is an almost direct proxy of the lifetime of a weak-heating phase, followed by more typical flare energy fluxes. Where we do not see ribbon fronts characteristics, two regimes exist. One is that there is a rather soft nonthermal electron distribution during the weak-heating phase, in which case we see a modest intensity increase in chromospheric and TR spectra that builds over some timeframe before the onset of explosive chromospheric evaporation during the strong-heating phase. The second is that in those locations there is no weak-heating phase, or a vanishingly small one, with only a large energy flux deposited into the chromosphere. Detailed study of the TR and coronal ribbon front spectral characteristics would likely provide important constraints on these different scenarios. Indeed, the width of the \ion{Si}{4} resonance lines (that form in the transition region) has been observed to increase prior to the increase of line intensity, from which it was inferred that turbulence was increasing prior to the main burst of flare energy deposition \citep{Jeffreyeaav2794}.

Our explanation as to the origin of both the \ion{He}{1} 10830~\AA\ and \ion{Mg}{2} ribbon front characteristics, determined by studying each line's formation properties, essentially boils down to the necessity to keep the upper chromosphere from getting too hot and dense. An alternative approach to forward modelling the flare chromosphere is to perform spectral inversions to obtain a model atmosphere that produces synthetic spectra consistent with the observations. This is a non-trivial challenge given the complexity of flare spectra, but some recent successful inversions of ribbon front \ion{Mg}{2} k line spectra using the \texttt{STiC} code \citep{2019A&A...623A..74D} found a local maximum of temperature in the mid-chromosphere \citep{2023FrASS..1033429S}. In bright ribbon \ion{Mg}{2} k line spectra, \cite{2023FrASS..1033429S} found a larger temperature in the upper chromosphere. These findings are consistent with our picture of the evolution of heating rates. 

What then causes this energy flux evolution, and can theories of flare energy release and energy transport explain this seemingly necessary pattern? We do not tackle this difficult question here, but do moot some speculative possibilities.

One is that two different particle acceleration mechanisms act in sequence, the first producing only a modest amount of nonthermal electrons with high average energies that produces the ribbon fronts, followed by a second mechanism some time later that accelerates a much greater number of electrons with a broader distribution of energies that produces the bright ribbons. 

Another is that there is some trapping of particles following reconnection that suppresses their transport for some time period, with some leakage of a small number of high energy particles before the bulk precipitate to the chromosphere. Mechanisms that could suppress transport include magnetic mirrors, magnetic traps \citep[e.g.][which are related to acceleration of particles by termination shocks]{2019ApJ...887L..37K}, Double Layers \citep[][]{2013ApJ...778..144L}, magnetic turbulence, or scattering by waves.

Indeed, this could be consistent with the fact that in the two-ribbon flares observed by \cite{2016ApJ...819...89X}, one of the ribbons showed only bright sources with no ribbon fronts. Similar observations of conjugate flare sources (including conjugate HXR footpoints) with asymmetric intensities could suggest asymmetric energy deposition \citep[e.g.][and references therein]{2009ApJ...693..847L, 2012ApJ...756...42Y, 2015ApJ...800...54G}. 

So far we have discussed evolving nonthermal electron distributions originating in the corona. \gsk{Though \cite{2021ApJ...912..153K} suggests that \ion{He}{1} 10830~\AA\ dimming demands nonthermal collisional ionisation, and hence energetic electrons, the possibility exists these could originate locally in the chromosphere}. \cite{2008ApJ...675.1645F} posited such a model, whereby Alfv\'en waves propagate downward from the reconnection region and accelerate chromospheric electrons. The role that this model could have in chromospheric heating has been explored \citep[e.g.][]{2013ApJ...765...81R,2016ApJ...818L..20R,2018ApJ...853..101R,2016ApJ...827..101K}, which did demonstrate that varying wave parameters could preferentially heat the mid-chromosphere rather than the upper chromosphere, but work needs to be done to determine its role in particle acceleration. See \cite{2023arXiv231102144R} for a recent review of the potential role of Alfv\'en waves in solar flares.


\section{Summary \& Conclusions}\label{sec:conc}
 
Solar flare ribbon fronts are the sites of initial flare energy deposition into the chromosphere, and as such are of great interest in understanding the energy release and transport following magnetic reconnection in the corona. We performed a series of numerical experiments in order to determine what we can learn about those processes from the lifetimes of ribbon fronts, which can range a few to many dozens of seconds. Our earlier experiments attempting to reproduce \ion{He}{1} 10830~\AA\ dimming \citep{2021ApJ...912..153K} and the characteristics of the \ion{Mg}{2} NUV, \ion{Fe}{21} and \ion{O}{4} spectra \citep{2023ApJ...944..104P}, though able to synthesising spectra consistent with ribbon fronts, were unable to reproduce the lifetimes. Both \ion{He}{1} 10830~\AA\ and \ion{Mg}{2} NUV rapidly transitioned to behaviours more consistent with the brighter main ribbon regions, with ribbon front lifetimes of only a few seconds. 

Here we addressed the question of lifetimes with a rather simple model in which an extended period of weak energy injection ($5\times10^{8}$~erg~s$^{-1}$~cm$^{-2}$), by a distribution of nonthermal particles with a range of properties, precedes the main energy injection phase. Those models were able to produce synthetic spectra which exhibited ribbon front behaviours for significantly longer than our original experiments. In fact, we found that the duration of the ribbon fronts was directly related to the duration of the weak-heating phase, so long as the injected nonthermal electron distribution was relatively hard (see Table~\ref{tab:rfsummary}). 

From a detailed analysis of the \ion{He}{1} 10830~\AA\  and \ion{Mg}{2} NUV spectra we were able to determine (1) what causes the ribbon front behaviour, (2) at what point the transition to bright ribbon regions occurs, and (3) why a relatively hard electron energy distribution is required. The answer is, in short, that in ribbon fronts sources the mid chromosphere is perturbed more than the upper chromosphere (which happens with a weak flux of higher energy electrons than can penetrate more deeply). This drives electron density and temperature enhancements, and introduces nonthermal collisional ionisation and gentle evaporation (upflowing plasma). \gsk{Once the upper chromosphere begins to be more strongly perturbed by the flare, the atmosphere becomes hotter and denser, and mass flows subside or become redshifts. This occurs either due to an increase of injected energy or because the nonthermal electron distribution is softer and begins to spread more evenly through the mid-upper chromosphere.  \ion{He}{1} 10830~\AA\ is then driven into emission, the depth of the \ion{Mg}{2} central reversal reduces and becomes redshifted.} The analysis of the \ion{Mg}{2} NUV formation further suggests that some of the results of \cite{2013ApJ...772...90L} could be broadly applicable during dynamic events also, in a qualitative sense at least.

All together, ribbon front observations in multiple wavelengths provide very strong constraints both on the plasma properties of the flare chromosphere and on the properties of the energy transport. They can act as a guide to the relative hardness and energy flux of populations of nonthermal particles on small spatial scales, and also as a sort of temperature and electron density gauge. 

Our results also demand an explanation as to why there is seemingly two stages of energy transport into single footpoints, which can in some sources be relatively long in duration. Does this imply trapping of particles with some leakage of high energy electrons that precede the bulk bombardment, or are there two distinct periods of energy release and particle acceleration?

In order to fully explore, and exploit, the potential diagnostics of flare energy deposition offered by flare ribbon fronts we suggest (1) that a detailed exploration akin to \cite{2023ApJ...944..104P} should be performed that includes additional spectral lines in order to tighten constraints (including on microturbulence through study of optically thin lines like \ion{O}{1}, \ion{O}{4}, or \ion{Si}{4});  (2) that we determine how \ion{He}{1} D3 behaves during ribbon fronts, and if its relative behaviour with \ion{He}{1} 10830~\AA\ can further constrain the nonthermal electron population; and (3) that our analysis should be repeated with a range of pre-flare atmospheres to better understand what impact that has on constraining the nonthermal electron beam properties (for example, does a different electron stratification mean that the delineation between electron beams capable or not capable of producing ribbon fronts shift?).  \\


\textsc{Acknowledgments:} \small{We appreciate the comments of the anonymous referee. GSK, VP, and YX acknowledge financial support from the NASA ROSES Heliophysics Supporting Research program (Grant\# 80NSSC19K0859). GSK also acknowledges the NASA ROSES Early Career Investigator Program (Grant\# 80NSSC21K0460). VP also acknowledges support from NASA ROSES Heliophysics Guest Investigator program (Grant \# 80NSSC20K0716) and from the NASA's IRIS mission (Grant \# NNG09FA40C). J.C.A. acknowledges funding from NASA’s Heliophysics Innovation Fund (part of the NASA Internal Scientist Funding Model), and NASA’s Heliophysics Supporting Research program.  Y.X also acknowledges NSF AGS 2228996, 2309939. This manuscript benefited from discussions held at a meeting of International Space Science Institute team: ``Interrogating Field-Aligned Solar Flare Models: Comparing, Contrasting and Improving,'' led by Dr. G. S. Kerr and Dr. V. Polito. IRIS is a NASA small explorer mission developed and operated by LMSAL with mission operations executed at NASA Ames Research Center and major contributions to downlink communications funded by the Norwegian Space Center (NSC, Norway) through an ESA PRODEX contract. Resources supporting this work were provided by the
NASA High-End Computing (HEC) Program through the
NASA Advanced Supercomputing (NAS) Division at Ames
Research Center.}

\bibliographystyle{aasjournal}
\bibliography{Kerr_etal_He10830_LongDur}


\newpage
\appendix
\section{\ion{He}{1} and \ion{Mg}{2} Spectra From the Main Grid of Flare Simulations}\label{sec:helium_profiles_maingrid}
\begin{figure}
	\centering 
	\vbox{
	\hbox{
	\subfloat{\includegraphics[width = 0.15\textwidth, clip = true, trim = 0.cm 0.cm 0.cm 0.cm]{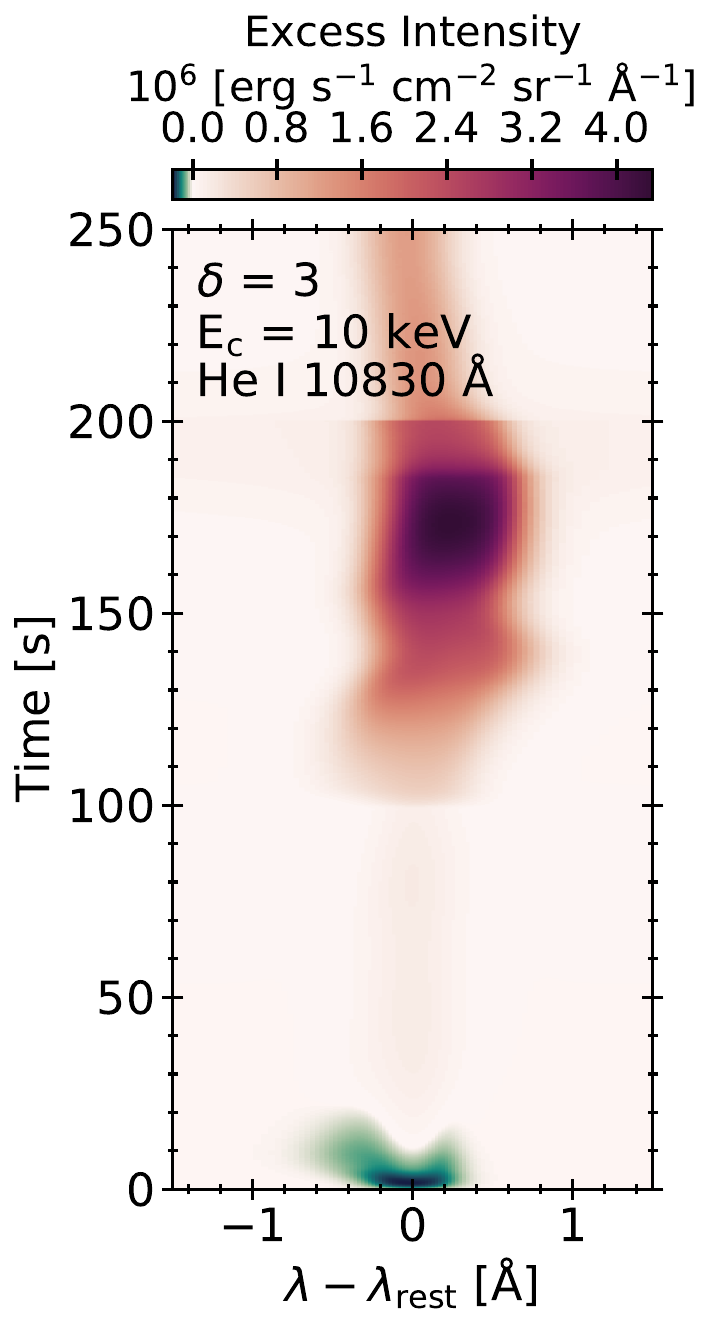}}
	\subfloat{\includegraphics[width = 0.15\textwidth, clip = true, trim = 0.cm 0.cm 0.cm 0.cm]{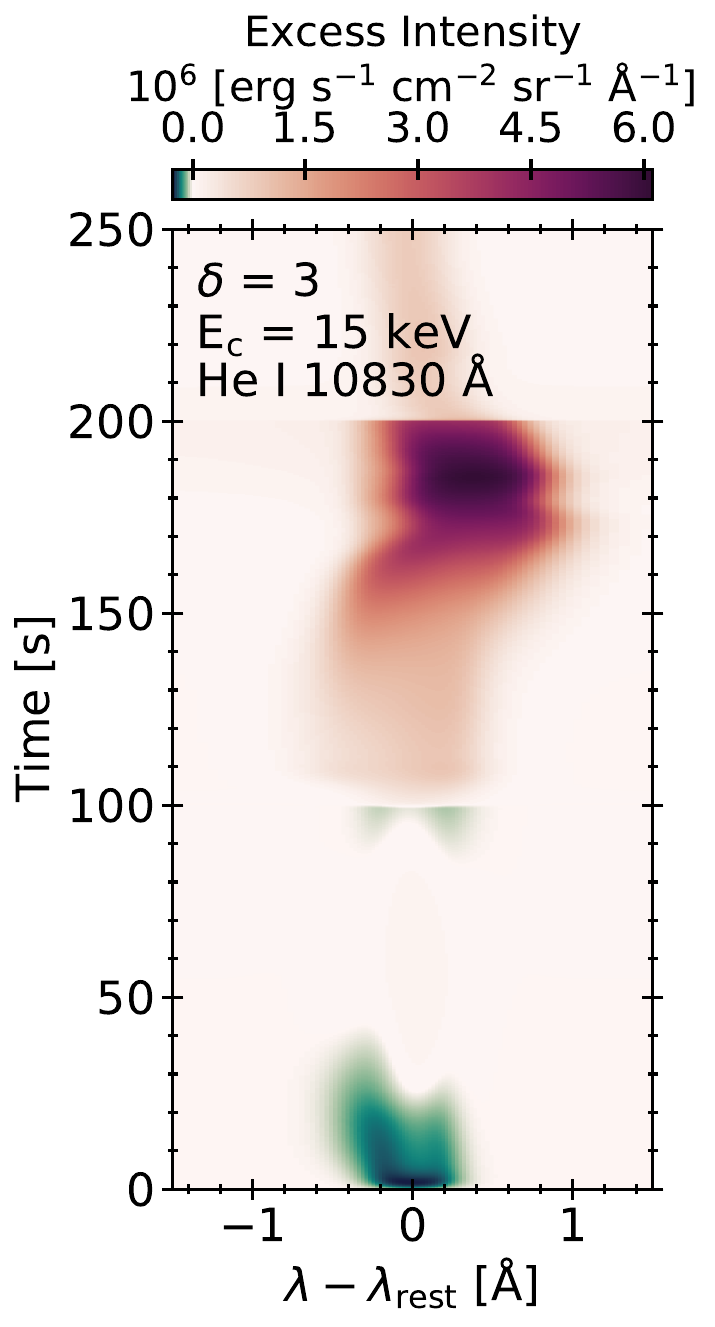}}
	\subfloat{\includegraphics[width = 0.15\textwidth, clip = true, trim = 0.cm 0.cm 0.cm 0.cm]{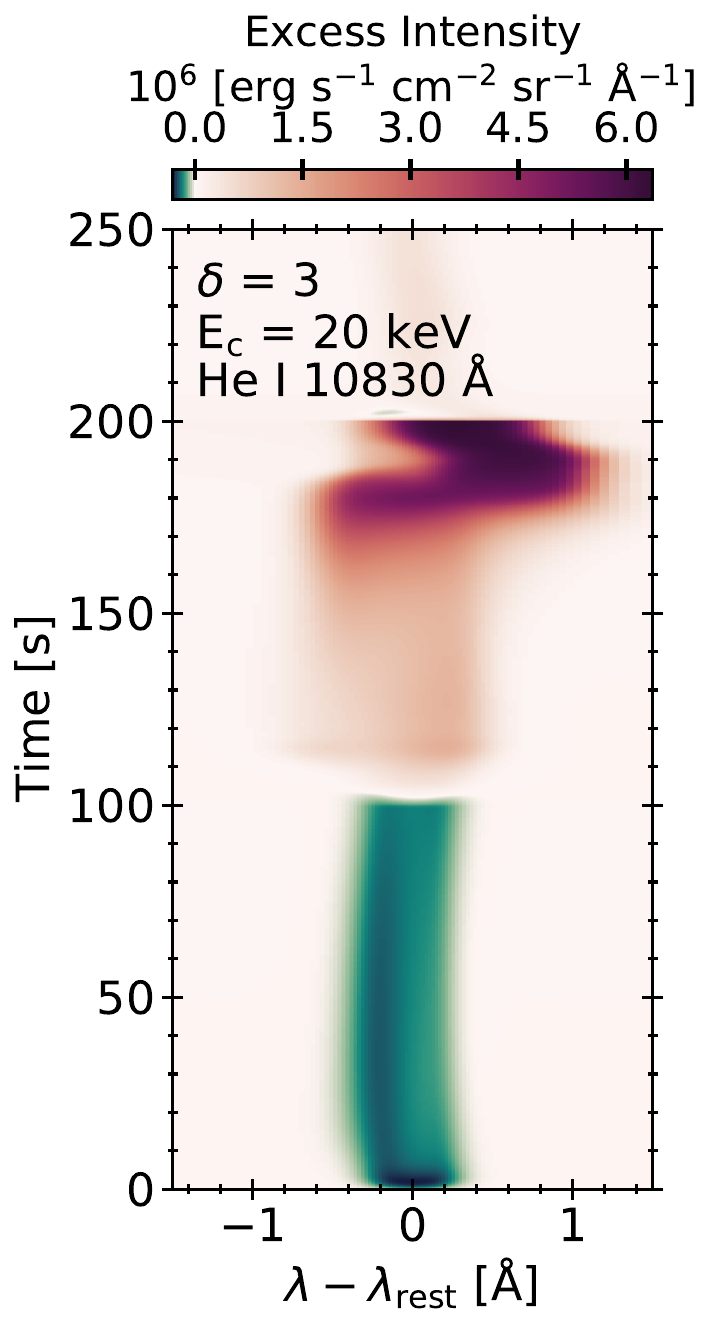}}
	\subfloat{\includegraphics[width = 0.15\textwidth, clip = true, trim = 0.cm 0.cm 0.cm 0.cm]{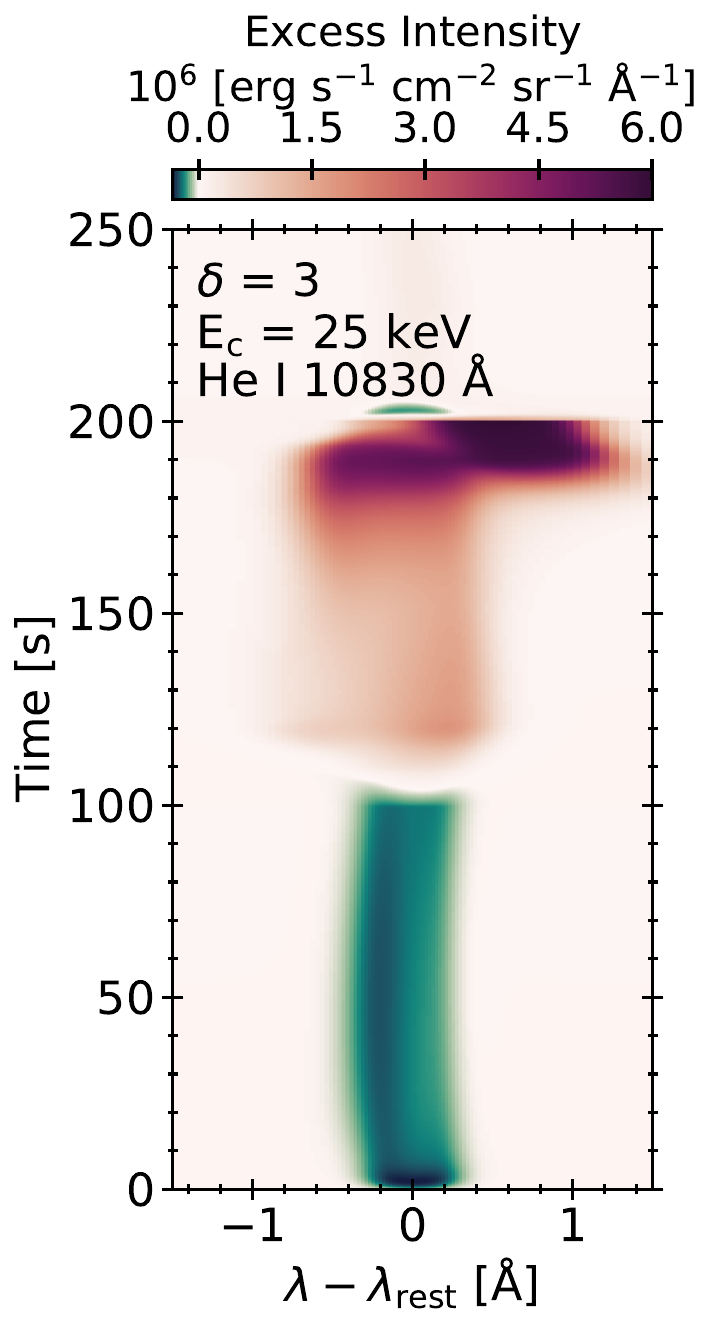}}
	\subfloat{\includegraphics[width = 0.15\textwidth, clip = true, trim = 0.cm 0.cm 0.cm 0.cm]{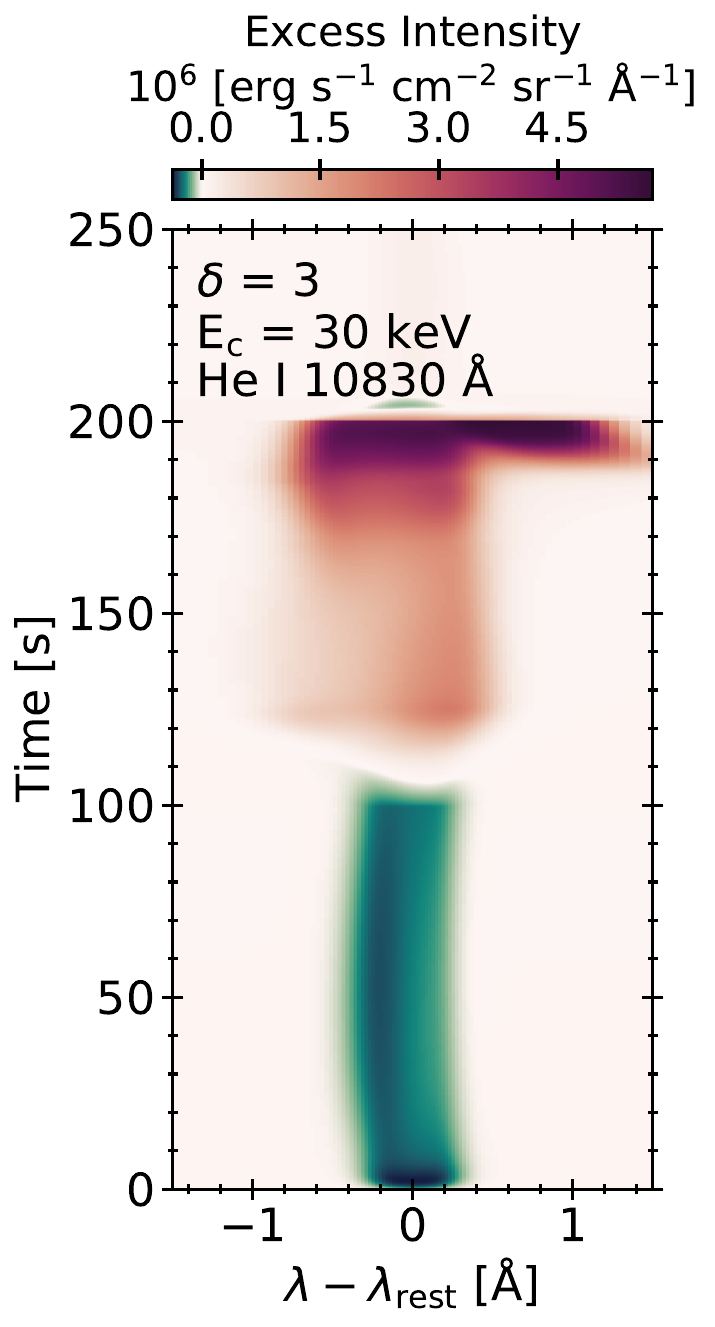}}
	}
	}
	\vbox{
	\hbox{
	\subfloat{\includegraphics[width = 0.15\textwidth, clip = true, trim = 0.cm 0.cm 0.cm 0.cm]{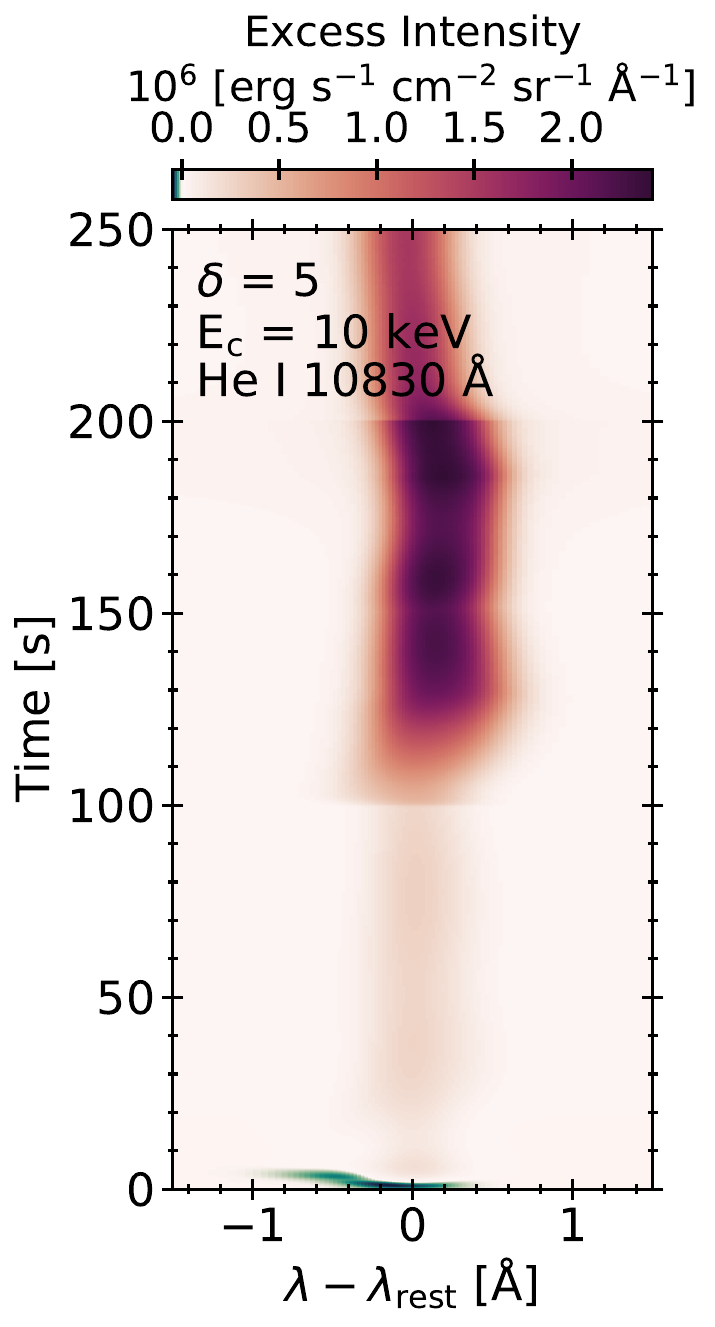}}
	\subfloat{\includegraphics[width = 0.15\textwidth, clip = true, trim = 0.cm 0.cm 0.cm 0.cm]{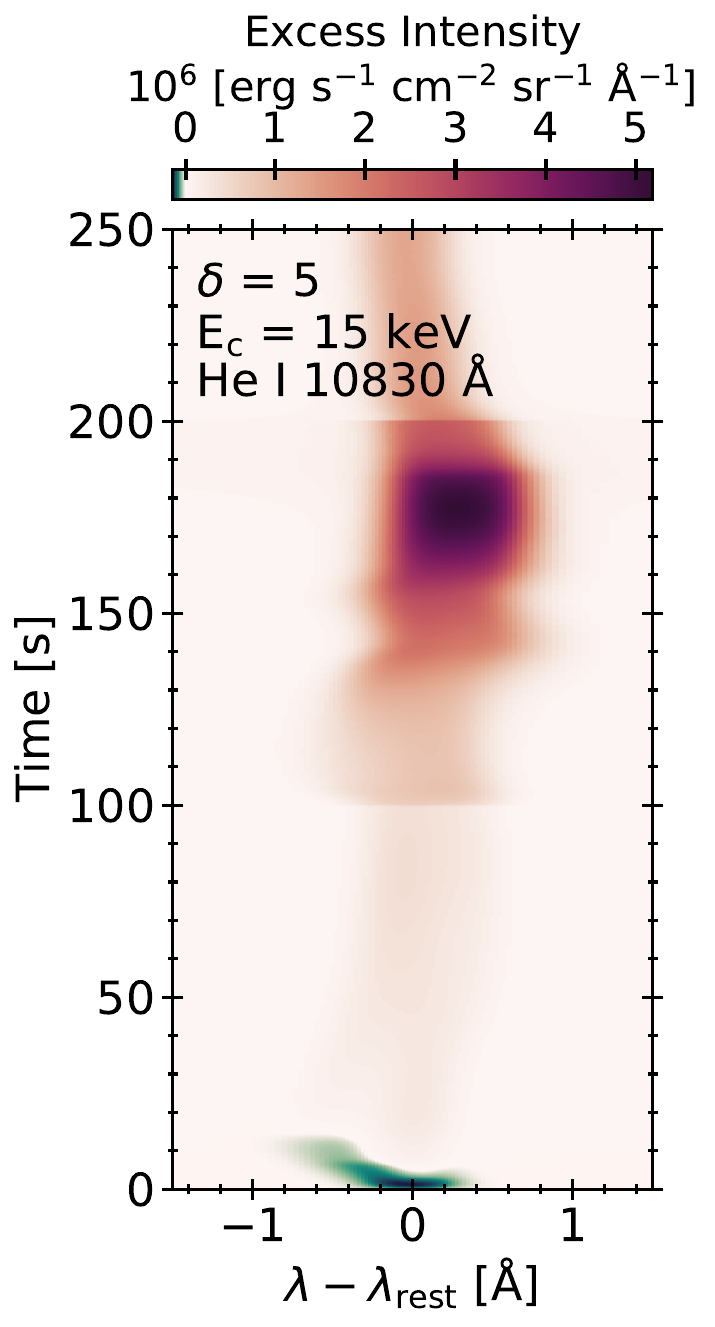}}
	\subfloat{\includegraphics[width = 0.15\textwidth, clip = true, trim = 0.cm 0.cm 0.cm 0.cm]{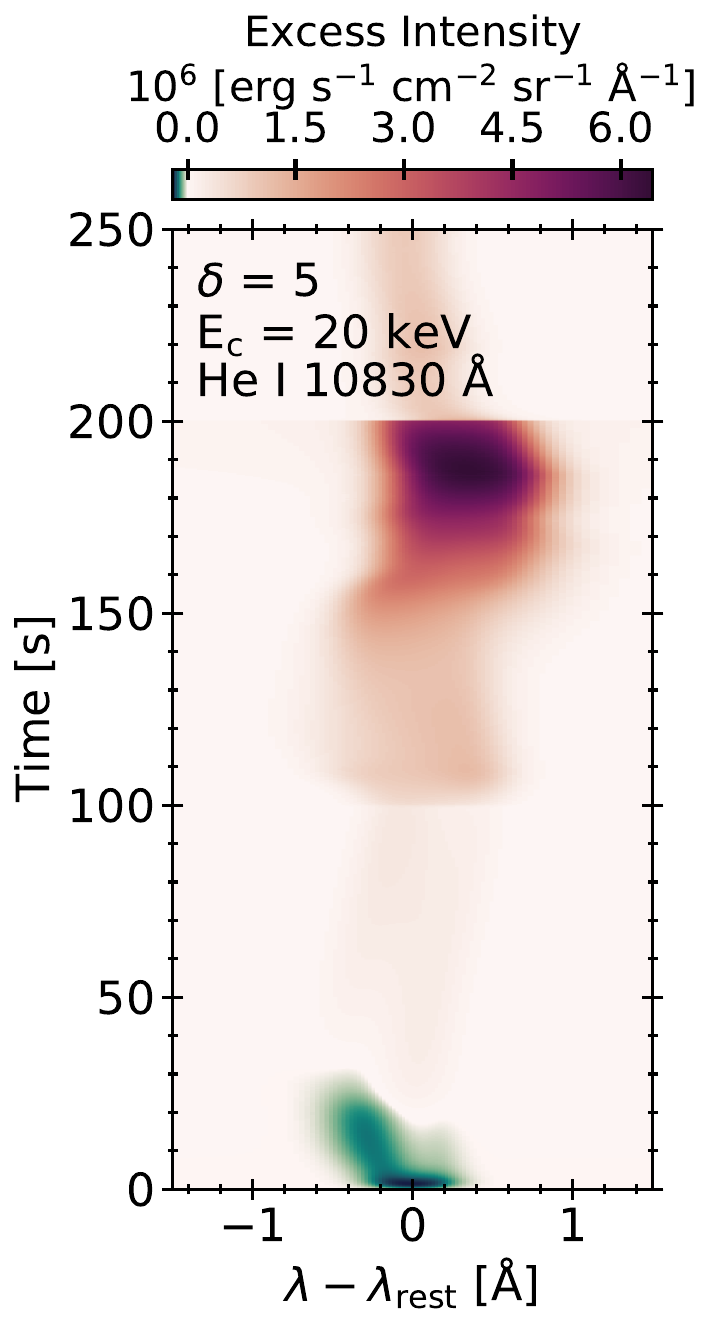}}
	\subfloat{\includegraphics[width = 0.15\textwidth, clip = true, trim = 0.cm 0.cm 0.cm 0.cm]{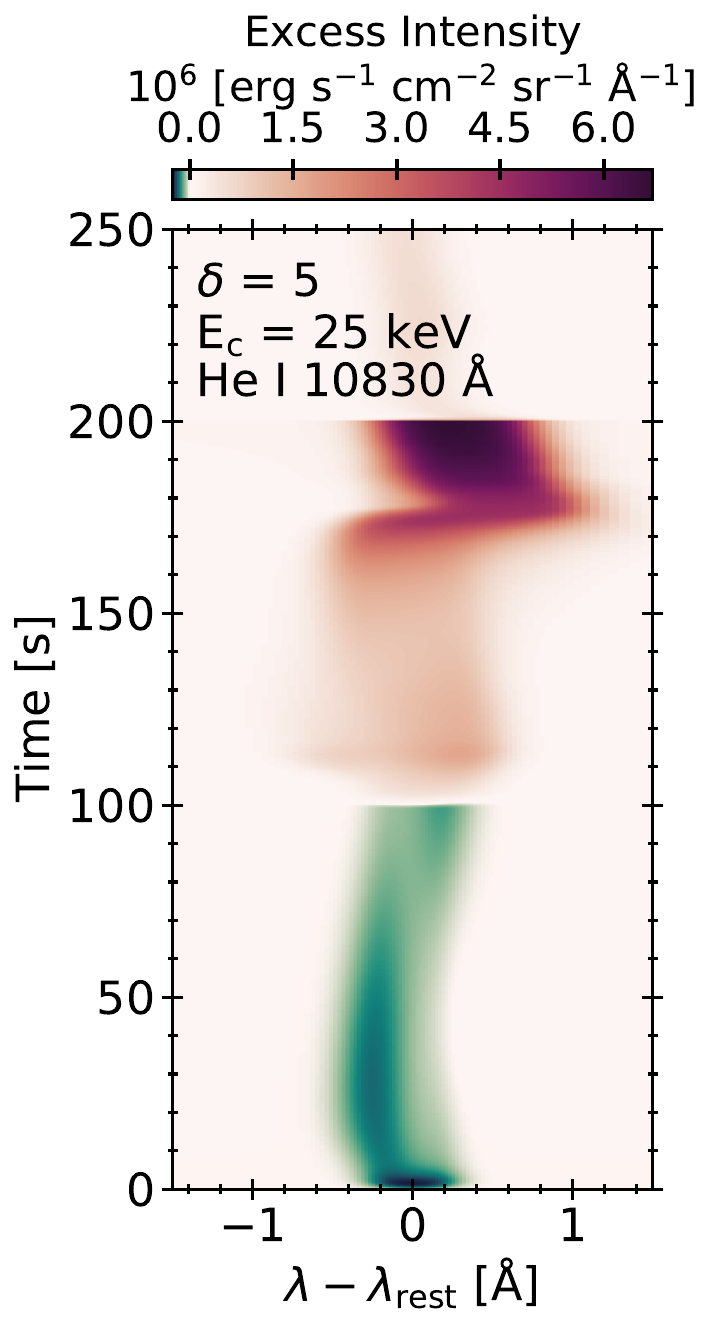}}
	\subfloat{\includegraphics[width = 0.15\textwidth, clip = true, trim = 0.cm 0.cm 0.cm 0.cm]{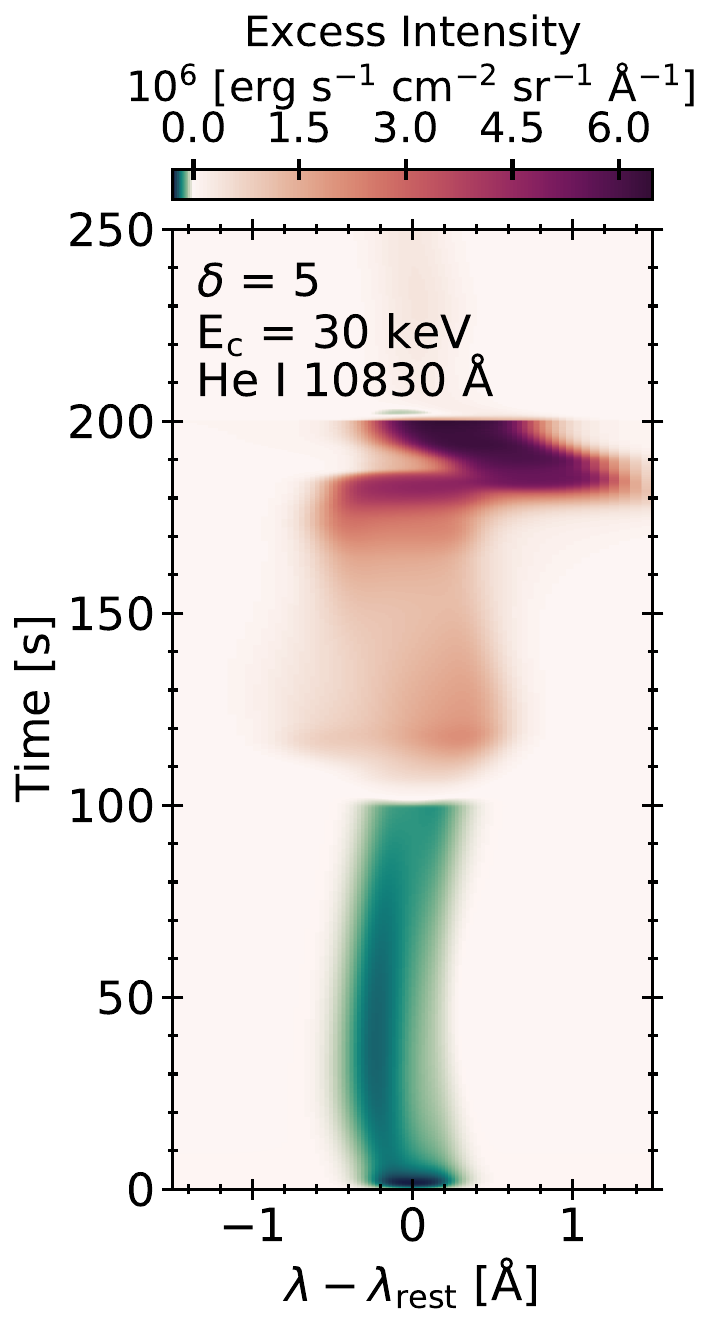}}
	}
	}
	\vbox{
	\hbox{
	\subfloat{\includegraphics[width = 0.15\textwidth, clip = true, trim = 0.cm 0.cm 0.cm 0.cm]{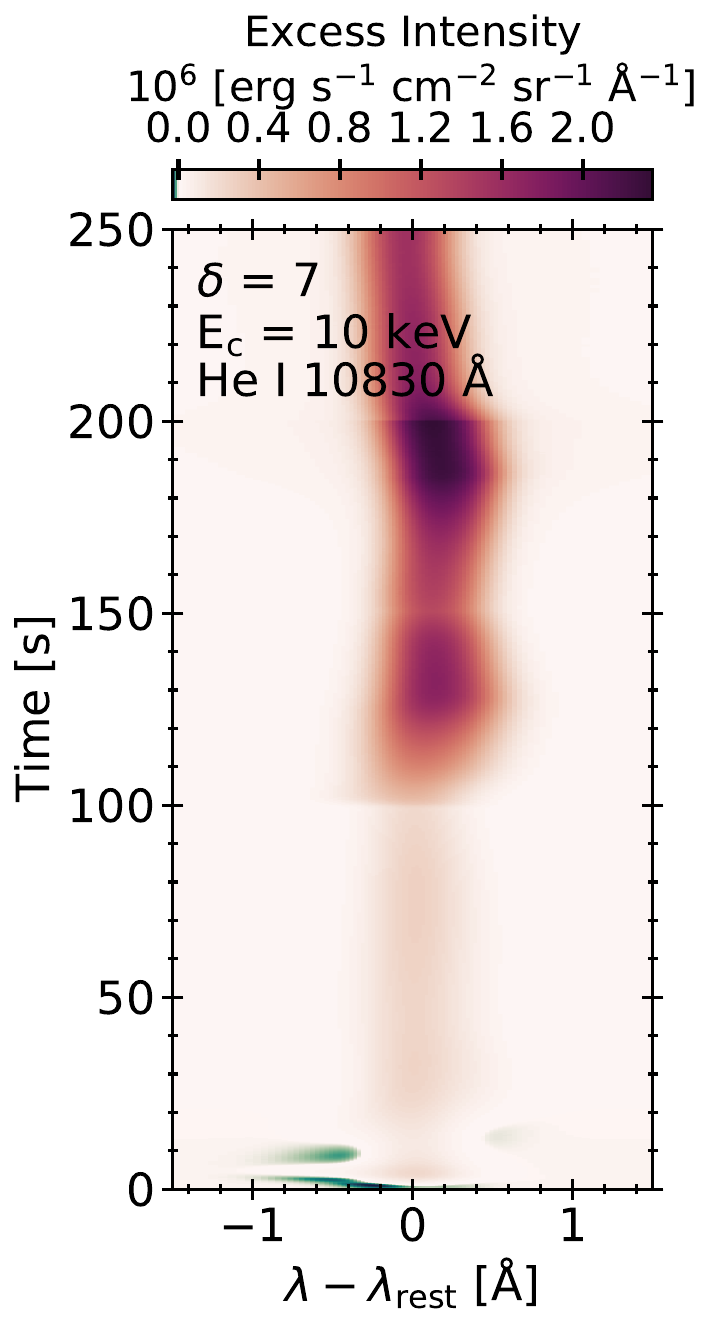}}
	\subfloat{\includegraphics[width =0.15\textwidth, clip = true, trim = 0.cm 0.cm 0.cm 0.cm]{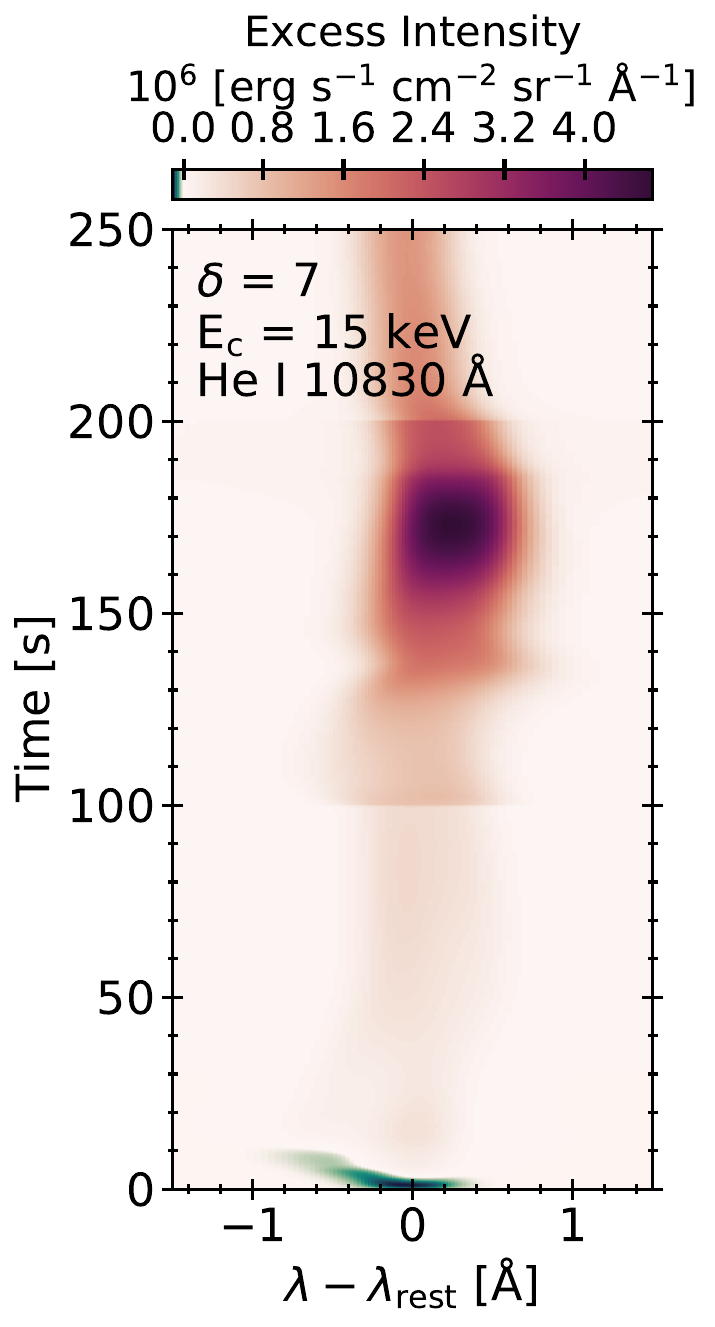}}
	\subfloat{\includegraphics[width = 0.15\textwidth, clip = true, trim = 0.cm 0.cm 0.cm 0.cm]{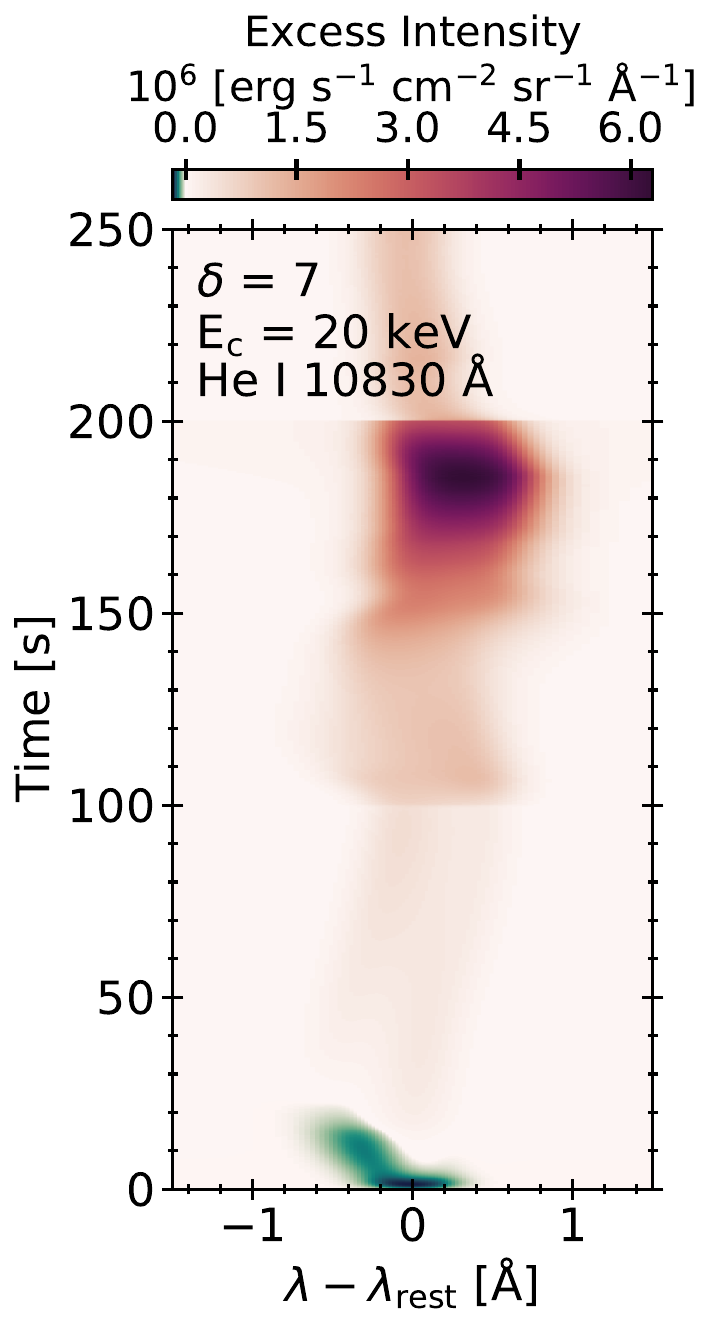}}
	\subfloat{\includegraphics[width = 0.15\textwidth, clip = true, trim = 0.cm 0.cm 0.cm 0.cm]{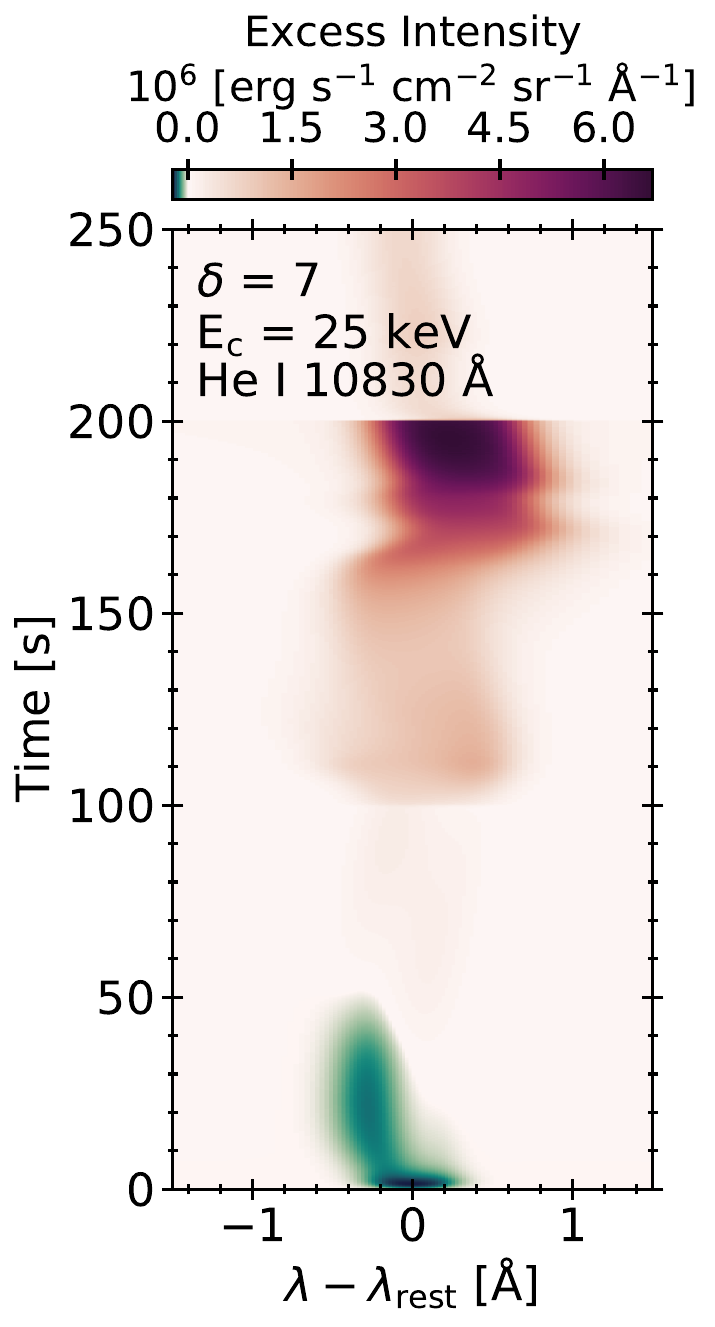}}
	\subfloat{\includegraphics[width = 0.15\textwidth, clip = true, trim = 0.cm 0.cm 0.cm 0.cm]{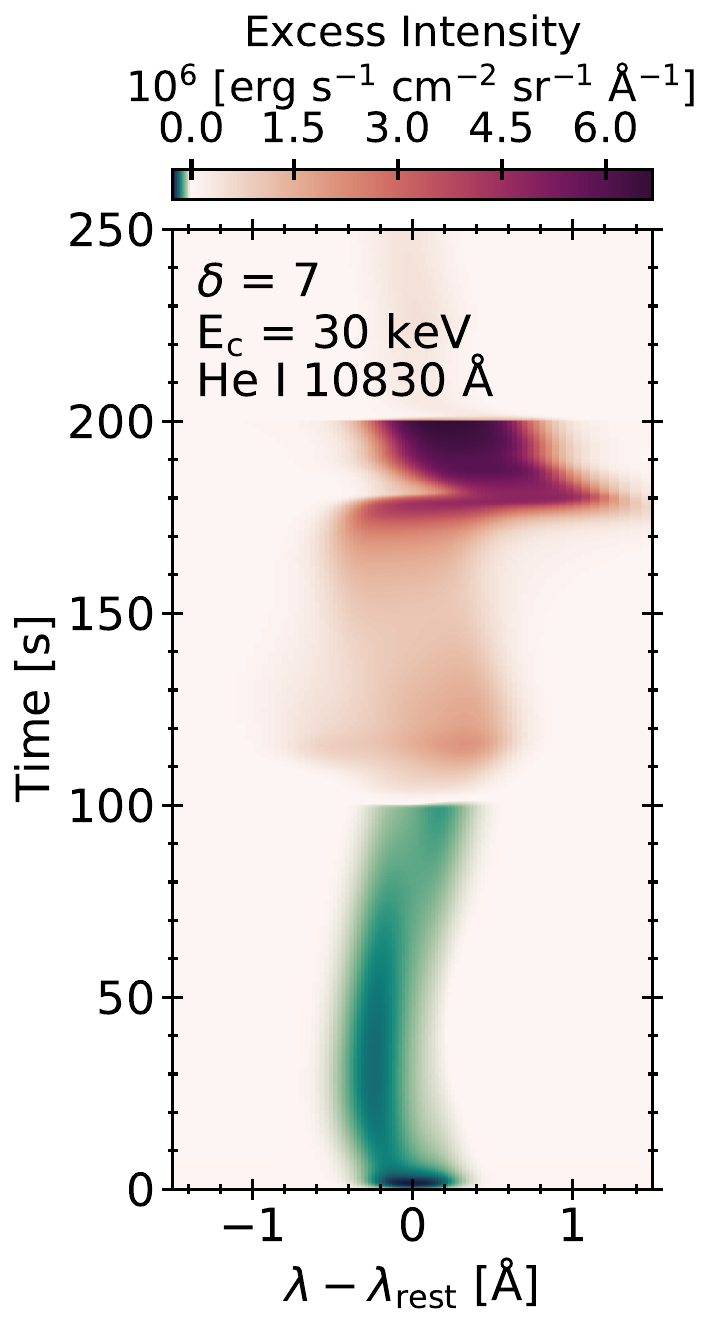}}
	}
	}
		\caption{\textsl{Synthetic pre-flare subtracted \ion{He}{1} 10830~\AA\ line profiles as a function of time in each flare simulation from the main parameter study. The top row is $\delta = 3$, the middle row is $\delta = 5$, and the bottom row is $\delta = 7$. From left to right the low energy cutoffs are $E_{c} = [10, 15, 20, 25, 30]$~keV, such that the hardest nonthermal electron distribution is the top right, and the softest is bottom left. Redder colours are positive (i.e. enhancements) and greener colours are are negative (i.e. dimmings), and note that the scales on each side of the zero line are not uniform, so that the range of negative values is very much smaller than the range of positive values.}}
	\label{fig:helium_profiles_maingrid}
\end{figure}

\begin{figure}
	\centering 
	\vbox{
	\hbox{
	\subfloat{\includegraphics[width = 0.15\textwidth, clip = true, trim = 0.cm 0.cm 0.cm 0.cm]{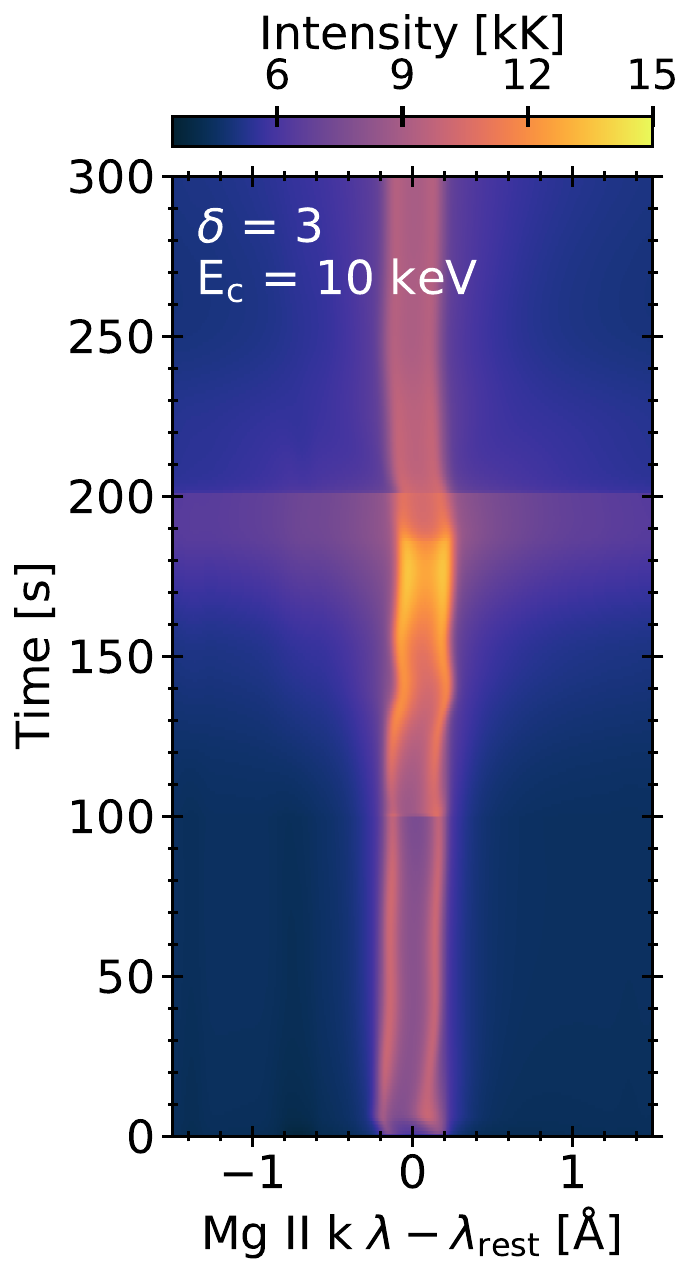}}
	\subfloat{\includegraphics[width = 0.15\textwidth, clip = true, trim = 0.cm 0.cm 0.cm 0.cm]{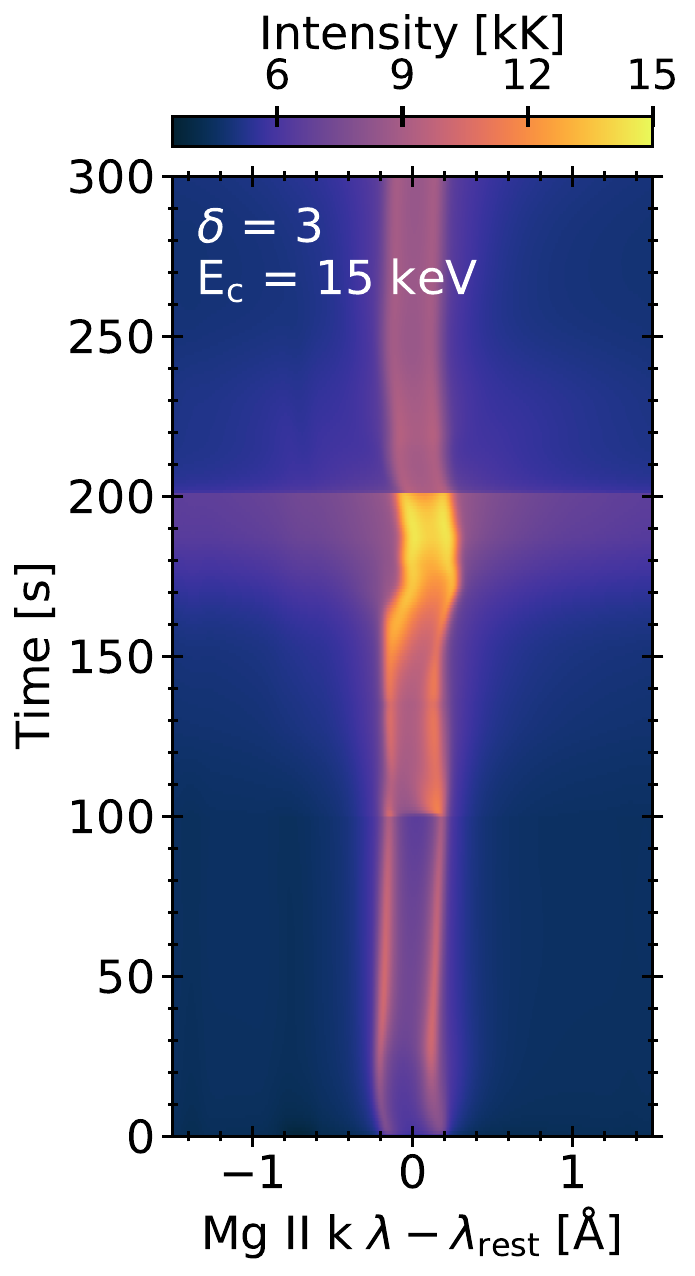}}
	\subfloat{\includegraphics[width = 0.15\textwidth, clip = true, trim = 0.cm 0.cm 0.cm 0.cm]{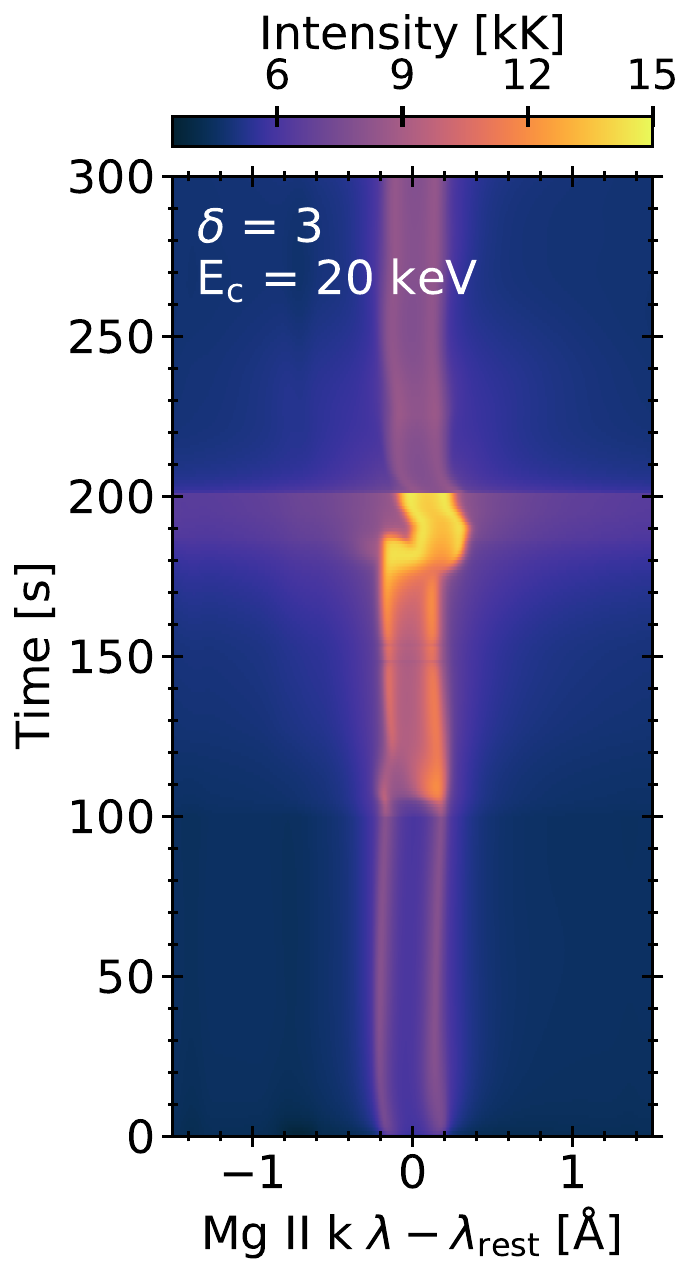}}
	\subfloat{\includegraphics[width = 0.15\textwidth, clip = true, trim = 0.cm 0.cm 0.cm 0.cm]{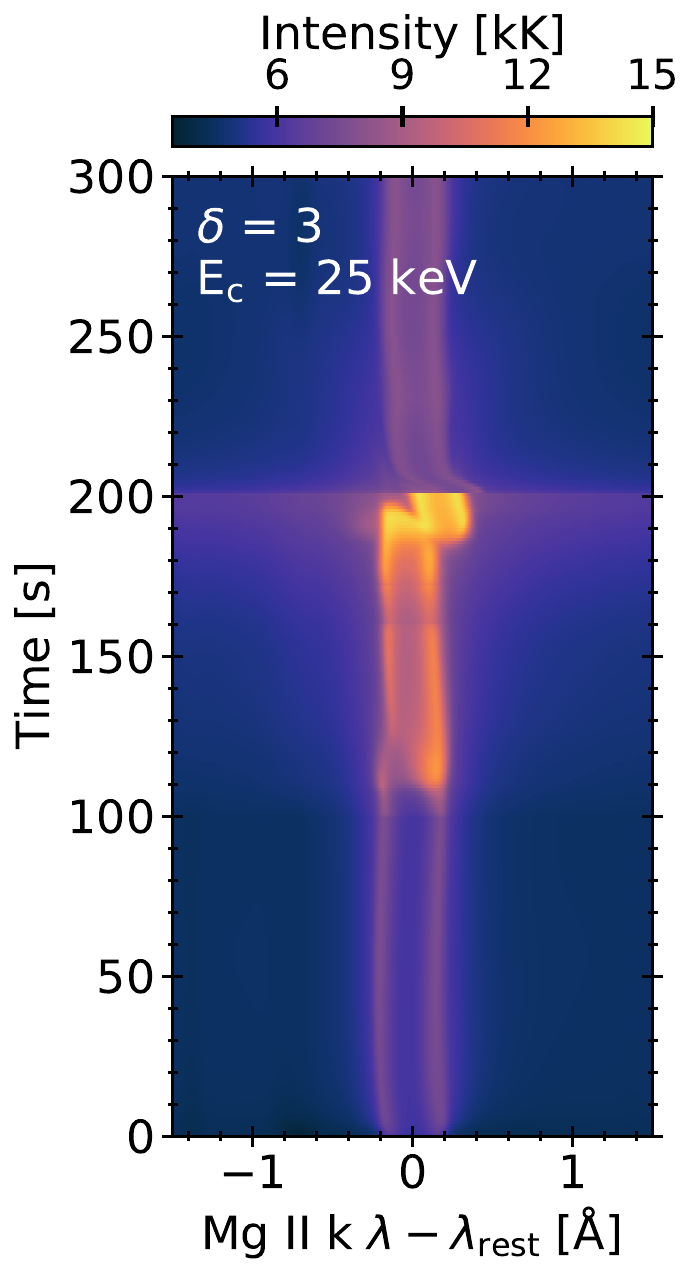}}
	\subfloat{\includegraphics[width = 0.15\textwidth, clip = true, trim = 0.cm 0.cm 0.cm 0.cm]{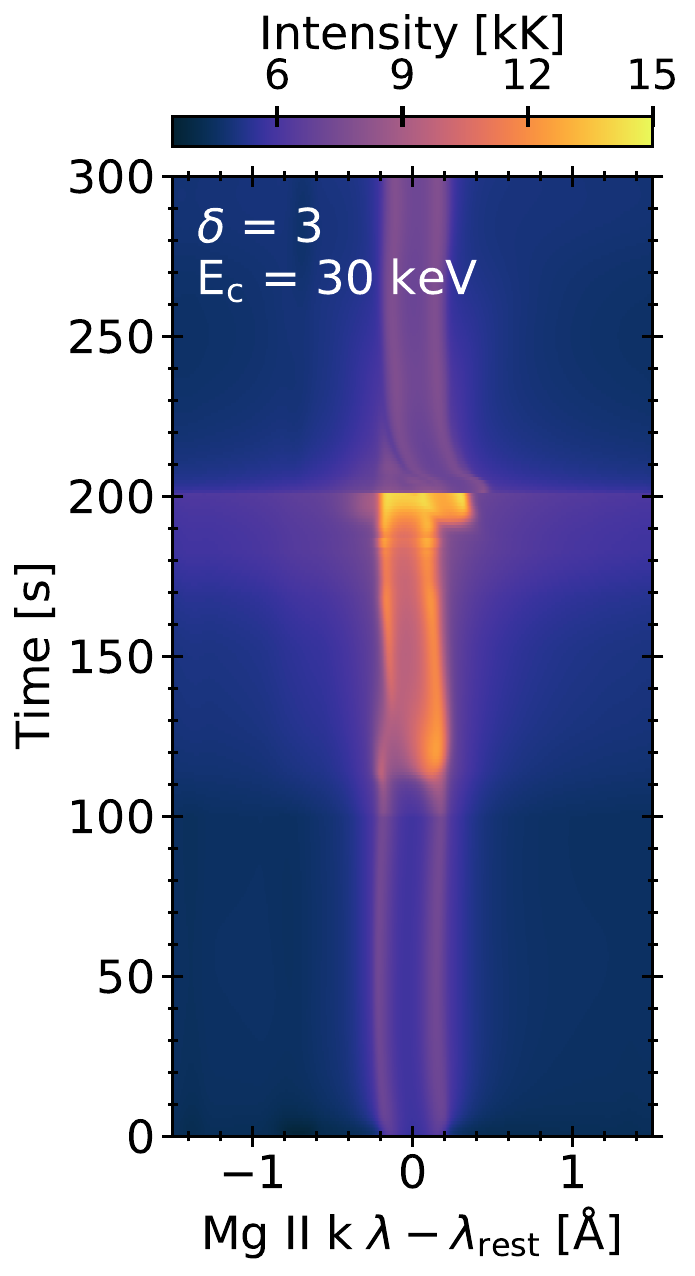}}
	}
	}
	\vbox{
	\hbox{
	\subfloat{\includegraphics[width = 0.15\textwidth, clip = true, trim = 0.cm 0.cm 0.cm 0.cm]{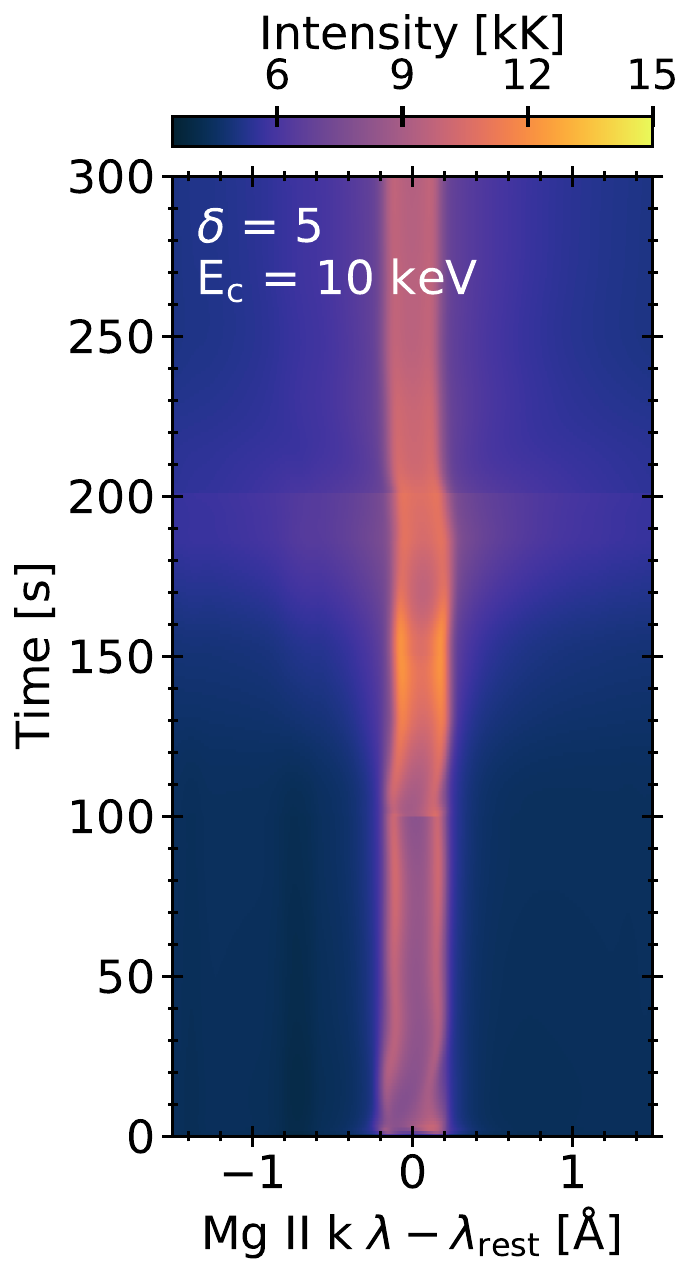}}
	\subfloat{\includegraphics[width = 0.15\textwidth, clip = true, trim = 0.cm 0.cm 0.cm 0.cm]{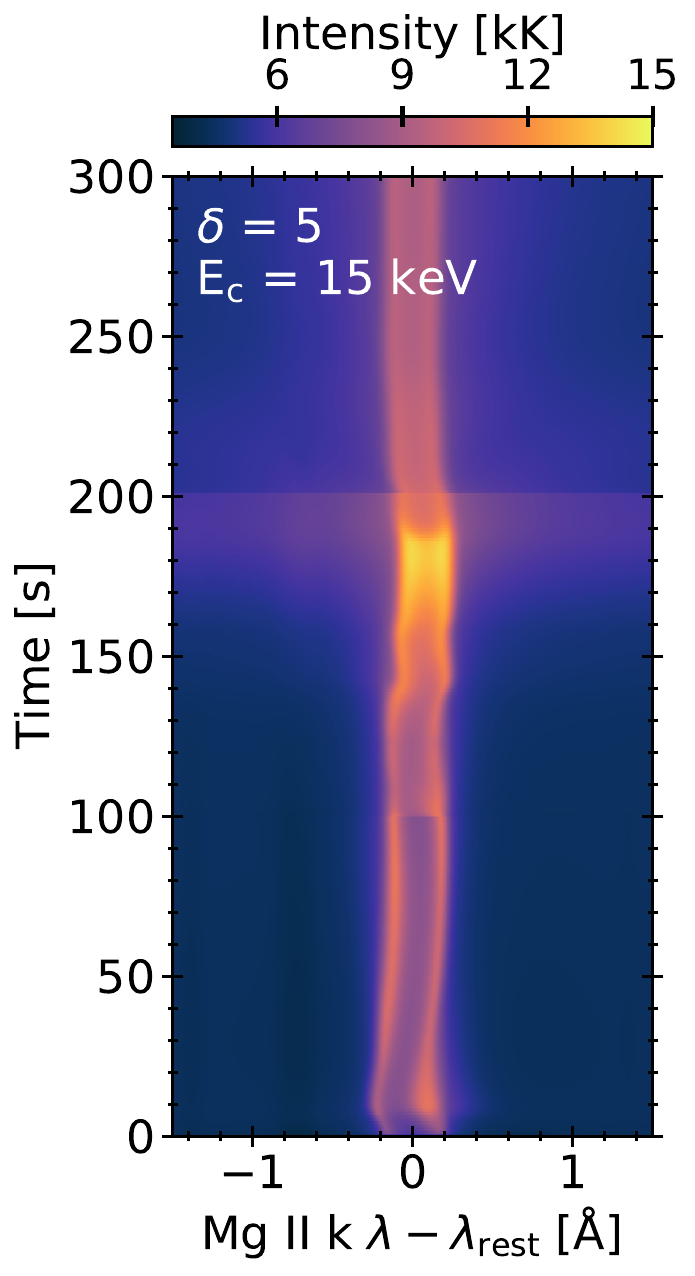}}
	\subfloat{\includegraphics[width = 0.15\textwidth, clip = true, trim = 0.cm 0.cm 0.cm 0.cm]{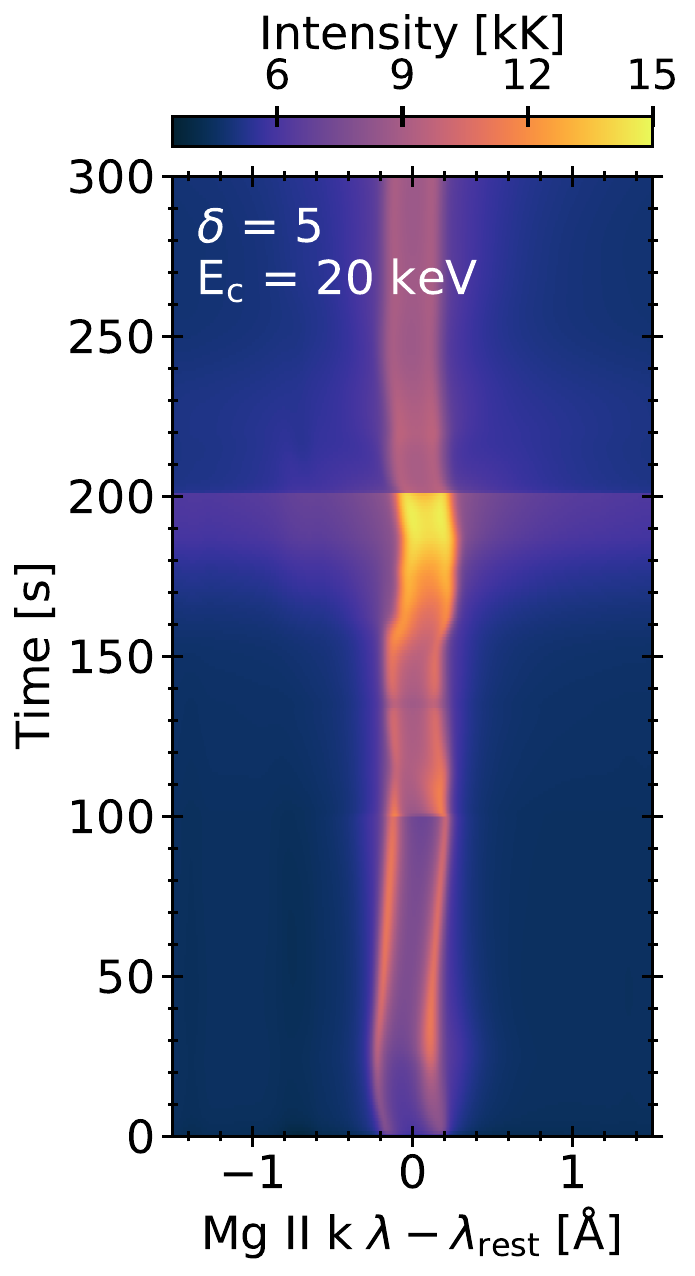}}
	\subfloat{\includegraphics[width = 0.15\textwidth, clip = true, trim = 0.cm 0.cm 0.cm 0.cm]{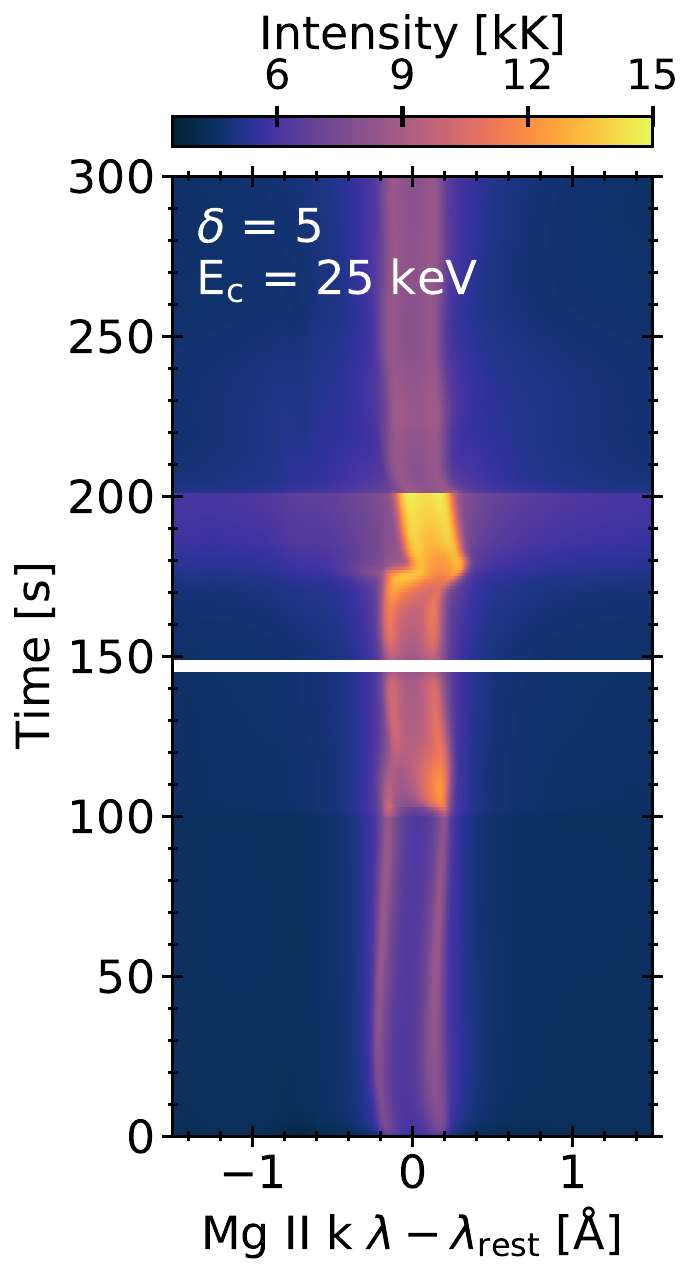}}
	\subfloat{\includegraphics[width = 0.15\textwidth, clip = true, trim = 0.cm 0.cm 0.cm 0.cm]{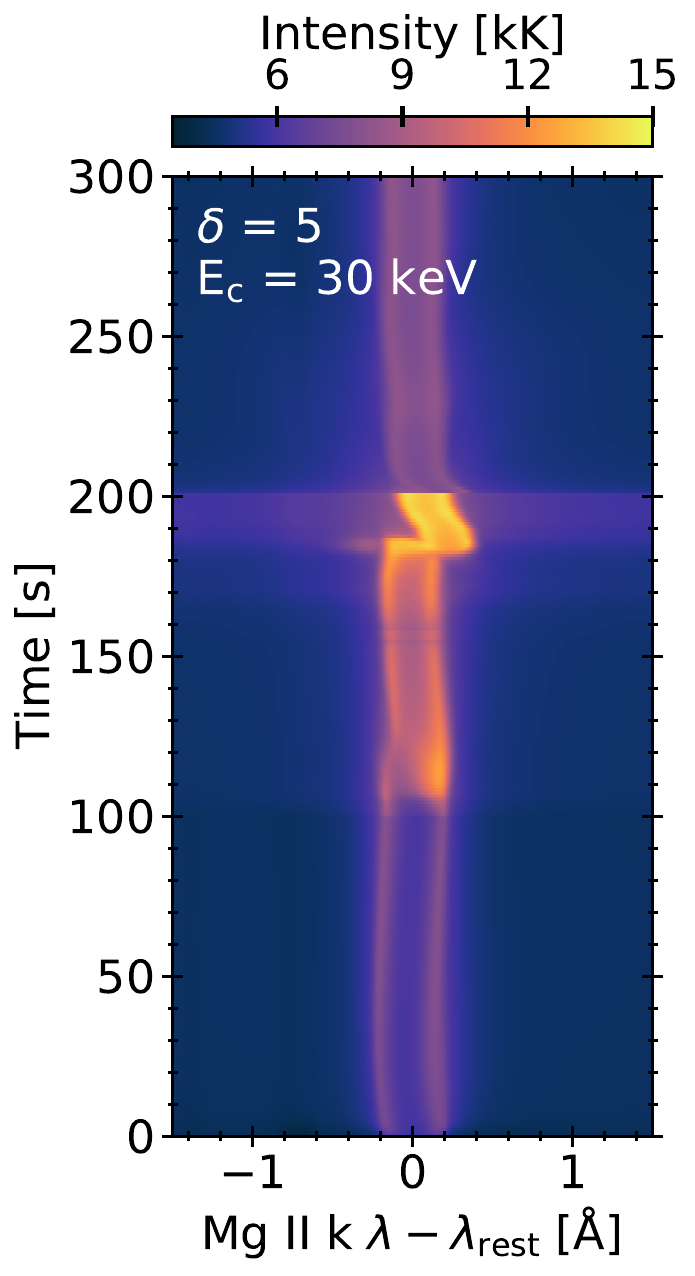}}
	}
	}
	\vbox{
	\hbox{
	\subfloat{\includegraphics[width = 0.15\textwidth, clip = true, trim = 0.cm 0.cm 0.cm 0.cm]{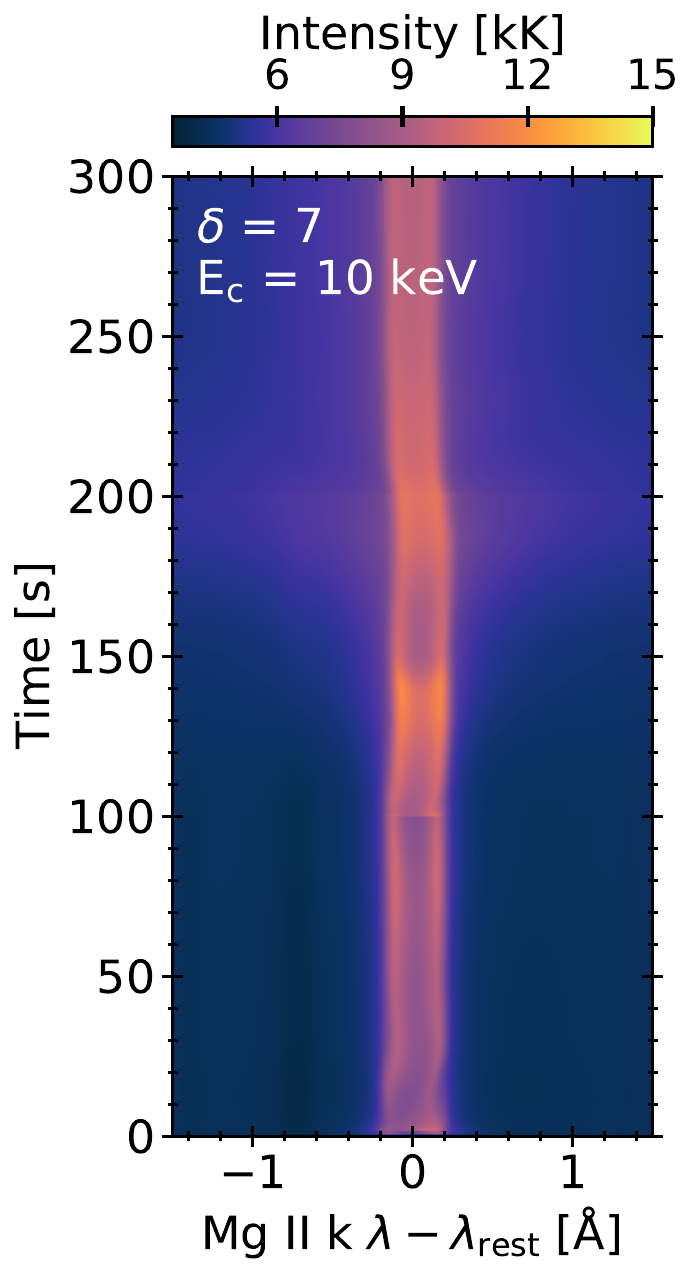}}
	\subfloat{\includegraphics[width =0.15\textwidth, clip = true, trim = 0.cm 0.cm 0.cm 0.cm]{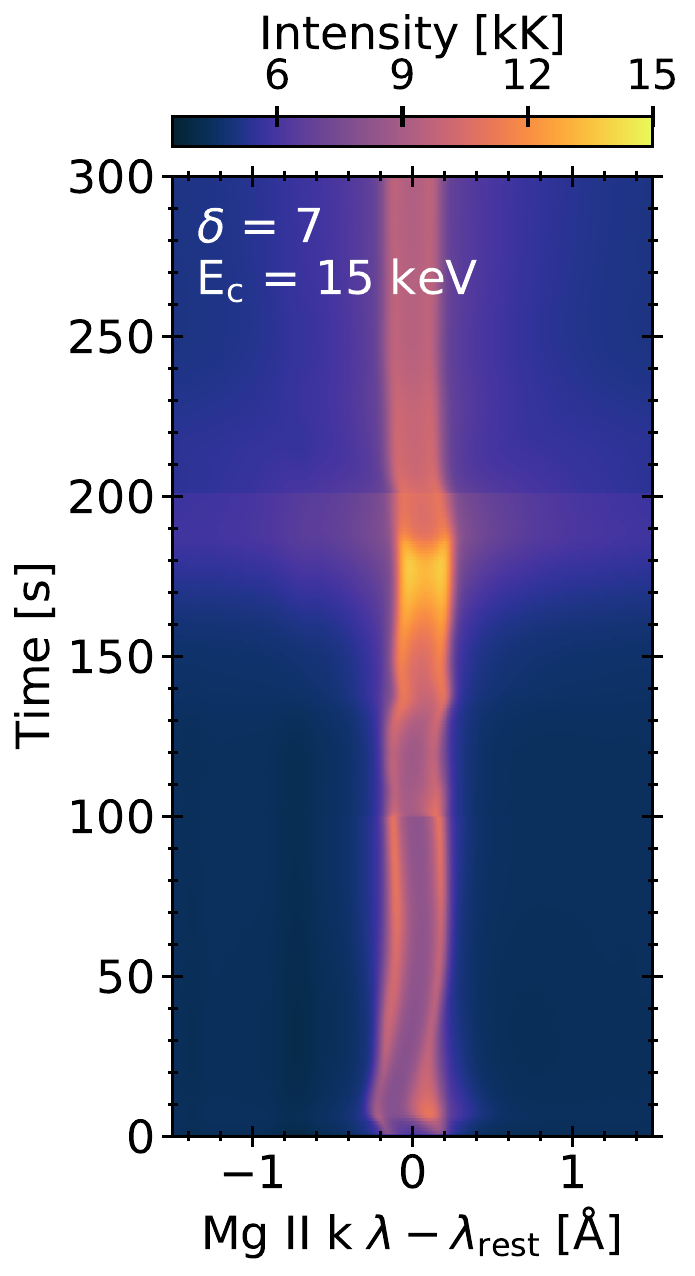}}
	\subfloat{\includegraphics[width = 0.15\textwidth, clip = true, trim = 0.cm 0.cm 0.cm 0.cm]{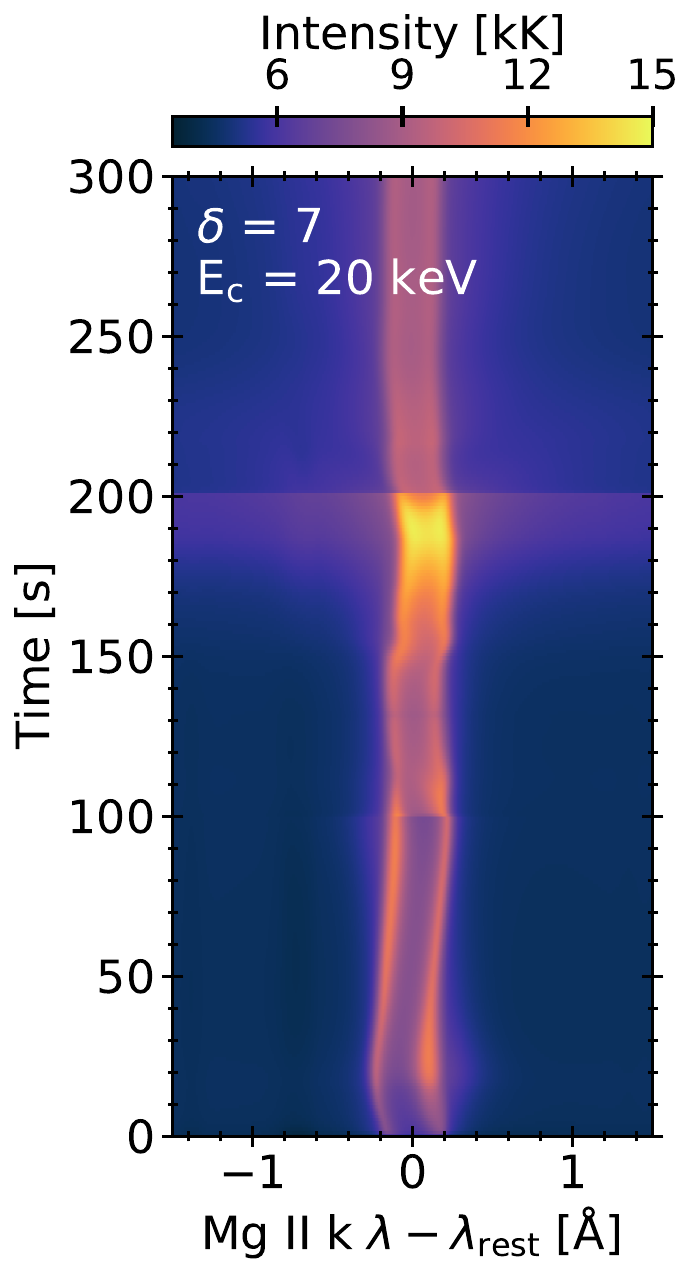}}
	\subfloat{\includegraphics[width = 0.15\textwidth, clip = true, trim = 0.cm 0.cm 0.cm 0.cm]{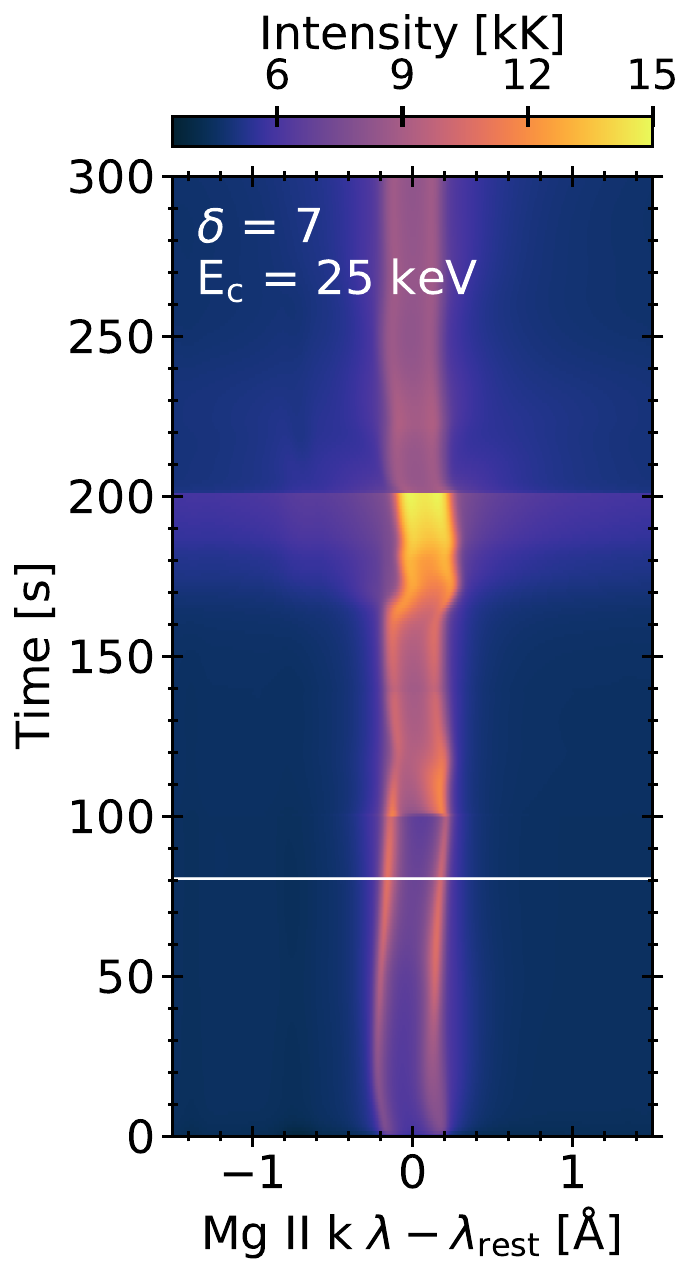}}
	\subfloat{\includegraphics[width = 0.15\textwidth, clip = true, trim = 0.cm 0.cm 0.cm 0.cm]{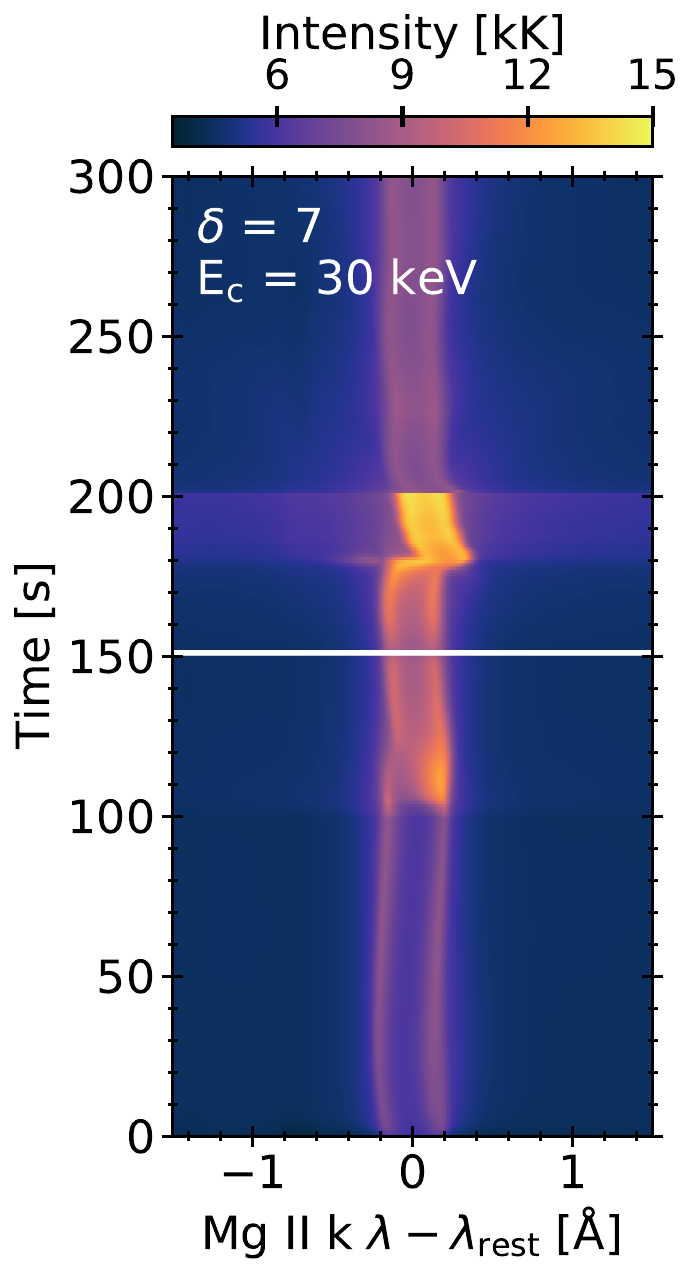}}
	}
	}
		\caption{\textsl{Synthetic \ion{Mg}{2} k line profiles as a function of time in each flare simulation from the main parameter study. The top row is $\delta = 3$, the middle row is $\delta = 5$, and the bottom row is $\delta = 7$. From left to right the low energy cutoffs are $E_{c} = [10, 15, 20, 25, 30]$~keV, such that the hardest nonthermal electron distribution is the top right, and the softest is bottom left. \gsk{For certain snapshots in  three of the simulations \rhpar\ was unable to obtain a solution, with the missing data represented as white horizontal bars.}}}
	\label{fig:mgii_profiles_maingrid}
\end{figure}

\begin{figure}
	\centering 
	\vbox{
	\hbox{
	\subfloat{\includegraphics[width = 0.15\textwidth, clip = true, trim = 0.cm 0.cm 0.cm 0.cm]{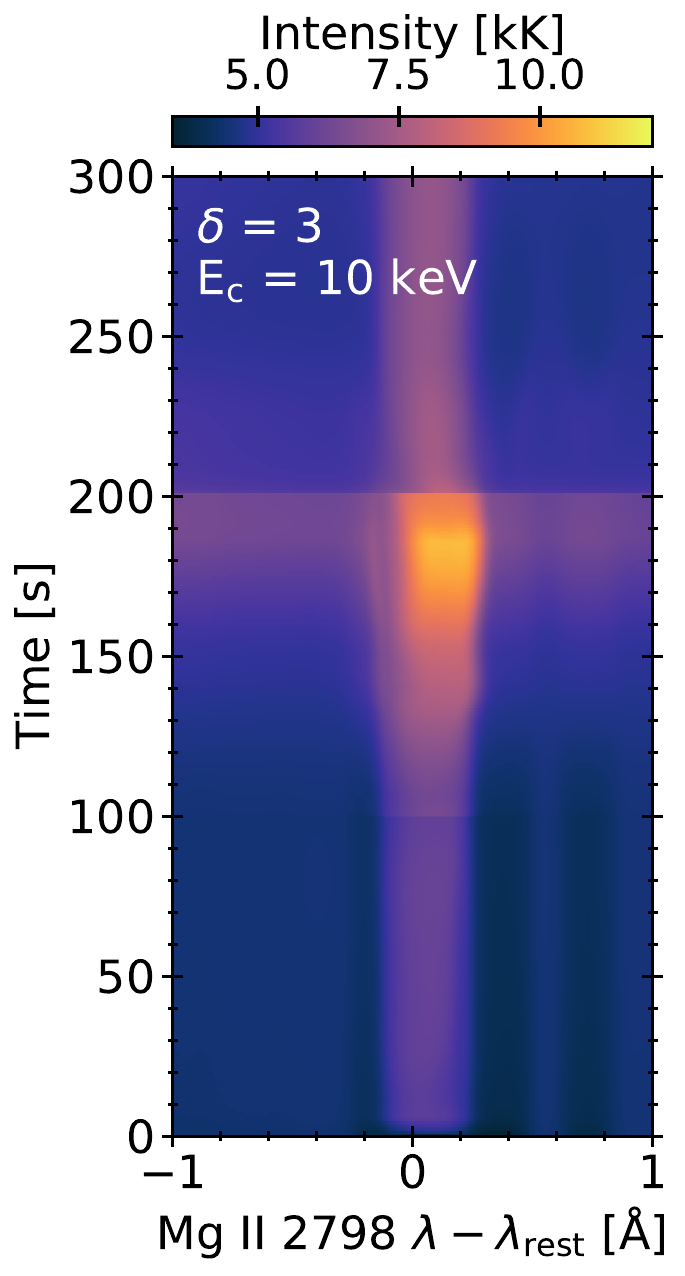}}
	\subfloat{\includegraphics[width = 0.15\textwidth, clip = true, trim = 0.cm 0.cm 0.cm 0.cm]{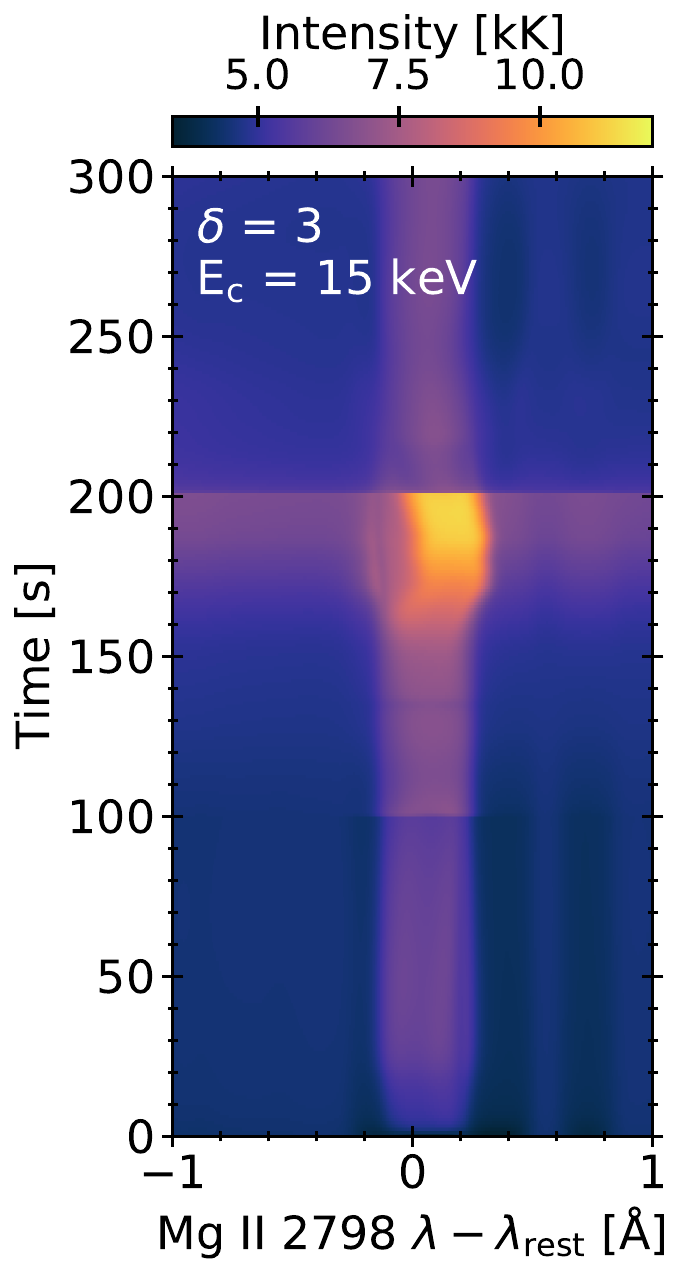}}
	\subfloat{\includegraphics[width = 0.15\textwidth, clip = true, trim = 0.cm 0.cm 0.cm 0.cm]{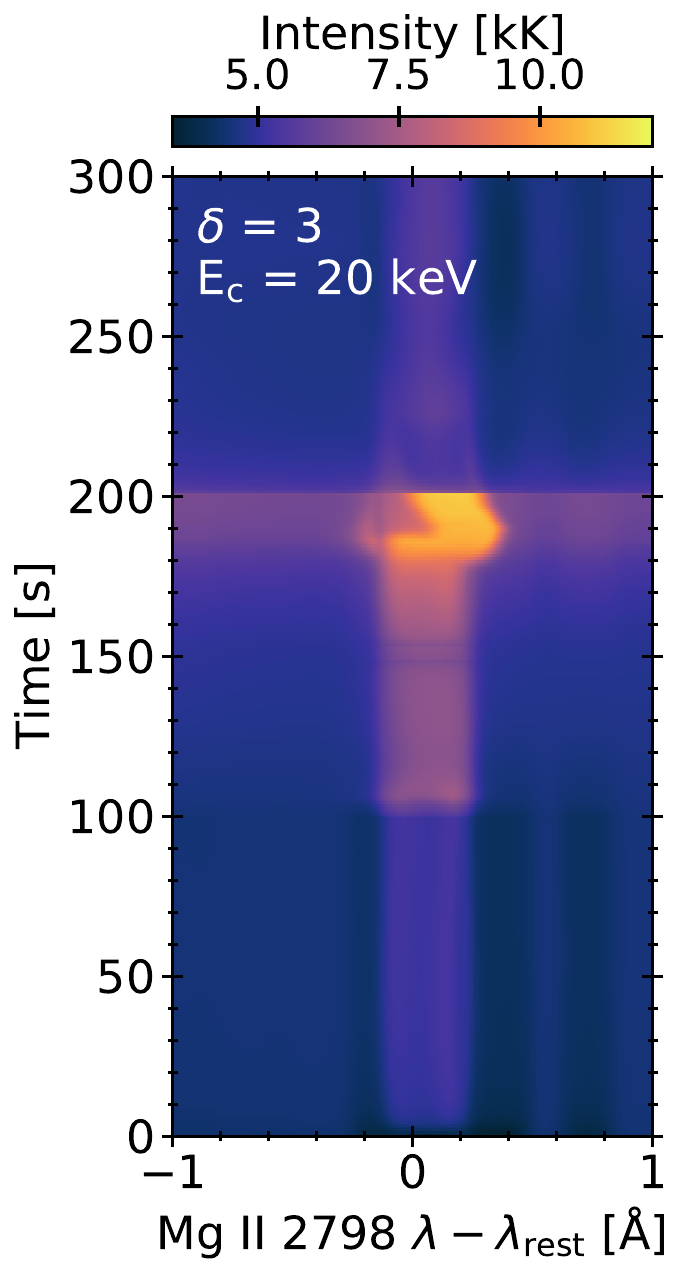}}
	\subfloat{\includegraphics[width = 0.15\textwidth, clip = true, trim = 0.cm 0.cm 0.cm 0.cm]{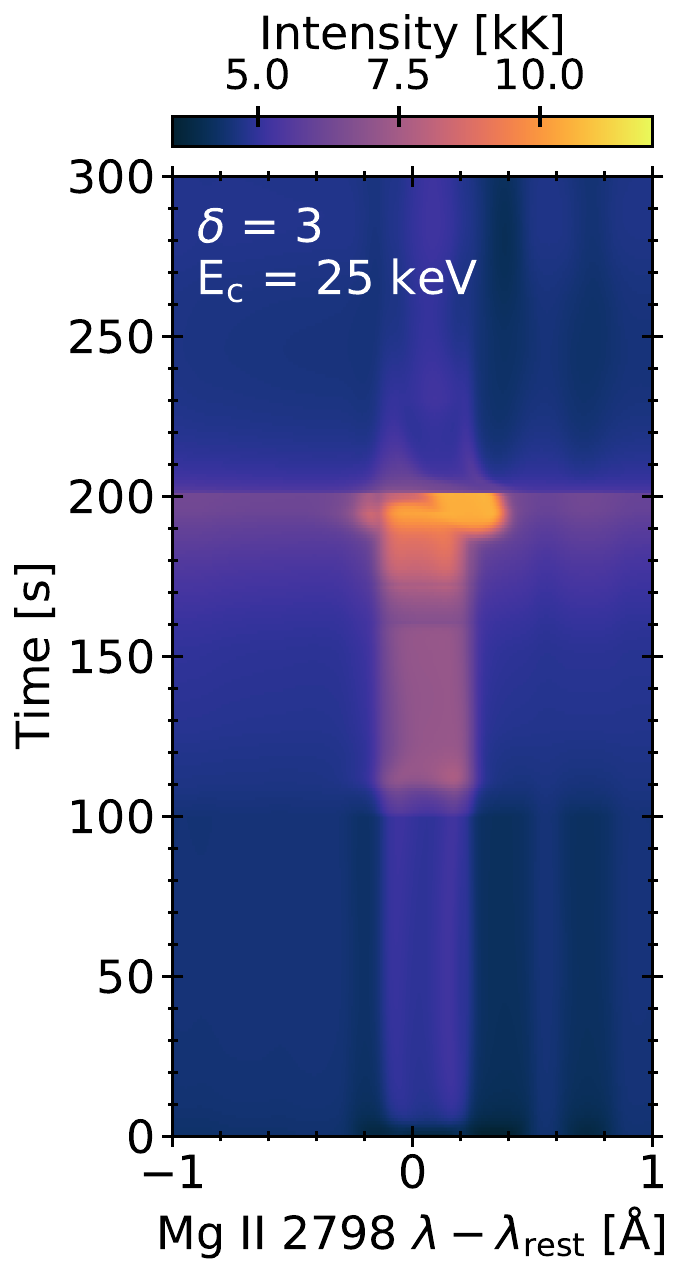}}
	\subfloat{\includegraphics[width = 0.15\textwidth, clip = true, trim = 0.cm 0.cm 0.cm 0.cm]{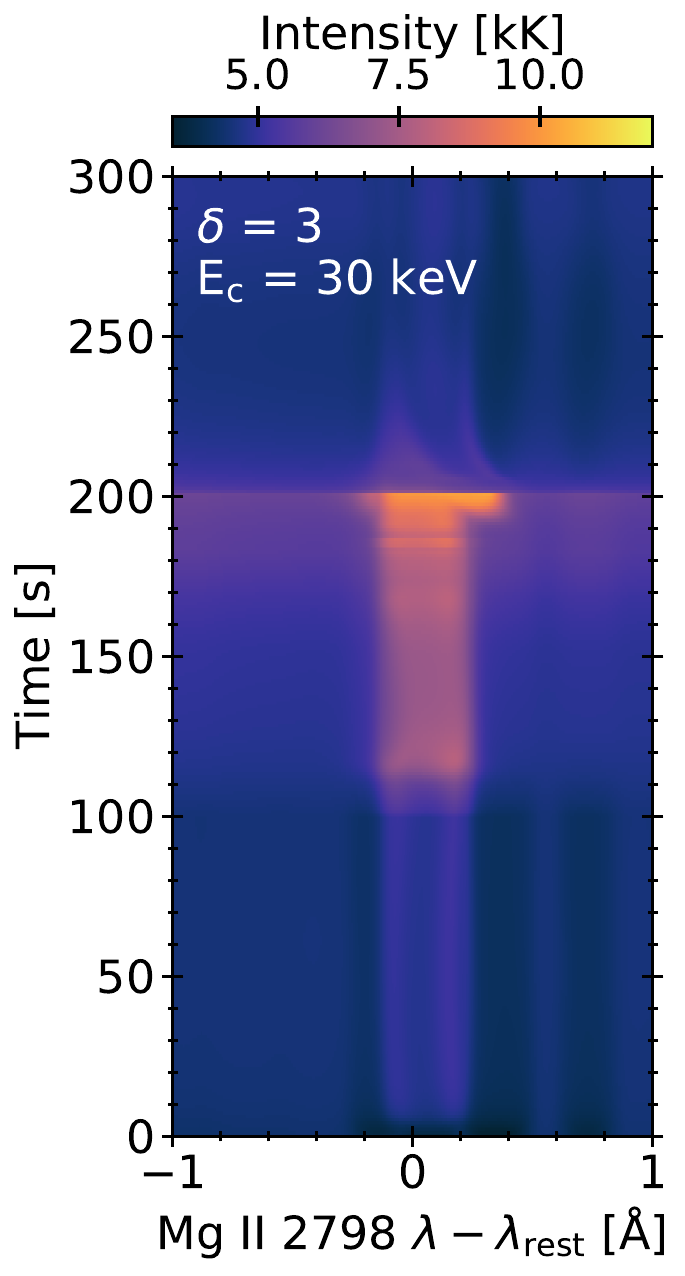}}
	}
	}
	\vbox{
	\hbox{
	\subfloat{\includegraphics[width = 0.15\textwidth, clip = true, trim = 0.cm 0.cm 0.cm 0.cm]{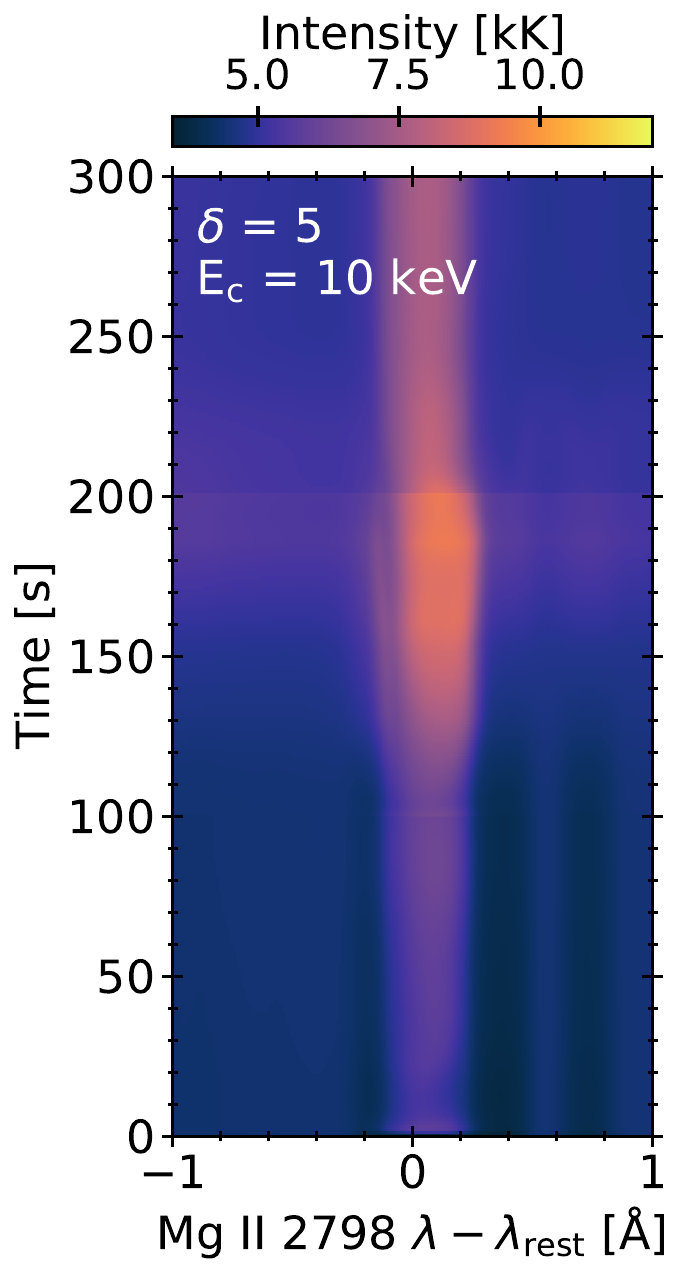}}
	\subfloat{\includegraphics[width = 0.15\textwidth, clip = true, trim = 0.cm 0.cm 0.cm 0.cm]{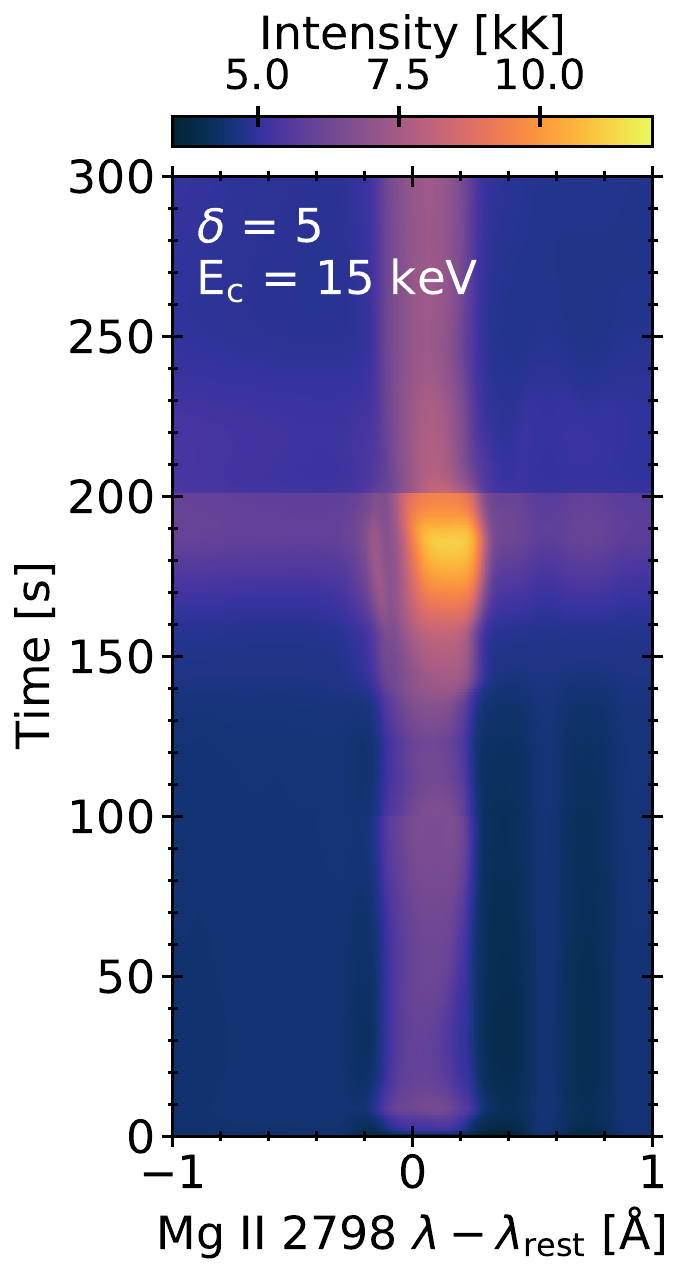}}
	\subfloat{\includegraphics[width = 0.15\textwidth, clip = true, trim = 0.cm 0.cm 0.cm 0.cm]{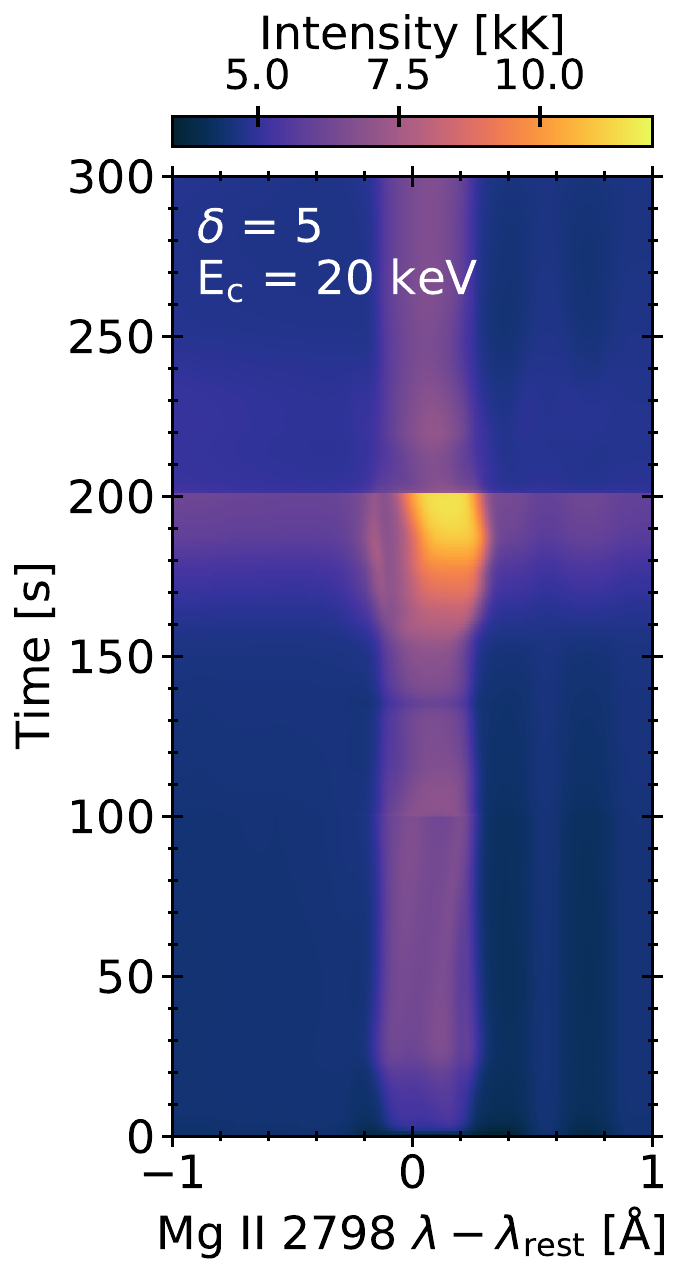}}
	\subfloat{\includegraphics[width = 0.15\textwidth, clip = true, trim = 0.cm 0.cm 0.cm 0.cm]{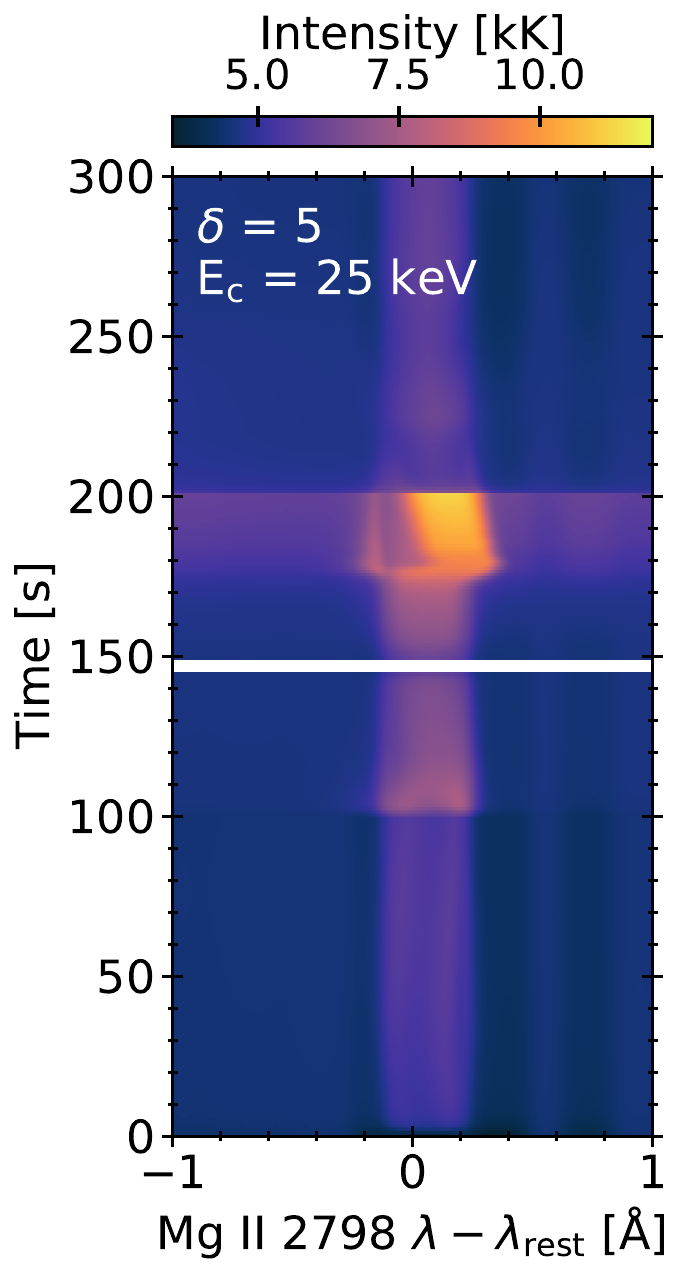}}
	\subfloat{\includegraphics[width = 0.15\textwidth, clip = true, trim = 0.cm 0.cm 0.cm 0.cm]{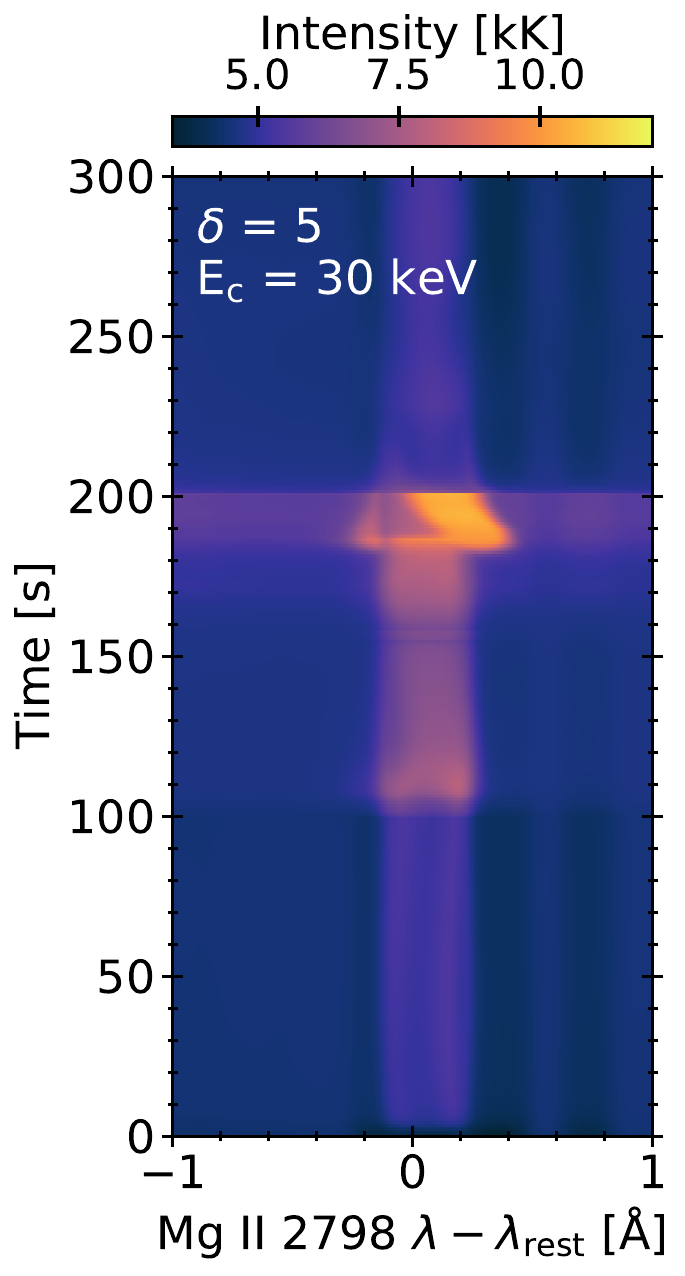}}
	}
	}
	\vbox{
	\hbox{
	\subfloat{\includegraphics[width = 0.15\textwidth, clip = true, trim = 0.cm 0.cm 0.cm 0.cm]{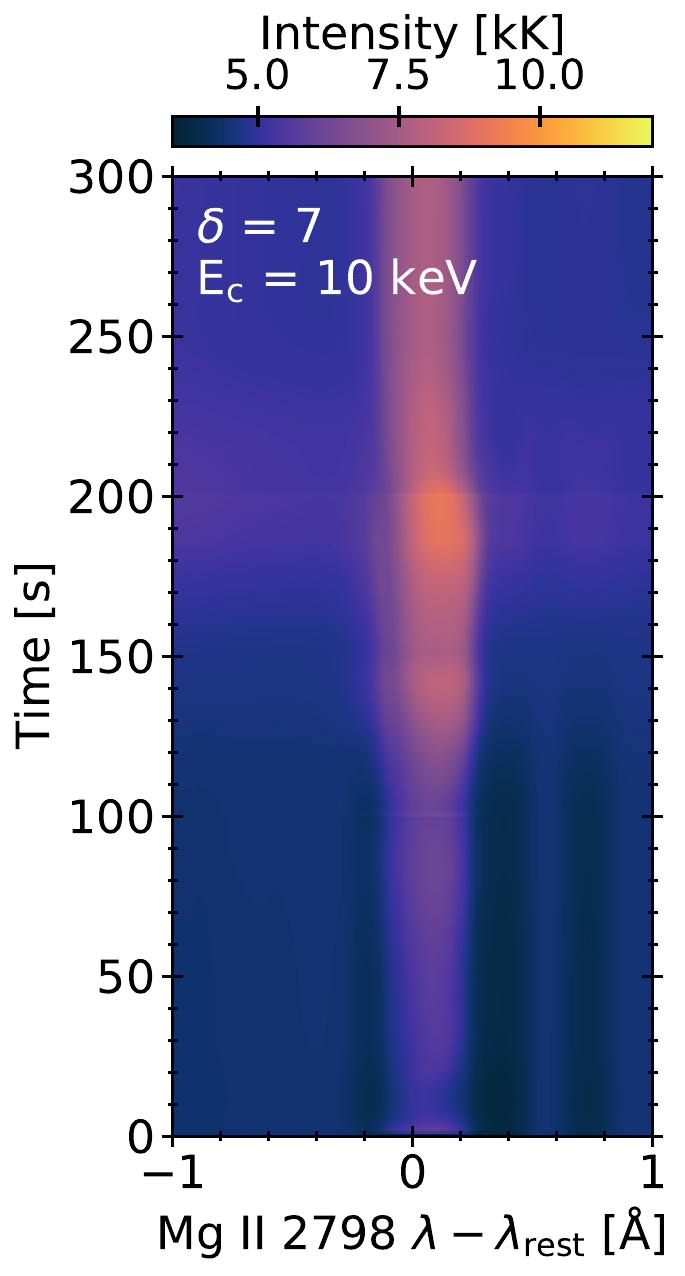}}
	\subfloat{\includegraphics[width =0.15\textwidth, clip = true, trim = 0.cm 0.cm 0.cm 0.cm]{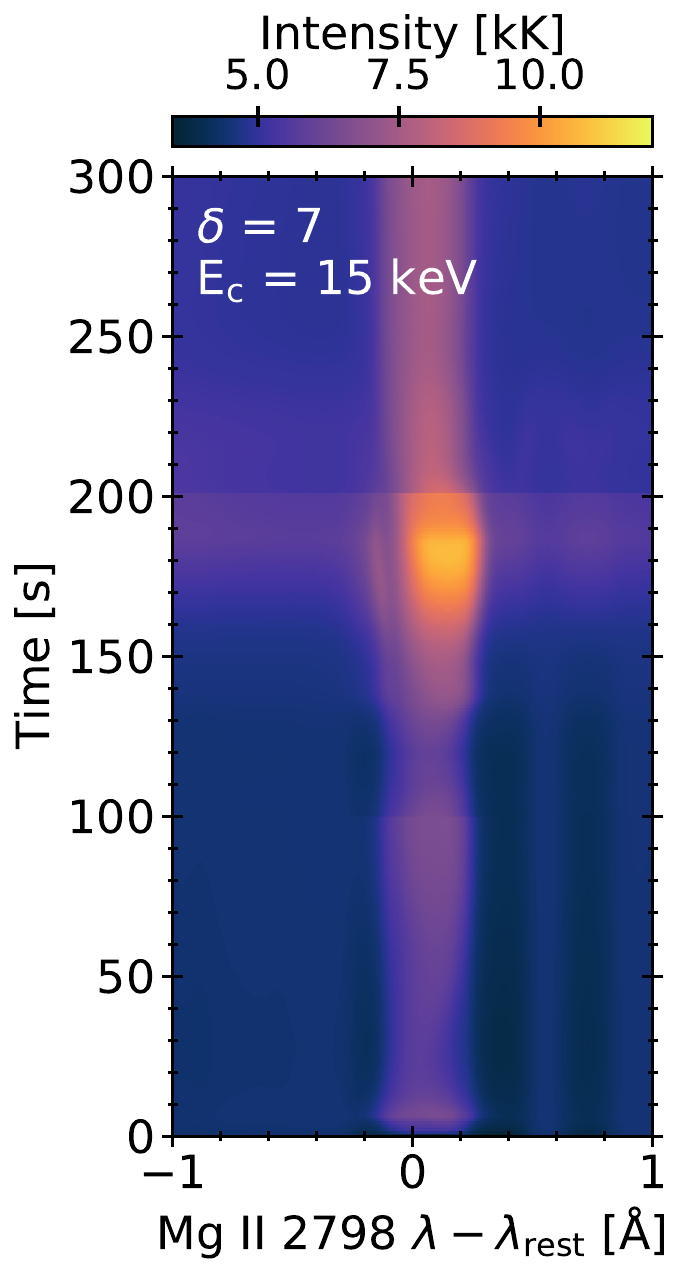}}
	\subfloat{\includegraphics[width = 0.15\textwidth, clip = true, trim = 0.cm 0.cm 0.cm 0.cm]{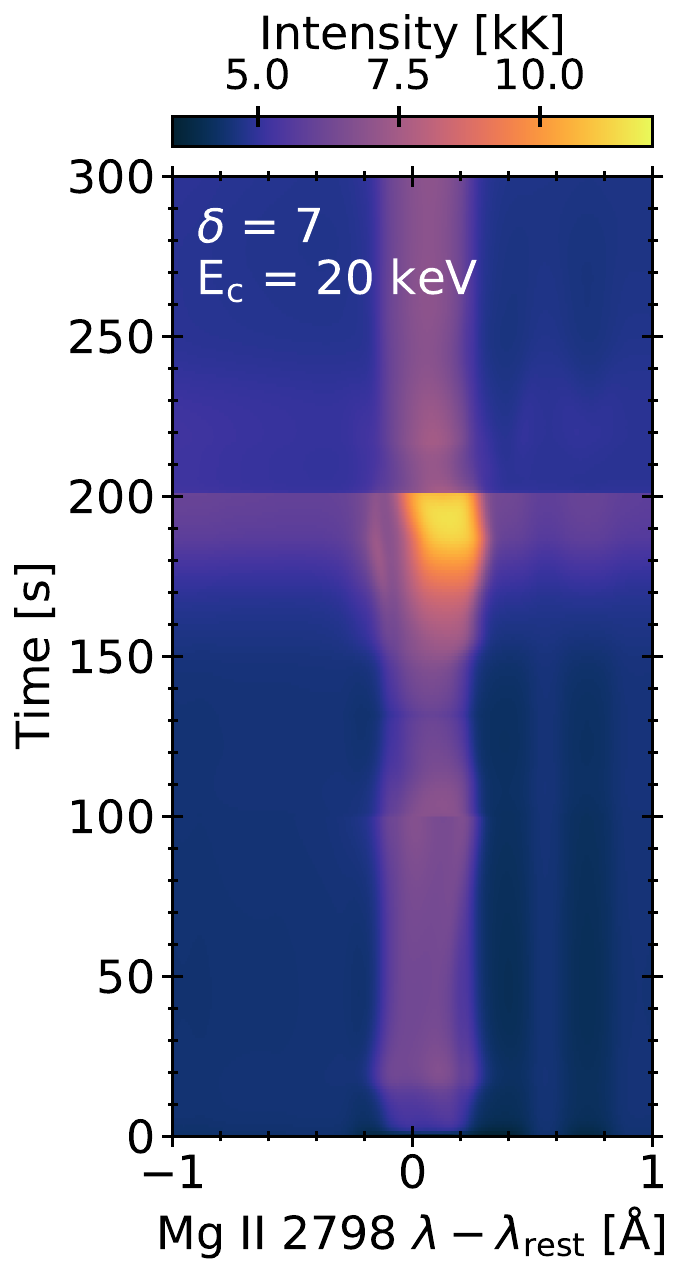}}
	\subfloat{\includegraphics[width = 0.15\textwidth, clip = true, trim = 0.cm 0.cm 0.cm 0.cm]{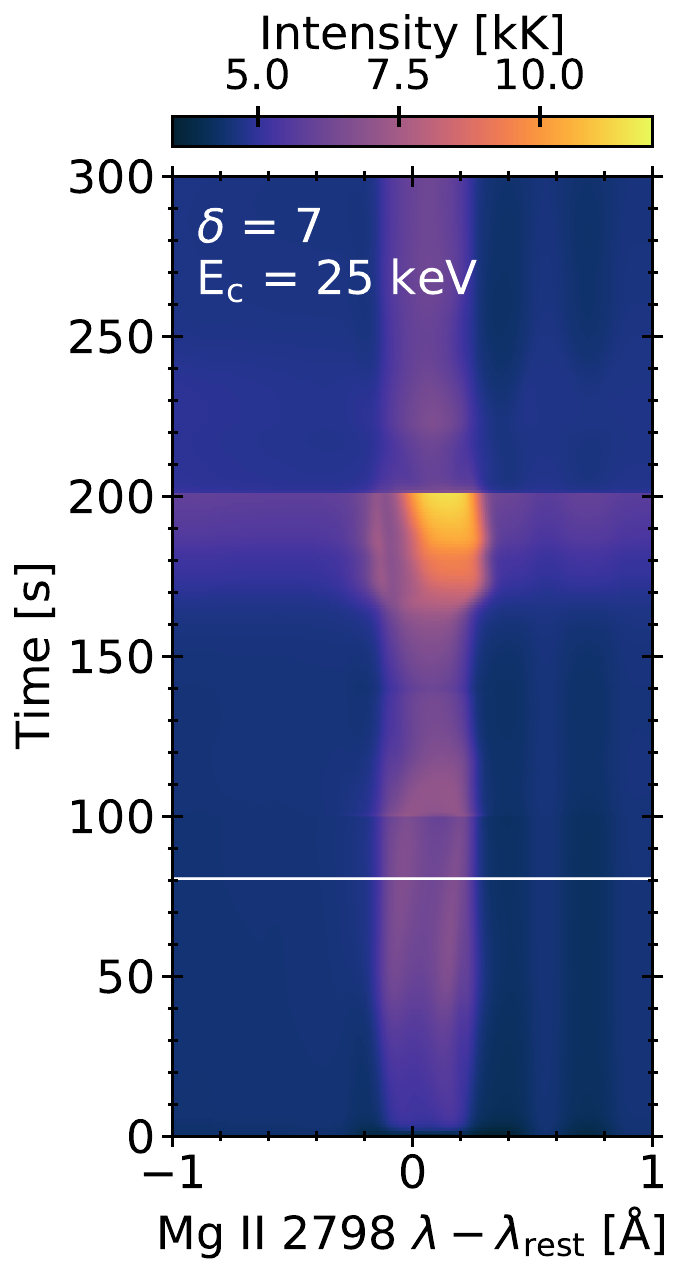}}
	\subfloat{\includegraphics[width = 0.15\textwidth, clip = true, trim = 0.cm 0.cm 0.cm 0.cm]{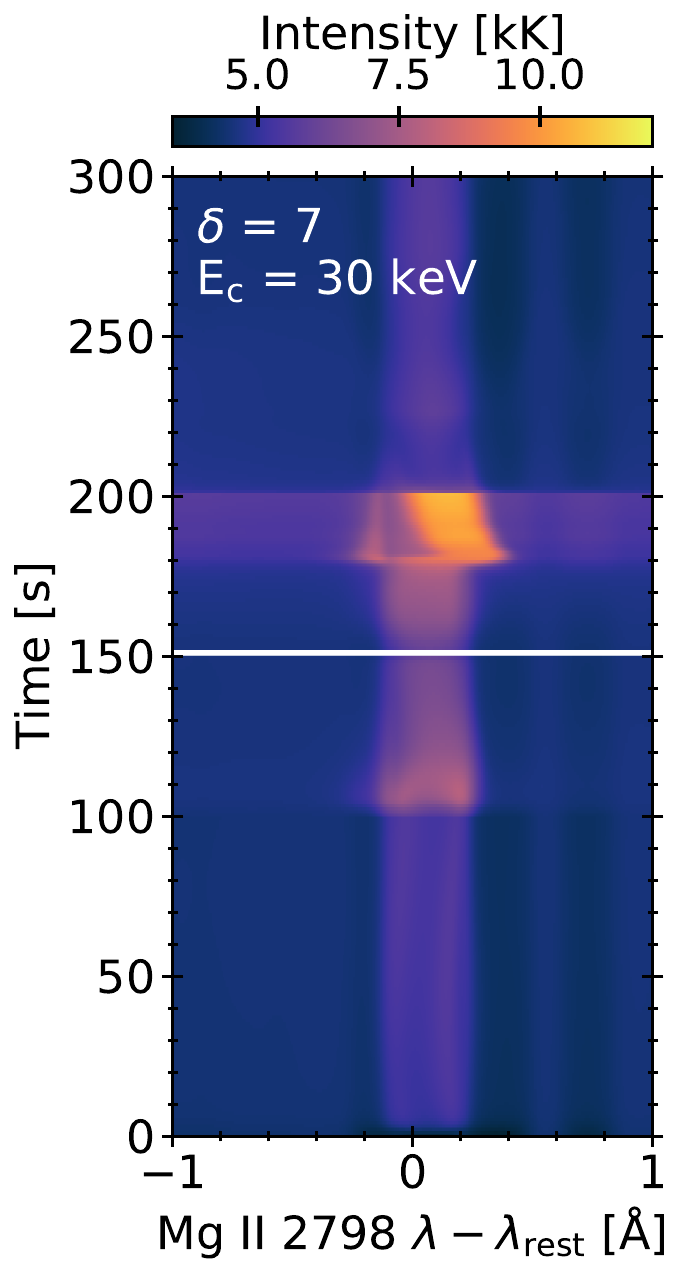}}
	}
	}
		\caption{\textsl{Same as Figure~\ref{fig:mgii_profiles_maingrid}, but showing the \ion{Mg}{2} blended subordinate lines near 2798~\AA.}}
	\label{fig:mgiis2_profiles_maingrid}
\end{figure}

Here we show the evolution of the \ion{He}{1} 10830~\AA\ spectra (Figure~\ref{fig:helium_profiles_maingrid}), \ion{Mg}{2} k spectra (Figure~\ref{fig:mgii_profiles_maingrid}), and \ion{Mg}{2} 2798~\AA\ subordinate blend (Figure~\ref{fig:mgiis2_profiles_maingrid}) from every simulation in our main parameter grid.  

\section{Impact of converting synthetic spectra to IRIS resolution}\label{sec:irisquality}

\begin{figure*}
	\centering 
	\vbox{
	\hbox{
	\hspace{0.25in}	
	\subfloat{\includegraphics[width = 0.4\textwidth, clip = true, trim = 0.cm 0.cm 0.cm 0.cm]{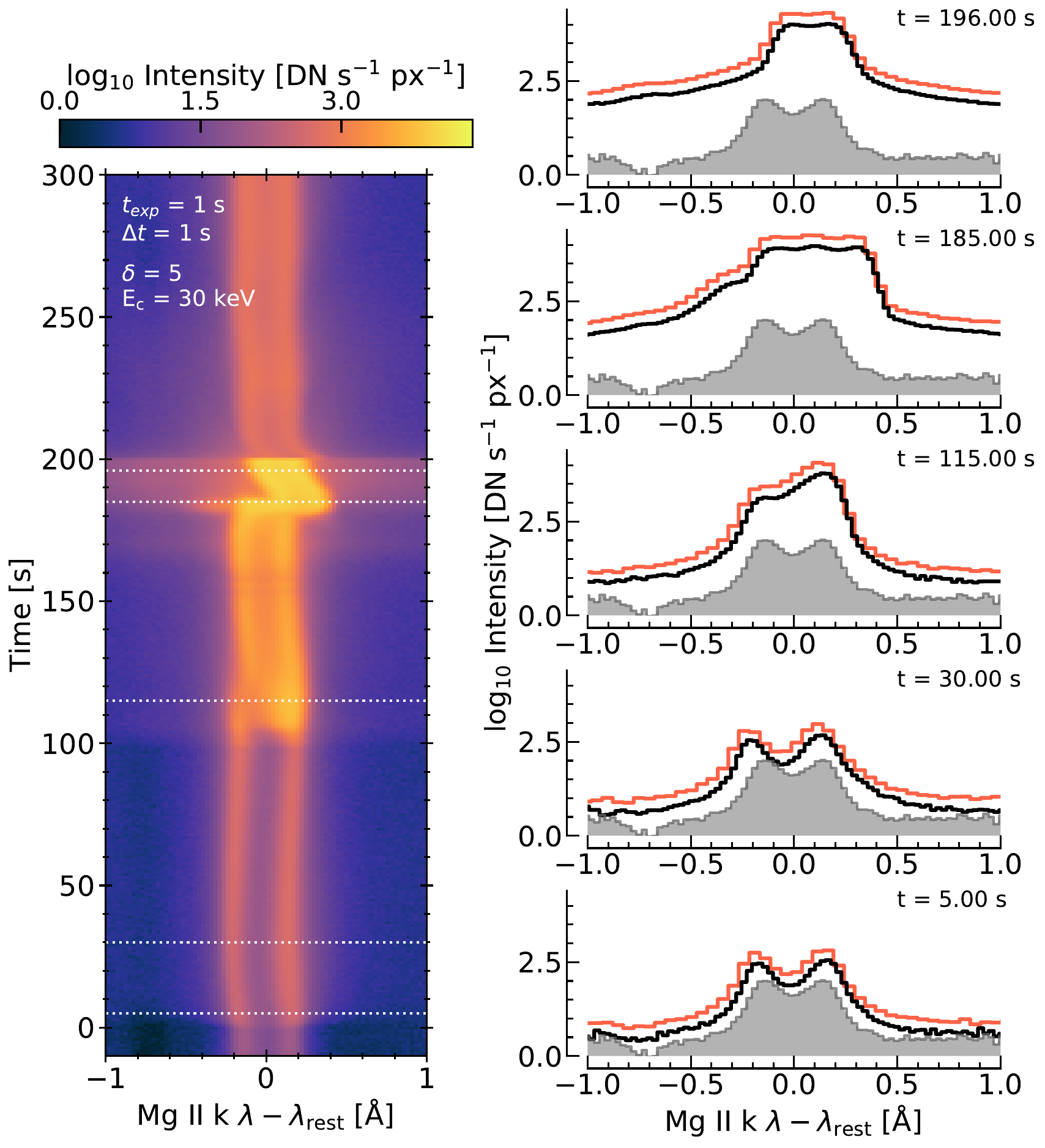}}	
		\subfloat{\includegraphics[width = 0.4\textwidth, clip = true, trim = 0.cm 0.cm 0.cm 0.cm]{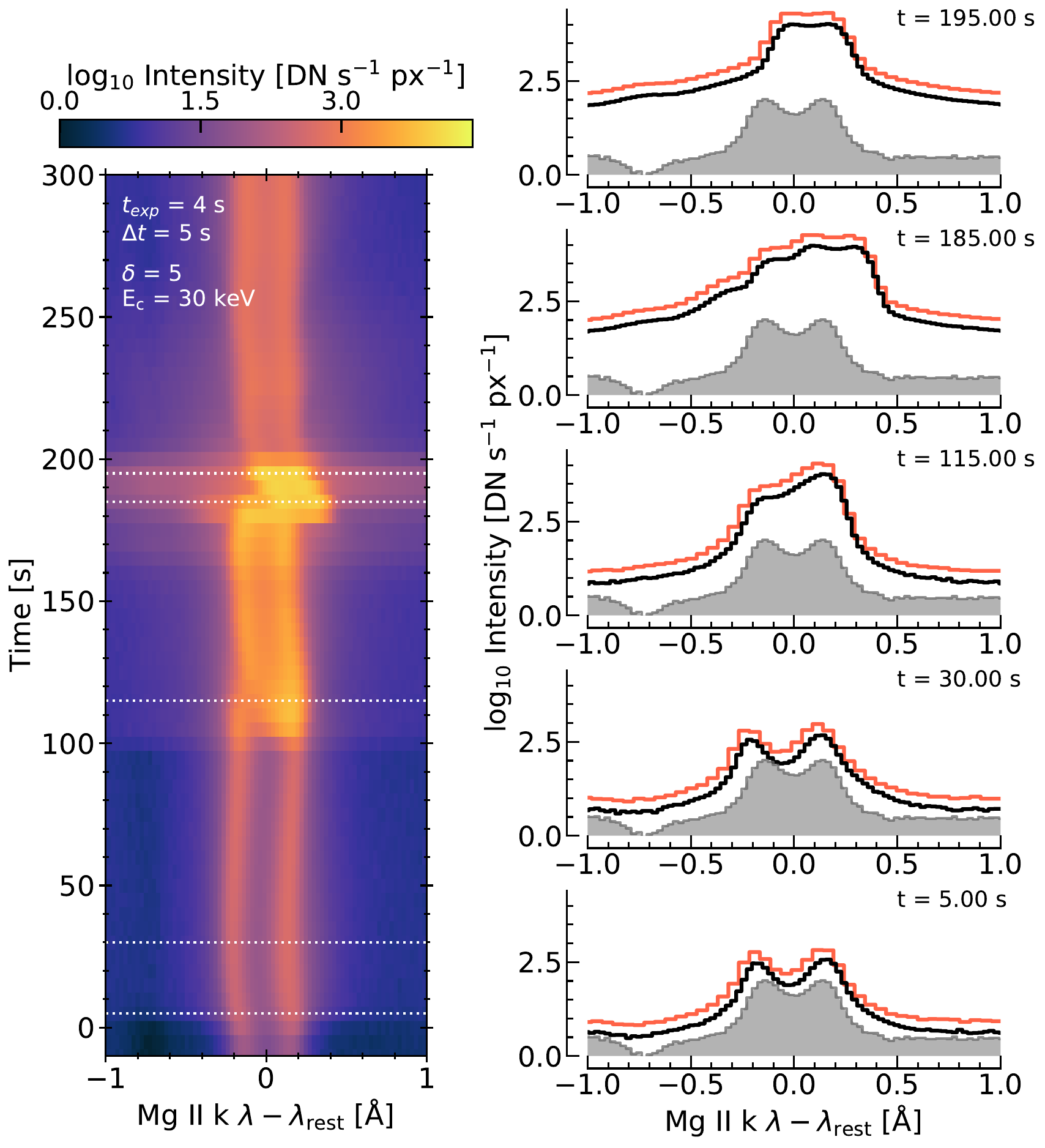}}	
	}
	}
	\vbox{
	\hbox{
	\hspace{0.25in}
	\subfloat{\includegraphics[width = 0.4\textwidth, clip = true, trim = 0.cm 0.cm 0.cm 0.cm]{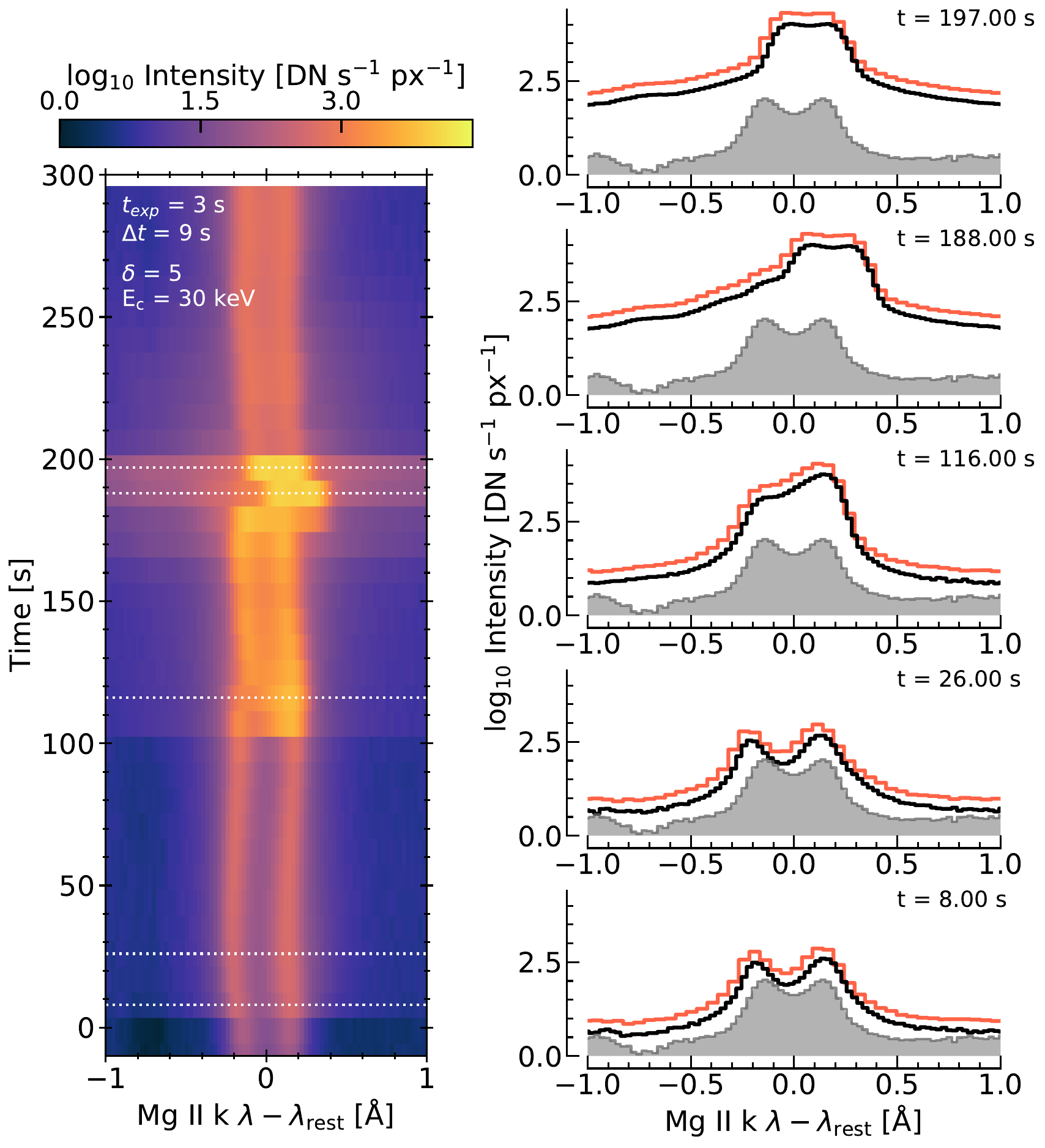}}	
	}
	}
	\caption{\textsl{Example \ion{Mg}{2} k line spectra from the $\delta = 5$, $E_{c} = 30$~keV simulation, converted to IRIS count rates. Each panel shows an image of the evolution of the spectra during the simulation, with cutouts at certain times. The setups are: $t_{\mathrm{exp}} = 1$~s, $\Delta t = 1$~s (top left); $t_{\mathrm{exp}} = 4$~s, $\Delta t = 5$~s (top right); $t_{\mathrm{exp}} = 3$~s, $\Delta t = 9$~s (bottom left). In each of those panels the black lines and greyscale (pre-flare) are un-summed data, and the red lines are $2\times2$ spectral summing. }}
	\label{fig:mgii_spectra_ex_iris}
\end{figure*}

Converting to IRIS count rates, spectral plate scales and resolution, and accounting for smearing over some typical exposure times \citep[using the same procedure as described in][]{2024MNRAS.527.2523K}, makes identifying multiple Doppler shifted components to the lines more difficult, and makes the central reversal appear shallower than in the `native' synthetic profiles. Indeed, at some points the \ion{Mg}{2} k lines appear almost single peaked, and the central reversal of the subordinate lines becomes very subtle or disappears. Figure~\ref{fig:mgii_spectra_ex_iris} compares the $\delta = 5$, $E_{c} = 30$~keV simulation including various IRIS observing setups with different exposure times ($t_{\mathrm{exp}}$), cadences ($\Delta t$), and spectral summing. These were non-exhaustive of the observing modes possible with IRIS, but provide a representative range: (1) $t_{\mathrm{exp}} = 4$~s, $\Delta t = 5$~s (similar to the 2014-Oct-25th X class flare); (2) $t_{\mathrm{exp}} = 3$~s, $\Delta t = 9$~s (similar to the 2014-Sept-10th X class flare\footnote{Here we have rounded to the nearest second to match the \rhpar\ output cadence.}); (3) $t_{\mathrm{exp}} = 1$~s, $\Delta t = 1$~s (to mimic high-cadence observations). For each, $2\times2$ spectral summing was performed in addition to no spectral summing. The effective areas were those from 2014-Oct-25th. These can be compared with Figure~\ref{fig:mgii_spectra_ex}.
	
\begin{figure*}
	\centering 
	\vbox{
	\hbox{
	\hspace{-0in}
	\subfloat{\includegraphics[width = 0.5\textwidth, clip = true, trim = 0.cm 0.cm 0.cm 0.cm]{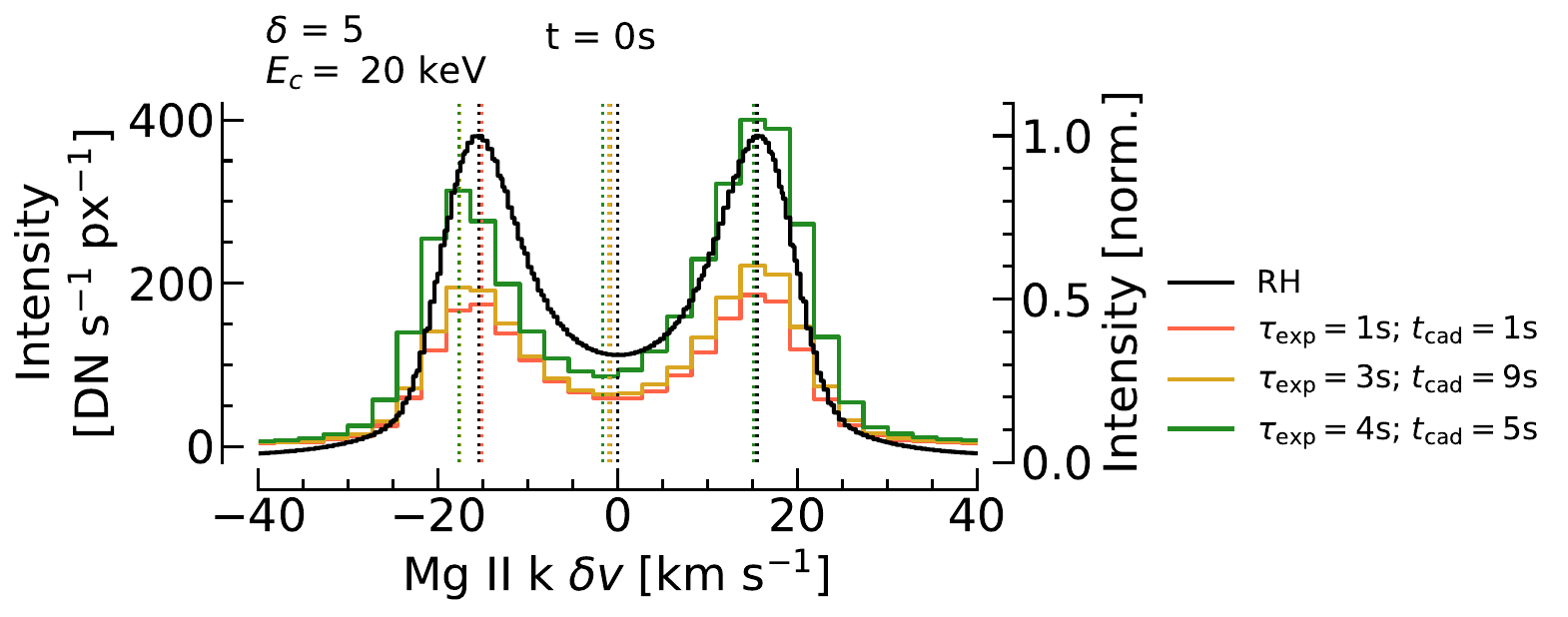}}	
	\subfloat{\includegraphics[width = 0.5\textwidth, clip = true, trim = 0.cm 0.cm 0.cm 0.cm]{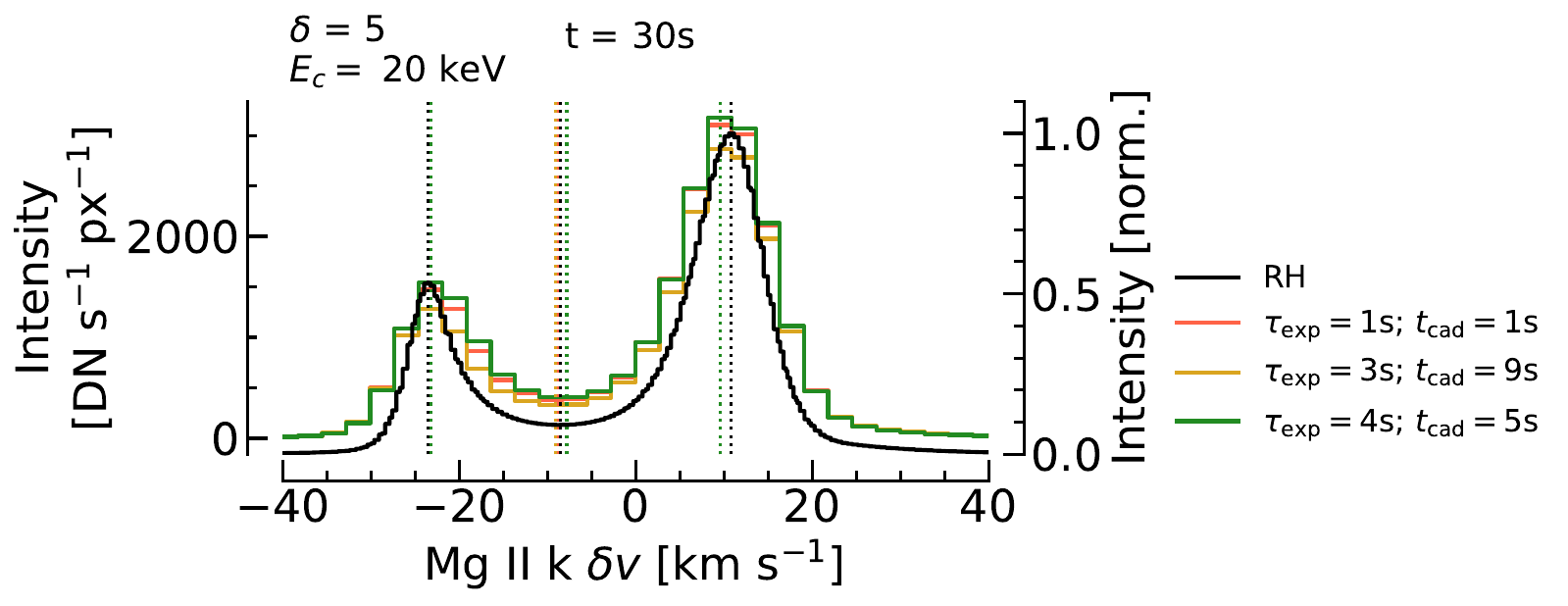}}	
	}
	}
	\vbox{
	\hbox{
	\hspace{0in}
	\subfloat{\includegraphics[width = 0.5\textwidth, clip = true, trim = 0.cm 0.cm 0.cm 0.cm]{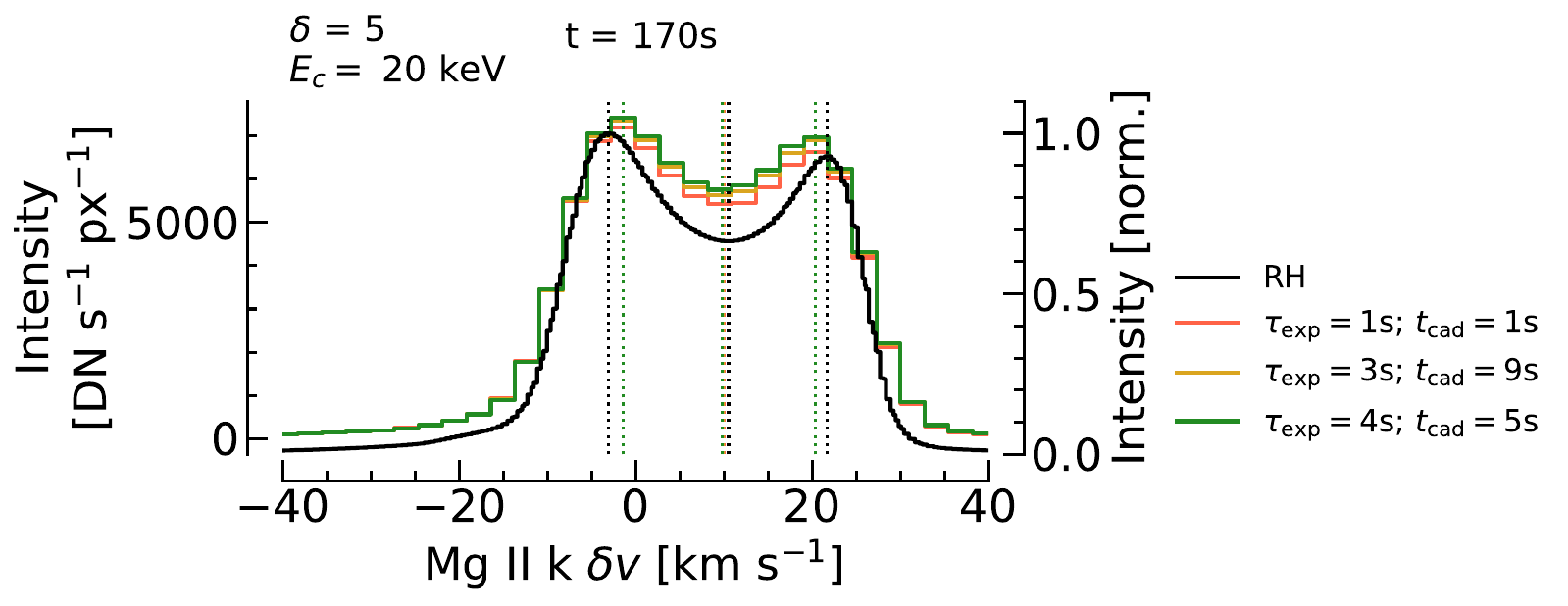}}	
	\subfloat{\includegraphics[width = 0.5\textwidth, clip = true, trim = 0.cm 0.cm 0.cm 0.cm]{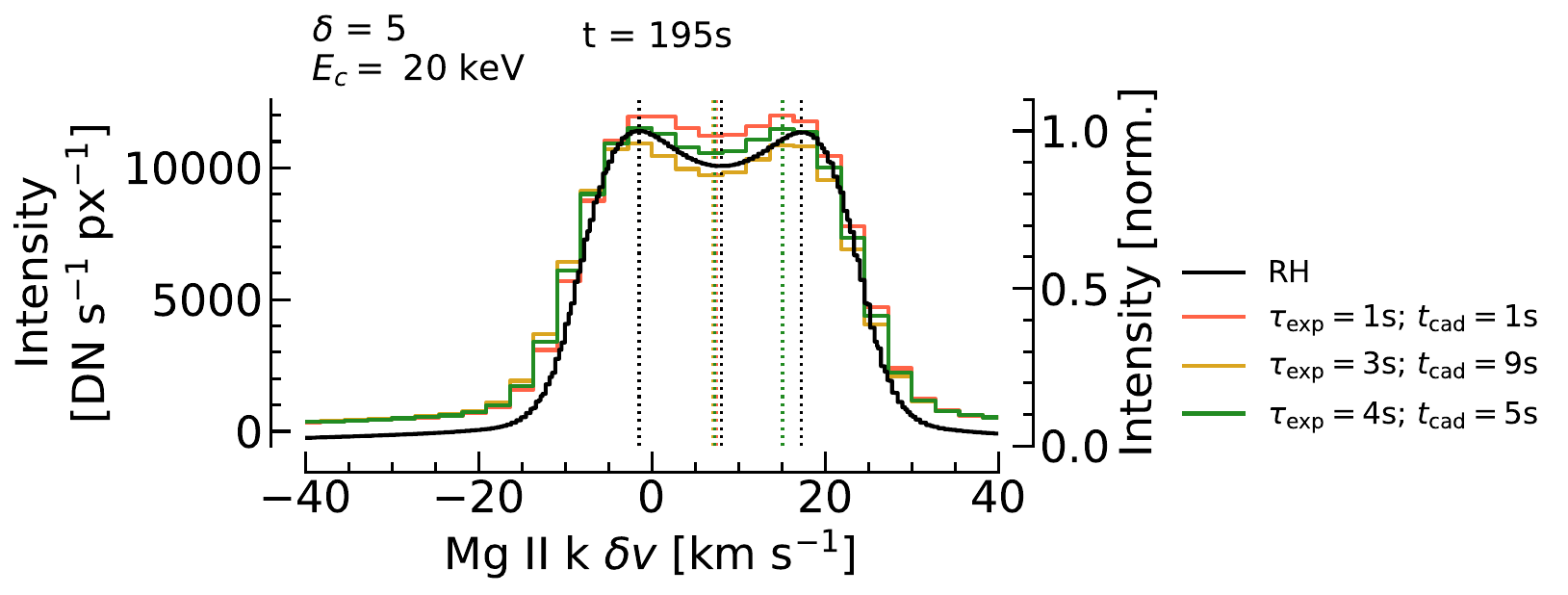}}	
	}
	}
	\caption{\textsl{Illustrating how the \ion{Mg}{2} k line differ when converting to IRIS count rates with different exposure times, in the $\delta = 5$, $E_{c} = 20$~keV simulation. Each panel is a different snapshot, and the vertical lines are the k2v, k2r, and k3 components where the colour indicates the exposure time and cadence applied when converting to IRIS resolution. Some vertical lines are hidden behind others where there is almost no difference in their wavelength positions. The native \rhpar\ profiles are normalised within each snapshot. }}
	\label{fig:mgii_peaklocs_ex}
\end{figure*}

\begin{figure*}
	\centering 
	\vbox{
	\hbox{
	\hspace{0.25in}
	\subfloat{\includegraphics[width = 0.45\textwidth, clip = true, trim = 0.cm 0.cm 0.cm 0.cm]{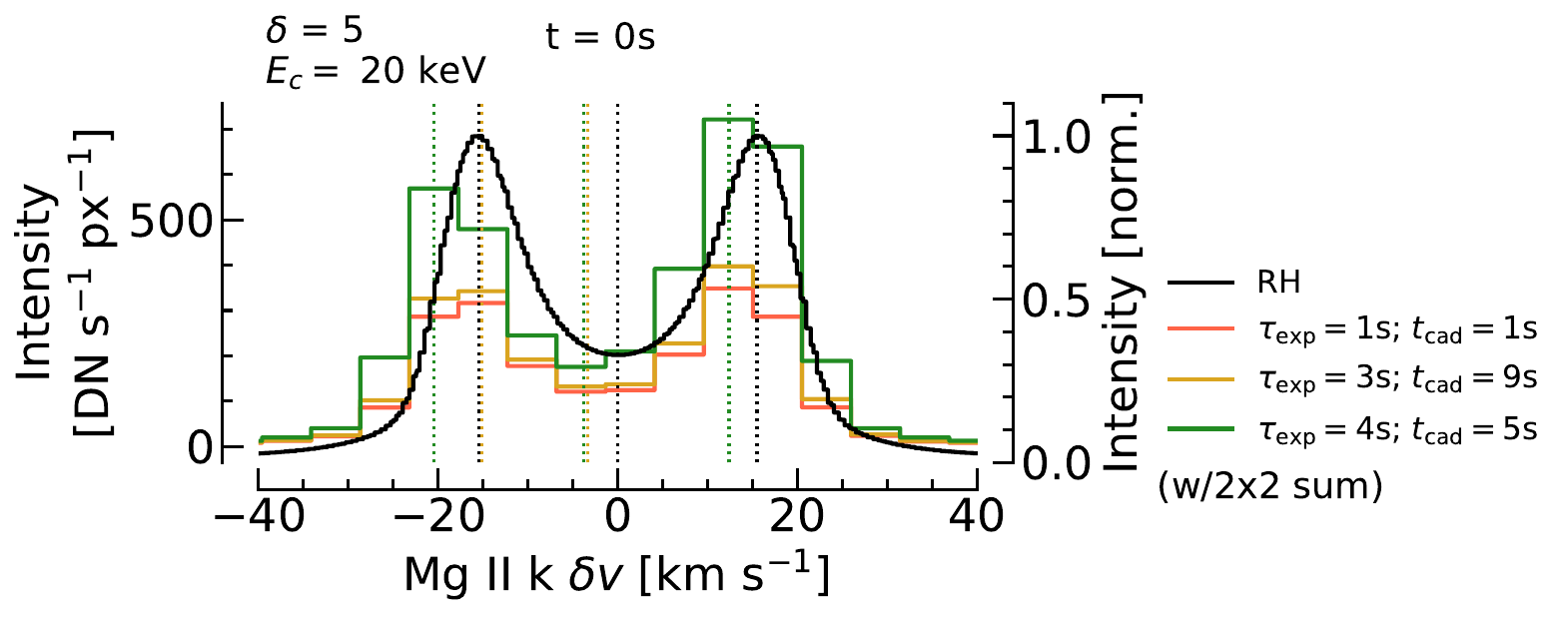}}	
	\subfloat{\includegraphics[width = 0.45\textwidth, clip = true, trim = 0.cm 0.cm 0.cm 0.cm]{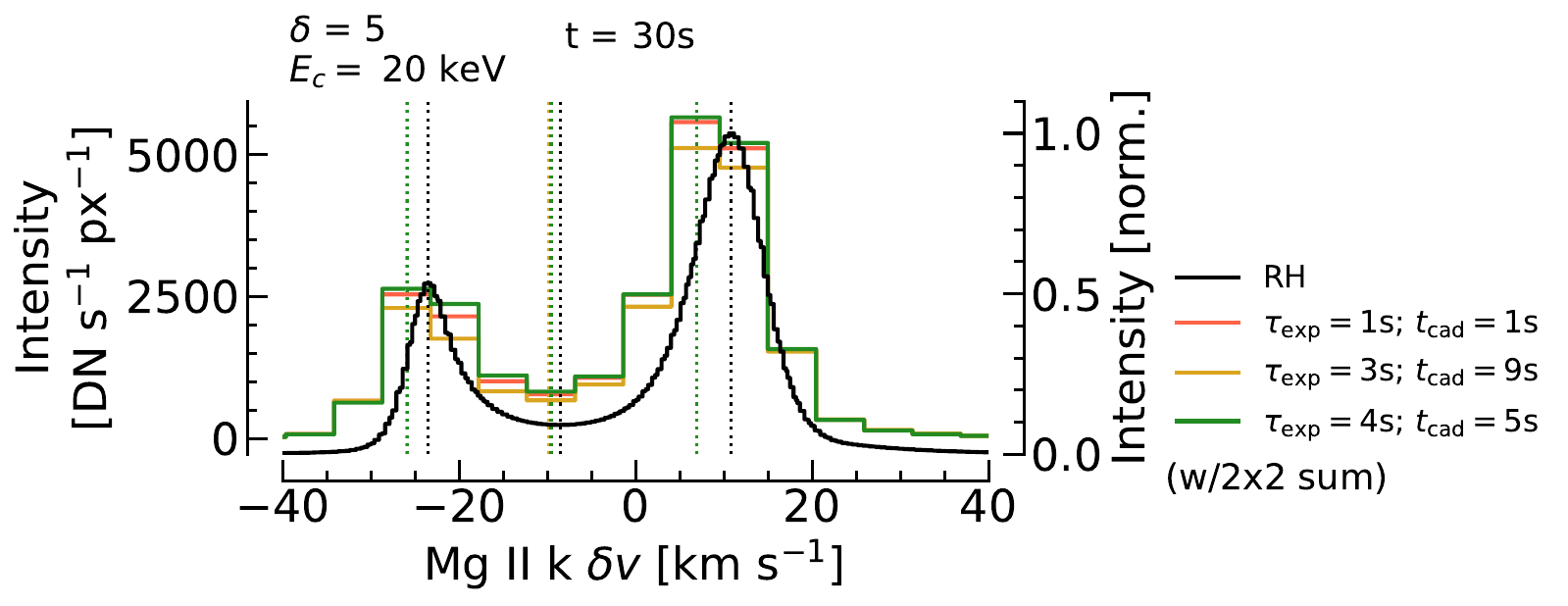}}	
	}
	}
	\vbox{
	\hbox{
	\hspace{0.25in}
	\subfloat{\includegraphics[width = 0.45\textwidth, clip = true, trim = 0.cm 0.cm 0.cm 0.cm]{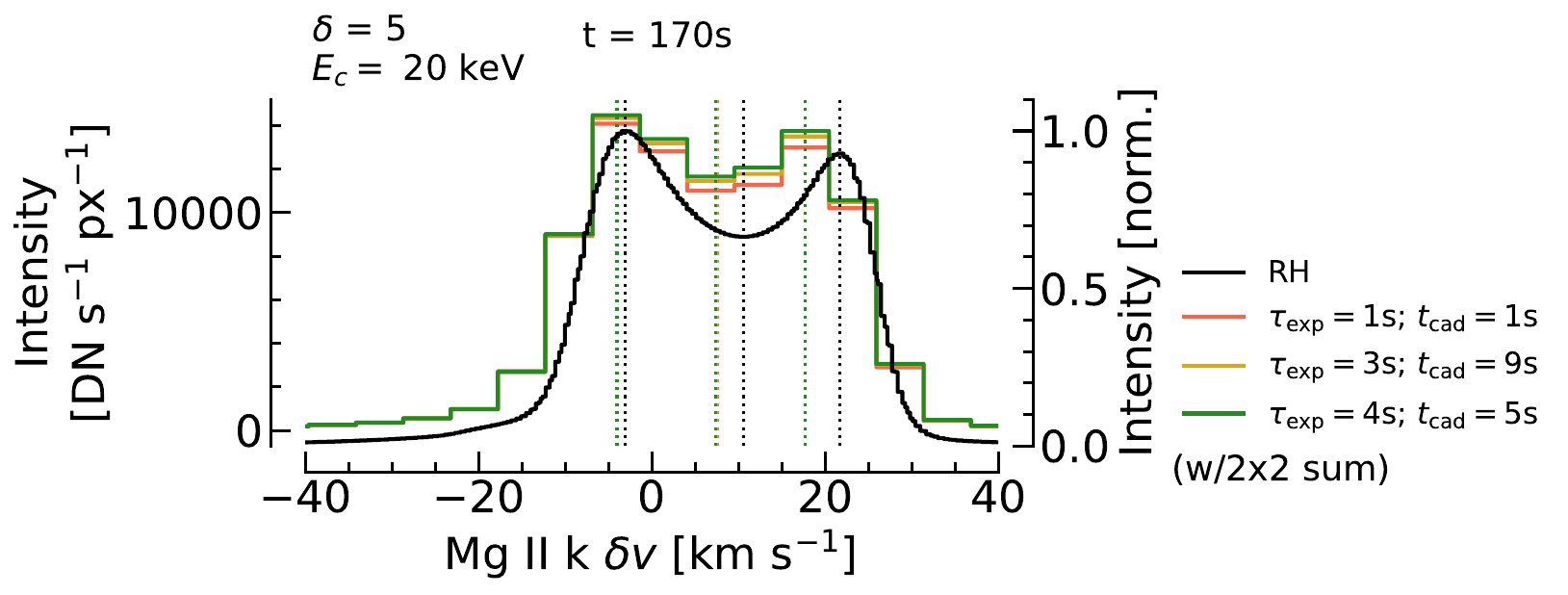}}	
	\subfloat{\includegraphics[width = 0.45\textwidth, clip = true, trim = 0.cm 0.cm 0.cm 0.cm]{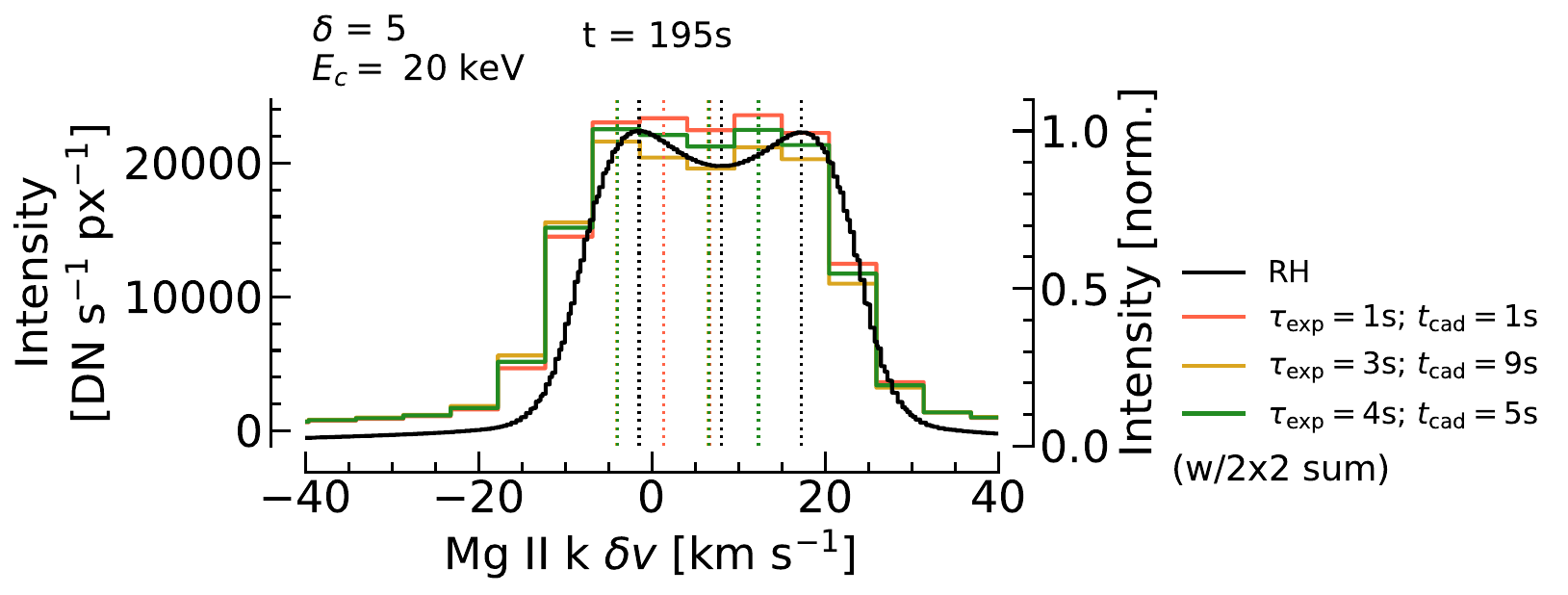}}	
	}
	}
	\caption{\textsl{Same as Figure~\ref{fig:mgii_peaklocs_ex} but including $2\times2$ spectral summing of synthetic IRIS \ion{Mg}{2} k line profiles.}}
	\label{fig:mgii_peaklocs_ex_sum}
\end{figure*}

	Since the identification of the line components may be affected by instrumental and observational effects we explored the differences that result from converting to IRIS-like spectra. Wavelength positions and relative intensities of the k2v, k2r, and k3 components were, broadly speaking, affected by the instrumental properties, the magnitude of which varied in time and with simulation. While these differences do not affect the trends in spectral characteristics, they did lead to some differences in the absolute magnitude of the various metrics. This is complicated by the fact that the mis-identification of the true k2v, k2r, and k3 components are not uniform. That is, in one snapshot the k2v may be mostly correctly identified whereas the k2r is not. These do not drastically affect the conclusions but should be borne in mind when making direct model-data comparisons. Figure~\ref{fig:mgii_peaklocs_ex} illustrates these impacts on identifying k2v, k2r, and k3 for various setups of the $\delta = 5$, $E_{c} = 20$~keV simulation. Figure~\ref{fig:mgii_peaklocs_ex_sum} is similar but includes $2\times2$ spectral summing of the synthetic IRIS spectra.

\section{\ion{O}{4} and \ion{Fe}{21} Spectra From the Main Grid of Flare Simulations}\label{sec:tr_profiles_maingrid}
\begin{figure}
	\centering 
	\vbox{
	\hbox{
	\subfloat{\includegraphics[width = 0.15\textwidth, clip = true, trim = 0.cm 0.cm 0.cm 0.cm]{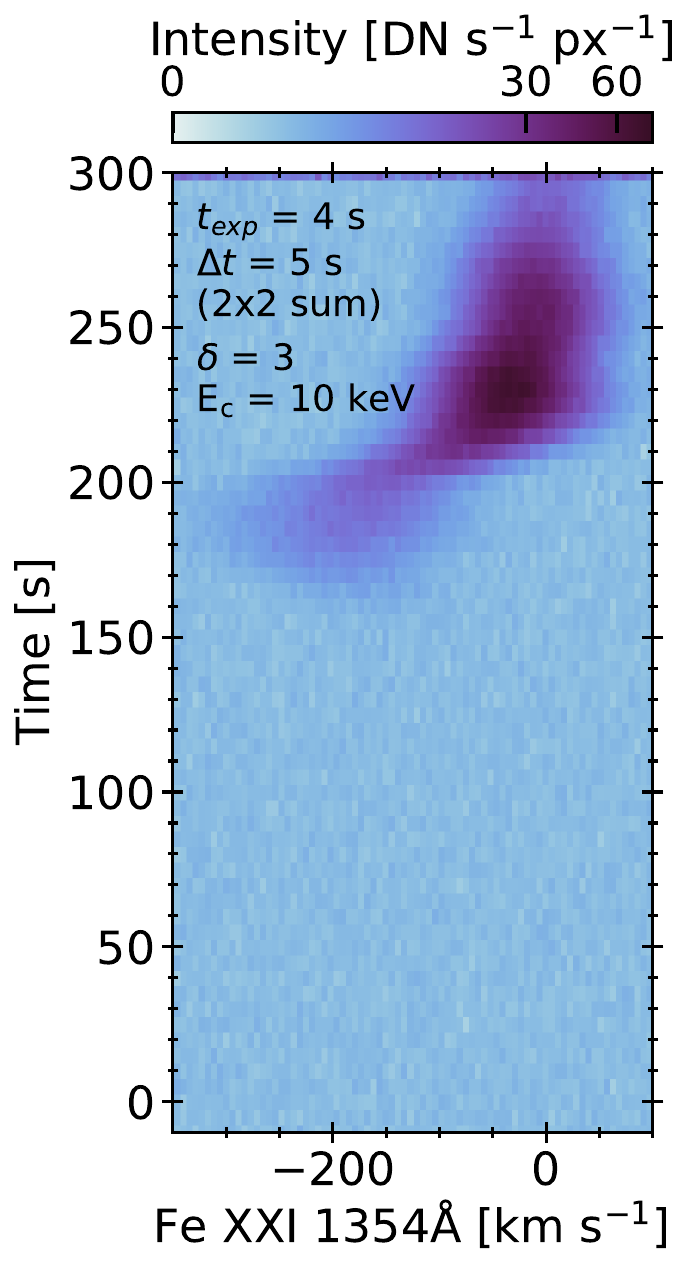}}
	\subfloat{\includegraphics[width = 0.15\textwidth, clip = true, trim = 0.cm 0.cm 0.cm 0.cm]{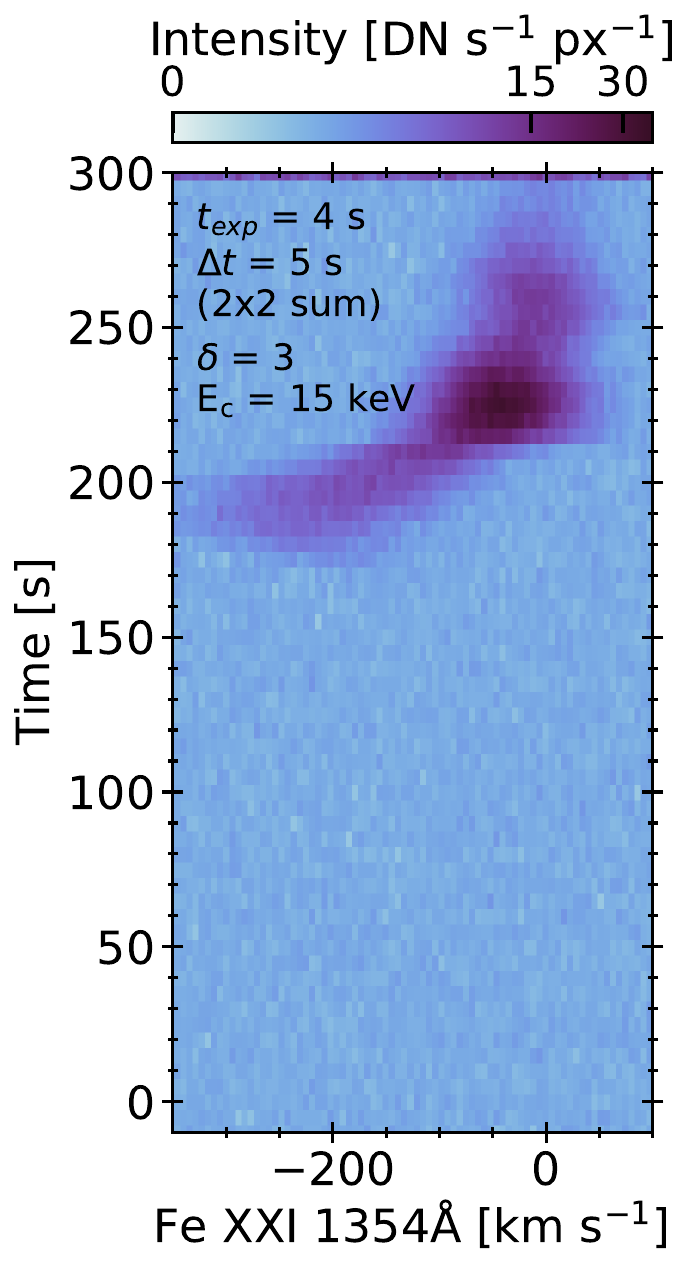}}
	\subfloat{\includegraphics[width = 0.15\textwidth, clip = true, trim = 0.cm 0.cm 0.cm 0.cm]{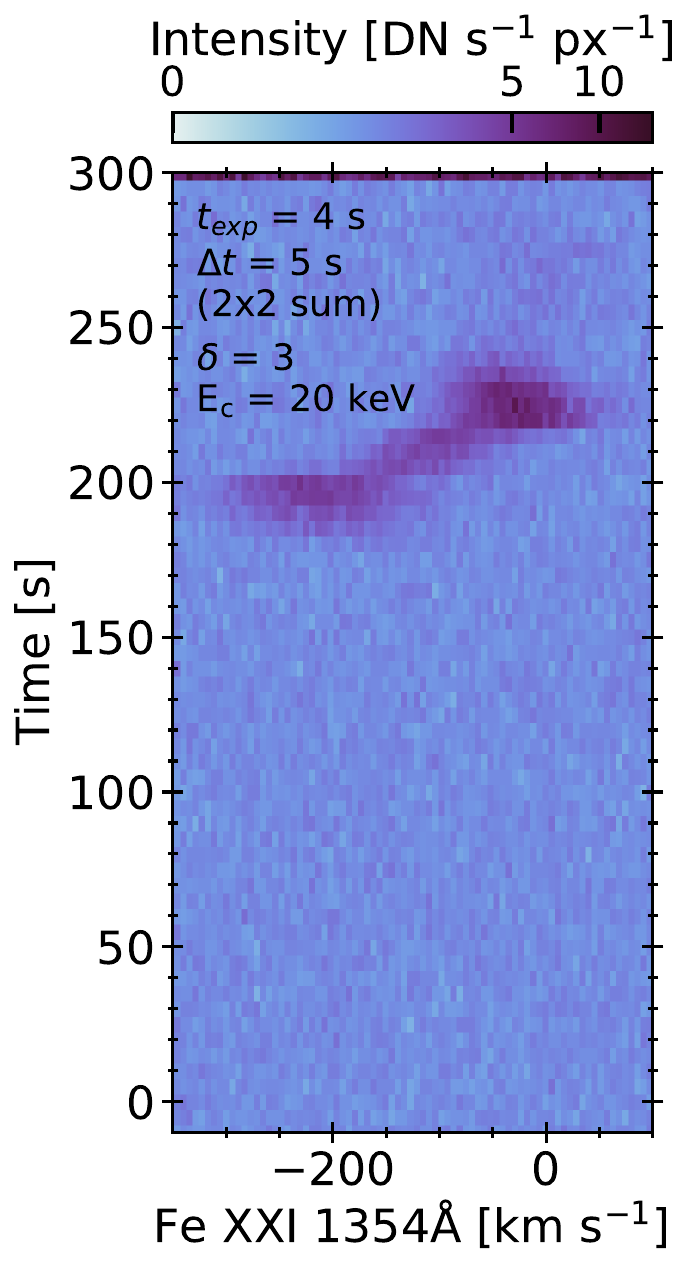}}
	\subfloat{\includegraphics[width = 0.15\textwidth, clip = true, trim = 0.cm 0.cm 0.cm 0.cm]{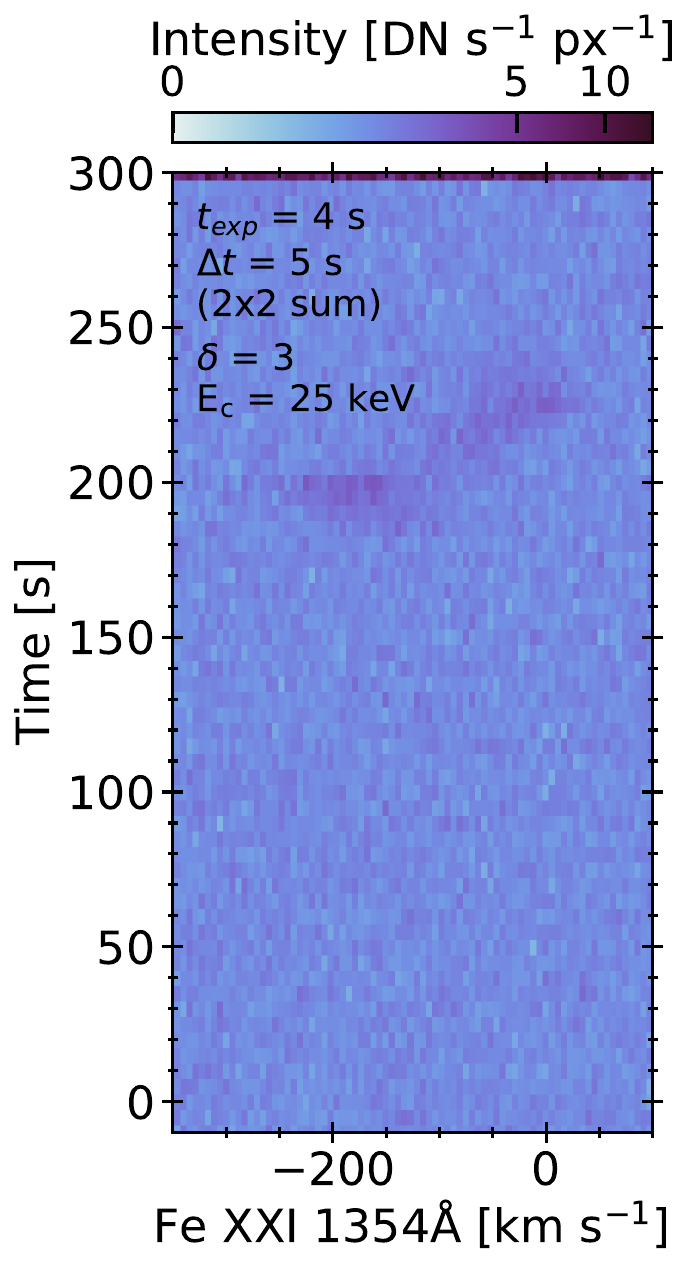}}
	\subfloat{\includegraphics[width = 0.15\textwidth, clip = true, trim = 0.cm 0.cm 0.cm 0.cm]{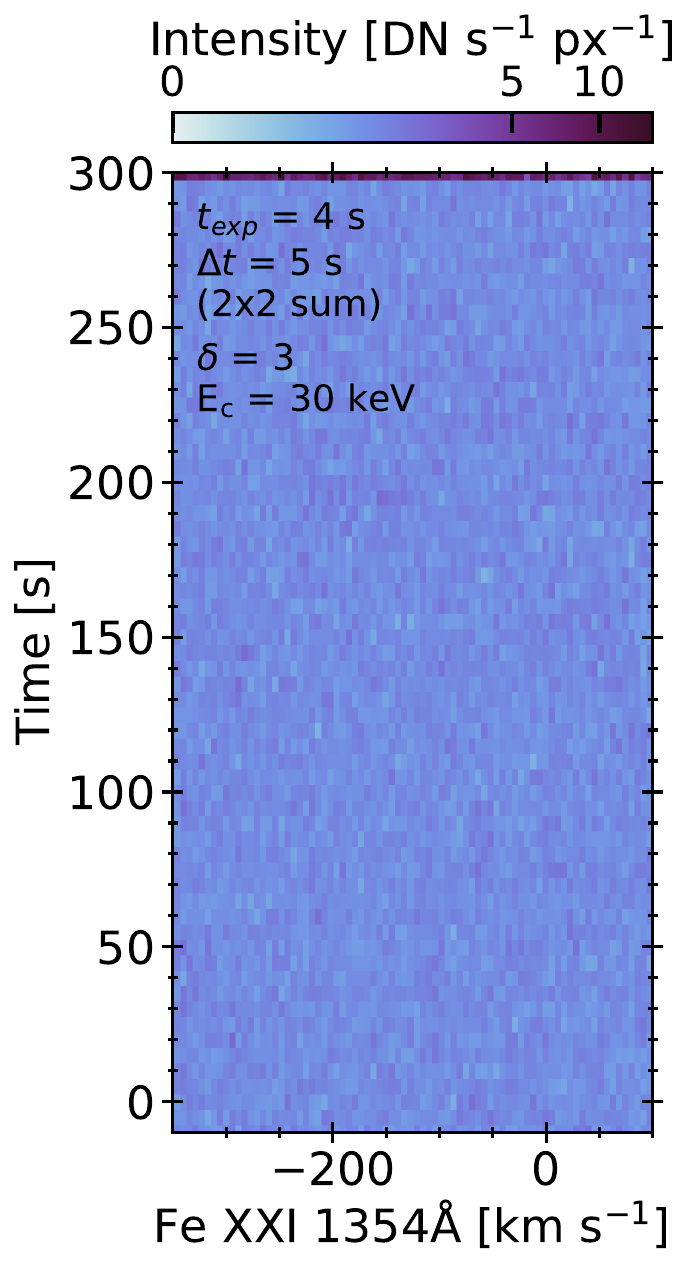}}
	}
	}
	\vbox{
	\hbox{
	\subfloat{\includegraphics[width = 0.15\textwidth, clip = true, trim = 0.cm 0.cm 0.cm 0.cm]{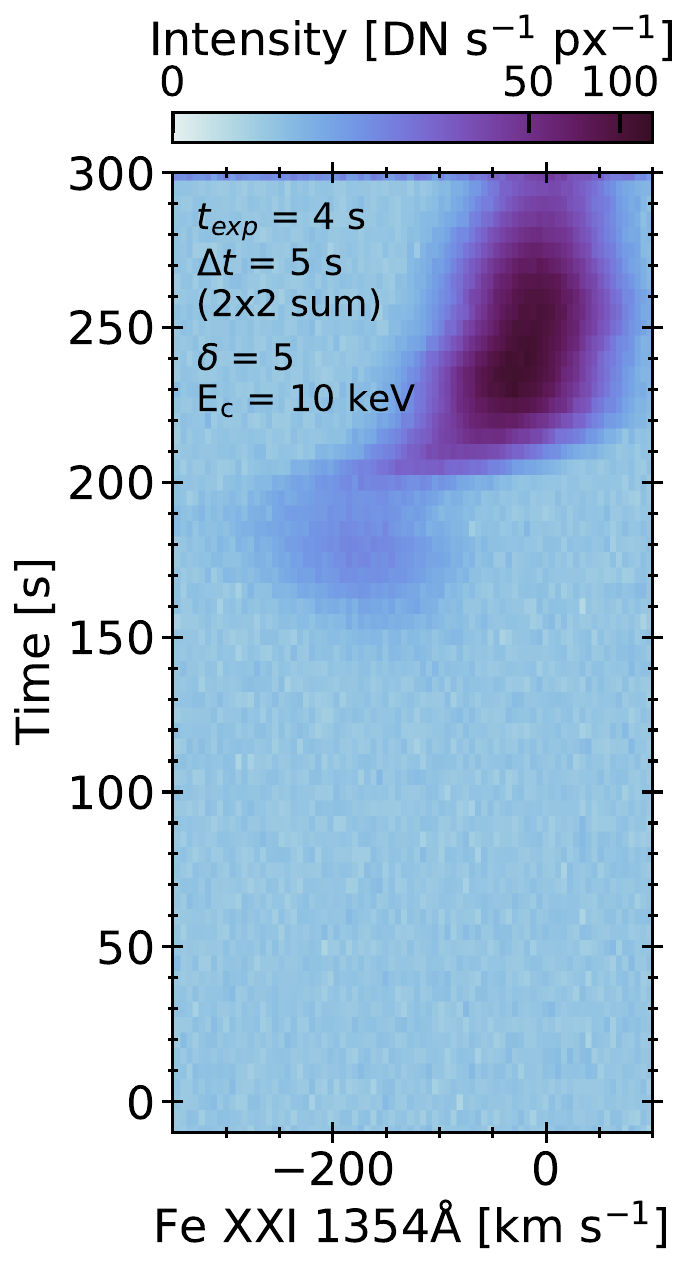}}
	\subfloat{\includegraphics[width = 0.15\textwidth, clip = true, trim = 0.cm 0.cm 0.cm 0.cm]{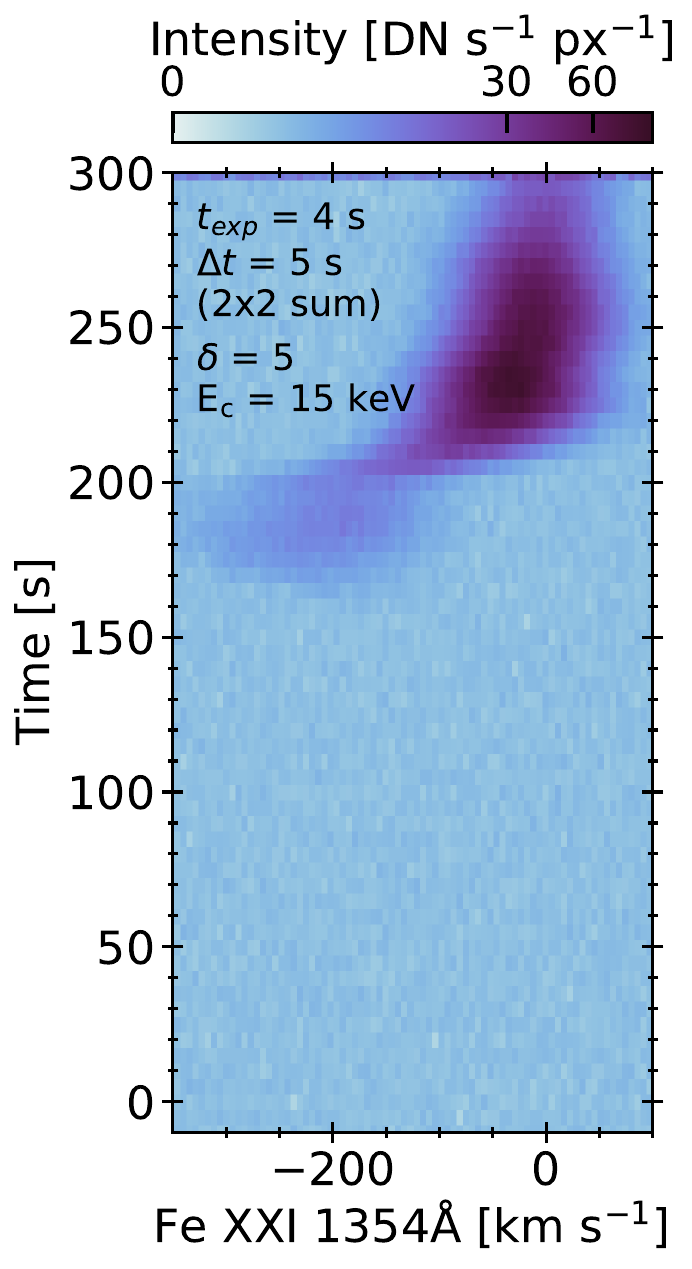}}
	\subfloat{\includegraphics[width = 0.15\textwidth, clip = true, trim = 0.cm 0.cm 0.cm 0.cm]{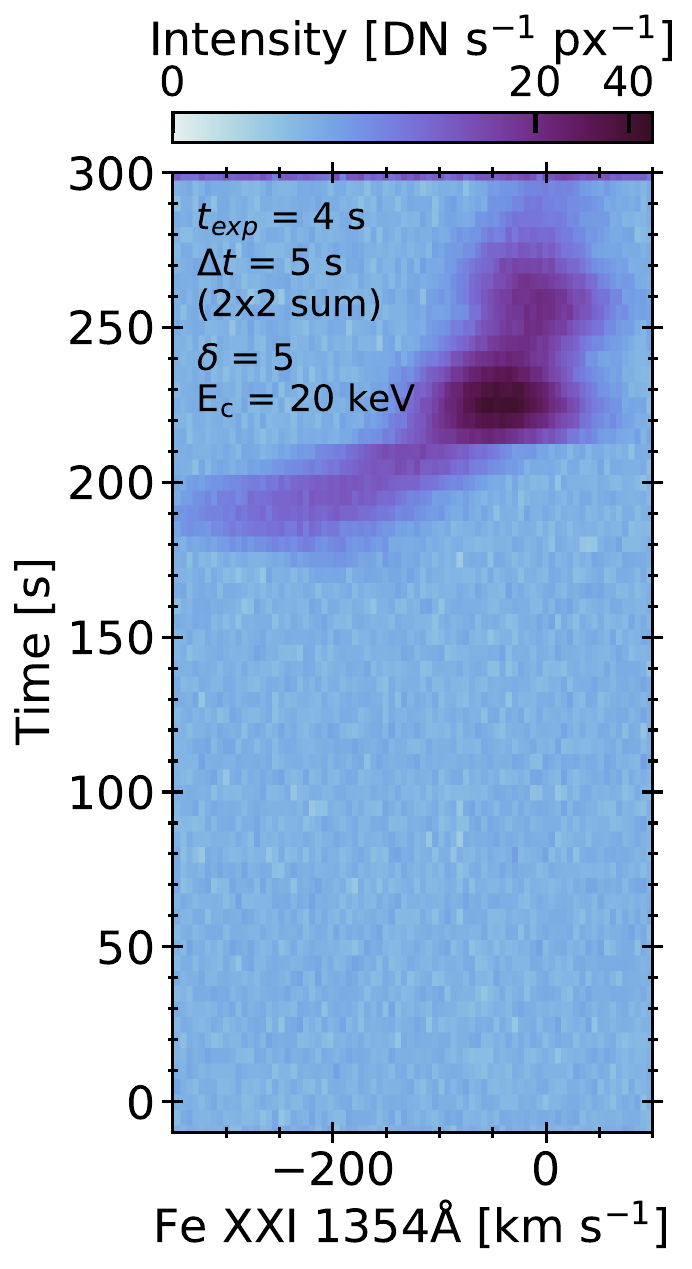}}
	\subfloat{\includegraphics[width = 0.15\textwidth, clip = true, trim = 0.cm 0.cm 0.cm 0.cm]{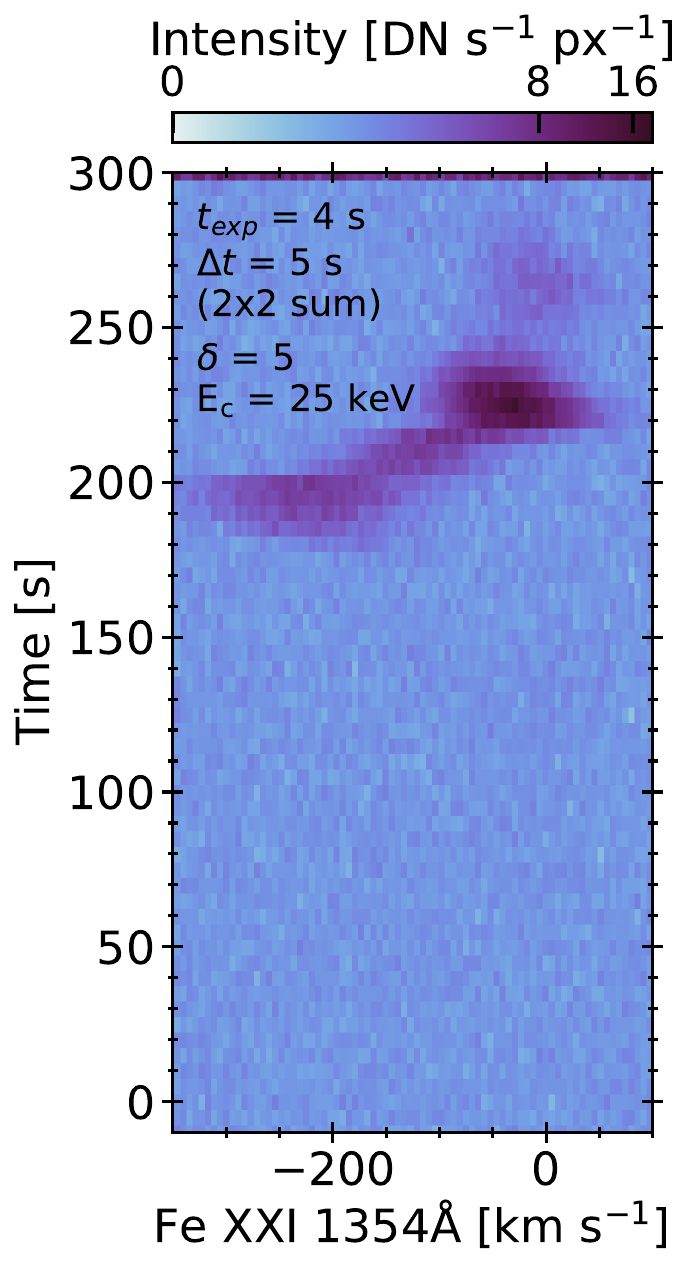}}
	\subfloat{\includegraphics[width = 0.15\textwidth, clip = true, trim = 0.cm 0.cm 0.cm 0.cm]{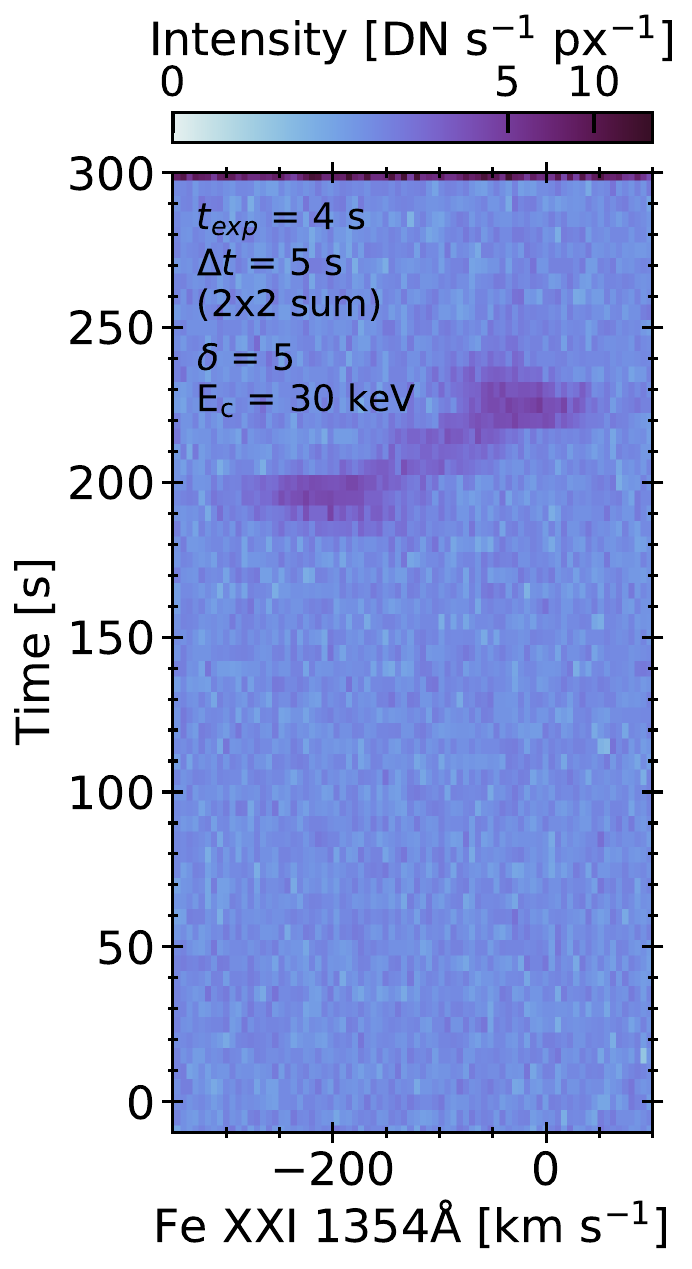}}
	}
	}
	\vbox{
	\hbox{
	\subfloat{\includegraphics[width = 0.15\textwidth, clip = true, trim = 0.cm 0.cm 0.cm 0.cm]{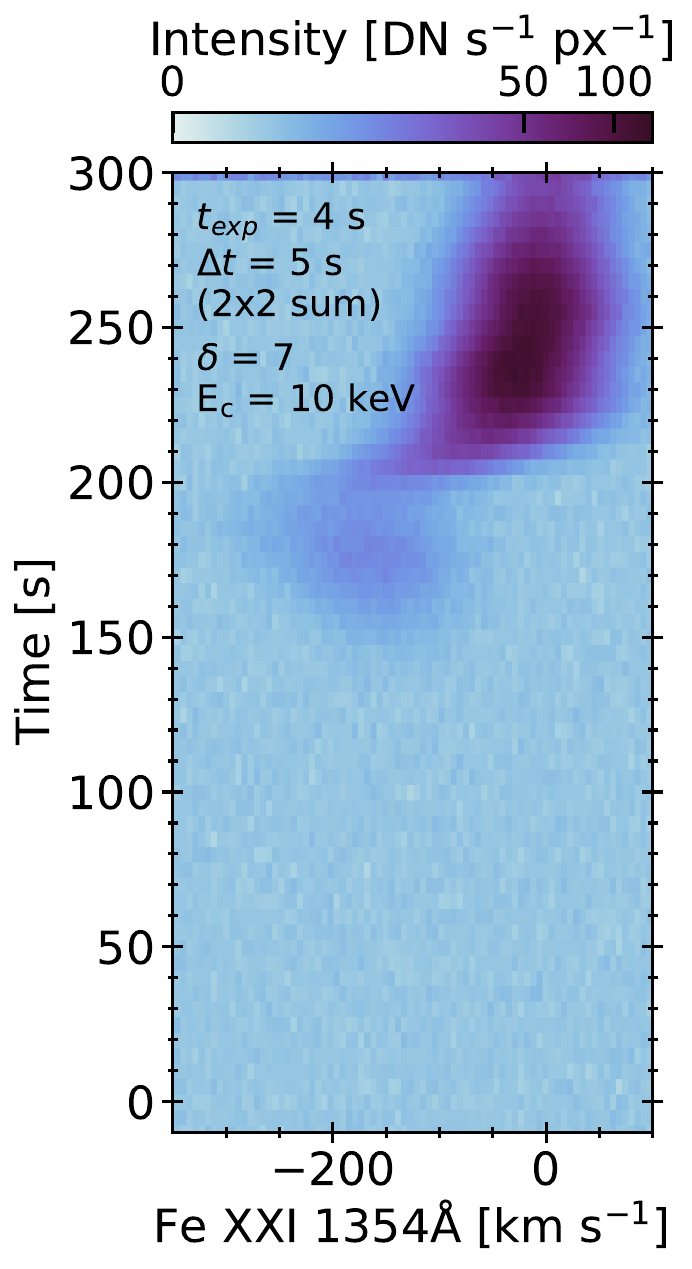}}
	\subfloat{\includegraphics[width =0.15\textwidth, clip = true, trim = 0.cm 0.cm 0.cm 0.cm]{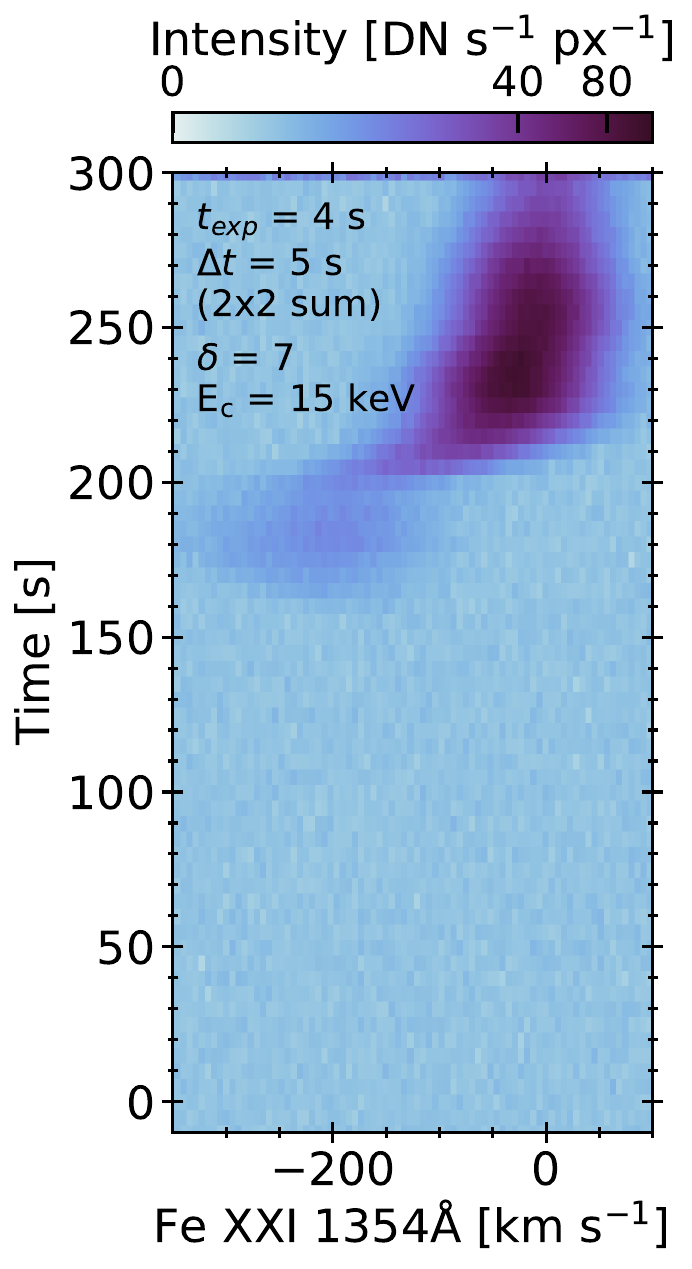}}
	\subfloat{\includegraphics[width = 0.15\textwidth, clip = true, trim = 0.cm 0.cm 0.cm 0.cm]{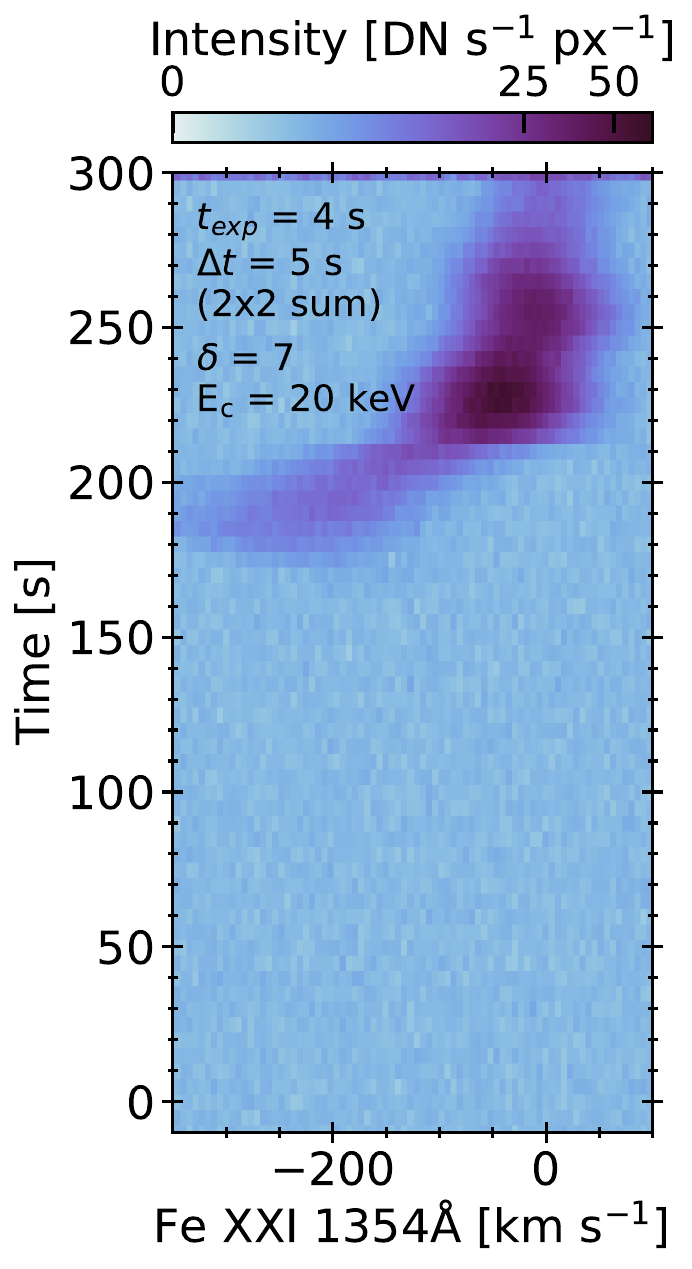}}
	\subfloat{\includegraphics[width = 0.15\textwidth, clip = true, trim = 0.cm 0.cm 0.cm 0.cm]{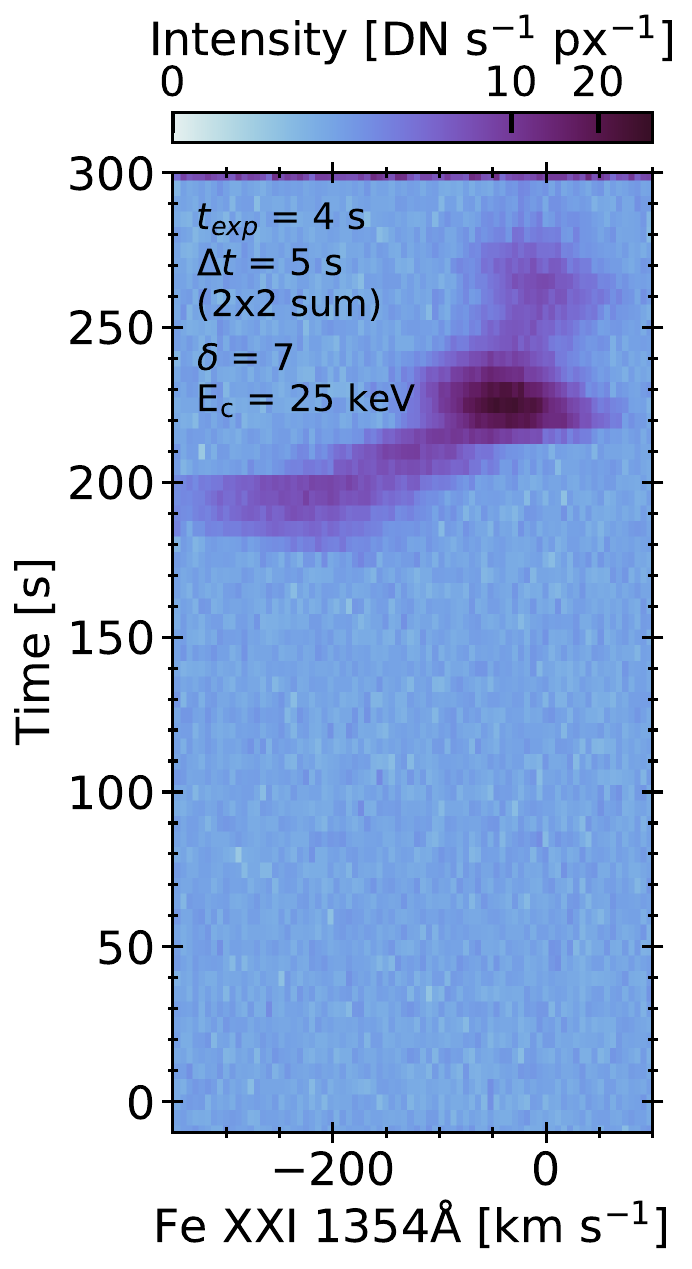}}
	\subfloat{\includegraphics[width = 0.15\textwidth, clip = true, trim = 0.cm 0.cm 0.cm 0.cm]{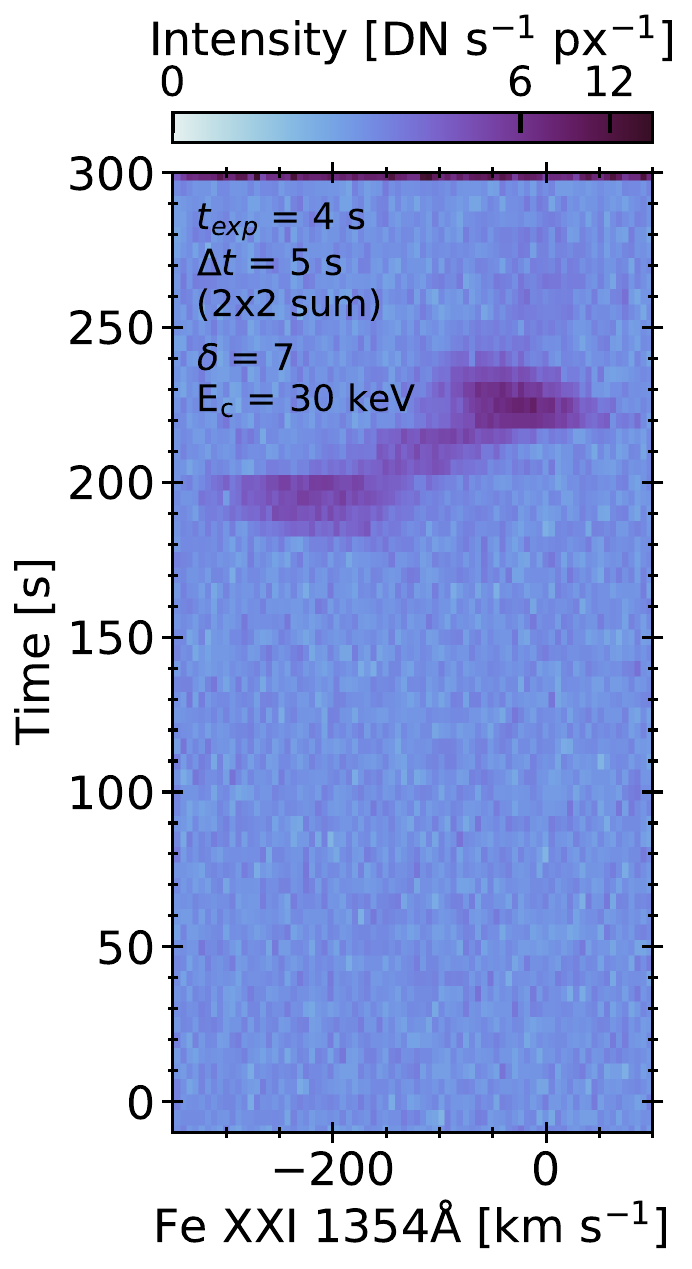}}
	}
	}
		\caption{\textsl{Synthetic \ion{Fe}{21} 1354.1~\AA\ line profiles as a function of time in each flare simulation from the main parameter study. The top row is $\delta = 3$, the middle row is $\delta = 5$, and the bottom row is $\delta = 7$. From left to right the low energy cutoffs are $E_{c} = [10, 15, 20, 25, 30]$~keV, such that the hardest nonthermal electron distribution is the top right, and the softest is bottom left. Spectra have been converted to IRIS quality, with $2\times2$ spectral summing, $t_{\mathrm{exp}} = 4$~s, and $\Delta t = 5$~s.}}
	\label{fig:fexxi_profiles_maingrid}
\end{figure}

\begin{figure}
	\centering 
	\vbox{
	\hbox{
	\subfloat{\includegraphics[width = 0.15\textwidth, clip = true, trim = 0.cm 0.cm 0.cm 0.cm]{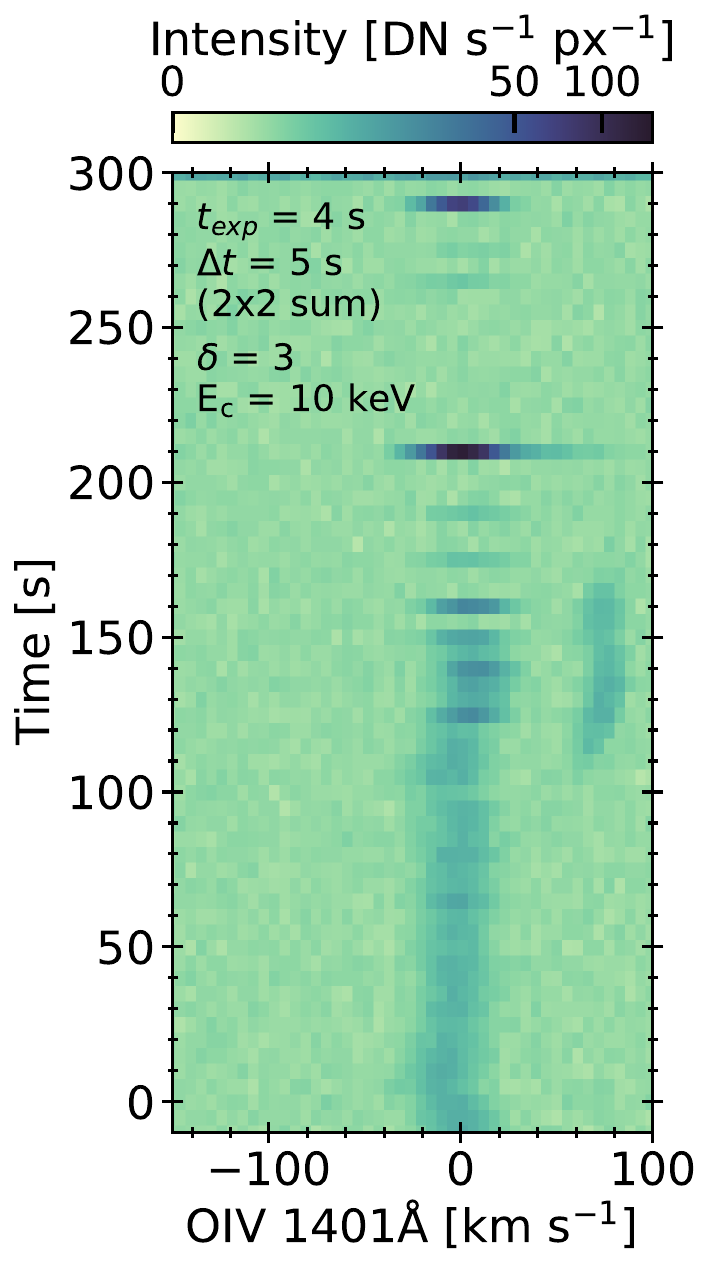}}
	\subfloat{\includegraphics[width = 0.15\textwidth, clip = true, trim = 0.cm 0.cm 0.cm 0.cm]{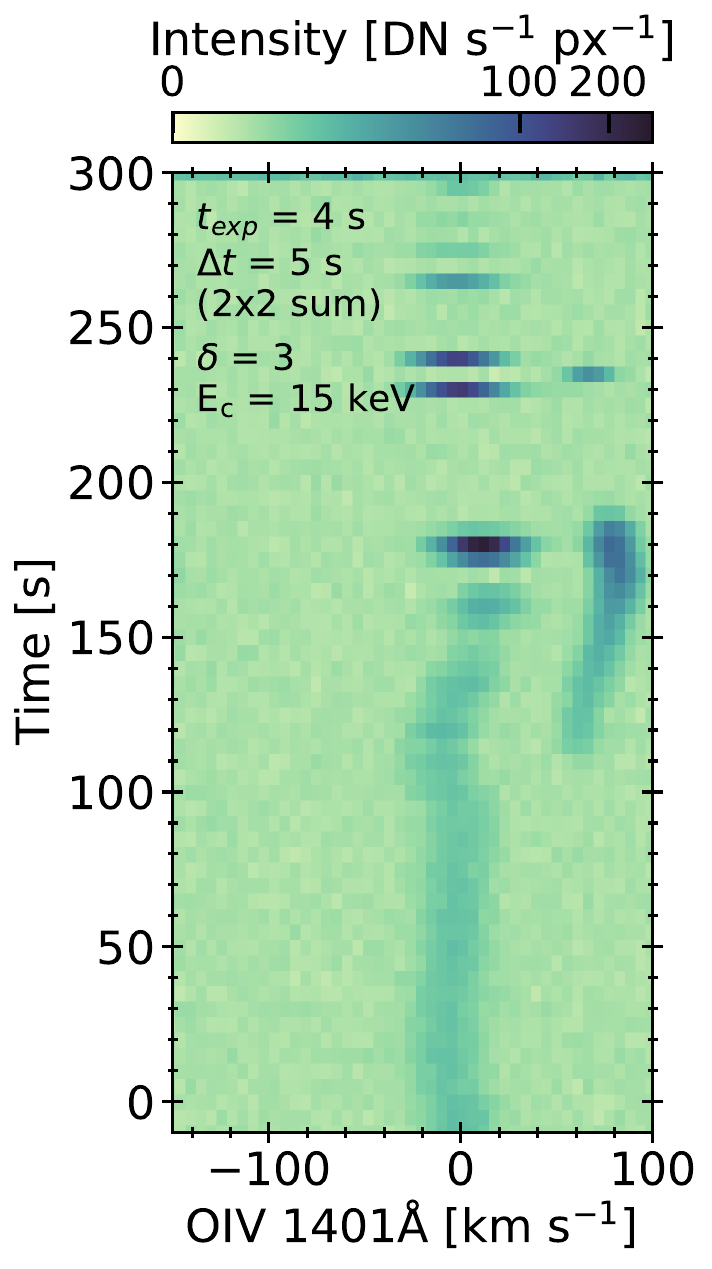}}
	\subfloat{\includegraphics[width = 0.15\textwidth, clip = true, trim = 0.cm 0.cm 0.cm 0.cm]{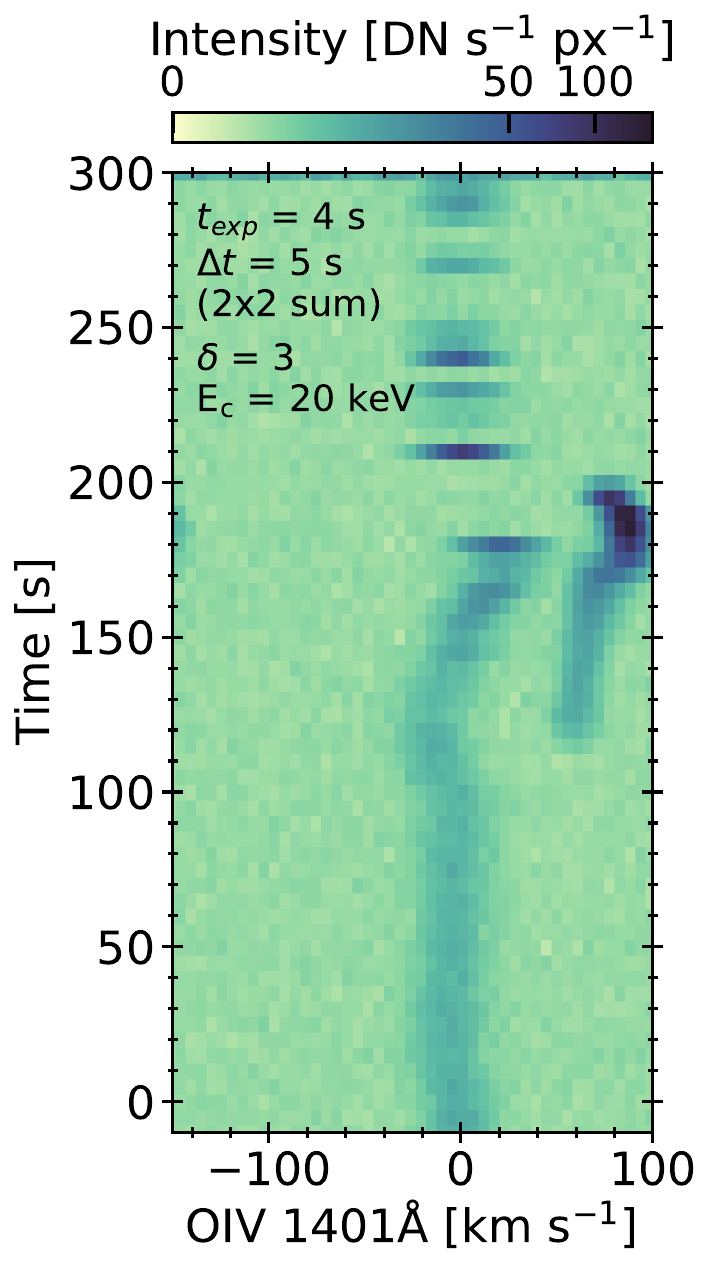}}
	\subfloat{\includegraphics[width = 0.15\textwidth, clip = true, trim = 0.cm 0.cm 0.cm 0.cm]{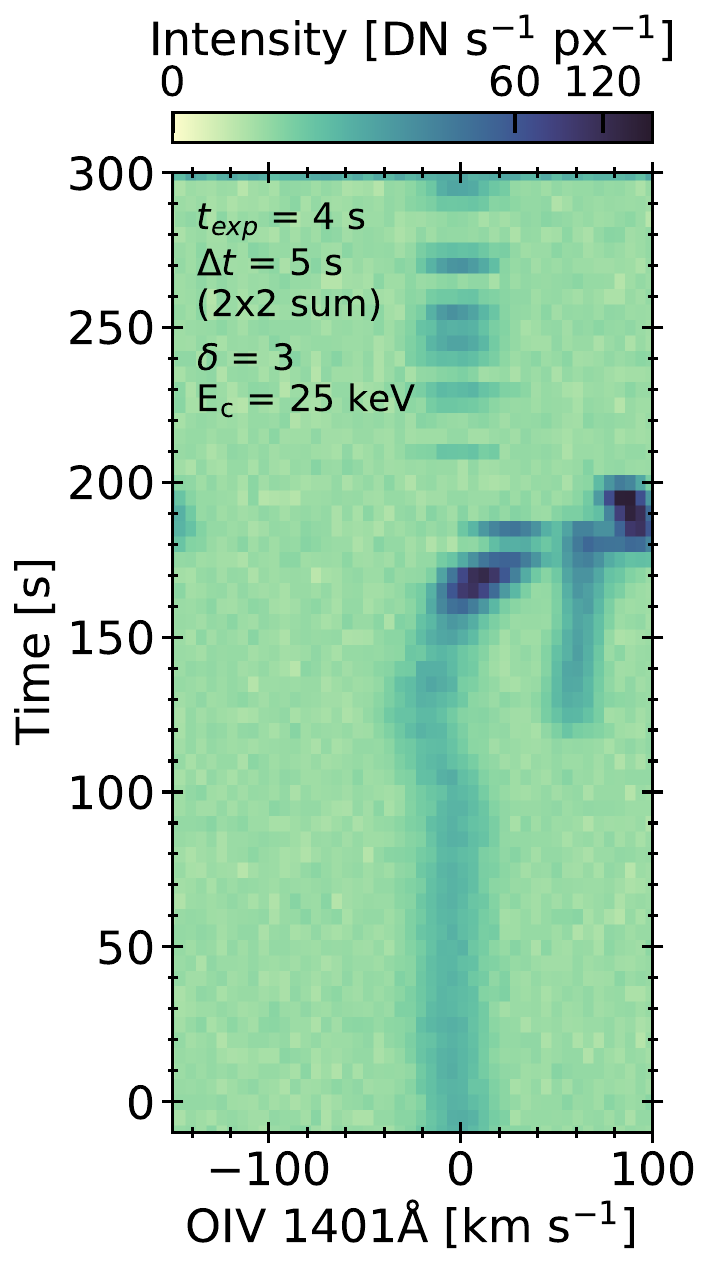}}
	\subfloat{\includegraphics[width = 0.15\textwidth, clip = true, trim = 0.cm 0.cm 0.cm 0.cm]{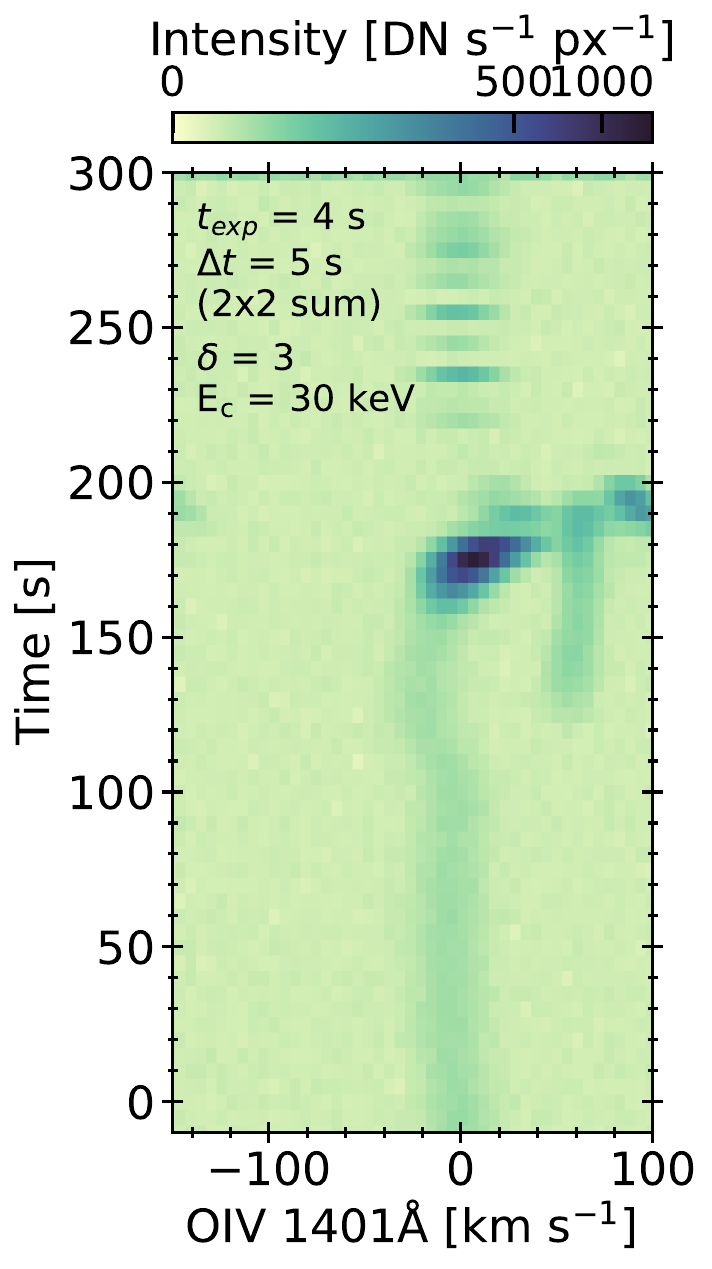}}
	}
	}
	\vbox{
	\hbox{
	\subfloat{\includegraphics[width = 0.15\textwidth, clip = true, trim = 0.cm 0.cm 0.cm 0.cm]{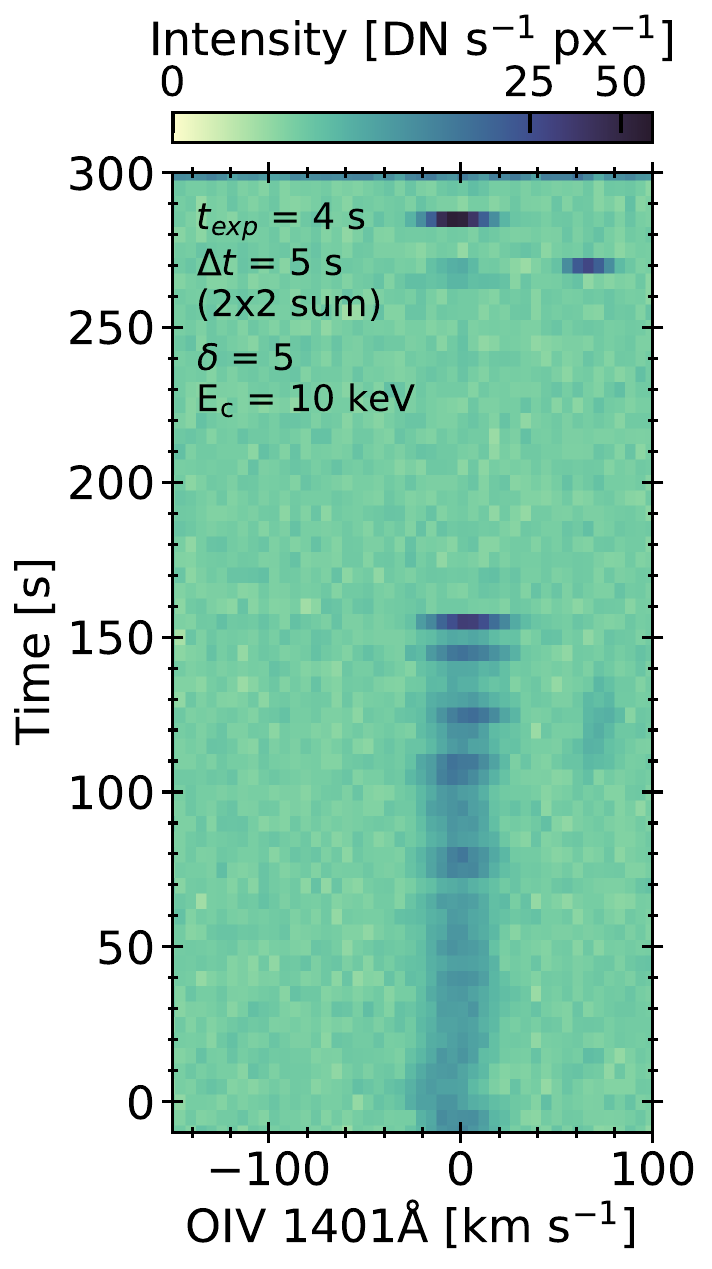}}
	\subfloat{\includegraphics[width = 0.15\textwidth, clip = true, trim = 0.cm 0.cm 0.cm 0.cm]{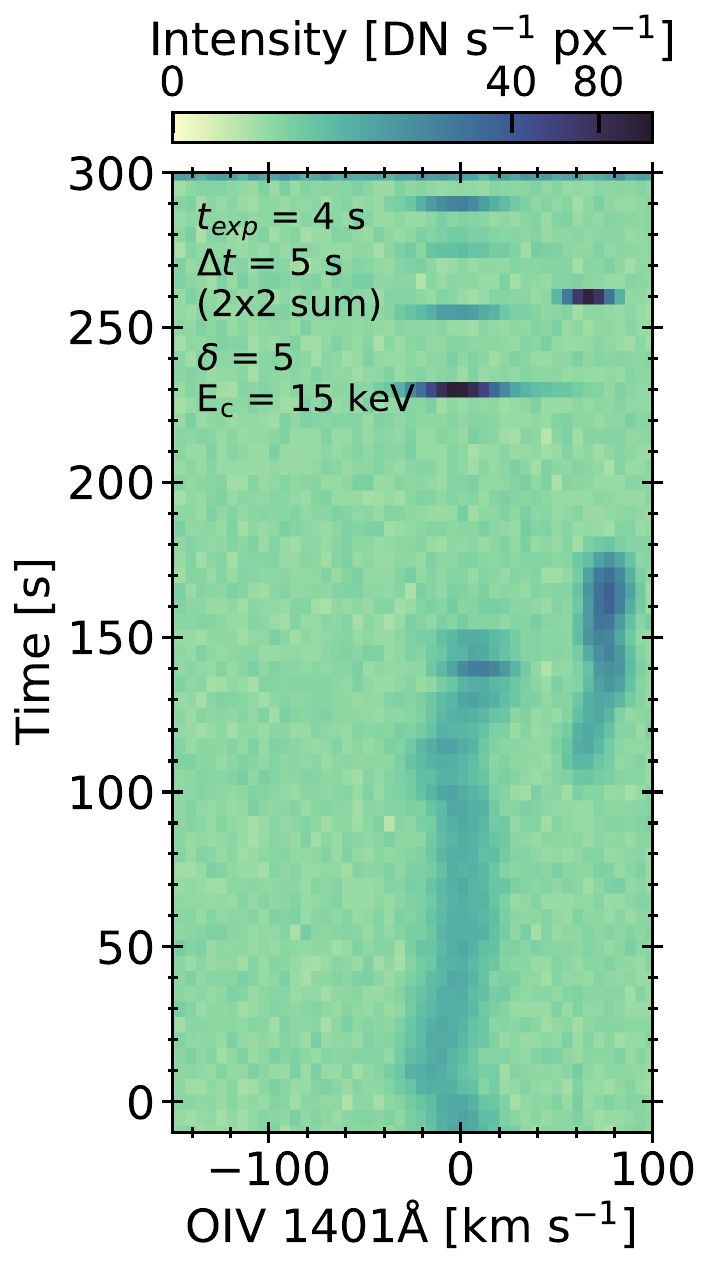}}
	\subfloat{\includegraphics[width = 0.15\textwidth, clip = true, trim = 0.cm 0.cm 0.cm 0.cm]{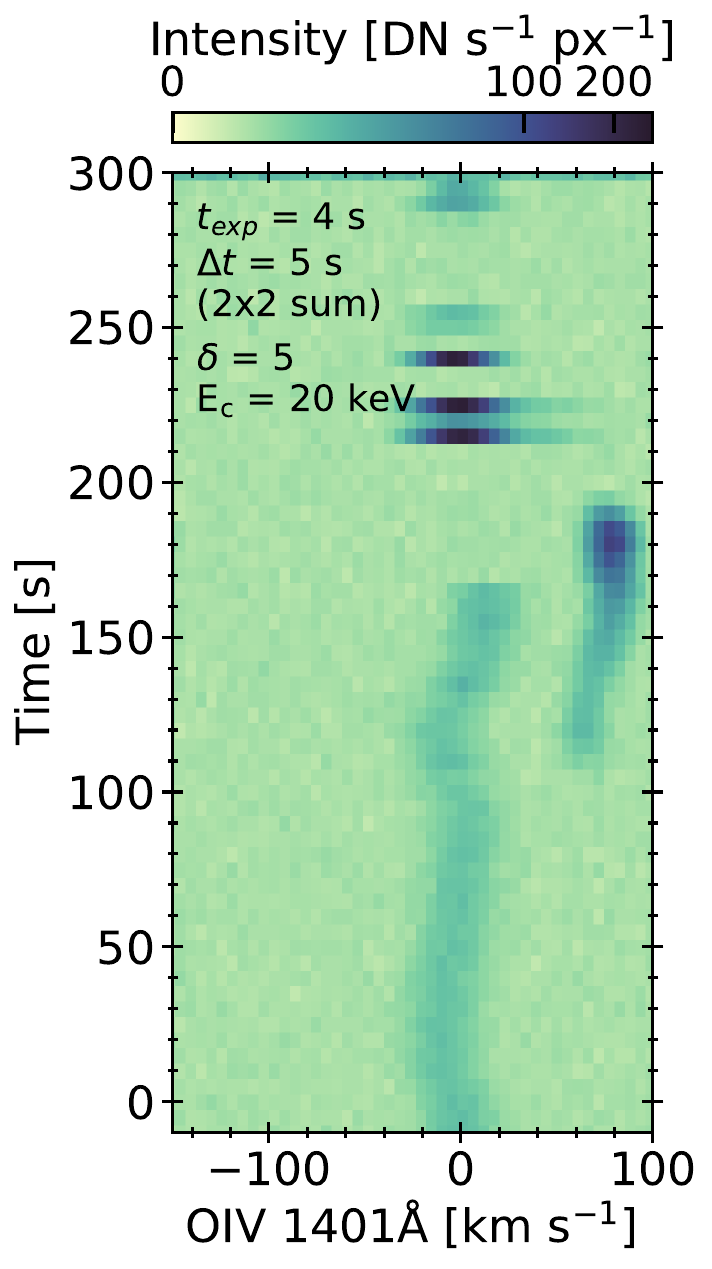}}
	\subfloat{\includegraphics[width = 0.15\textwidth, clip = true, trim = 0.cm 0.cm 0.cm 0.cm]{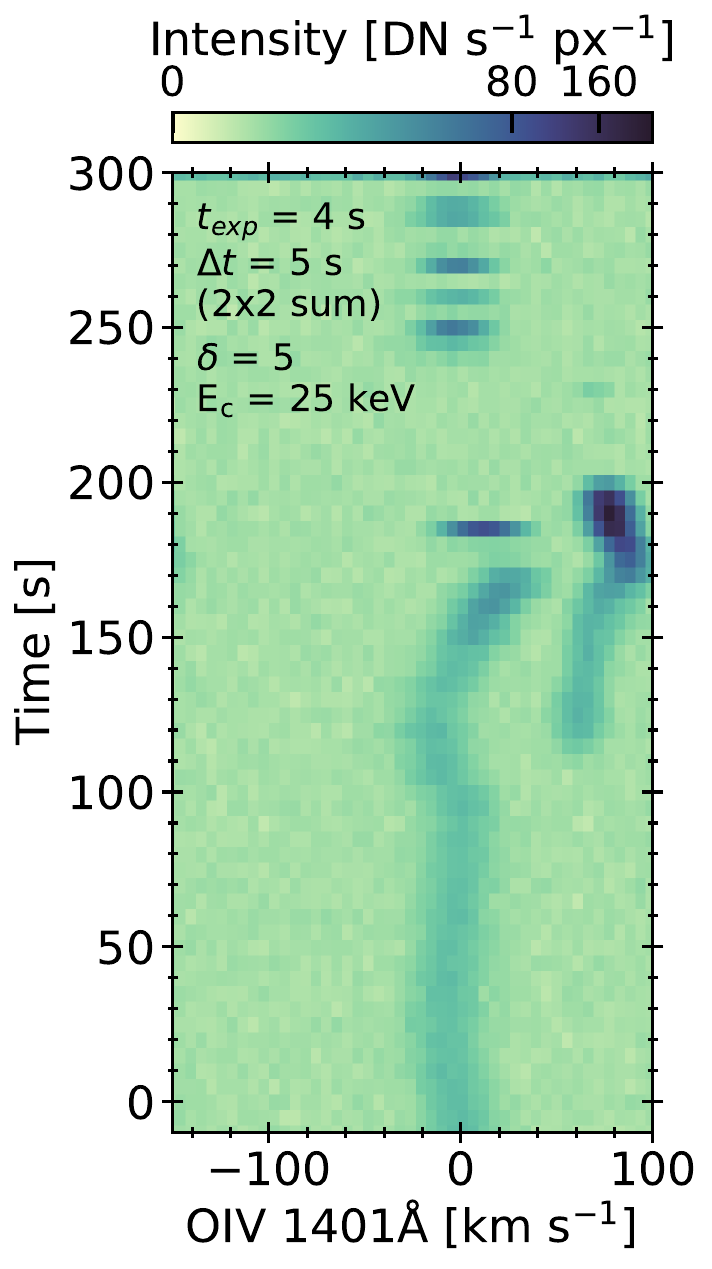}}
	\subfloat{\includegraphics[width = 0.15\textwidth, clip = true, trim = 0.cm 0.cm 0.cm 0.cm]{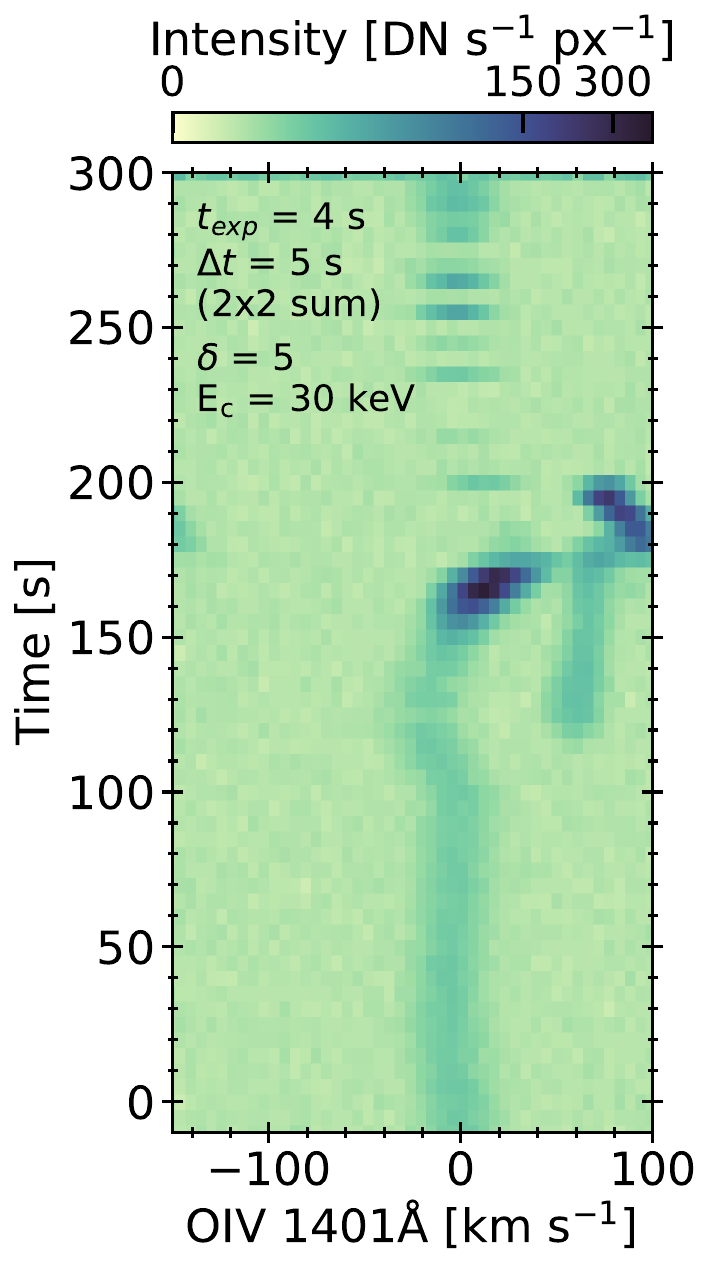}}
	}
	}
	\vbox{
	\hbox{
	\subfloat{\includegraphics[width = 0.15\textwidth, clip = true, trim = 0.cm 0.cm 0.cm 0.cm]{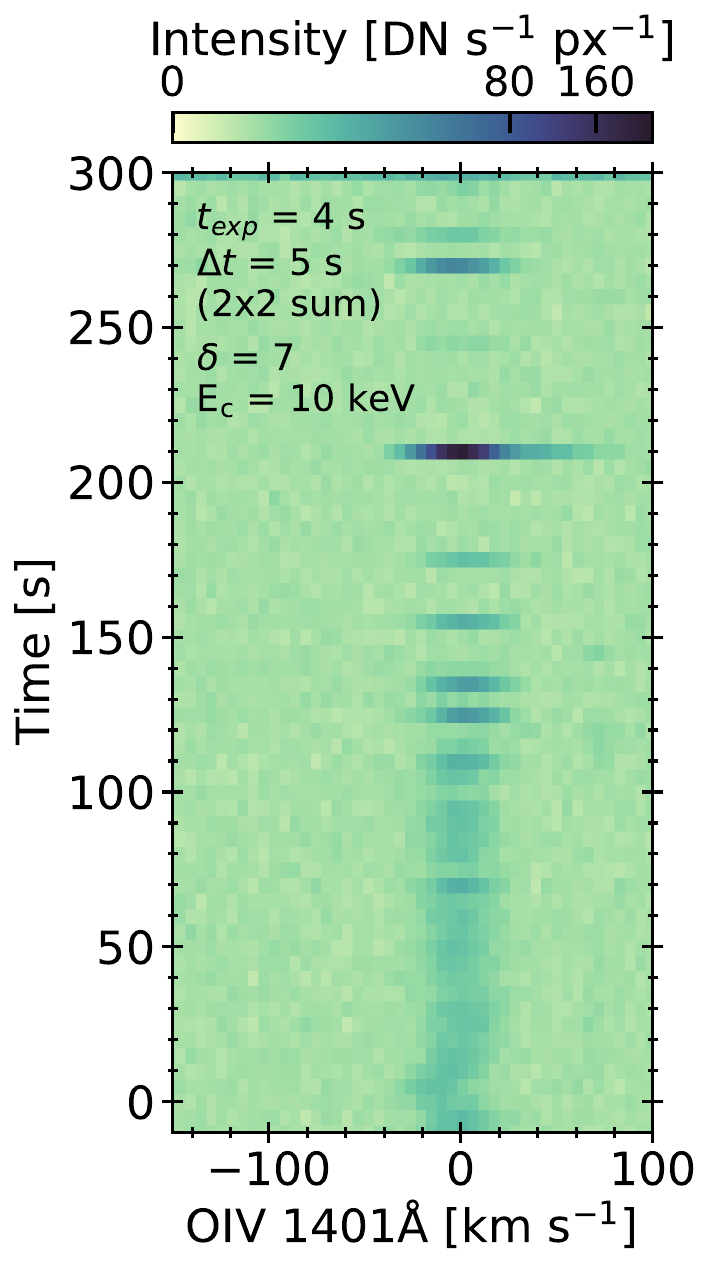}}
	\subfloat{\includegraphics[width =0.15\textwidth, clip = true, trim = 0.cm 0.cm 0.cm 0.cm]{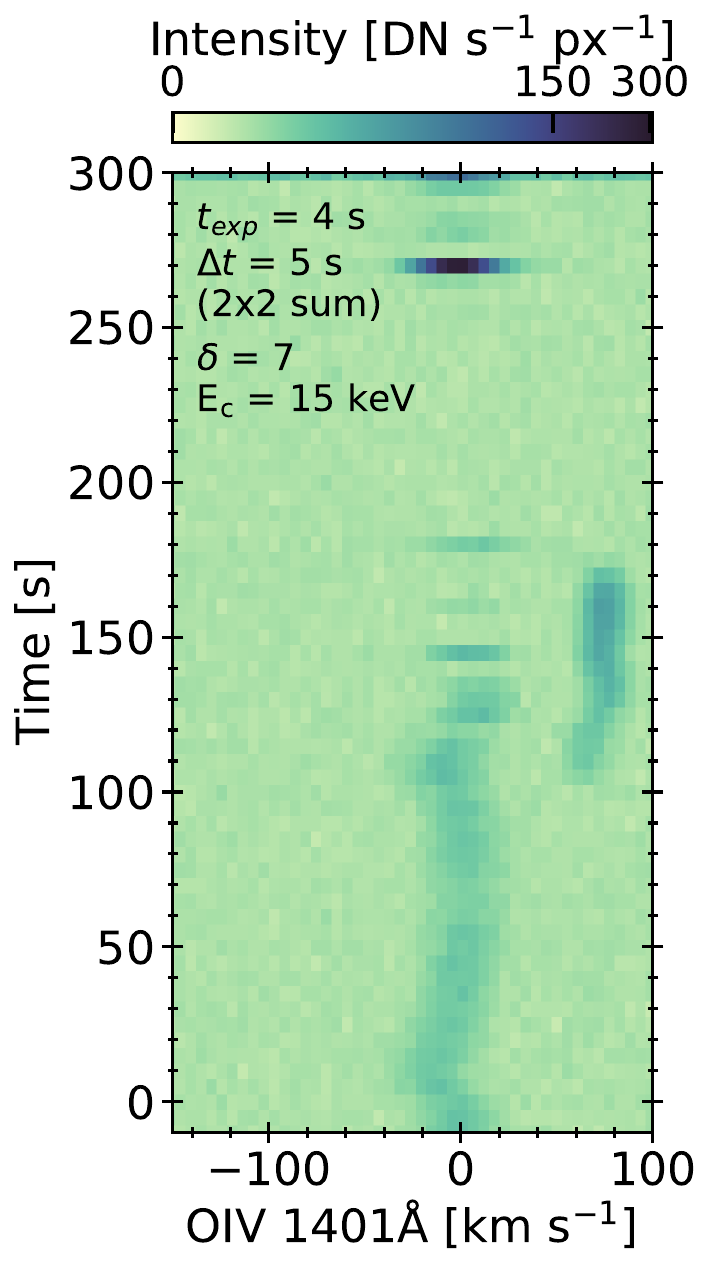}}
	\subfloat{\includegraphics[width = 0.15\textwidth, clip = true, trim = 0.cm 0.cm 0.cm 0.cm]{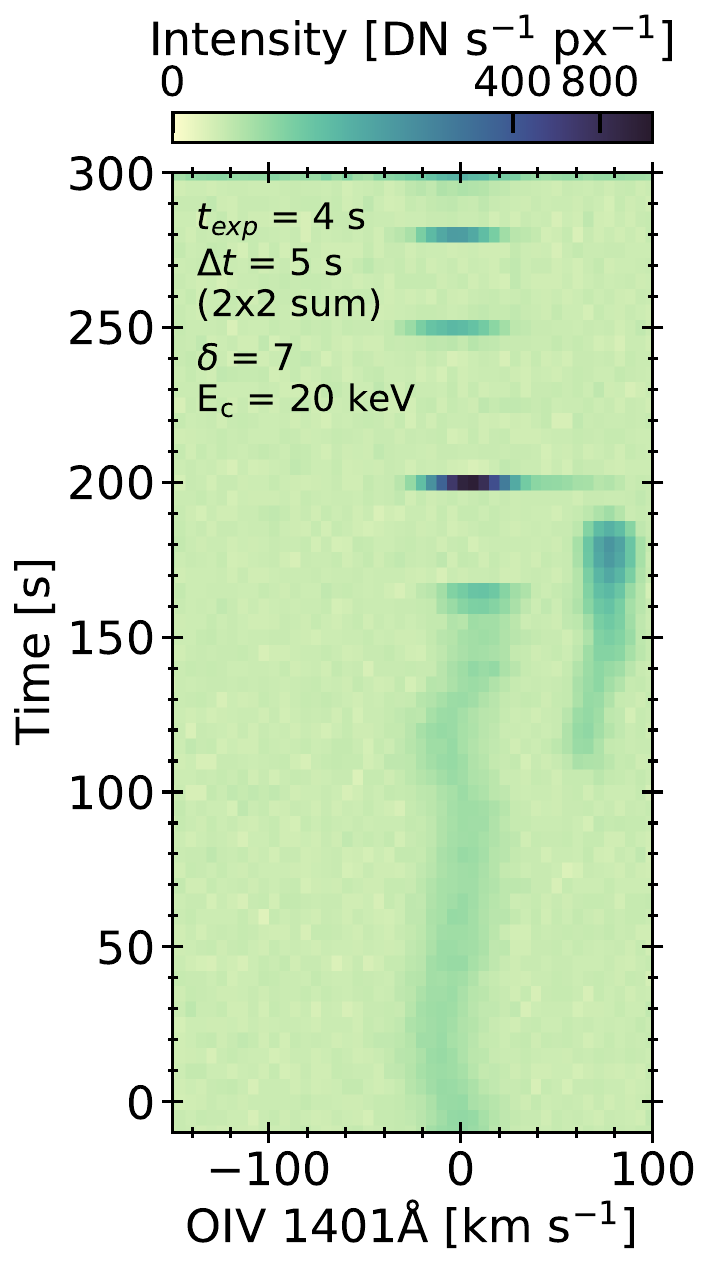}}
	\subfloat{\includegraphics[width = 0.15\textwidth, clip = true, trim = 0.cm 0.cm 0.cm 0.cm]{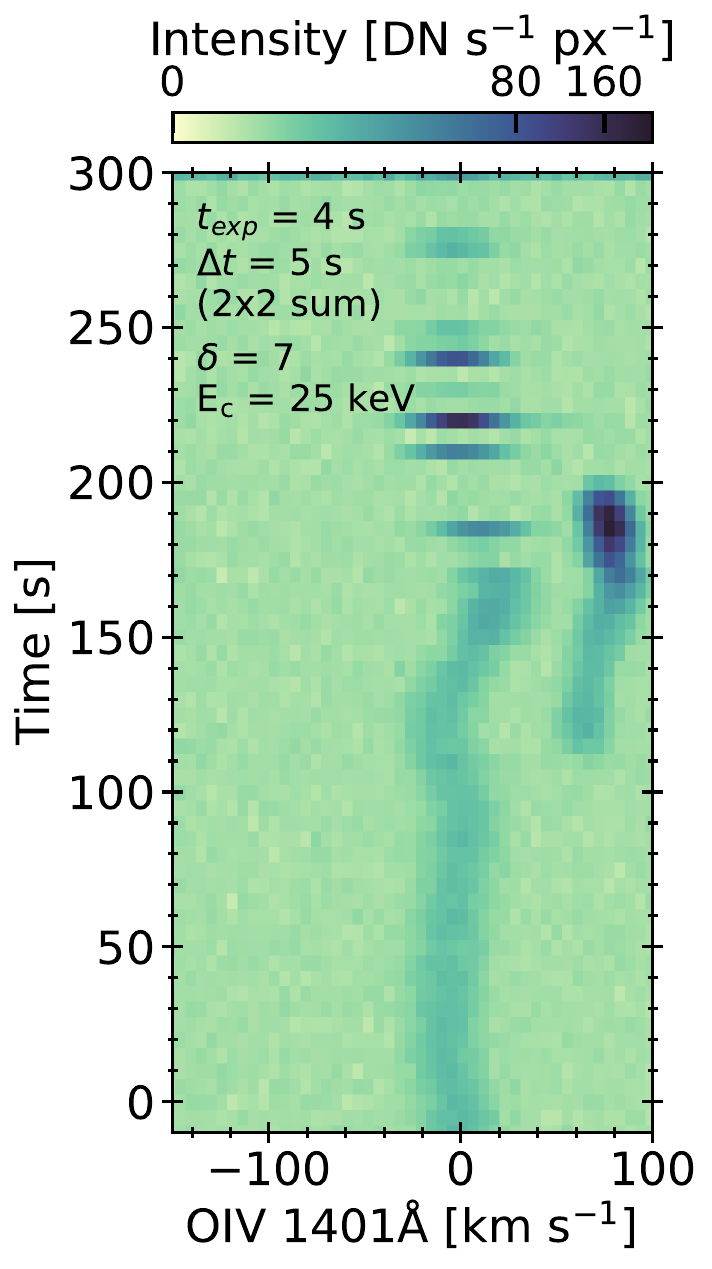}}
	\subfloat{\includegraphics[width = 0.15\textwidth, clip = true, trim = 0.cm 0.cm 0.cm 0.cm]{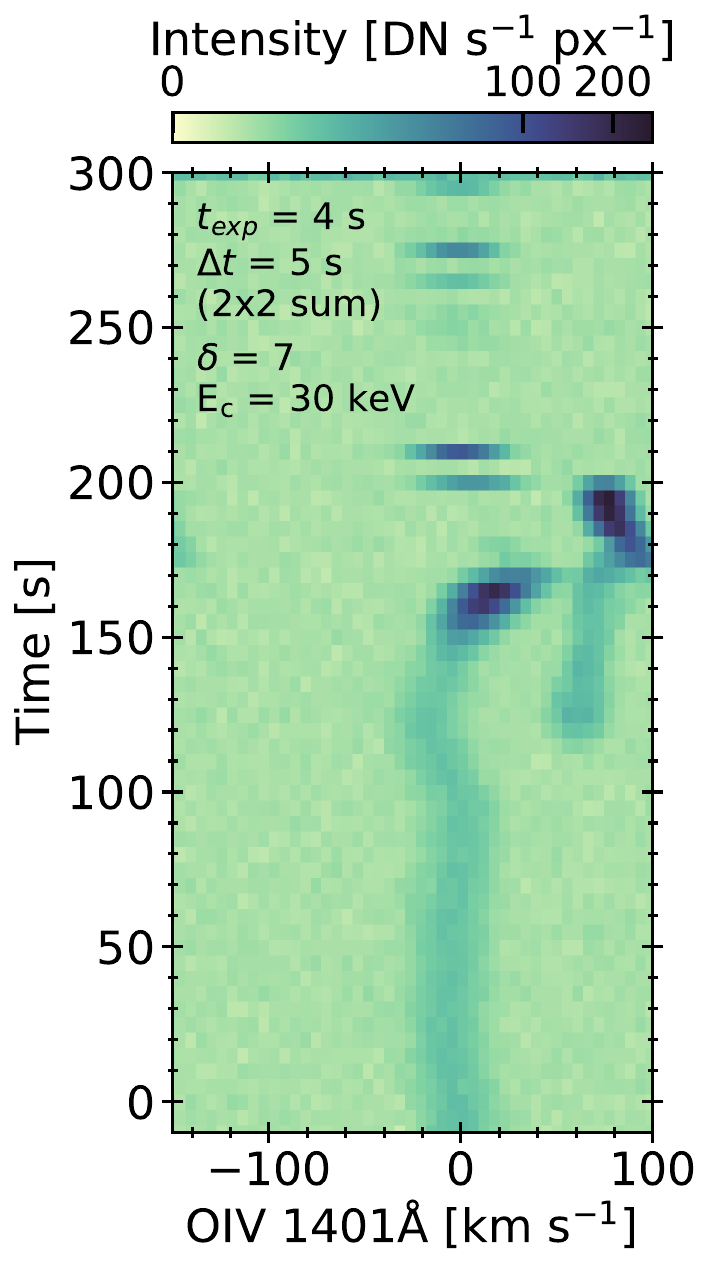}}
	}
	}
		\caption{\textsl{Same as Figure~\ref{fig:fexxi_profiles_maingrid} but showing \ion{O}{4} 1401~\AA.}}
	\label{fig:oiv_profiles_maingrid}
\end{figure}

Here we show the evolution of the \ion{Fe}{21} 1354.1~\AA\ spectra (Figure~\ref{fig:fexxi_profiles_maingrid}), that forms at $T\sim11$~MK, illustrating that in most simulations explosive chromospheric evaporation occurs in the latter phase. These spectra were synthesised assuming optically thin formation using atomic data from the CHIANTI database \citep[][]{1997A&AS..125..149D,2015A&A...582A..56D} in the same way as described in \cite{2019ApJ...871...23K}, integrating the intensity in each \radyn\ grid cell from $z = 500-3000$~km. Coronal abundances were assumed for these illustrative figures. The spectra were converted to IRIS count rates assuming $t_{\mathrm{exp}} = 4$~s, $\Delta t = 5$~s, and $2\times2$ spectral summing. Similarly, we synthesised \ion{O}{4} 1401~\AA, a transition region line (Figure~\ref{fig:oiv_profiles_maingrid}; in that window there is also another line to the red of \ion{O}{4}).

\end{document}